%% file: OSETPaper.tex
\documentclass[preprint,11pt]{article}
\usepackage{latexsym}
\usepackage{amsthm}
\usepackage{graphics}
\usepackage{epstopdf}
\usepackage{epsfig}
\usepackage{bbm}
\usepackage{amsmath}
\usepackage{amssymb}
\usepackage{pstricks,pst-node,pst-tree}

\def \CalcHEP {{\sc CalcHEP}}
\def \CompHEP {{\sc CompHEP}}
\def \Herwig {{\sc Herwig}}
\def \Sherpa {{\sc Sherpa}}
\def \Pandora {{\sc Pandora}}
\def \Whizard {{\sc Whizard}}
\def \Grace {{\sc Grace}}
\def \StdHEP {{\sc StdHEP}}
\def \Bard {{\sc Bard}}
\def \Vista {{\sc Vista}}
\def \Sleuth {{\sc Sleuth}}
\def \Quaero {{\sc Quaero}}
\def \MadGraph {{\sc MadGraph}}
\def \MadEvent {{\sc MadEvent}}
\def \Pythia {{\sc Pythia}}

\newcommand{\f}[2]{\frac{#1}{#2}}
\newcommand{\sh}{\hat{s}}
\newcommand{\that}{\hat{t}}
\newcommand{\uh}{\hat{u}}

\newcommand{\Msq}{|\mathcal{M}|^2}

\def\slashchar#1{\setbox0=\hbox{$#1$}           
   \dimen0=\wd0                                 
   \setbox1=\hbox{/} \dimen1=\wd1               
   \ifdim\dimen0>\dimen1                        
      \rlap{\hbox to \dimen0{\hfil/\hfil}}      
      #1                                        
   \else                                        
      \rlap{\hbox to \dimen1{\hfil$#1$\hfil}}   
      /                                         
   \fi}                                         %
\catcode`@=11
\long\def\@caption#1[#2]#3{\par\addcontentsline{\csname
  ext@#1\endcsname}{#1}{\protect\numberline{\csname
  the#1\endcsname}{\ignorespaces #2}}\begingroup
    \small
    \@parboxrestore
    \@makecaption{\csname fnum@#1\endcsname}{\ignorespaces #3}\par
  \endgroup}
\catcode`@=12

\begin{document}

\baselineskip=18pt

\setcounter{footnote}{0}
\setcounter{figure}{0} \setcounter{table}{0}

\begin{titlepage}
\begin{center}

{\LARGE \bf \sc MARMOSET: The Path from LHC Data to the New Standard
Model via On-Shell Effective Theories}

\vspace{0.5cm}

{\bf Nima Arkani-Hamed\footnote{Contact and supplemental information
available at: \texttt{http://marmoset-mc.net}}, Philip
Schuster, Natalia Toro}\\
{\it Jefferson Laboratory, Physics Department, Harvard University, Cambridge, MA 02138}\\
\vspace{.2cm}
{\bf Jesse Thaler}\\
{\it Department of Physics, University of California, Berkeley, CA 94720}\\
{\it Theoretical Physics Group, Lawrence Berkeley National Laboratory, Berkeley, CA 94720}\\
\vspace{.2cm}
{\bf Lian-Tao Wang}\\
{\it Department of Physics, Princeton University, Princeton, NJ 08544}\\
\vspace{.2cm}
{\bf Bruce Knuteson}\\
{\it Department of Physics, Massachusetts Institute of Technology, Cambridge, MA 02139}\\
\vspace{.2cm}
{\bf Stephen Mrenna} \\
{\it Fermi National Accelerator Laboratory, Batavia, IL 60510}

\vspace{.3cm}

\end{center}

\begin{abstract}
\medskip
We describe a coherent strategy and set of tools for reconstructing
the fundamental theory of the TeV scale from LHC data. We show that
On-Shell Effective Theories (OSETs) effectively characterize hadron
collider data in terms of masses, production cross sections, and decay
modes of candidate new particles. An OSET description of the data
strongly constrains the underlying new physics, and sharply motivates
the construction of its Lagrangian. Simulating OSETs allows efficient
analysis of new-physics signals, especially when they arise from
complicated production and decay topologies. To this end, we present
MARMOSET, a Monte Carlo tool for simulating the OSET version of
essentially any new-physics model.  MARMOSET enables rapid testing of
theoretical hypotheses suggested by both data and model-building
intuition, which together chart a path to the underlying theory. We
illustrate this process by working through a number of data
challenges, where the most important features of TeV-scale physics are
reconstructed with as little as $5$ fb$^{-1}$ of simulated LHC
signals.

\end{abstract}

\end{titlepage}
\tableofcontents \vfill\eject


\input{intro.tex}

\input{ProductionParameterization.tex}

\input{UsersGuide.tex}

\input{MARMinPractice.tex}

\input{endmatter.tex}

\appendix
\input{AppendixProduction.tex}
\input{AppendixExampleFits.tex}
\input{AppendixDecay.tex}

\bibliographystyle{JHEP}
\bibliography{OSETPaper}
\end{document}

%% file: intro.tex
\section{Anticipating the New Standard Model}

With the upcoming turn-on of the Large Hadron Collider (LHC), high
energy physics is on the verge of entering its most exciting period in
a generation. How will we reconstruct the fundamental theory of the
TeV scale from LHC data?  The major discoveries at hadron
colliders in the last thirty years---those of the $W$ and $Z$ bosons
\cite{Arnison:1983rp,Banner:1983jy,Arnison:1983mk,Bagnaia:1983zx} and
the top quark \cite{Abe:1995hr,Abachi:1995iq}---were of particles
whose properties were exactly predicted by the Standard Model (for the
$W$ and $Z$) or characterized by a single unknown parameter ($m_t$ for
the top quark). By contrast, the LHC is so exciting precisely because
the answer to the question---what will we see?---has never been more
uncertain.

The questions about new physics that must be answered first are
therefore big-picture, structural ones: What kinds of particles are
being made with what sorts of rates? What pattern of decays do they
exhibit? Only with this information can we tackle the fundamental
questions: What new physical principles are being manifested at the
TeV scale? Is nature supersymmetric? Is the electroweak scale natural?
Given the tremendous range of possibilities, a coherent strategy for
going from data to a still-unknown theory is necessary.

\begin{figure}[tbp]
\begin{center}
\includegraphics[width=5in]{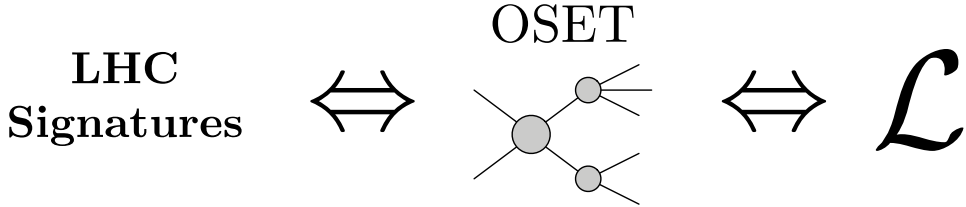}
\end{center}
\caption{\label{fig:datatoLagrangian} Instead of attempting to
reconstruct the TeV-scale effective Lagrangian directly from LHC data,
we propose an On-Shell Effective Theory (OSET) characterization of the new physics in terms of new
particle masses, production cross sections, and branching ratios as a
crucial intermediate step.  }
\end{figure}

We propose and develop such a strategy, using On-Shell Effective
Theories (OSETs) as an intermediary between LHC data and the
underlying Lagrangian as in Figure \ref{fig:datatoLagrangian}. An OSET is a model-independent characterization
of new-physics processes in terms of their dominant kinematic
structure---the masses, production cross sections, and decay modes of
candidate new particles.  The success of this approach relies on three
interrelated facts that we establish in this paper:
\begin{enumerate}
\item Only a few dynamical variables control the essential phenomenology
of new physics at hadron colliders. By characterizing new physics
directly in terms of these variables, OSETs permit a simple, accurate
parametrization of almost {\it any} model of new physics.
\item OSET parameters can be scanned efficiently through
re-weighting of post-detector-simulated Monte Carlo.  By contrast, the
relation between Lagrangian parameters and observables is often
obscure, and Monte Carlo must be generated separately for every point
in the Lagrangian parameter space.  Thus, OSETs can be simulated more
rapidly than Lagrangians, and are more readily interpreted.
\item An
OSET concisely describes many distinct event topologies with
correlated rates.  Therefore, simple observables and correlations
between them place powerful constraints on OSET descriptions of
new-physics signals; an OSET in turn sharply motivates the
construction of the underlying new-physics Lagrangian.  Of course,
Lagrangian consistency is a more stringent condition than OSET
consistency, and feeds back into this analysis.
\end{enumerate}

The rest of this paper is devoted to an exposition of these three
ideas; Sections \ref{sec2}, \ref{sec:tool}, and
\ref{sec:MARMinPractice} are each self-contained and logically
independent. Section \ref{sec2} contains the heart of the physics of
OSETs; we show that OSETs are an effective, accurate, and physical way
of parametrizing new physics at Hadron Colliders, with far fewer
parameters than a full Lagrangian description. A number of important
related physics issues are further discussed in the
appendices. Section \ref{sec:tool} presents a step-by-step
introduction to MARMOSET (Mass And Rate Matching In On-Shell Effective
Theories) as a Monte Carlo tool using a variety of examples. In
Section \ref{sec:MARMinPractice}, MARMOSET swings into action; we let
it loose on the ``Michigan Black Box'' and show how it naturally leads
us down a path to reconstructing the underlying model. The analysis
used in Section \ref{sec:MARMinPractice} is simple-minded, and we
expect any fully realistic analysis to be more sophisticated.

The remainder of this introduction summarizes the main ideas developed
in Sections \ref{sec2}, \ref{sec:tool}, and \ref{sec:MARMinPractice},
and the interested reader is encouraged to refer to the appropriate
sections of the paper.

\subsection{The New Standard Model at Low Luminosity}
As is well-known, the first step in uncovering the New Standard
Model at the LHC will be the rediscovery of the old Standard Model
at $14$ TeV \cite{Green:2006fa}.  Many discrepancies between data
and Standard Model ``predictions'' will likely be uncovered in the
early running phases of the LHC, most of which will not be new
physics signals. The difficult problem of describing the Standard
Model accurately will necessarily occur in parallel with the attempt
to interpret discrepancies as they arise. Consequently, a simple,
transparent, and model-independent characterization of big-picture
features of discrepancies is crucial. This is the context in which
OSETs can be useful, depending, of course, on what Nature has in
store for us.

In extreme cases, new physics stands out and the appropriate
characterization is obvious.  For example, a prominent $Z^\prime$
resonance, quasi-stable charged particles, or decays happening at a
macroscopic separation from the beam pipe are unmistakable signals for
new physics with clear interpretations \cite{ATLASTDR1,ATLASTDR2,CMSTDR1,CMSTDR2}.

But the more challenging scenario is equally exciting: if, as is
widely expected, there is a natural solution to the hierarchy problem
at the weak scale, its phenomenology will probably be quite rich.  New
colored particles lighter than $\sim 2$ TeV could be produced in the
thousands to millions within the first $1$ to $5$ fb$^{-1}$ of
accumulated data.  But what is the new physics, and what information
should we try to extract from LHC data?

Ultimately, we seek the effective Lagrangian at the TeV scale, but
this goal will not be immediately attainable.  The effective
Lagrangian compatible with early data may not be unique; for instance,
there could be particles that can be kinematically produced, but
aren't made at sufficiently high rates to play any role. More
importantly, typical models of new physics have scores of new
parameters and particles, of which only a few combinations affect the
broad features of physical observables.  It seems imprudent to
reconstruct an under-constrained Lagrangian from early signals.  Is there
a simpler and more invariant characterization of the data?

In some very straightforward cases, a more physical characterization
of the data is already aimed for and obtained by experimental
collaborations. LEP II Higgs search results are presented as exclusion
plots in the $m_{\rm Higgs}$ vs. $\sigma \times \mbox{Br}$ planes for
various Higgs decay channels \cite{Barate:2003sz,Carena:2000yx}; searches for singly-produced
resonances at the Tevatron report limits in the same way \cite{Melnitchouk:2005xv,Abulencia:2005nf,Abazov:2006nr,Abazov:2006hn,CDFExotic,D0Exotic}.

Model-independent limits are more difficult to present in theories,
such as low energy supersymmetry (SUSY), with many production channels
and long cascade decay chains.  To deal with these complications,
exclusion limits on these theories are often presented on the
Lagrangian parameters for simplified ``benchmark'' models, such as the
constrained MSSM \cite{Allanach:2002nj}. But these limits lack a transparent interpretation
in a broader parameter space, and are often highly misleading. In
attempting to directly reconstruct the parameters of a highly
simplified underlying Lagrangian, important physics can be lost.

In this paper, we propose a characterization of new physics in an
extension of what is done in resonance and Higgs searches.  This is a
first step towards the ultimate goal of reconstructing the TeV-scale
effective Lagrangian, and is more practical than trying to do so
directly, especially when many new processes are operative near the
same energy scale. Such a characterization has transparent physical
meaning and is practical and effective {\it especially} when the new
physics has many complicated production and decay topologies. We call
such a characterization of the data an ``On-Shell Effective Theory'',
or OSET.

\subsection{Introducing On-Shell Effective Theories}
An OSET is specified by the gauge quantum numbers and masses of new
particles, their production cross sections, and branching ratios for
their decays to other new and/or Standard Model particles. An OSET can
be associated with a set of Feynman-like diagrams for production and
decay of new particles, but every particle in the diagram is
on-shell.

\begin{figure}[tbp] \begin{center}
\includegraphics[width=5in]{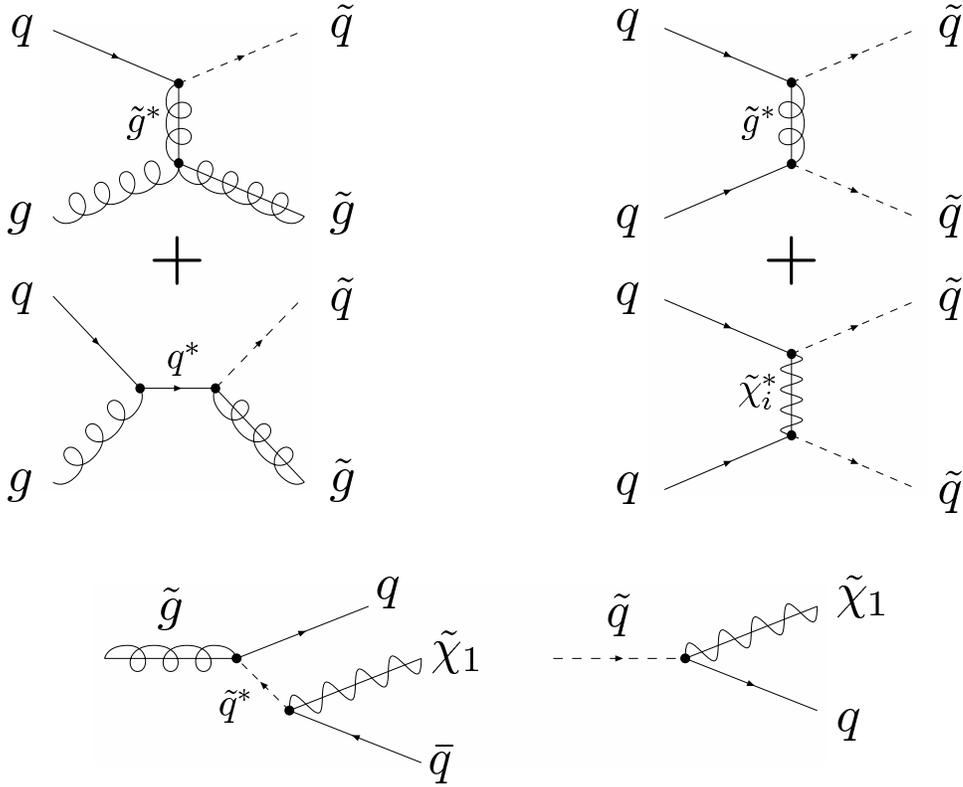}
\caption{A full Lagrangian
process where on- and off-shell masses and several combinations of couplings
control both kinematics and rates. The squark-gluino associated production receives contributions from intermediates quarks and gluinos; for same-sign squark production, the four neutralino states propagate in the $t$-channel and the vertices involve the neutralino mixing matrices. Squarks only appear off-shell in the gluino decays. The OSET analogue is
shown in Figure \ref{fig:SampleOSETDiagrams}.}\label{fig:SampleSUSY}
\end{center}
\end{figure}

\begin{figure}[tbp] \begin{center}
\includegraphics[width=5in]{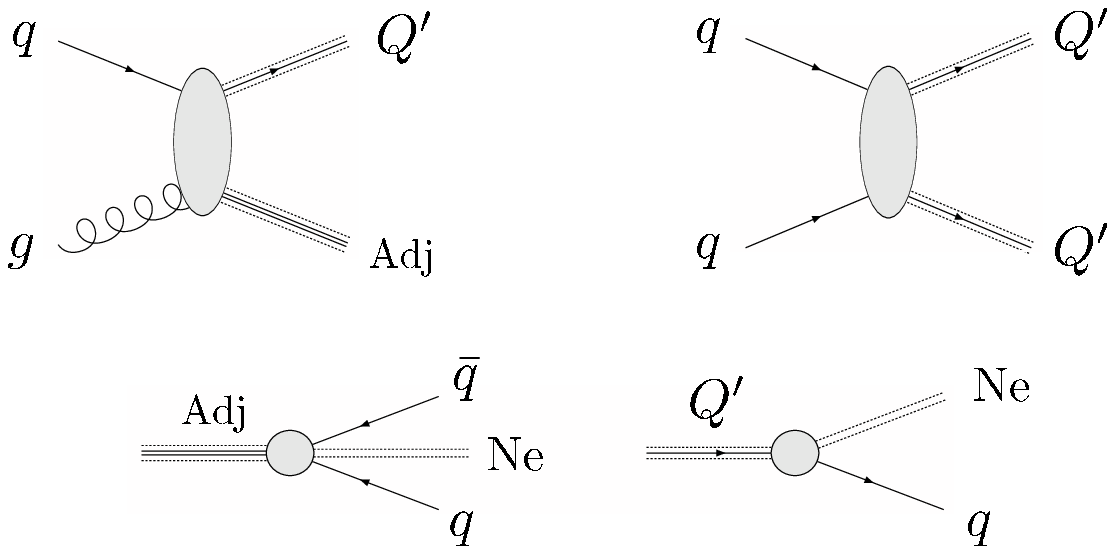}
\caption{The OSET analogue of the diagrams in Figure
  \ref{fig:SampleSUSY}.  Gray circles or ``blobs'' represent dynamics
  that are parameterized by one rate and possibly an additional shape
  parameter, as discussed in Section \ref{sec2}.  Together with the
  spectrum of particles that appear on-shell, these rates fully
  specify the OSET.  For example, the gauginos appear only off-shell in
  same-sign squark production, so they do not appear in the OSET
  containing only this process.  Their indirect effects are instead
  captured in overall rate parameters.  In the case of gluino-squark
  associated production, the squark appears on-shell on one side of
  the diagram, but on the other side, where it appears off-shell in the
  SUSY description, it is absent from the
  OSET.}\label{fig:SampleOSETDiagrams}
\end{center}
\end{figure}

As examples, we consider two processes in a supersymmetric theory with
squarks $\tilde{q}$ and a gluino $\tilde g$ with $m_{\tilde q} >
m_{\tilde g}$: associated production of
$\tilde g$ and $\tilde q$ and same-sign squark production, as in Figure \ref{fig:SampleSUSY}.  The
squark decays directly to a quark and the LSP, while the gluino decays
through an off-shell $\tilde{q}$ to two quarks and the LSP. The full
quantum amplitudes for these processes depend on many Lagrangian
parameters of the MSSM.  The OSET description of Figure
\ref{fig:SampleOSETDiagrams} is, however, very simple---there are three
new particles $\tilde{g}$, $\tilde{q}$, and LSP, two production modes,
and two decay modes. We emphasize two features of the OSET
description.  Though the same-sign squark production amplitude depends
on the gaugino spectrum, the gauginos do not appear on-shell anywhere in
this process, and are excised from the OSET description. In SUSY
associated production, the squark appears both on-shell (directly
produced) and off-shell (in gluino decay); the OSET analogue keeps the
squark where it appears on-shell, and ``integrates it out'' where it
is off-shell.

As we will show in Section \ref{sec2} and the appendices,
OSET production and decays can be simply and accurately parameterized
with no more than an overall (integrated) cross section and, in some
cases, one or two ``shape'' variables. This simple parameterization
reproduces features such as the locations of invariant mass edges and
endpoints, and in most cases the locations and widths of peaks in
$p_T$ distributions.  In our figures, we represent the parameterized
components of amplitudes as shaded circles, or ``blobs''.  More or
less detail can be incorporated as appropriate, for example the
addition of spin correlations or simplification of cascade decays into
a single blob.

An OSET that describes a particular topology of production and decay
will almost always make predictions for the existence and relative
strengths of other topologies. Thus, a consistent picture is tightly
constrained.  The OSET has transparent physical meaning. An OSET that
fits the data well will strongly constrain the structure of the
underlying new physics, and a model inspired by the OSET can suggest
new channels to be be included in a refined OSET description.  Thus,
the OSET framework allows us to describe the new physics in stages
rather than getting the full effective Lagrangian in one go.
Consistency with an underlying Lagrangian
also implies basic guidelines
that should be considered throughout the process of
interpretation---for example, a 1 TeV weakly interacting particle
should not be produced with pb cross section!  As this feedback loop
converges on a model or set of models for new physics, detailed
simulation may be required to verify features that are
\emph{not} incorporated in the OSET.

\subsection{Inclusive Data Analysis with OSETs}
With a well-defined OSET parametrization scheme in place, the field is
open for applying powerful analysis techniques to OSET-based searches
and data characterization. We fully expect that the experimental
community will be adept at analysis; our remarks here and in
Section \ref{sec:MARMinPractice} are intended to illustrate potential
benefits of OSET-based analysis, not to advocate a particular analysis
strategy.

\begin{figure}[tbp] \begin{center}
\includegraphics[width=4in]{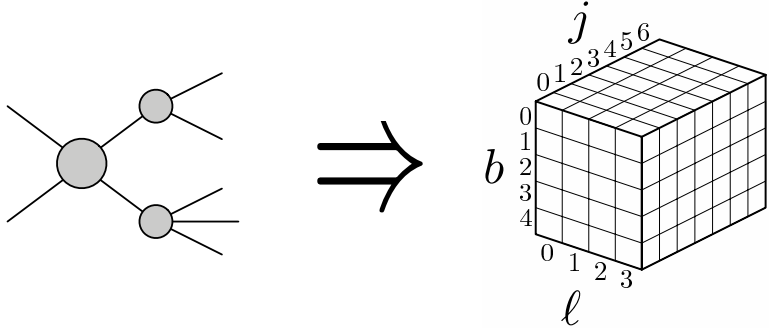}
\end{center}
\caption{\label{fig:topotorate} Each process populates a particular
  set of final states.  Though they may overlap, we can constrain OSET
  parameters by exploiting the differences between processes and
  correlations over multiple final states.  When distinguishing
  similar event topologies, the inclusive count data we use here can
  and should be replaced with more carefully designed
  characterizations of final states.}
\end{figure}

\begin{figure}[tbp]
\begin{center}
\includegraphics[width=6in]{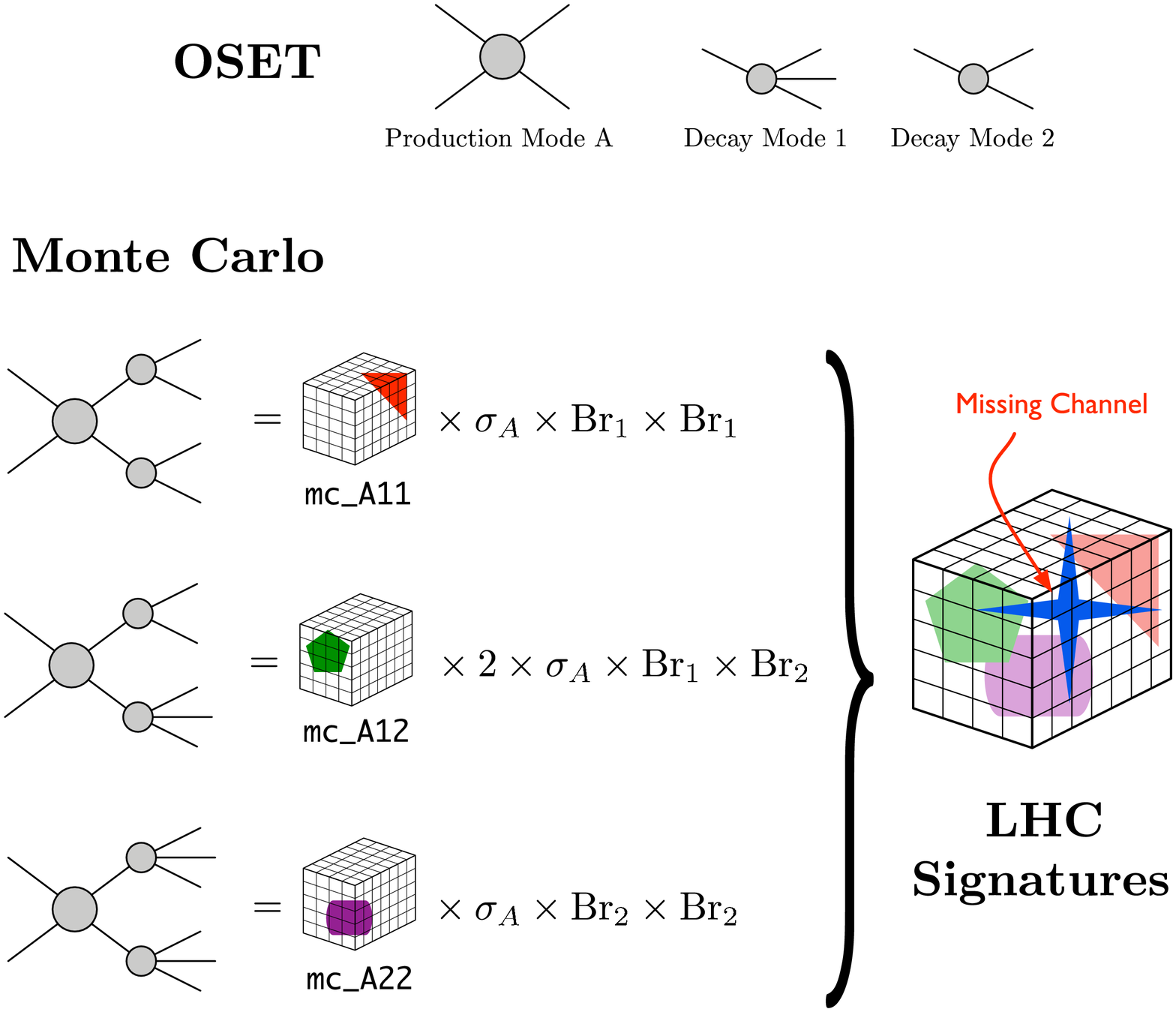}
\end{center}
\caption{\label{fig:ratetodata}An illustration of an OSET and the
accompanying rate parametrization scheme MARM (Mass And Rate Modeling). At top is the
production mode and two decay modes defining an OSET process. Each
production mode receives a cross section weight.  Each decay vertex
receives a branching ratio weight.  For each allowed topology that can
be derived from the OSET, a weight factor consistent with the topology
is assigned. A good fit to the data consistent with the OSET would
suggest that the underlying topological interpretation is
consistent. Scanning the rate variables is efficient as it only
involves re-weighting the Monte Carlo, or alternatively re-weighting
the signature templates constructed from each process.  In the above
example, the failure of the OSET to capture a missing signature
channel motivates the addition of a new process to the OSET.}
\end{figure}

Compared to Lagrangian parameters, OSET parameters are relatively easy
to extract from data.  Object multiplicities and the shapes of
kinematic distributions motivate a set of OSET diagrams. Approximate
mass scales can be inferred from standard kinematic distributions, and
a simple set of inclusive signatures---for instance counts for
leptons, jets and $b$-jets binned in $H_T$ ranges as in Figure
\ref{fig:topotorate}---constrain cross sections and branching
ratios \cite{Baer:1995va,Baer:1995nq,Mrenna:1995ax,Hinchliffe:1996iu}. These signatures should be supplemented with more carefully
designed discriminators to distinguish processes with similar
topologies from one another and from Standard Model processes.  An
OSET will populate this space of final states with a particular
pattern; a pattern in the data not captured by the OSET then suggests
new ``blobs'' to be included in a better OSET description. In
practice, this process has been most effective on examples when the
OSETs are well-motivated by top-down theory considerations.  As a
first step towards designing more sophisticated analysis techniques,
we show that OSET parameters can be determined by crude but effective
Mass And Rate Modeling (MARM) analysis techniques, shown schematically
in Figure \ref{fig:ratetodata}. This will be the main topic of Section
\ref{sec:MARMinPractice} using the ``Michigan Black Box'' as an
example.

We have implemented OSET-level Monte Carlo generation in a
\Pythia-based \cite{Sjostrand:2006za} analysis package called MARMOSET, described in detail in
Section \ref{sec:tool} and available for download at \texttt{http://marmoset-mc.net/}. MARMOSET is a Monte Carlo tool allowing direct
simulation of OSETs for new physics, in place of an effective
Lagrangian.  The machinery to Monte Carlo such models has been
available in event generators for years.  What has been missing is an
interface for simple specification of OSETs with minimal vertex
structure and for bookkeeping of rate correlations.  The organization
of Monte Carlo in MARMOSET allows efficient comparison of OSET
hypotheses to Hadron Collider data.

We use this in the simple analysis of Section
\ref{sec:MARMinPractice}.  This strategy is analogous to the template
method familiar from top mass measurements \cite{Freeman:2004tx}, but focuses on gross observables, explicitly treating
kinematics and rates independently.  In principle, best fit OSET
parameters can be obtained through a simultaneous fit to signal and
various background channels weighted independently, reducing the
possibilities of ``fake'' discoveries through background normalization
errors.  Of course, the optimal analysis path will depend on the
details of the new physics, the important backgrounds, and detector
resolution.

Because Monte Carlo generation and detector simulation are
time-consuming, MARMOSET is organized to exploit the economy of OSETs
for computational efficiency. This economy is achieved through two
simplifications.  In the underlying Lagrangian description, rate
variables are strongly dependent on masses, so that a high resolution
in scanning many masses is needed to adequately scan rates.  By
separating rate information from masses and rendering masses of
off-shell particles altogether irrelevant, OSETs greatly reduce the
dimensionality of parameter space and the resolution required in mass
scanning; crude mass estimators can often be used to approximately fix
mass scales without compromising the accuracy of branching ratio and
cross-section fits. Moreover, because the Monte Carlo for different
processes is additive and rates are floating parameters, these rates
can be varied without re-generating Monte Carlo by simply re-weighting
events.

As discussed in Section \ref{sec:UGfuture}, this MARMOSET analysis
strategy is analogous to that taken in \Bard\ \cite{Knuteson:2005ev,Knuteson:2006ha}, while the organization of the physics characterization is
complementary.  MARMOSET is
aimed at the problem of characterizing mass scales and rate
correlations among many competing topologies, mainly with LHC
scenarios in mind.  \Bard\ approaches the problem of characterizing
new physics by systematically generating and evaluating Lagrangian
explanations of discrepancies.

\subsection{SUSY Examples with OSETs}

\begin{figure}[tbp]
\begin{center}
\includegraphics[width=2.2in]{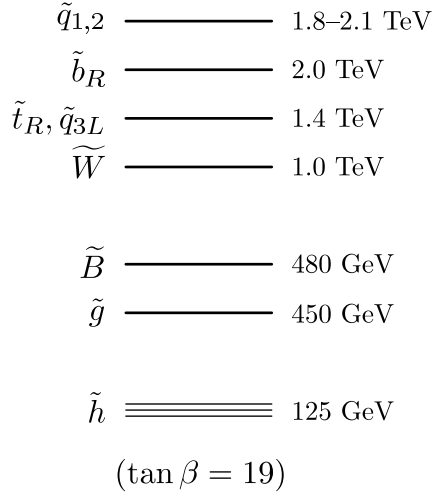}
\caption{Spectrum for the Michigan Black Box.  Note the unconventional
  electroweak-ino hierarchy and decoupled squarks mediating 3-body
  decays of the gluino.  The low-energy features of this model are
  controlled by a few combinations of these soft
  masses}\label{fig:MichSpectrum}
\end{center}
\end{figure}
\begin{figure}[tbp]
\begin{center}
\includegraphics[width=4.5in]{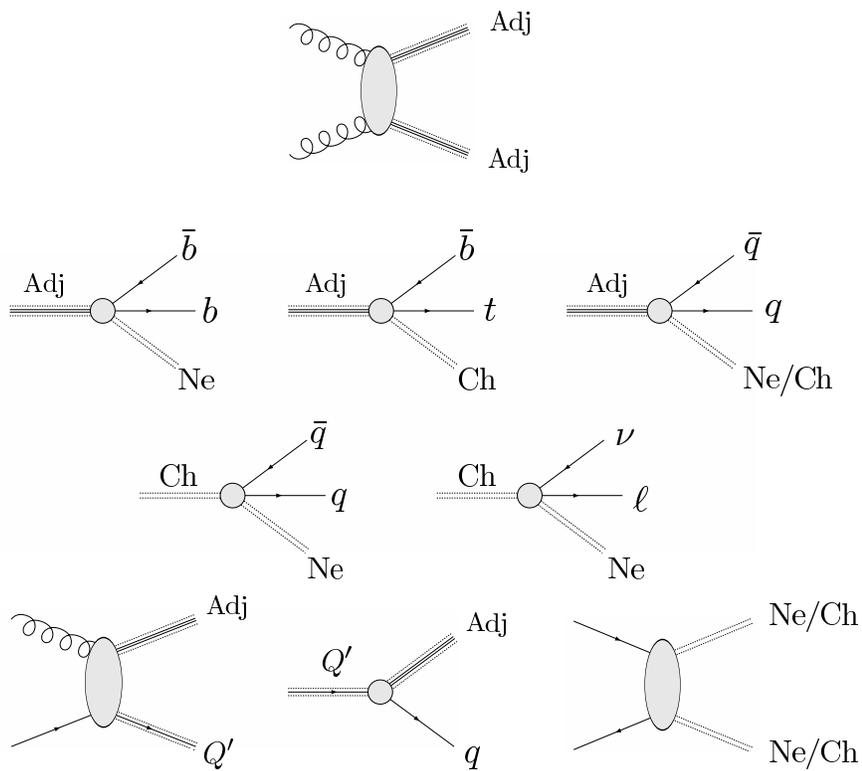}
\caption{\label{fig:IntroMichOSETDiagrams} Diagrams in the OSET corresponding to the
    Michigan Black Box of Figure \ref{fig:MichSpectrum}.  The bottom row of diagrams correspond to processes that are difficult to observe at low luminosity, but which might be inferred from theoretical considerations and are ultimately discoverable at high luminosity.}
\end{center}
\end{figure}
\begin{figure}[tbp]
\begin{center}
\includegraphics[width=6in]{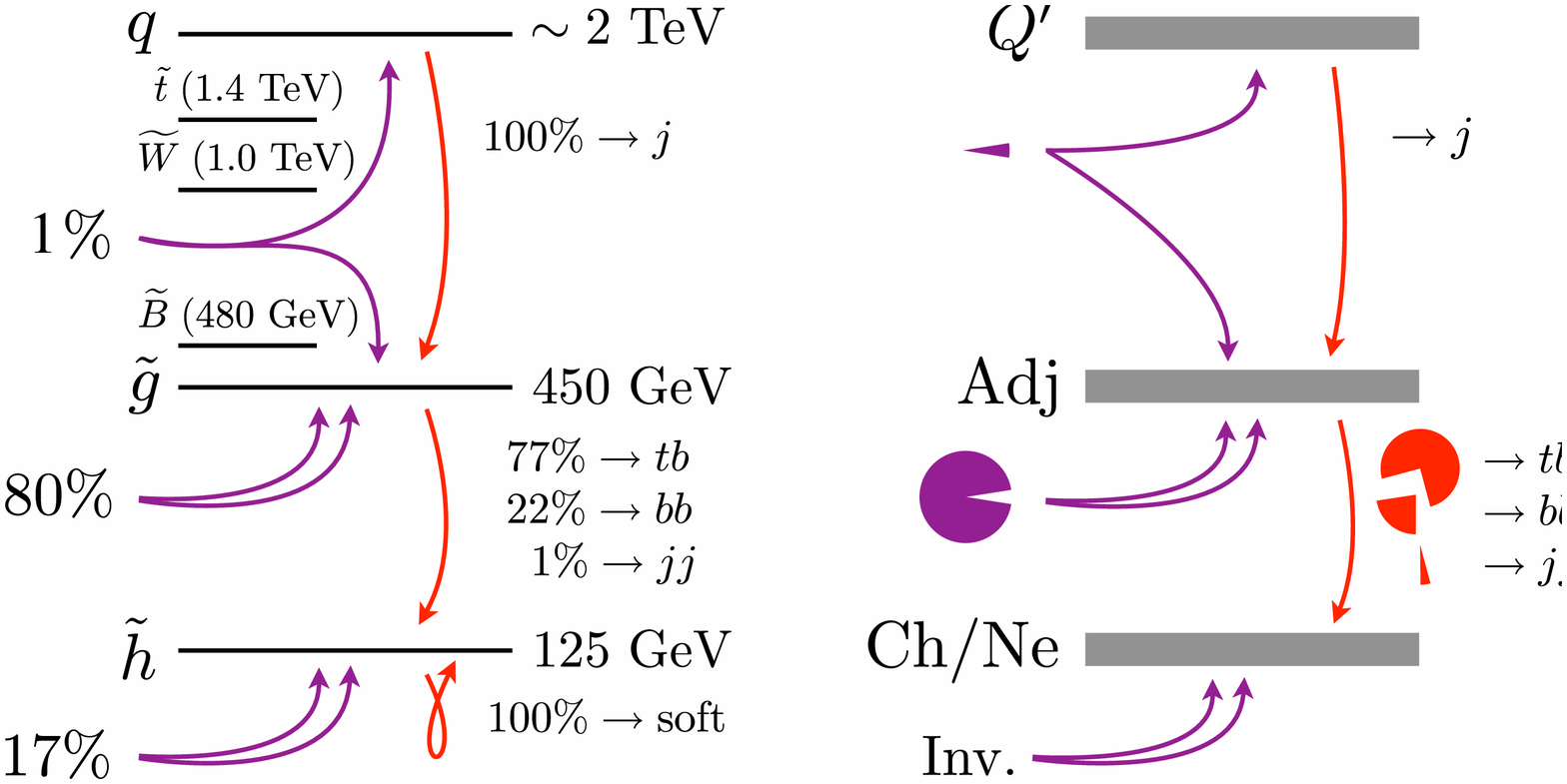}
\caption{Left: A summary of the dominant production and decay modes in
  the Michigan Black Box; Higgsino pair production does not trigger.
  Right: An OSET analysis readily suggests an accurate
  characterization of the model: we obtain the correct hierarchy of
  states, and dominant decay modes.  We do \emph{not} try to
  explicitly determine the identity of the LSP from data, nor to
  constrain the masses of particles using cross section or branching
  ratio information.}\label{fig:MichComparison}
\end{center}
\end{figure}

The need for an intermediate characterization of new physics becomes
clear in attempting to solve ``black box data challenges''; we
illustrate the use of OSETs in this context.  A ``black box'' is a
simulated data set for some theory of new physics---unknown to
us---with $5 \mbox{ fb}^{-1}$ of integrated luminosity at the LHC.
The simulations are done as realistically as reasonably possible, with
fast detector simulation and in some cases relevant Standard Model
backgrounds included.  The ``data challenge'' is to determine the
underlying model from the ``black box'' data.  In all examples,
unsophisticated analysis of simple kinematic distributions allow us to
build OSETs that, in turn, guide us rapidly to the ``right'' TeV-scale
Lagrangian, often in surprisingly interesting ways.

We begin with two SUSY examples for which the general structure of the
theory is already widely familiar. The first example is the
supersymmetric ``Michigan Black Box'' \cite{LHCOlympics}. The spectrum of this model is
shown in Figure \ref{fig:MichSpectrum}; the dominant production and
decay rates are shown in Figure \ref{fig:MichComparison}. The spectrum
is evidently unconventional, with a very heavy Wino, a light gluino
and a Higgsino LSP.  The top and bottom squarks, which mediate gluino
decays and so are essential in a renormalizable field theory
description of the LHC signals, are produced with tiny cross-sections
$\lesssim 1 \mbox{ fb}$, so their properties are not directly
measurable.  In the absence of electroweak cascades, the LSP identity
is difficult to ascertain. Thus, the MSSM Lagrangian is complicated
and is impossible to deduce experimentally.  As we show in Section
\ref{sec:MARMinPractice}, however, the OSET description found in
Figure \ref{fig:IntroMichOSETDiagrams} is quite simple. Moreover, the
OSET fit we have obtained from limited signal information, as
illustrated in Figure \ref{fig:MichComparison}, demonstrates a
striking agreement between OSET and model.

\begin{figure}[tbp]
\begin{center}
\includegraphics[width=6in]{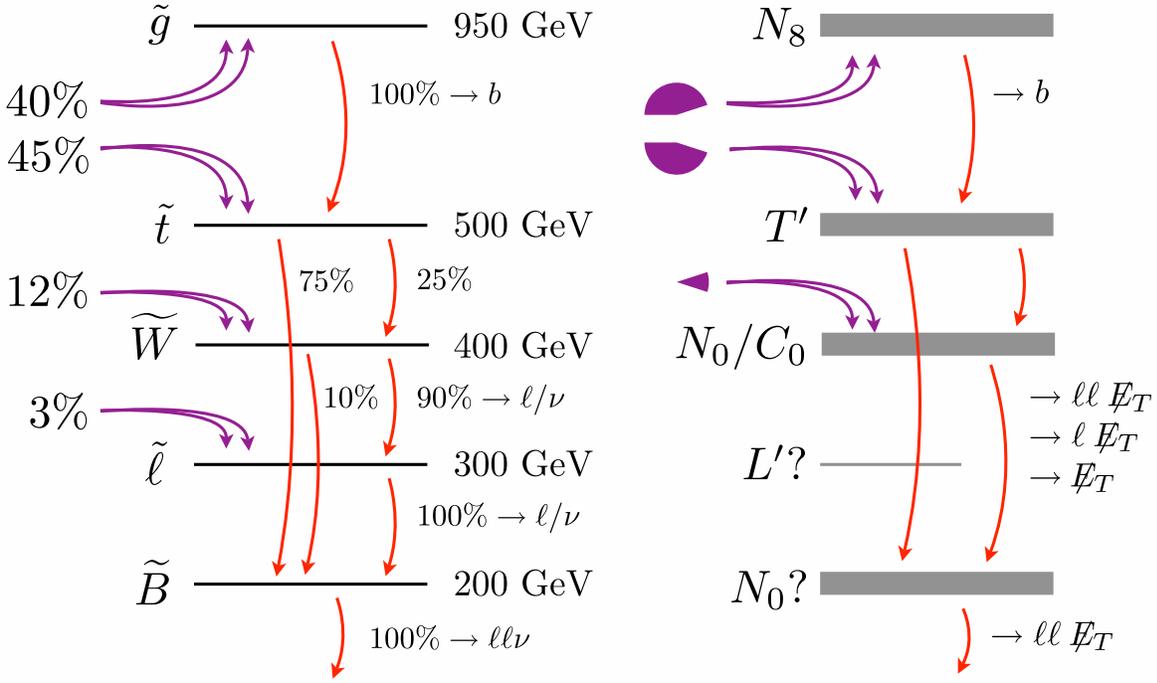}
\caption{Left: A summary of the dominant production and decay modes
in the Rutgers Black Box, an R-parity-violating SUSY model with
cascade decays involving up to 4 leptons on either side of a decay
chain. Right: Here, an OSET analysis suggests a description with
much of the underlying structure; though we are not directly
sensitive to the presence of a slepton-like state ($L'$), it is
suggested by the need to generate many leptons in cascade
decays.}\label{fig:AboxComparison}
\end{center}
\end{figure}

Though the ``Michigan Black Box'' was peculiar from a theoretical
perspective, it was dominated by a small number of production and
decay modes; the advantage of an OSET parameterization is in
representing these few processes more simply.  OSETs are also useful
in describing models with numerous production modes and more
complicated decay topologies.  This is illustrated by our second SUSY
example, the model of Figure \ref{fig:AboxComparison}, and its
accompanying OSET.  The underlying theory is a supersymmetric theory
with R-parity violation, the ``Rutgers Black Box'' \cite{LHCOlympics}---the absence of
certain details in the OSET (e.g.\ the on-shell slepton) reflects the
difficulty of resolving this aspect of the physics in the first stages
of analysis with only 4 fb$^{-1}$.

\begin{figure}[tbp]
\begin{center}
\includegraphics[width=4in]{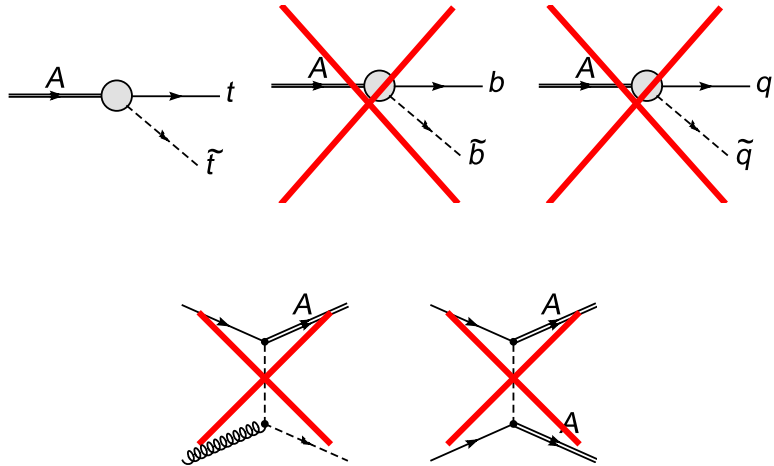}
\caption{Top: The SUSY model with a chiral adjoint appears at first
to be garden-variety MSSM, but the branching ratios of the adjoint
quickly reveal that it cannot be. Although adjoint decays to the
$\tilde b$ and first- and second-generation squarks are
kinematically allowed, neither occurs at an appreciable rate, with
$Adj$ instead dominantly decaying to $t \tilde t$. Bottom: The
apparent absence of $q \tilde q\, Adj$ couplings is confirmed in
production: the expected associated production channel is absent,
and the rate for $Adj$ pair-production is consistent with QCD
production with no $t$-channel squark contribution. These two facts,
readily visible in an OSET analysis, point immediately to the need
to consider models beyond the MSSM.}\label{fig:no12decays}
\end{center}
\end{figure}
\begin{figure}[tbp]
\begin{center}
\includegraphics[width=4in]{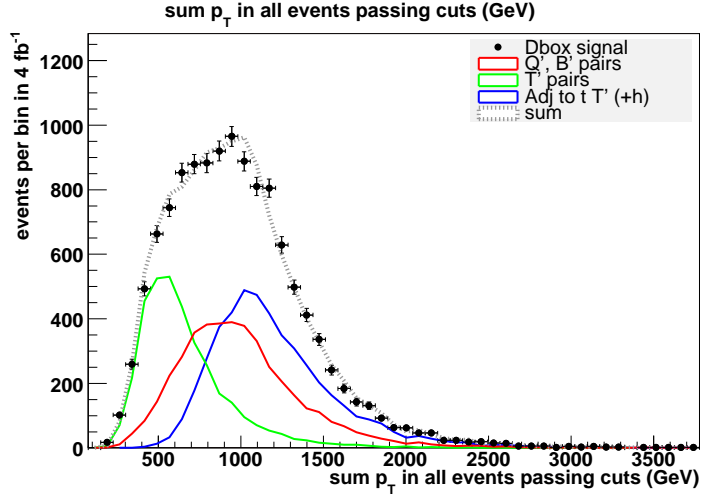}
\caption{Three processes dominate new-physics production in the SUSY
black box with a chiral adjoint: light-generation squark pair
production, top squark pair production, and adjoint pair-production.
They populate different regions of $\sum p_T$ as shown here, and
different final states.  Analysis of final-state reconstructed object
multiplicities reveals the striking absence of $Adj \rightarrow q
\tilde q$ decays, though they are kinematically allowed.}\label{fig:DboxFit}
\end{center}
\end{figure}
\begin{figure}[tbp]
\begin{center}
\includegraphics[width=6in]{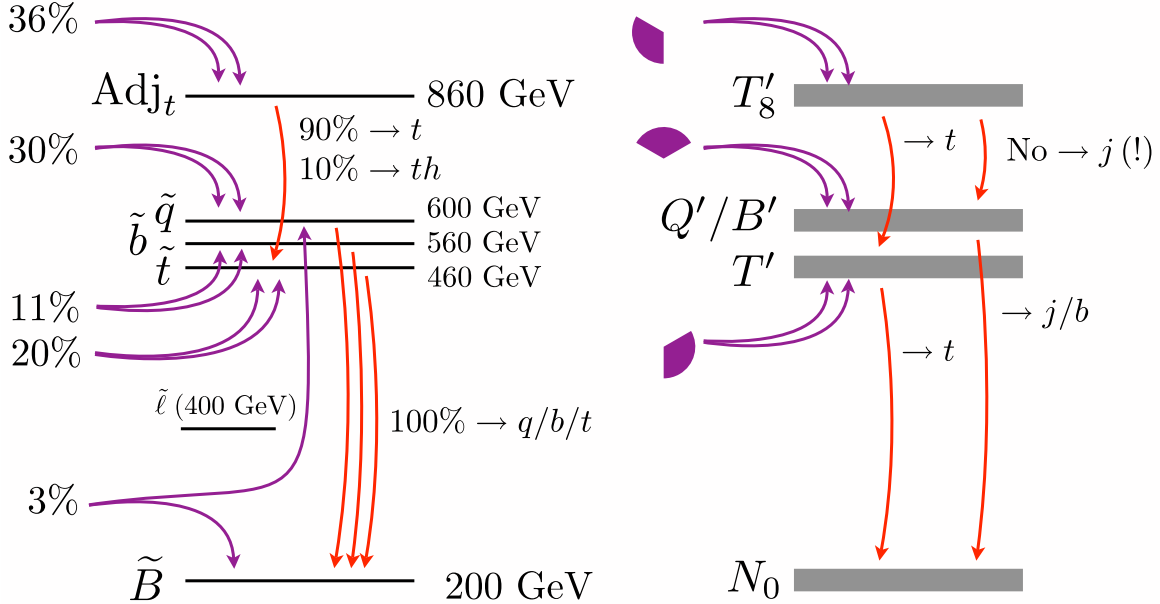}
\caption{Left: The spectrum for the modified SUSY black box with a
  chiral adjoint.  Right: An OSET characterization for this model.
  The unusual (non-gluino-like) decay properties of $Adj$ are readily
  incorporated in an OSET simulation.}\label{fig:DboxComparison}
\end{center}
\end{figure}

Our third example is another supersymmetric model, but with a twist.
Suppose that gauginos are very heavy, say all heavier than $5$ TeV,
so aren't directly produced at the LHC, while the scalars are light
and accessible as usual. However in addition, there are light chiral
superfields in the adjoint representation of the standard model;
this can easily occur in a wider variety of well-motivated theories
of SUSY breaking. In particular there is colored adjoint field
$\Sigma$. The only way this particle can decay is via
higher-dimension operators in the superpotential of the form $W
\supset \Sigma f f^c H$ where the $f$, $f^c$ are Standard Model fields.  Since these
couplings are necessarily flavor violating they are naturally
largest for the top so the leading decay is via the operator $W
\supset \frac{1}{M} Q_3 \Sigma U_3 H_u$. Let's suppose that $\Sigma$
is heavier than both the third generation as well as first-two
generation squarks. Then the leading $\Sigma$ decays are $\Sigma \to
t \tilde{t}$, as well as a three-body decay to $t \tilde{t}h$.

A simple rate analysis of the signal isolates three production and
decay modes, all consistent with a garden-variety MSSM:
pair-production of colored adjoints and of colored triplets decaying
to the third generation as well as first two generation jets.  The
surprise is in the preferred branching ratios for adjoint decays in
the OSET---though they are kinematically allowed, decays $Adj
\rightarrow q \tilde q$ to the first two generations are absent or
highly suppressed as in Figure \ref{fig:no12decays}.  This colored
adjoint can't be a usual gluino! This interpretation is further
confirmed by the absence of an adjoint-triplet associated production
mode, and by the production cross-section for Adj, which is
consistent with QCD production through an $s$-channel gluon, but is
missing the expected $t$-channel contribution from the
first-generation squark.

The peculiar properties of the OSET
immediately suggest the central interesting feature of the
underlying model---that the colored adjoint is not a gluino but a
new field decaying through a higher-dimension operator!  In turn,
this model suggests an additional channel for the decay of the
adjoint---the three body decay also including the Higgs, that can be
added to the OSET and searched for. Indeed, even with $5$ fb$^{-1}$ of
data, there are hints for this rare decay in events with many jets +
$\gamma \gamma$. The final OSET fit to this theory, again with the
zeroth order approximations to production and decay, is shown in
Figure \ref{fig:DboxFit}. Again, the agreement is striking, and the
OSET description in Figure \ref{fig:DboxComparison} is a clear
characterization of the physics.

\subsection{OSETs at the Energy Frontier}
The above examples illustrate that when faced with signal, attempting
to describe hypotheses in the OSET language of particle production and
decay modes is second-nature; the ability to verify hypotheses at this
OSET level dramatically simplifies the interpretation process. As we
will argue in Section \ref{sec:MARMinPractice}, an OSET description is
readily attainable once the new-physics signal has been isolated from
Standard Model background.  This OSET is determined using simple shape
and object-multiplicity observables, analyzed with theoretical
constraints and prejudices in mind.

Certain aspects of data analysis can and should be automated,
including the generation and testing of hypotheses.  At the same time,
the development of the ``big picture'' needs a human operator to guide
it.  In the examples above, the initial guesses for the OSET structure
are made by a global survey of gross features of the data, using
relatively simple tools. The interpretation of these results involves
guesswork and theoretical intuition, allowing humans to do what humans
do best---recognize broad patterns and make models.  MARMOSET
organizes the ``grunt work'' of simulating arbitrary hypotheses and
comparing them to data.

MARMOSET represents a coarse-grained approach to LHC signal
interpretation.  The ability to accurately simulate general, complete
Lagrangians is clearly also important, and tremendous progress has
been made in implementations such as {\sc Bridge} \cite{Meade:2007js}, \CompHEP\ \cite{Pukhov:1999gg}/\CalcHEP\ \cite{Pukhov:2004ca},
\Grace\ \cite{Ishikawa:1993qr}, \Herwig\ \cite{Corcella:2002jc}, \MadGraph\ \cite{Maltoni:2002qb}, \Pandora\ \cite{Peskin:1999hh}, \Sherpa\ \cite{Gleisberg:2003xi}, and \Whizard\ \cite{Kilian:2001qz}.  An interface between MARMOSET and these tools would also be useful. In the opposite
direction, \Bard\ can be further developed as an automated tool for
searching through tree-level renormalizable Lagrangians matching
excesses.  A successful Lagrangian characterization may be derived
from the top down or from the bottom up, and will be most convincing
if understood in both ways.

%% file: ProductionParameterization.tex
\section{The Physics of On-Shell Effective Theories}\label{sec2}

An OSET is a description of new physics processes in terms of just
on-shell particles.   The kinematics---masses and phase space---of
production and decay vertices are treated exactly, while the highly
model-dependent dynamics appearing in quantum amplitudes are
approximated using a simple parametrization of $\Msq$.  In this
section, we develop this broadly applicable parametrization for
dynamics, and show that it accurately reproduce a large class of
observables. The behavior of parton distribution functions
(PDFs)---and in some cases the inability to fully reconstruct events
with missing energy---will facilitate this task immensely.

Of course, the complexity of the parametrization required depends on
the desired accuracy; our goal is to enable the self-consistent
determination of an OSET from new physics signals.  We will argue in
Section \ref{sec:MARMinPractice} that relatively uncorrelated
observables, such as single-object $p_T$'s, $E_T^{\rm miss}$, $H_T$
distributions, and object multiplicities, are sufficient to shed
light on the topology of new physics processes as well as constrain
mass differences and rates. Therefore, these are the distributions
we will try to reproduce faithfully.  A useful OSET must also
reproduce rapidity and relative angular variables on which data is
frequently cut to reduce Standard Model backgrounds, so that the
sculpting of new physics distributions by these cuts is reasonably
modeled.  In this discussion, where no detector simulator is used,
we will use the rapidity $y$ of a hard object in the lab frame as a
proxy for the reconstructed detector $\eta$ on which cuts are
applied.

We consider production dynamics in Section \ref{sec2:Production},
and introduce a leading-order parametrization for $2 \rightarrow 2$
hard scattering matrices. The brief discussion here is supplemented
by a more thorough one in Appendix \ref{sec2:APP}. After introducing
an expansion for $|\mathcal{M}|^2$ in $2\rightarrow 2$ production in
Appendix \ref{sec2:2to2}, we note useful approximate
factorization/shape-invariance results in Appendix
\ref{sec:empirical2to2}, and demonstrate analytically the
PDF-suppression of the higher-order pieces in our expansion in
Appendix \ref{sec2:ShapeInv}.  In Appendix \ref{app:ExampleFits}, we
present a large number of examples illustrating the success of the
two-parameter models of $\Msq$ considered in this section.

Decay dynamics are discussed in Section \ref{sec2:Correlations} and
elaborated upon in Appendix \ref{app:decay}, where we find that some
 care must be taken in interpreting di-object invariant mass distributions
in an OSET framework.  On the other hand, because the axis of the
initial production has been lost, single object distributions are
reproduced with reasonable accuracy by decaying particles simply via phase space.
These discussions of production and decay dynamics provide the
theoretical foundation for Section \ref{sec2:osetdef}, where we give
a definition of an OSET suitable for the Monte Carlo program we
introduce in Section \ref{sec:tool}.

We will see that dynamics play a secondary role to kinematics in
predicting LHC signatures, such that precise modeling is less
important than gross structural properties.  The reasons are that
long decay chains wash out detailed correlations, low luminosity
measurements have insufficient statistics to resolve theoretical shape
differences, and detector effects smear out all distributions.

Having emphasized the importance of kinematics over dynamics, a
caveat is in order.  The degree of coarse-graining appropriate to an
analysis is determined by experimental resolution---i.e. data set
size, detector performance, understanding of Standard Model
backgrounds, and the whims of a given experimentalist. In some
cases, kinematics are not completely resolvable, while in others it
may be desirable to include more detailed dynamics in matrix element
structure for production or decays.    Though we will focus on the
middle ground of accurate kinematic modeling and coarse dynamic
modeling, the MARMOSET tool we develop in Section \ref{sec:tool} is
organized in anticipation of a progressive analysis that
interpolates between approximate OSETs and actual Lagrangians.

\subsection{Particle Production Without a Lagrangian}\label{sec2:Production}
Our first goal is to develop a parametrization scheme for describing
particle production at a Hadron Collider when we have no, or
only partial, information about the underlying theory or Lagrangian.
The production and decay of a single, narrow resonance ($2\rightarrow
1$) is fully specified by the resonance mass $M$, production
cross section $\sigma$, the branching ratio $\mbox{Br}_i$ to the final state of
interest, and a finite polynomial in $\cos\theta^*$ determining the
angular distribution of decay products.

It is clear that no such exact, finite scheme exists for
parameterizing non-resonant $2 \rightarrow 2$ production.  These
processes must be described over a finite range of partonic
center-of-mass energy, so full $s$ dependence is important.  Also,
$\cos\theta^*$-dependence doesn't truncate at finite order as
evidenced by a $t$-channel propagator that yields an infinite power
series in $\cos\theta^*$.  Both $s$-dependence and angular
dependence of $\Msq$ would have to be expanded in infinite series to
capture the full $2 \rightarrow 2$ matrix element.

\begin{figure}[tbp]
\begin{center}
\includegraphics[width=6in]{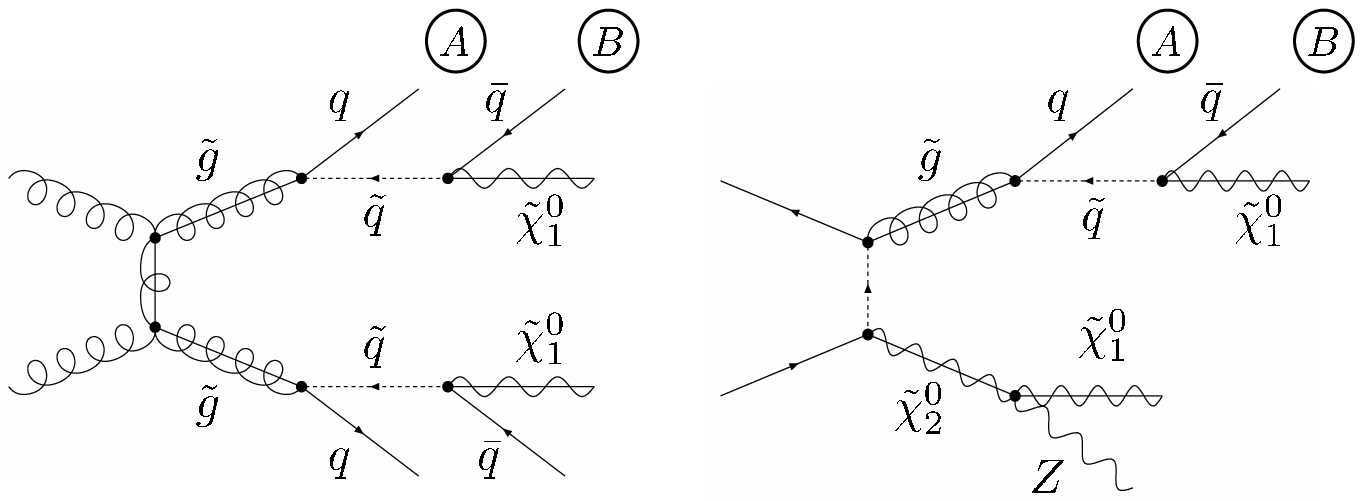}

\vspace{0.1in}
\begin{tabular}{|c|c|c|}
\hline Particle & Mass (GeV)\\
\hline
$\tilde g$ & 992 \\
$\tilde q$ & 700 \\
$\tilde{\chi}^0_2$   & 197 \\
$\tilde{\chi}^0_1$   & 89  \\
\hline
\end{tabular}
\caption{The left figure shows the topology and labeling scheme for
the $\tilde g\tilde g$ production example. There is also a $q\bar{q}
\rightarrow \tilde g \tilde g$ production mode not illustrated, but
included in our analysis. Note that the intermediate left handed
squarks appear on-shell. The right figure illustrates the
$\tilde \chi_2 \tilde g$ associated production example. The spectrum
for both of these examples is given in the table.}
\label{fig:ex1and2}
\end{center}
\end{figure}

Rather than a power series, we could use tree-level field theory to
constrain the form of matrix elements.  For example, in Figure
\ref{fig:ex1and2} we illustrate gluino pair
production through the gluon PDFs \cite{Martin:2002aw}, which has the matrix element \cite{Dawson:1983fw}
\def\gl{\tilde{g}}
\def\mg{m_{\tilde g}}
\begin{eqnarray}\label{eq:SUSYgluinoPairMEFull}
|{\cal M}(gg\to\gl\gl)|^2   \propto  \left(1
  -\frac{t_g u_g}{s^2}\right) \left[\frac{s^2}{t_g\,u_g}
  -2
  +4\,\frac{\mg^2 s}{t_g u_g} \left (1-\frac{\mg^2 s}{t_g u_g}\right)
  \right],
\end{eqnarray}
with $t_g = (p_{g,1}-p_{\gl,1})^2 -\mg^2$, $u_g = (p_{g,1}-p_{\gl,2})^2 -\mg^2$.
There is also a contribution from a $q \bar{q}$ initial state
\begin{equation}
|{\cal M}(q \bar{q}\to\gl\gl)|^2 \propto \bigg[
 \frac{t_g^2 + \mg^2 \,s}{s^2}+ \frac{4}{9} \frac{t_g^2}{t_q^2}
 + \frac{t_g^2 + \mg^2 \,s}{s\, t_q}+ \f{1}{18}\f{\mg^2 s}{t_g\, u_g} +
(t \leftrightarrow u) \bigg],
\end{equation}
with $t_q$ ($u_q$) $= (p_{g,1}-p_{\gl,1(2)})^2 -m_{\tilde q}^2$
dependent on the squark mass.  Within a given model, both expressions
involve a finite number of pieces, and we can add terms appropriate to
different spin possibilities. But the choice of matrix element depends
on, say, knowing how many states can propagate in the $t$-channel, and
one must specify all of their masses to parameterize the matrix
element fully.

\begin{figure}[tbp]
\begin{center}
\includegraphics[width=2.2in,angle=-90]{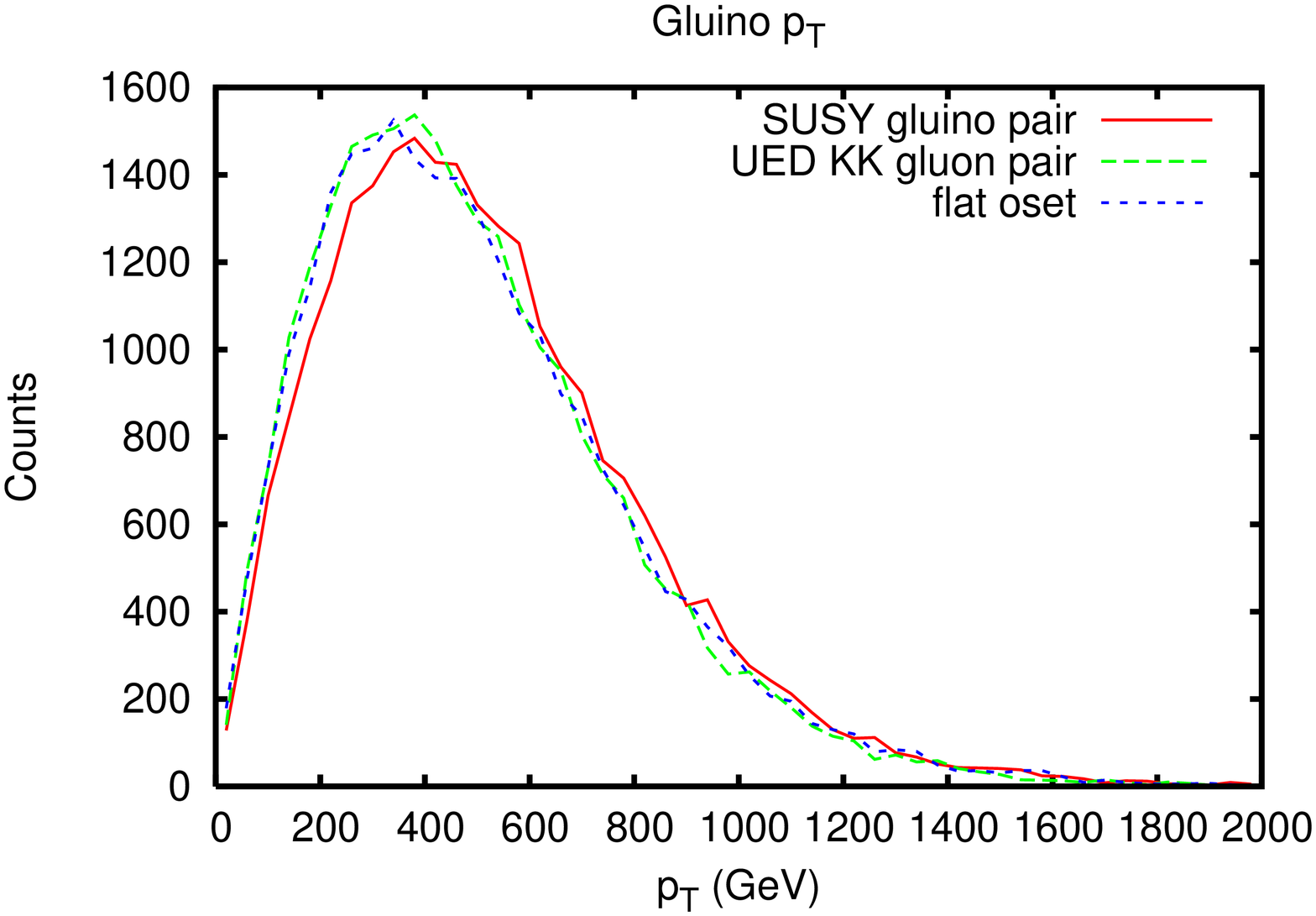}
\includegraphics[width=2.2in,angle=-90]{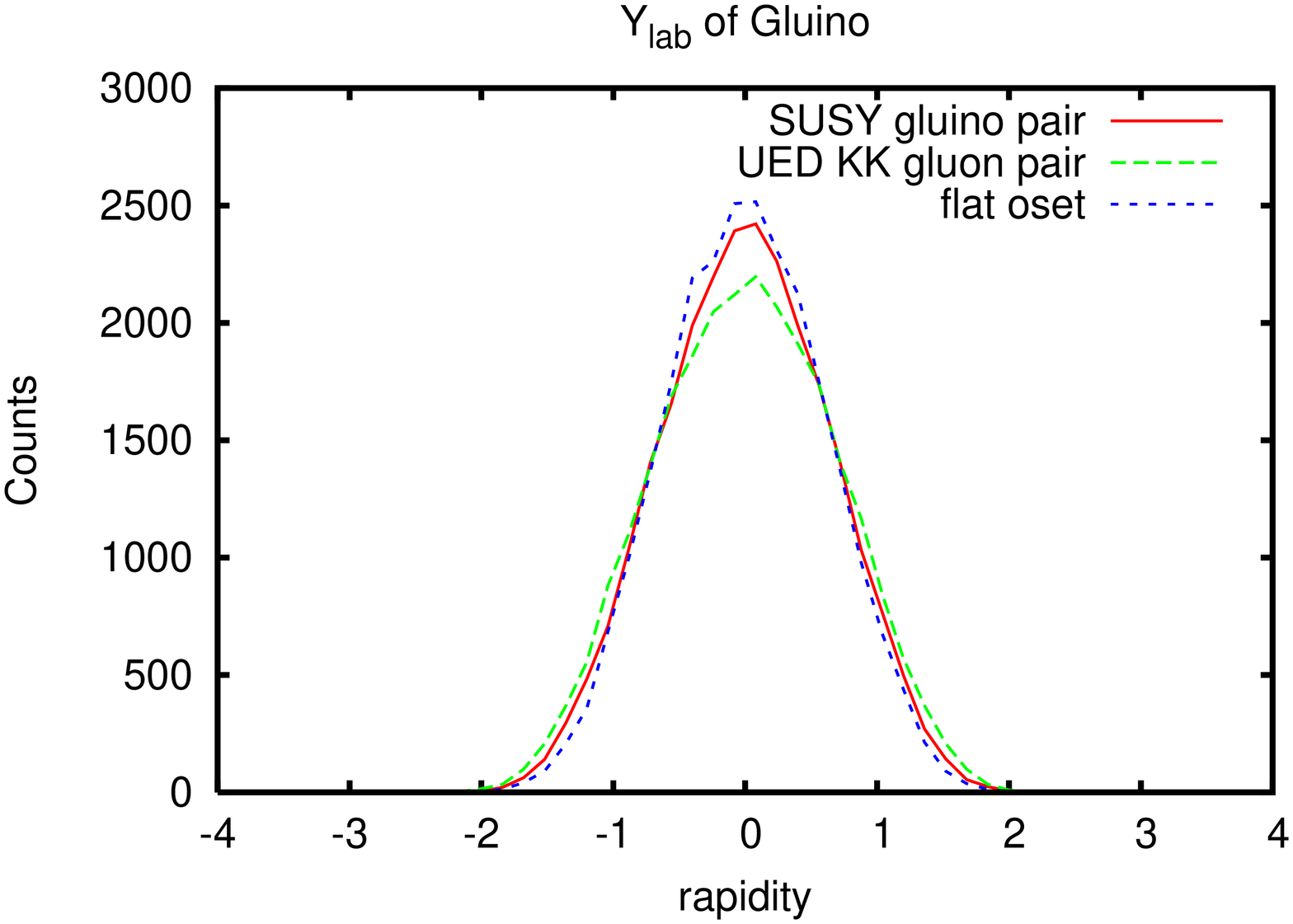}
\includegraphics[width=2.2in,angle=-90]{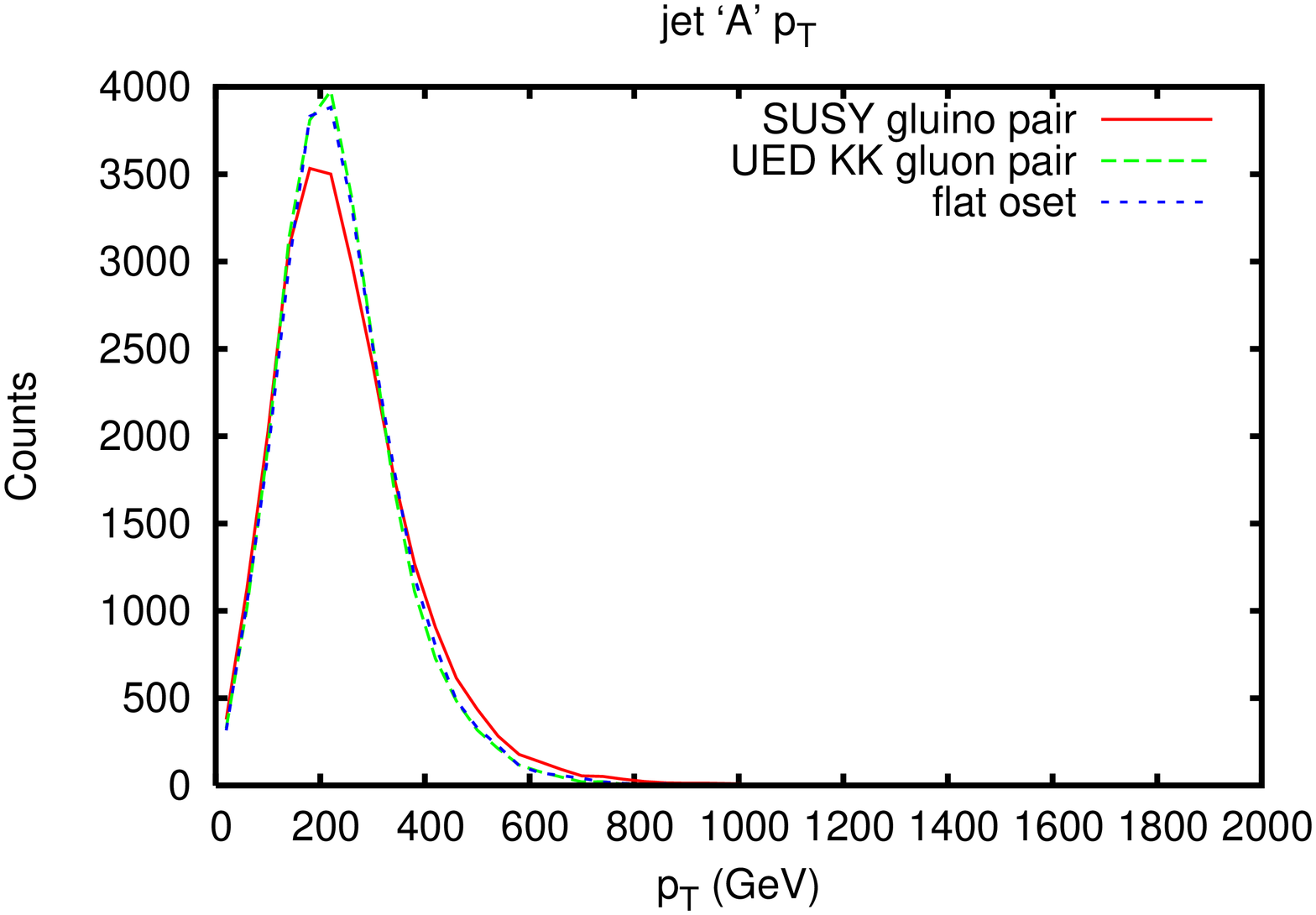}
\includegraphics[width=2.2in,angle=-90]{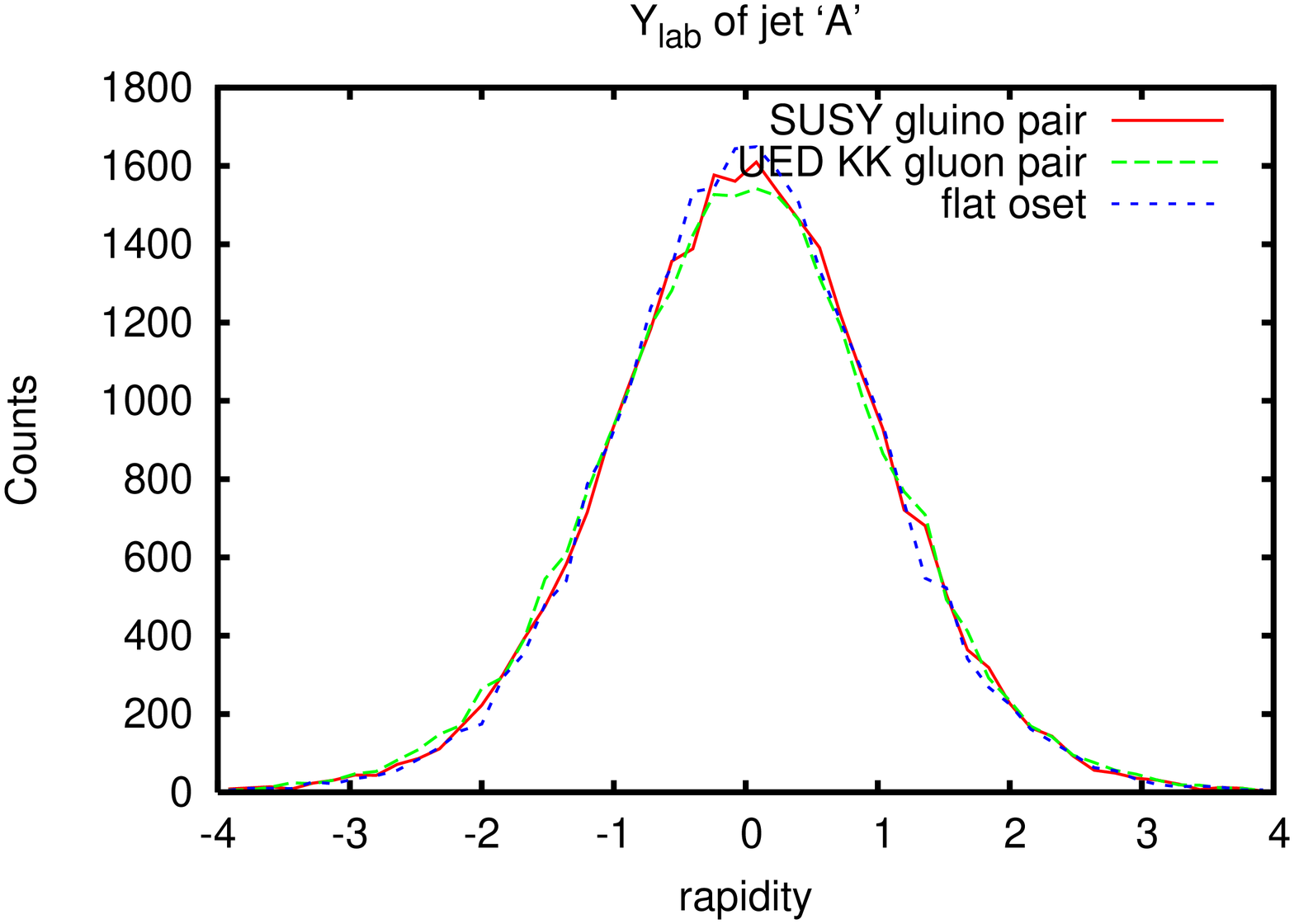}
\caption{Inclusive $p_T$ and $\eta$ distributions for the SUSY $\tilde
g\tilde g$ pair production process in Figure \ref{fig:ex1and2}
compared with a flat matrix element $\Msq \propto 1$ for both $u \bar u$ and $g g$ initial
states. Top: $p_T$ (left) and $\eta$ (right) of partonic
gluino.  Bottom: same distributions for the quark labeled ``A'' in
Figure \ref{fig:ex1and2}.}
\label{fig:FlatGlPairPt}
\label{fig:gluinoPairIntro}
\end{center}
\end{figure}

There is reason to hope, however, that hadronic production is
insensitive to the details of this structure.  The matrix element
$\Msq$ varies smoothly over energy, whereas parton luminosities fall
rapidly about threshold.  The kinematics of the gluino
pair-production process of Eq.~\eqref{eq:SUSYgluinoPairMEFull} is
quite well reproduced by the truncated expression
\begin{equation}\label{eq:SUSYgluinoPairMEApprox}
|{\cal M}|^2 = {\rm constant},
\end{equation}
with a value chosen to reproduce the total hadronic production cross
section.  Of course, this constant $\Msq$ has no Lagrangian
interpretation, but it serves as an effective leading order
parametrization for various kinematic variables.  The $p_T$ and
rapidity distributions for a gluino produced according to the full
matrix element of Eq.~\eqref{eq:SUSYgluinoPairMEFull} and the
approximation of Eq.~\eqref{eq:SUSYgluinoPairMEApprox} are shown in
Figure \ref{fig:gluinoPairIntro}. The agreement is striking! The
distributions for a visible decay product have been convolved with
decay kinematics, and are even less sensitive to detailed production
dynamics.

The success of this approximation relies dominantly on the
approximately polynomial fall-off of parton luminosities. When the
most naive constant approximation fails, it does so for the clear,
physical reason that the threshold- or high-energy scaling of $\Msq$
is extreme in one limit or the other. For example, $p$-wave
scattering amplitudes vanish at threshold, and four-fermion contact
interactions grow up to a cutoff scale $M_* \gg \sqrt{s_0}$
($\sqrt{s_0}$ is the threshold scale). In parameterizing corrections
to constant $\Msq$, it is convenient to introduce dimensionless
energy and angular variables related to the Mandelstam parameters
$\hat s$, $\hat t$, and $\hat u$ of the hard subprocess. We define
\begin{eqnarray}
X & \equiv & \hat s/s_0, \\
\xi & \equiv & \f{\hat t-\hat u}{\hat s} = \beta \cos\theta^{*},
\end{eqnarray}
where $s_0$ is the minimum possible value of $\hat s$ (for production of
species of masses $m_c$ and $m_d$, $s_0 \equiv (m_c+m_d)^2$),
\begin{equation}
\beta^2 =
\left(1-\f{m_c^2}{\sh}-\f{m_d^2}{\sh}\right)^2-4\f{m_c^2}{\sh}\f{m_d^2}{\sh}
\end{equation}
is the relative velocity of the products, and $\theta^{*}$ is the
scattering angle in the center-of-mass frame.  The variable $\xi$ is
the $z$-component of momentum of the particles in the center-of-mass system
scaled by half the center-of-mass energy.

\begin{figure}[tbp]
\begin{center}
\includegraphics[width=4in,angle=0]{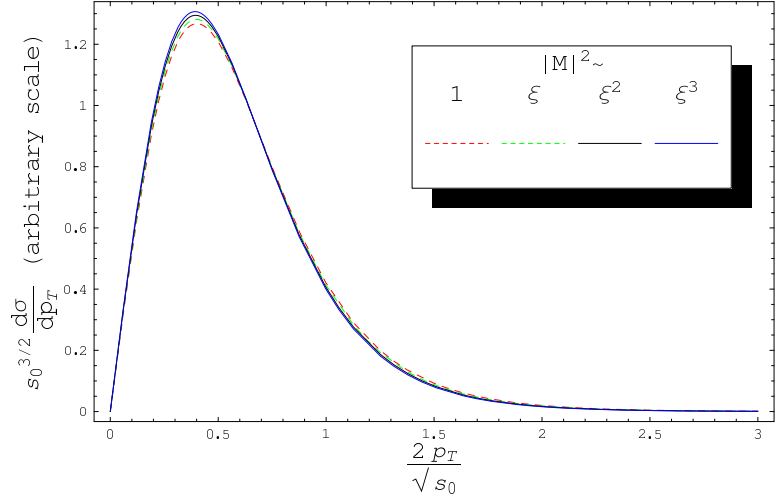}
\end{center}
\caption{Invariance of the shape of $\f{d\sigma}{dp_T}$ under changes
  in the $\xi$-scaling of the production matrix element.  All shapes
  have been normalized to unity.  The shape-invariance is also valid
  for $\Msq \propto X^p \xi^q$---at fixed $p$, the power of $q$ does
  not affect the shape of $\f{d\sigma}{dp_T}$.  Small deviations from
  this shape-invariance when the two final-state masses are not equal
  are discussed in Appendix \ref{sec2:APP}.}
\label{fig:shapesec2}
\end{figure}

In this basis, parameterizations that reproduce $p_T$ distributions
simplify remarkably because $\Msq \propto f(X)$ and $\Msq \propto
f(X) \xi^p$ produce nearly identical shapes.  This equivalence is
exhibited in one example in Figure \ref{fig:shapesec2}, but is quite
general and will be discussed further in Appendix \ref{sec2:APP}.
Though $\xi$-dependence does affect final-state rapidities, the
leading $\xi^0$ term usually dominates. Moreover, PDFs and cascade
decays tend to wash out $\xi$ effects on individual decay product
rapidities, as discussed further in Appendix \ref{app:ExampleFits}.
If events cannot be fully reconstructed, and $\eta$ distributions
matter mostly for their systematic effects on analysis cuts,
modeling $\Msq$ by a function that depends only on $X$ is
sufficient. Motivated by extreme scaling behaviors of $\Msq$ with
center-of-mass energy, we will try to find a small but sufficient
basis for parameterizing $X$-dependence.

\begin{figure}[tbp]
\begin{center}
\includegraphics[width=2.2in,angle=-90]{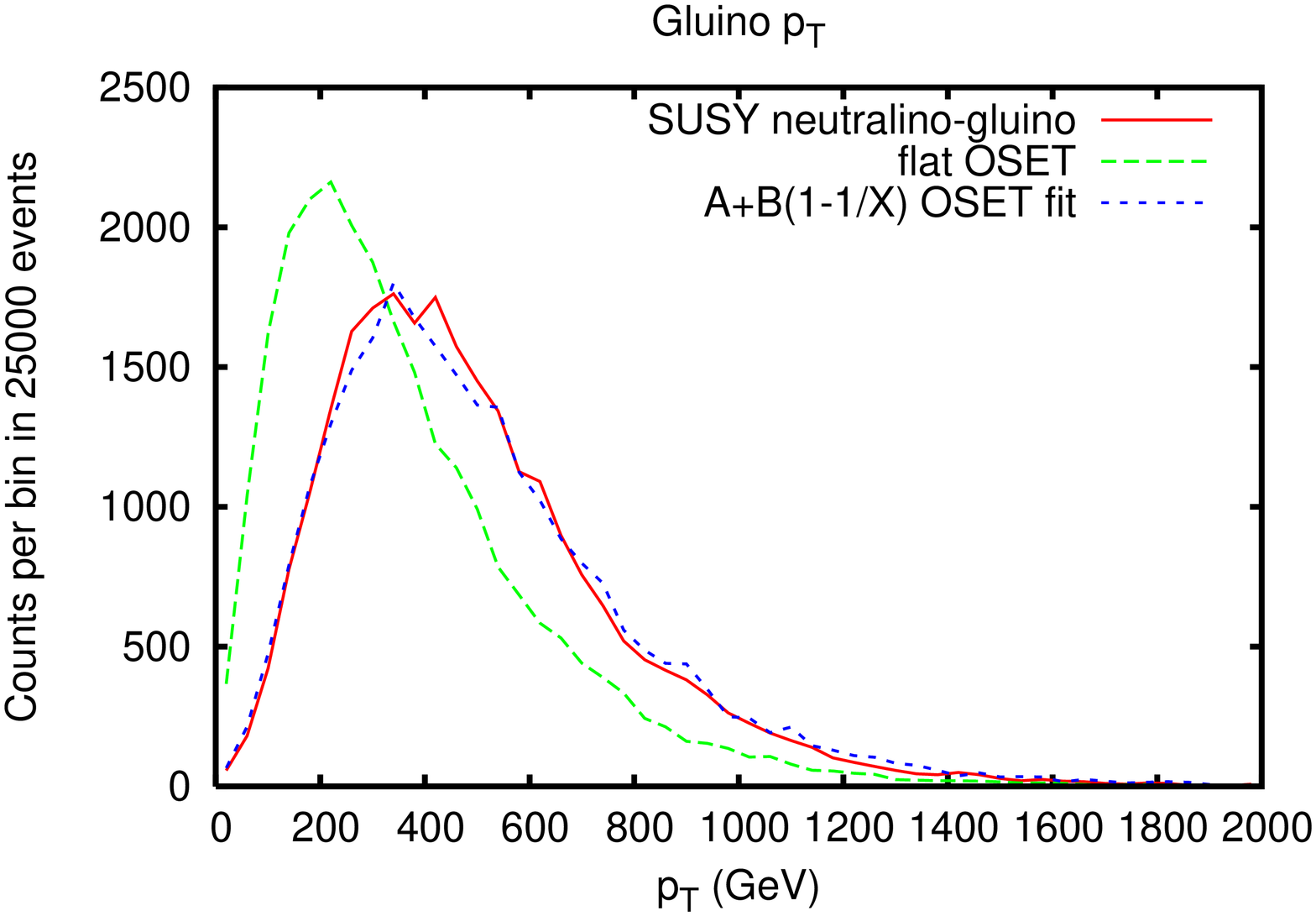}
\includegraphics[width=2.2in,angle=-90]{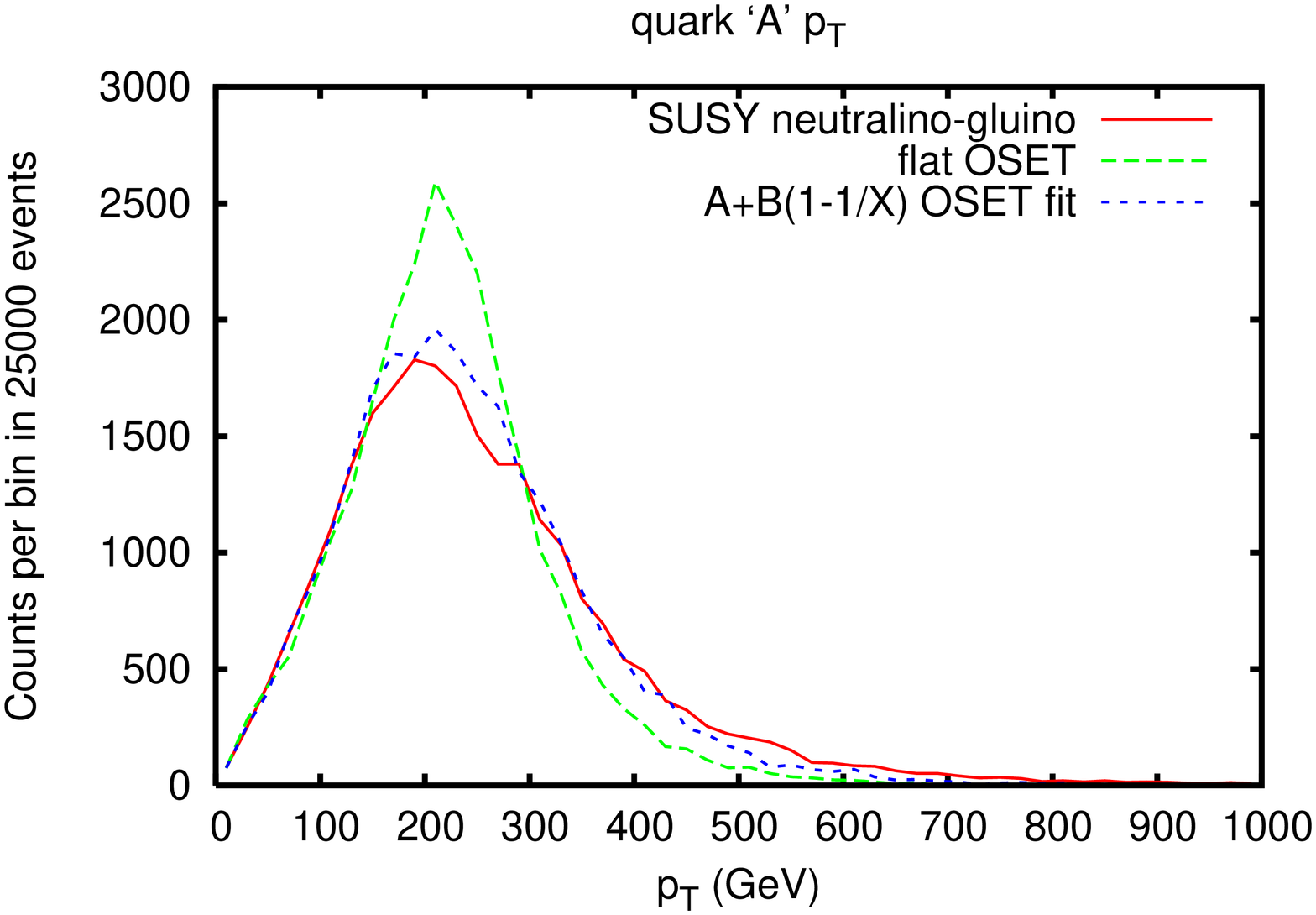}
\caption{Inclusive $p_T$ distributions for $\tilde g$ in the SUSY
$\tilde \chi_2 \tilde g$ associated production process in Figure
\ref{fig:ex1and2} compared with an OSET with a flat matrix element
$\Msq \propto 1$ and a threshold-corrected matrix element $\Msq \rightarrow A + B (1-1/X)$. Left:
$p_T$ of partonic gluino. Right: $p_T$ of quark labeled ``A'' in
Figure \ref{fig:ex1and2}.}
\label{fig:chi2gIntro}
\end{center}
\end{figure}

As an example of threshold-suppressed production, we consider the
associated production process $u\bar{u}\rightarrow \widetilde g
\widetilde \chi_2$ in the model of Figure \ref{fig:ex1and2}.  Unlike gluino
production, it is $p$-wave dominated and $\Msq \propto \beta^2$
near threshold.  As shown in Figure
\ref{fig:chi2gIntro}, a flat $\Msq$ ansatz does not reproduce the gluino
decay product $p_T$. The function $(1-1/X)$ reproduces this
threshold behavior (for $m_c=m_d$, $1-1/X = \beta^2$), and a
linear combination
\begin{equation}
\Msq \rightarrow A + B (1-1/X),
\end{equation}
with $A$ and $B$ obtained by a $\chi^2$ fit reproduces the SUSY
distributions quite well in Figure \ref{fig:chi2gIntro}.  We
emphasize that the $A$ and $B$ that reproduce overall rates and
$p_T$ distributions are \emph{not} the coefficients one would find
in a Taylor expansion of $\Msq$!  Rather, because of the $\Msq
\rightarrow \Msq \xi^p$ shape invariance of the transverse
structure, the coefficients $A$ and $B$ are only an effective
parametrization after integrating over PDFs. The $A$ term in this
case arises from $\Msq \propto \xi^p$ pieces in the underlying matrix
element.
%

\begin{figure}[tbp]
\begin{center}
\includegraphics[width=2.2in,angle=-90]{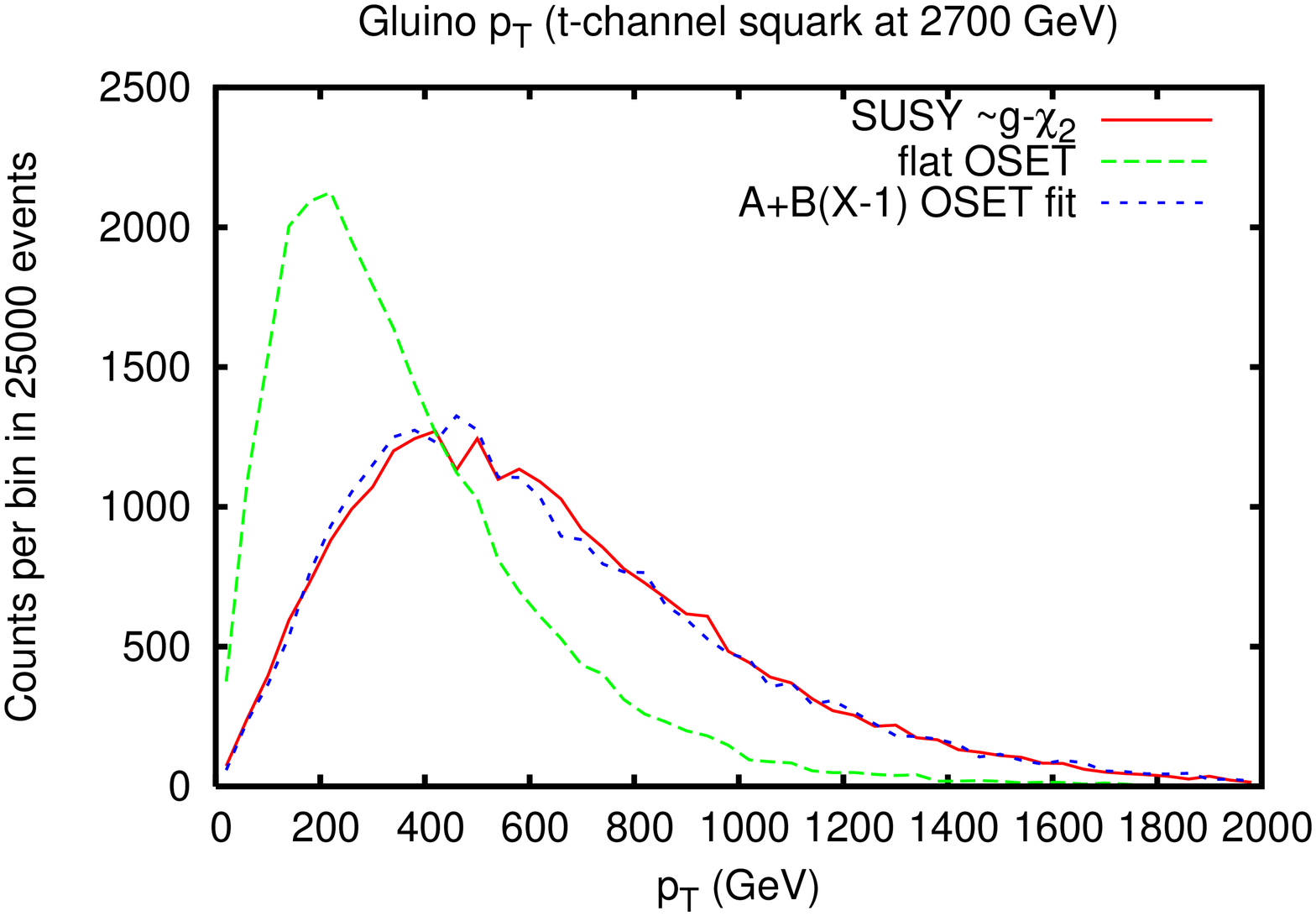}
\includegraphics[width=2.2in,angle=-90]{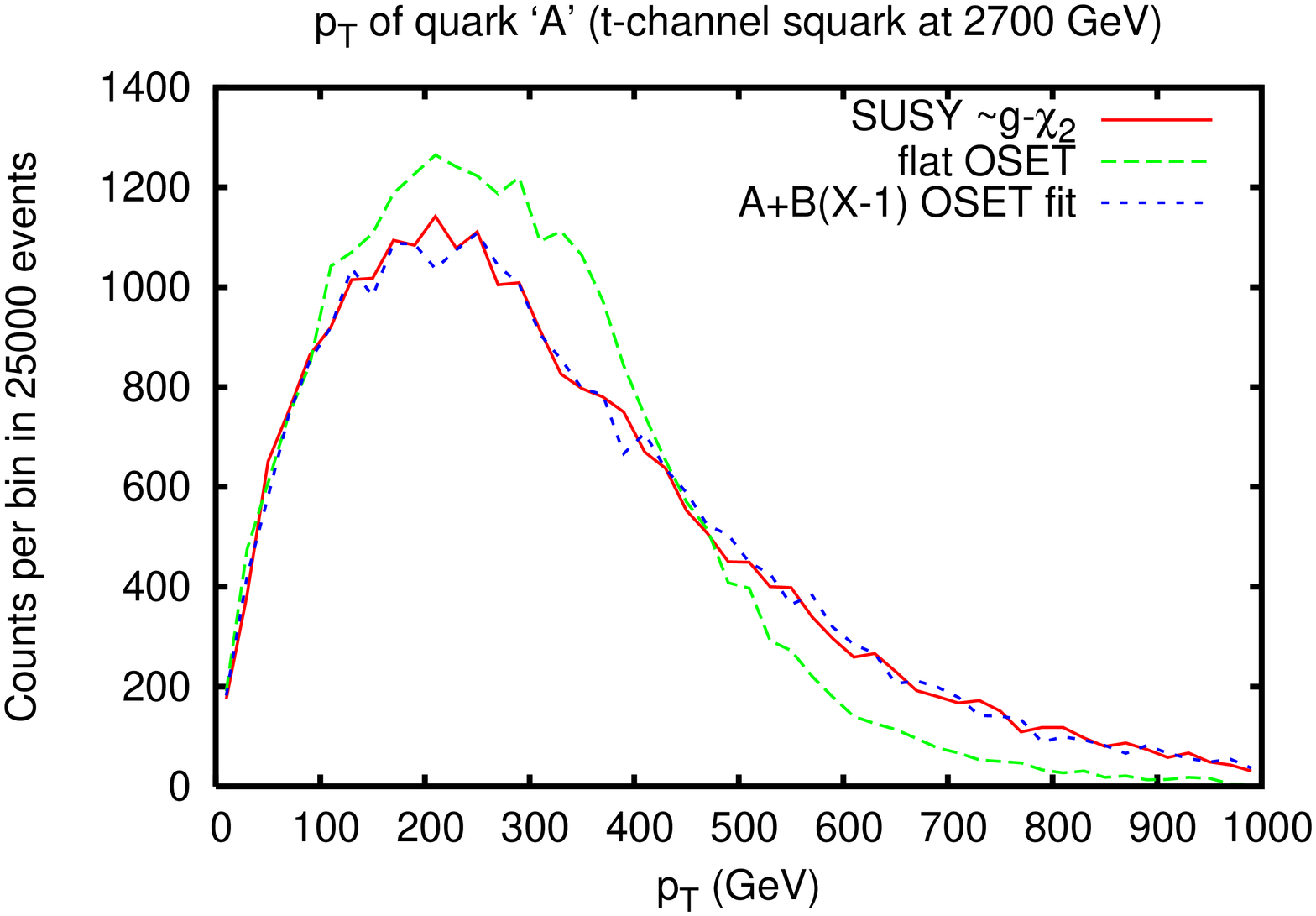}
\caption{Inclusive $p_T$ distributions for $\tilde g$ in the SUSY
$\tilde \chi_2 \tilde g$ associated production process mediated by an
off-shell $t$-channel squark.  The masses are $m(\tilde g)=991$ GeV
and $m(\tilde \chi_2)=197$ GeV as in Figure \ref{fig:ex1and2}, but
$m(\tilde q) = 2700$ GeV significantly exceeds the threshold
center-of-mass energy $\sqrt{s_0}$, and a contact interaction is an
appropriate description of this operator.  We compare the $p_T$ of the
gluino (red) to the prediction from an OSET with a flat matrix element
$\Msq \propto 1$ and corrected matrix element with $\Msq =A+B (X-1)$,
with the parameters $A$ and $B$ fit to this $p_T$ distribution. Left:
$p_T$ of partonic gluino. Right: $p_T$ of a final-state quark.  It is
striking that, although the flat matrix element misestimates the peak
of the gluino $p_T$ distribution by nearly a factor of $2$, it
introduces only $\sim 15\%$ inaccuracy in the $p_T$ of a quark
produced in the decay.}
\label{fig:chi2gl2700}
\end{center}
\end{figure}

When $m_{\tilde q} \gg m_{\tilde g}$, $|\mathcal{M}(q \bar q
\rightarrow \gl \tilde \chi)|^2$ displays qualitatively different energy
scaling.  Integrating out the $t$-channel squark, we see that at
energies well below $m_{\tilde q}$, the interaction is modeled by a
four-fermion operator, with $|\mathcal{M}|^2 \sim \f{t^2}{m_{\tilde
q}^4}$.  This growth of $\Msq$ is, of course, cut off for $t \sim
m_{\tilde q}^2$, but PDFs will cut off the $p_T$ distributions,
anyway. With contact-like interactions, a good fit to $p_T$'s is achieved with
\begin{equation}
\Msq \rightarrow A + B (X-1).
\end{equation}
For $t$-channel intermediate squark mass $m(\tilde q) = 2.7$ TeV, the
results of constant and linear parameterizations for $\Msq$ are shown
in Figure \ref{fig:chi2gl2700}; we note that $A$ is quite small,
contributing $\sim 1\%$ to the total cross-section. At higher squark
masses (where the cross-section for this process is also lower), a
quadratic term $C (X-1)^2$ may be added for a more accurate fit.

This example illustrates another reason that simple matrix elements
are often quite effective: when decay products are boosted, their
relative $p_T$ in the rest frame of the parent particle washes out
inaccuracies in modeling of the $p_T$ distribution of this parent
particle.  For instance, whereas $\Msq \propto 1$ misestimates
the gluino $p_T$ by nearly a factor of 2 in Figure
\ref{fig:chi2gl2700}, the $p_T$ of a quark into which it decays is
only inaccurate at the $\sim 15\%$ level.

Although angular dependence is the only important structure of $\Msq$
in the $2 \rightarrow 1$ resonant production case, we have said nothing about its role in
non-resonant pair production.  Indeed, we achieved a good fit to $p_T$
and $\eta$ in the previous example by fitting a function of $\hat s$,
when $\Msq$ is manifestly a function of $\hat t$.  In Appendix
\ref{sec2:ShapeInv}, we note the striking result that, when integrated
over PDFs, $p_T$ distributions are independent of the $\xi$-dependence
of $\Msq$ to a very good approximation.  Moreover, modeling $p_T$
shape with $\Msq = f(X)$ has little effect on the rapidity
distribution $\f{d\sigma}{dy}$, which is controlled dominantly by PDFs
and secondarily by the $\xi$-dependence of $\Msq$.  Therefore, the
problem of modeling $p_T$'s and $y$'s is largely factorizable; in
practice, if events cannot be fully reconstructed and one seeks only
to reproduce single-object $p_T$ and lab-frame $\eta$ distributions,
$\xi$-dependence of the amplitude can be ignored.

Deviations from this factorizable structure will give rise to theory
systematic errors; for example, the $X$-dependent parameterization
of $\Msq$ obtained by fitting to momentum distributions of central
objects differs from what one would expect from theory if all events
were treated equally.  Therefore, when using the approximate
parametrization method, experimentalists should be aware of these
systematic effects, and try to estimate them.

Then again, the theory systematics of different parametrizations for
$\Msq$ are not very different in magnitude from the kinds of theory
systematics one encounters in comparing LO and NLO results for a
given theoretical model.  In the context of a model with real matrix
elements, the largest NLO systematic is on overall normalization,
whereas in an OSET parametrization, the total cross section is a
free parameter that is matched to the data anyway.  The advantage of
a OSET parametrization is that systematic effects reflect our actual
uncertainty about the underlying model while still allowing
quantitative comparisons between models and data.

Finally, in some cases, $2\rightarrow 3$ production modes can be key to
discovery of new physics.  Notable examples include vector boson and
top fusion in production of Higgs-like resonances.  In these examples,
Standard Model dynamics has enormous effect on event shapes, but
new physics internal lines are absent.  Extending the matrix elements
for Standard Model Higgs production to more general new physics
scenarios is a reasonable approximation; we will not discuss it further.

\subsection{Decay Kinematics and Final State Correlations}\label{sec2:Correlations}

So far, when presenting $p_T$ and rapidity distributions for decay
products, we have assumed that all decays are weighted only by phase
space.  In this section, we examine the structure of more general
decays matrix elements, and their effects on variables of interest.
Though the current implementation of MARMOSET only allows for phase space decays,
we discuss systematic ways of including these effects and
improving our approximation.

We choose to focus on single- and two-object observables that are
relevant to determining the gross structure of a model and could be
measurable with low luminosity---$p_T$'s of single objects (or
uncorrelated sums of several of them) and pairwise invariant masses
such as $m_{\ell^+ \ell^-}$ are useful variables for characterizing
decay chains, $\delta \phi(j,E_T^{\rm miss})$
is frequently employed in cuts to reduce Standard Model background, and object $\eta$'s are used both to determine signal acceptance and background reduction.
Though invariant mass observables involving more than two objects can
be useful \cite{Miller:2005zp}, interpreting them will probably
require higher statistics and we will not discuss them here.

Correlations in the rest frame of a decaying particle tend to be
washed out after boosting to the lab frame.  For this reason, when
events cannot be fully reconstructed, the phase space approximation
reproduces single-object $p_T$ and $\eta$ distributions quite well.
Spin correlations do induce observable effects on $p_T$ distributions
of decay products, such as the lepton $p_T$ in top quark decay. Such
effects can be exploited for precision analysis (for example, the
determination of $W$ helicity and top polarization in top decays
\cite{Hubaut:2005er,Abulencia:2005xf,Abulencia:2006ei,Lillie:2007yh}),
but are subleading for discovery.

Inaccuracies in $\delta \phi(j, E_T^{\rm miss})$ would lead to
systematic errors in measuring the cross section of new physics when
$\delta\phi$ cuts are used to remove QCD background from searches
for new physics in hadronic channels.  In principle, the directions
of a jet and the missing particle in the same decay chain are
correlated. But averaging over both decay chains in $E_T^{\rm miss}$
makes this variable fairly insensitive to spin correlations.
Partonic $\delta \phi(j, E_T^{\rm miss})$ distributions for SUSY, UED,
and OSET processes with the same topologies are visually identical
(see Figure \ref{fig:FitGlPairPT}, again for the $\tilde g$
pair-production process).

Next, we discuss the distributions of invariant mass observables. We
should stress that phase space, together with on-shell
conditions, reproduces positions of kinematic end-points, but do
\emph{not} reproduce their shapes.  Unlike the transverse variables,
invariant mass distributions are invariant under the boost to the
lab frame, and spin correlation information is not diluted. Exploring
such information to measure spin has been studied recently
\cite{Barr:2004ze,Battaglia:2005zf,Smillie:2005ar,Datta:2005zs,Datta:2005vx,Barr:2005dz,Alves:2006df,Athanasiou:2006ef,Wang:2006hk,Smillie:2006cd,LW-spin2}.

\begin{figure}[tbp]
\begin{center}
\includegraphics[scale=0.4]{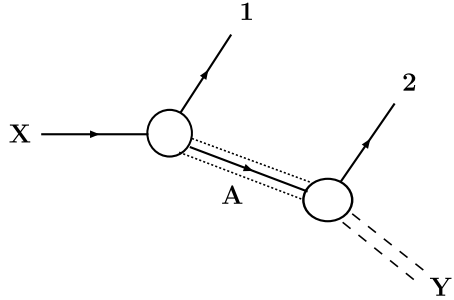}
\end{center}
\caption{A decay process in which $X\rightarrow 1 + A(\rightarrow
  2+Y)$. Depending on the mass spectrum, $A$ could be either on or off
  the mass-shell. From observable particles $1$ and $2$, we could form
  the invariant mass combination $m_{12}^2=t_{12}=(p_1+p_2)^2$.
  \label{decay}}
\end{figure}

Factors that can change the shape of the invariant mass distribution
significantly include, spin correlations, effects of off-shell
propagators, as well as interference between several processes. We consider the decay process shown in Figure~\ref{decay}. We can systematically improve our approximation by expanding
the differential decay width in a single pair's invariant mass
$t_{12}=(p_1 + p_2)^2$ as
\begin{equation}
\label{minv-general} \frac{\mbox{d}}{\mbox{d} t_{12}} \Gamma= a_0 +
a_1 t_{12} + ... + a_{n} (t_{12})^{n}.
\end{equation}
Keeping only $a_0$ corresponds to our original flat matrix element
approximation.

We begin by considering the case that the intermediate particle $A$
is on-shell. In this case, we can consider a single channel shown
in Figure~\ref{decay} since interference effects are typically small
in on-shell decays. In this case, the shape is almost completely
determined by spin correlations. In particular, the degree of the
polynomial satisfies $n=2J_{\rm A}$ \cite{Wang:2006hk}, where $J_{\rm
  A}$ is the spin
of the intermediate particle $A$. Constants $a_0,..., a_{J_{\rm A}}$
depend on the masses of particles in the decay chain, the
chiralities of the couplings, and the masses and spins of
final-state particles $X$ and $Y$.  If, for instance, $A$ is
fermion, the sign and value of $a_1$ depends on whether $X$
and/or $Y$ are vector or scalar, as well as mass differences such as
$M_A^2 - 2 M_Y^2$ \cite{LW-spin2}.

\begin{figure}[tbp]
\begin{center}
\includegraphics[angle=270,scale=0.4]{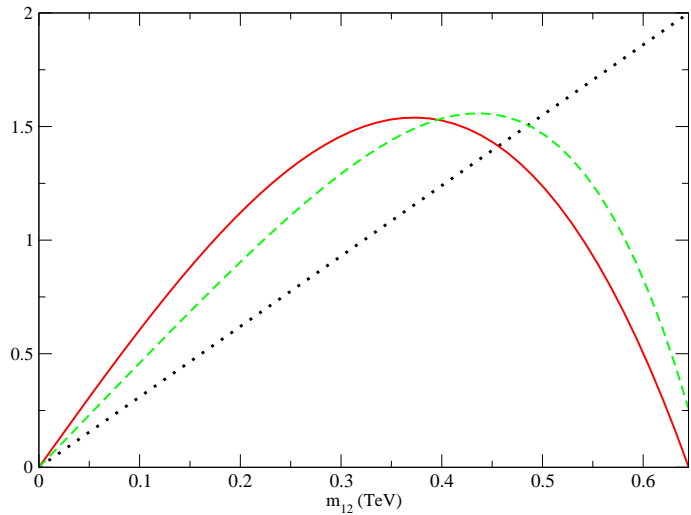}
\end{center}
\caption{Comparison of shapes for di-object invariant mass distributions.
The solid (red)
  curve is the case that $X$ and $Y$ are scalars. The dotted curve is
  the result of decaying through pure phase space. The dashed (green)
  curve results from a decay chain in which $A$ is a vector while
  $X$ and $Y$ are fermions. \label{minv_comp_twotwo}}
\end{figure}

The dependence of the parameters $a_i$ on masses demonstate
perfectly the need to characterize mass hierarchies before
extracting spin information from decays.  But particularly when
decays can be cleanly isolated (e.g. di-lepton invariant masses),
the $a_i$ cannot be forgotten.  An edge that appears clearly in an
OSET with the correct spectrum may be invisible in the data if
angular momentum conservation forces the distribution to vanish at
the edge!  If both $X$ and $Y$ are scalars, angular momentum
conservation force the distribution to vanish exactly at the edge as
in Figure \ref{minv_comp_twotwo}.  In other cases correlations may
reduced the edge so much that we cannot tell it apart from
statistical fluctuations \cite{Smillie:2005ar}. For comparison, we
have also included a UED-like case where the intermediate particle
is a vector and $X$ and $Y$ are fermions. Given fluctuations and
combinitorics, it will be challenging to resolve the end point
structure of the solid curve. Therefore, the absence of an edge in
the data \emph{should not} rule out an OSET---it should instead
warrant of a thorough study of the $a_i$ dependence of
Eq.~(\ref{minv-general}).

\begin{figure}[tbp]
\begin{center}
\begin{tabular}{ccc}
\parbox{5cm}{\includegraphics[scale=0.3]{sec2plots/3body-decay}} & \parbox{2cm}{ \hspace{0.5cm} \huge $+$}& \parbox{5cm}{\includegraphics[scale=0.3]{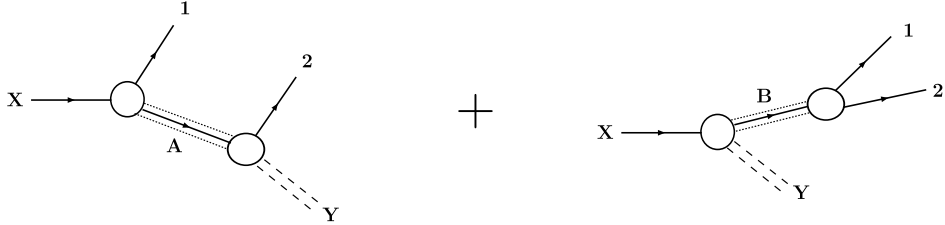} }
\end{tabular}
\end{center}
\caption{A more general decay topology where two channels contribute
to the same final states. The parameterization of Eq.~(\ref{minv-general}) is still valid even in this case where there is the possibility of interference between the channels.  \label{decay-2}}
\end{figure}

When an intermediate particle is off-shell, $\f{d\Gamma}{dt_{12}}$
is modified by both spin correlations \emph{and} the
momentum-dependence of the propagator, which is more significant
when the intermediate state is close to on-shell.  Both deformations
can be parameterized by a polynomial in $t_{12}$.  So long as we
consider only a single pairwise invariant mass distribution, the
presence of multiple decay modes as in Figure \ref{decay-2} and the
resulting interference can also be modeled by a polynomial in
$t_{12}$.  This improved treatment of decay matrix elements based on
Eq.~(\ref{minv-general}) is particularly useful in the cases when an
OSET analysis reveals a large deviation in invariant mass
distributions while other observables show good fits.

A qualitatively different case is when a new resonance decays into
several visible Standard Model particles, allowing full
reconstruction and facilitating spin determination. More complete
angular information in the rest frame of the decaying particle would
have to be included. For instance, in a two body decay, angular
information could be systematically included by expanding
$\mathcal{M} = \sum a_\ell \cos\theta^\ell$. Conservation of total
angular momentum restricts the allowed values of $\ell$.

\subsection{Definition of an OSET}
\label{sec2:osetdef} The preceding discussion allows us to define an
OSET more precisely than the preceding intuitive descriptions.  It
is specified, first, by a spectrum of new particles with given
masses and possibly observable widths (if instrumental widths
dominate, unknown physical widths can be set to arbitrarily small
values).  Though not strictly speaking necessary, the $U(1)_{\rm
EM}$ and $SU(3)_C$ gauge quantum numbers for new particles are also
specified, both for reasons of theoretical consistency as well as to
enable OSETs to be simulated using Monte Carlo tools like \Pythia\
that model initial/final state radiation and parton showering.

\begin{figure}
\begin{center}
\includegraphics[width=4in]{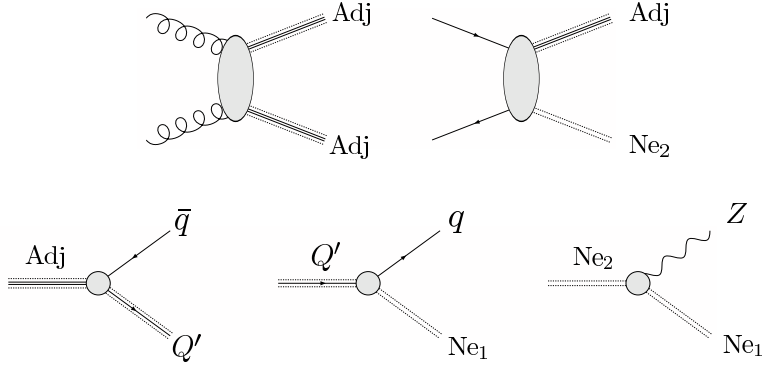}
\caption{Production and decay elements in an OSET describing the
  processes of Figure \ref{fig:ex1and2}.}\label{fig:OSETelements}
\end{center}
\end{figure}

The remaining content of an OSET concerns the observable production
and decay modes, in terms of only the particles that appear on-shell
in each mode.  For example, the processes of Figure
\ref{fig:ex1and2} would be described by an OSET containing the
production and decay elements shown in Figure
\ref{fig:OSETelements}. Each OSET vertex is represented by a gray
blob, which reminds us that we are choosing an approximate,
parameterized matrix element $\Msq$. The type of structure depends
on the context in which a blob appears (e.g.\ production vs.\ decay)
and the amount of structure is chosen such that terms that do not
produce resolvably different observables are not included.  In
practice, more detail is required in modeling production than decay.

In our leading-order OSET parameterization, the blob structures are as
follows:

\begin{itemize}
\item \textbf{$2 \rightarrow 1$ Resonant Production} from a given
partonic initial state is specified by an overall hadronic cross
section, integrated over the appropriate PDFs.  Together with the
mass and width of the resonance, this fully determines the resonance
production from unpolarized initial states.  In this case, it is
possible to extract from the hadronic cross section and PDFs a
partonic cross section, and from this a squared coupling. This
near-equivalence between an OSET process and a renormalizable
Lagrangian is particular to resonant production, and not generic.
Of course, without specifying the decay modes of the resonance,
observables such as forward/backward asymmetries will be lost.

\item \textbf{$2 \rightarrow 2$ Pair or Associated Production} from a
given partonic initial state can typically be approximated by
\begin{equation}
\Msq = A + B \begin{cases} (1-1/X) \\ (X-1) \end{cases}.
\end{equation}
The two forms are appropriate for threshold-suppressed and
contact-like interactions, as discussed in Section
\ref{sec2:Production}.  When a $B$ term is needed to match $p_T$
distributions, the preferred form is usually clear from the shape of
the tail.  In many cases---production of heavy particles of comparable
masses, with poor statistics, or in generating exclusions on processes
that have \emph{not} been observed---a parameterization in terms of
$A$ alone is most useful.  In cases of high-precision measurement, the
expansion can be readily expanded to include higher powers of $X$
and/or leading $\xi$-dependence (this could be linear or quadratic,
depending on symmetries).  We refer the reader to Appendix
\ref{sec2:APP} for more detailed discussion of $2 \rightarrow 2$
matrix element structures.

As free rate parameters for $2 \rightarrow 2$ production, we can use
either the partonic $A$ and $B$ or the more directly observable
PDF-integrated hadronic cross-sections for each term.  If $A$ and
$B$ are used, the OSET expression for $\Msq$ \emph{looks} like an
effective Lagrangian, but it cannot be interpreted as such---$A$ is
not the coefficient of $\Msq \propto 1$ in an effective Lagrangian,
but a linear combination of coefficients for $\xi$-dependent
contributions to $\Msq$.  For example, $p$-wave production does
\emph{not} mean that $A=0$, but only that the dominant contribution
to $A$ has non-trivial $\xi$-dependence.

\item \textbf{$2 \rightarrow 3$ Production} is typically very far from
uniformly filling out phase space.  For example, in weak boson
fusion into a Higgs, the scattered quarks are typically quite
forward, and we can think of the Higgs as being resonantly produced
from off-shell weak bosons.  Thus, Standard model dynamics affects
phase space considerably, whereas the dynamics of new physics is
quite simple. The same is true of $\bar{t} t h$ production, though
this includes radiated Higgs production from off-shell top quarks in
$t\bar t$ final states.  In our current implementation, we will only
include these two classes of $2 \rightarrow 3$ production in OSETs,
with the same dynamics as for the Standard Model Higgs processes.
Such processes would be characterized by one overall rate.  For most
intents and purposes, $2 \rightarrow 3$ processes with two new
particle final states and one QCD jet are adequately described by $2
\rightarrow 2$ production with parton showering.

\def\Msqd{|\mathcal{M}_\text{decay}|^2}
\item \textbf{$1 \rightarrow n$ Decay} is modeled at leading-order by
$\Msqd = $ constant. This is currently the only decay incorporated
in MARMOSET. Two simple situations require more detail: if the decay
is of a singly-produced resonance or if events can be fully
reconstructed, it may be desirable to incorporate angular
dependence. Likewise, angular and (for three-body decays) energy
dependence may be needed to reproduce the shapes of kinematic edges
and endpoints.  The ability to handle the decay parameterization of Eq.~(\ref{minv-general}) will be added to future versions of MARMOSET.

When total decay widths cannot be measured, only
products of branching fractions and cross sections are observable.
However, when more than one process has been observed, it is often easier and instructive to
parametrize rates in terms of branching fractions instead of overall rates for individual processes.  This is because branching fractions properly account for combinatoric factors in decays, so correlated final state multiplicity information is retained. This is the MARMOSET default.

\end{itemize}

We note that in this parameterization, spin is not explicitly
included.  Nonetheless, non-trivial matrix elements for various processes
can incorporate the \emph{leading} effects of spin on threshold behavior
and angular distributions.

Now that we have defined an OSET parametrization, we have a choice
for how to generate OSET Monte Carlo. We can imagine two ways of
implementing a $2\rightarrow 2$ production matrix element $\propto
A+B(1-1/X)$ with unknown $A$ and $B$.  Currently in MARMOSET, one
generates separate samples with $\Msq \propto$ 1 and $1-1/X$, and
mixes them together with appropriate weights.  Though simple, this
approach does not generalize well to more complicated expressions,
particularly if the formulas in question are not additive.  A more
robust alternative is to generate and simulate Monte Carlo for a
flat matrix element, keeping both detector-simulated event records
and partonic kinematic information. Then, a given choice of $A$ and
$B$ determines a weight associated with each event.  These weights
can be stored, or events can be unweighted by keeping only those
that satisfy $(A+B(1-1/X))^2 > (A+B) r$ for a random number $r\in
[0,1]$. The difficulty with this method is that unweighting can be
extremely inefficient for certain matrix elements.  In practice, we
imagine that some combination of preweighting and reweighting will
be the most efficient use of OSET Monte Carlo.

The matrix element parameterizations we have outlined are useful when
we are in the dark about the underlying physics.  One would hope,
however, that an OSET analysis (to be discussed in Section
\ref{sec:MARMinPractice}) combined with theoretical understanding will
quickly lead us to a correct framework for describing the new physics
observed at the LHC.  In this case, a parameterization in terms of
on-shell masses and the efficient process-by-process Monte Carlo
generation that it permits will continue to be useful as an efficient
means of scanning parameter space, but naive rate parameterizations
will have outlived their usefulness.  In this scenario, it may well be
more reasonable to parameterize dynamics through a full set of model
parameters, sliced into surfaces of fixed masses for species produced
on-shell.  As one moves along such a slice, changes in $\Msq$ can be
incorporated by re-weighting of saved Monte Carlo as discussed above.

\begin{figure}
\begin{center}
\includegraphics[width=4in]{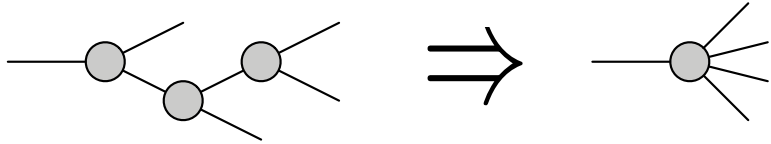}
\end{center}
\caption{When mass scales are neither well-separated nor squeezed, it
  may be difficult to determine that a state is on-shell; in this
  case, it may be convenient to characterize a cascade decay as a
  single $1 \rightarrow {\it  many}$ decay instead.}\label{fig:cascade->many}
\end{figure}

The converse situation can also arise, in which an OSET contains too
much kinematic detail.  When mass ratios are neither very large nor
very small, on-shell particle masses may not be clearly resolvable
and it may not be clear which states in a decay chain are on-shell.
It may then be useful to parameterize a chain of decays as a single
$1 \rightarrow \mathit{many}$ decay, as in Figure \ref{fig:cascade->many}.
This is readily done in an OSET framework.


%% file: UsersGuide.tex
\section{MARMOSET: A Tool for OSET Monte Carlo}
\label{sec:tool} The OSET parametrization introduced in the previous
section allows us to describe a wide class of TeV-scale models quite
simply, in terms of the masses, quantum numbers, production modes,
and decay modes of new particles, at the level of cross sections and
branching ratios. We have seen that even with leading-order coarse
approximations, an OSET reproduces single-object kinematics and some
features of correlated kinematic variables (edge and endpoint
locations in invariant masses) with reasonable accuracy, and a more
detailed correspondence can be achieved systematically. One key
feature of an OSET is that mass parameters that affect kinematics
are disentangled from the specification of rates for new physics
processes.  In this way, an OSET can be simulated \emph{additively}
process by process. In this section, we introduce MARMOSET (Mass and
Rate Modeling in On-Shell Effective Theories), currently a
\Pythia-based tool for generating Monte Carlo for any model that is
well characterized by a finite set of particles. More information
about installing and using MARMOSET can be found at
\texttt{http://marmoset-mc.net}.

MARMOSET manipulates \Pythia's behavior in two ways: we have extended
\Pythia's production modes to include parameterized production of the
kind described in Section~\ref{sec2:osetdef}, and we have built a user
interface and organization scheme for generating separate Monte Carlo
for individual decay topologies while keeping track of the
appropriate relative weightings.  Thus, MARMOSET provides a
straightforward means of event generation for a broad variety of
models for which no full Monte Carlo implementation exists.   Development of an
alternative \MadGraph/\MadEvent\ \cite{Maltoni:2002qb} back-end for MARMOSET is ongoing,
and it would be straightforward to modify MARMOSET to interface with other Monte Carlo tools.

For studying single topologies, MARMOSET can generate exclusive
processes that can be used for setting experimental limits on the
overall rates as functions of the physical mass scales involved.
For studying correlated topologies, MARMOSET allows for an
efficiently scanning of production cross sections and branching
ratios even when this scanning demands consistency relations between
the rates for different processes.  This is accomplished by
reweighting Monte Carlo events \emph{after} they are generated.  A
simple analysis technique that makes use of the property will be
discussed in Section \ref{sec:MARMinPractice}.

Currently, MARMOSET simulation of a new physics model is achieved in
three steps:
\begin{enumerate}
\item {\bf To simulate an OSET, a user specifies the masses and $SU(3)$
and $U(1)_{\rm EM}$ quantum numbers of new particles, as well as
their production and decay modes.}  The specified production and
decay modes are sewed together in all possible ways to generate a
list of {\it consistent topologies}.  Each topology is assigned a
weight function $w = \sigma \times \prod_i \mbox{Br}_i$, where
numerical values of $\sigma$ and $\mbox{Br}_i$ will be determined
through later analysis. The $2 \rightarrow 2$ production matrix
elements can be specified using the leading-order parameterization
of Section \ref{sec2:osetdef}.  In the current implementation,
decays always proceed according to phase space.
\item {\bf Monte Carlo events are generated separately for each
topology.}  \StdHEP\ and PGS4-simulated output are currently
supported. This stage is highly parallelizable.  Exclusion limits on
$\sigma \times \prod_i \mbox{Br}_i$ for a given topology can be set
directly using the Monte Carlo for a single process.  For each
topology, Monte Carlo is implemented by creating a special-purpose
branching ratio file in \Pythia\ native format.
\item {\bf The appropriate weights for each topology are uniquely
determined by specifying hadronic cross sections for each production
mode and branching fractions for each decay mode} and the individual
Monte Carlo can be readily combined. Alternatively, each process can
be treated as a template with a constrained rate, and an
optimization technique can be used to quickly find the ``best fit''
values of OSET rate parameters.  A simple version of this approach
is illustrated in Section \ref{sec:MARMinPractice}.
\end{enumerate}
In the following subsections, we describe the input format and give more
 details on the workflow for Monte Carlo generation.

\subsection{Building a Simple OSET}
\label{sec:simpleOSET}
The process of building an OSET input file is best illustrated by example.  A new $Z'$ with a mass of
900 GeV and a width of 50 GeV would be specified as:
\begin{equation}\begin{minipage}{5.5in}\begin{verbatim}
Zprime : charge=0 color=0 mass=900.0 width=50.0
\end{verbatim}\end{minipage}\end{equation}
The variable \verb=charge= specifies $3 Q_{EM}$, and \verb=color= is 0
for a color singlet, 3 ($-3$) for a color (anti-)fundamental,
and $8$ for an adjoint.  Note that the spin of the $Z$ is ignored,
but we plan to add spin-related correlation functionality in future implementations. We
have not yet given MARMOSET production or decay modes for the $Z'$,
which we do as follows:
\begin{equation}\begin{minipage}{5.5in}\begin{verbatim}
u ubar > Zprime
Zprime > e- e+
\end{verbatim}\end{minipage}\end{equation}
The first line defines production of $Z'$ from $u$ and $\bar u$ PDFs
using a Breit-Wigner parametrization, while the second line permits $Z'$
decays to $e^+e^-$. MARMOSET knows about these Standard Model
particles from definitions such as:
\begin{equation}\begin{minipage}{5.5in}\begin{verbatim}
u ubar : pdg=2 charge=2 color=3 mass=0.33
e- e+  : pdg=11 charge=-3 color=0 mass=0.00051
\end{verbatim}\end{minipage}\end{equation}
The \verb=pdg= code is the Particle Data Group's standard Monte Carlo reference number \cite{Yao:2006px}. Note that the \verb=charge=, \verb=color=, and \verb=mass=
specifications are used by MARMOSET to check consistency, but \Pythia\
standard values are used in Monte Carlo generation.

\begin{figure}[tb]
\begin{center}
\begin{minipage}{5.5in}\begin{verbatim}
# Standard Model Particles
u ubar : pdg=2 charge=2 color=3 mass=0.33
e- e+  : pdg=11 charge=-3 color=0 mass=0.00051

# New Particles
Zprime : charge=0 color=0 mass=900.0 width=50.0

# Decay Modes
Zprime > e- e+
Zprime > u ubar

# Production Modes
u ubar > Zprime
\end{verbatim}\end{minipage}
\caption{\label{fig:simpleOSET}A simple MARMOSET input file to generate
$u \bar{u} \rightarrow Z'$, where the $Z'$ can decay to $u \bar{u}$ or $e^+ e^-$.
In the current implementation, the spin of the $Z'$ is ignored.  At the level
of specifying the OSET, the user need not give explicit rate information, because
rates can be chosen after Monte Carlo generation.}
\end{center}
\end{figure}

\begin{figure}
\begin{center}
\begin{minipage}{5.5in}\begin{verbatim}
# Processes
p001 * s1 b1 $ u ubar > (Zprime > e+ e-)
p002 * s1 b2 $ u ubar > (Zprime > u ubar)

# Rates
s1 = 1000 $ Sigma( u ubar > Zprime )
b1 = 0.5 $ Br( Zprime > e+ e- )
b2 = 0.5 $ Br( Zprime > u ubar )
@ b1 b2
\end{verbatim}\end{minipage}
\caption{\label{fig:simpleOSETprocess}The two production topologies
specified by the OSET in Figure~\ref{fig:simpleOSET}.  Monte Carlo for
$u \bar{u} \rightarrow Z' \rightarrow e^+ e^-$ and $u \bar{u} \rightarrow Z' \rightarrow u \bar{u}$
are generated independently and will be combined according the weight formulas on the first two lines.
The cross section and branching ratio default values can be modified by the user,
and the \texttt{@} symbol indicates rates that are constrained to sum to 1.  No relation is assumed between the $u \bar{u} \rightarrow Z'$ total cross section and the $Z' \rightarrow u \bar{u}$ partial width.}
\end{center}
\end{figure}

There is a crucial difference between OSET vertices and Lagrangian
vertices: even though the $Z'$ can be produced by $u \bar{u}$, it
does not necessarily need to decay to $u \bar{u}$. An OSET is
organized with \emph{experimentally resolvable} branching ratios in
mind, so only if the decay channel $Z'\rightarrow u \bar{u}$ is
visible would the user be interested in adding this mode. Adding it
to the OSET is simple using:
\begin{equation}\begin{minipage}{5.5in}\begin{verbatim}
Zprime > u ubar
\end{verbatim}\end{minipage}\end{equation}
The two processes $u \bar u \rightarrow Z' \rightarrow e^+ e^-$
and $u \bar u \rightarrow Z' \rightarrow u \bar u$
will be generated independently, so the OSET input file does not need to
supply any coupling strength or rate information.

Using the complete input file from Figure~\ref{fig:simpleOSET},
MARMOSET constructs two processes \verb=p001= and \verb=p002= shown
in Figure~\ref{fig:simpleOSETprocess}. Between the index
(\verb=pXXX=) and the \verb=$= sign is the rate formula for each
process in terms of the overall cross section \verb=s1= and two
branching ratios \verb=b1= and \verb=b2=. The rate parameters are
listed alongside default values which can be changed by the user.
The \verb=@= symbol indicates which branching ratios are constrained
to sum to 1. Since $\verb=b1= + \verb=b2= = 1$, there are two free
rates, one for each process.  This counting always holds for
resonant production with decays to the Standard Model.  For longer
decay chains, there are typically more processes than free
parameters, and correlations between the process weights are
important.

\subsection{Pair Production and Stable New Particles}

Let us add to our $Z'$ OSET a new process, perhaps to account for an
observed 4 jet $+$ missing $E_T$ excess.  The syntax is given in
Figure~\ref{fig:pairOSET}.  We have added a neutral particle
$\mathit{Ne}$ and a color adjoint particle $\mathit{Adj}$.  The new
particle \verb=Ne= is stable, so we give it a \texttt{pdg} number
1000022, so that it is properly interpreted as a stable, neutral
particle by detector simulators.\footnote{In the current
implementation, there can only be one kind of neutral stable
particle and it must have \texttt{pdg=1000022}.  In a future
release, the user will be able to define arbitrary ``stable''
particles by assigning them non-zero \texttt{pdg} numbers, and it
will be the users responsibility to make sure those particles are
handled properly by Pythia and detector simulators.}  The new
particle \verb=Adj= can be pair produced through $gg\rightarrow
\mathit{Adj}\,\mathit{Adj}$ and has two decay modes
$\mathit{Adj}\rightarrow c\bar{c}\,\mathit{Ne}$ and
$\mathit{Adj}\rightarrow b\bar{b}\,\mathit{Ne}$.

\begin{figure}
\begin{center}
\begin{minipage}{5.5in}\begin{verbatim}
# More Standard Model Particles
g      : pdg=21 charge=0 color=8 mass=0
c cbar : pdg=4 charge=2 color=3 mass=1.5
b bbar : pdg=5 charge=-1 color=3 mass=4.8

# More New Particles
Ne  : pdg=1000022 charge=0 color=0 mass=250.0
Adj : charge=0 color=8 mass=600.0

# More Decay Modes
Adj > c cbar Ne
Adj > b bbar Ne

# More Production Modes
g g > Adj Adj : matrix=1
g g > Adj Adj : matrix=2
\end{verbatim}\end{minipage}
\caption{\label{fig:pairOSET} Additional processes added to the OSET
of Figure~\ref{fig:simpleOSET} to generate colored adjoint pair production.
The adjoint particle $\mathit{Adj}$ decays via two different three-body modes,
each which involves a stable neutral particle $\mathit{Ne}$.  The variable
\texttt{matrix} chooses from among a set of predefined matrix elements, and
in this example, we can reweight Monte Carlo events to get the matrix element
$\Msq = A + B (1+1/X)$ for arbitrary $A$ and $B$.}
\end{center}
\end{figure}

\begin{figure}
\begin{center}
\begin{minipage}{5.5in}\begin{verbatim}
# More Processes
p003 * b3 b3 s2 $ g g > (Adj > Ne bbar b) (Adj > Ne bbar b) : matrix=1
p004 * b3 b4 s2 $ g g > (Adj > Ne bbar b) (Adj > Ne cbar c) : matrix=1
p005 * b4 b3 s2 $ g g > (Adj > Ne cbar c) (Adj > Ne bbar b) : matrix=1
p006 * b4 b4 s2 $ g g > (Adj > Ne cbar c) (Adj > Ne cbar c) : matrix=1
p007 * b3 b3 s3 $ g g > (Adj > Ne bbar b) (Adj > Ne bbar b) : matrix=2
p008 * b3 b4 s3 $ g g > (Adj > Ne bbar b) (Adj > Ne cbar c) : matrix=2
p009 * b4 b3 s3 $ g g > (Adj > Ne cbar c) (Adj > Ne bbar b) : matrix=2
p010 * b4 b4 s3 $ g g > (Adj > Ne cbar c) (Adj > Ne cbar c) : matrix=2

# More Rates
s2 = 1000 $ Sigma( g g > Adj Adj : matrix=1 )
s3 = 1000 $ Sigma( g g > Adj Adj : matrix=2 )
b3 = 0.5 $ Br( Adj > Ne bbar b )
b4 = 0.5 $ Br( Adj > Ne cbar c )
@ b3 b4
\end{verbatim}\end{minipage}
\caption{\label{fig:pairOSETprocess} Eight new processes defined by the
OSET in Figure~\ref{fig:pairOSET}.  The combinatorics of decays are handled
by generating many symmetry-equivalent topologies.  In an experimental environment,
these topologies can either be used independently for dedicated searches or
combined to find best fit cross sections and branching ratios.  Note that
production modes with the same initial and final states but different matrix
elements are treated completely independently.}
\end{center}
\end{figure}

Note that we have specified \emph{two} different matrix elements for the $2
\rightarrow 2$ production of $\mathit{Adj}$ fields.  The variable \verb$matrix=i+10*j$
chooses from a set of predefined matrix elements  $\Msq = f_i(X) \xi^j$, where
\begin{equation*}
f_1(X) = 1, \quad f_2(X) = \left(1 - \frac{1}{X}\right),\quad f_3(X) = \left(1 - \frac{1}{X}\right)^2,
\end{equation*}
\begin{equation}
\quad f_4(X) = \left(X-1\right), \quad f_5(X) = \left(X-1\right)^2.
\end{equation}
We proposed in Section \ref{sec2} that nearly all processes are well modeled by a matrix
element of the form $\Msq = A + B (1-1/X)$.  Therefore in Figure~\ref{fig:pairOSET}, we
have selected two processes, one with $\Msq \propto 1$ (\verb!matrix=1!) and one with
$\Msq \propto (1-1/X)$ (\verb!matrix=2!).  By reweighting each Monte Carlo set from
these two production modes, we can  separately vary the values of $A$ and $B$.

The OSET input file in Figure~\ref{fig:pairOSET} generates 8
distinct new processes as in Figure~\ref{fig:pairOSETprocess}. Four
are symmetry-equivalent to another process, for example \verb=p004=
and \verb=p005=.  For technical reasons, it is much more convenient
to handle combinatorics by generating lots of equivalent topologies
instead of keeping track of numerical
coefficients.\footnote{Currently, separate Monte Carlo files are
generated for symmetry-equivalent processes.   While this limitation
will be improved in a future release, in practice, when two branching
ratios are comparable, one wants to generate events pre-weighted
according to combinatoric factors anyway.}  As before, weight and
rate information is stored separately such that the same OSET input
file can accommodate many different choices of cross sections and
branching rates.

\subsection{Searches Using Single Processes and Measurements Using Many}

Any one of the processes in Figure~\ref{fig:pairOSETprocess} could
be used as signal Monte Carlo for a topology-based dedicated search.
The natural product of such a search would be a limit on the rate
($\sigma \times \mbox{Br}_1 \times \mbox{Br}_2 \times \cdots$) for
that process as a function of the masses of the states involved.
For example, exotics searches at the Tevatron are often presented as
limits on overall rates as a function of masses \cite{CDFExotic,D0Exotic}.   An OSET
topology can be used to place limits on models for which reliable
Monte Carlo tools do not exist or for which the translation between
Lagrangian parameters and mass/rate information is tedious.

Deviations in shape between different theoretical models---or
equivalently, between different hard scattering matrix
elements---introduce systematic uncertainties, which can be
accounted for by placing limits separately on particular production
matrix elements for the same topology.  The residual
parameterization error can compare favorably to the typical 5-10\%
theoretical uncertainties in overall cross sections for a given
assumed model.  In cases where differences in matrix elements affect
experimental cuts or efficiencies, the OSET parametrization can aid
in understanding those systematics.

If a signal is observed, the question of interest will shift from a
limit on the combined product of cross section and branching ratios to a
parameterization of the individual rates that control all observed processes.  Although we
have introduced six symmetry-inequivalent new processes in Figure~\ref{fig:pairOSETprocess},
they are controlled by only four rate parameters \verb=s3=, \verb=s4=, \verb=b3=, and \verb=b4=.
Of these, the two branching ratios are constrained to sum to 1.  Thus, the rates for the six
processes of interest are not independent, and provide a non-trivial verification of the
description!  In Section~\ref{sec:MARMinPractice}, we outline an analysis strategy for
measuring these rates from data.

If we wanted to change the masses of $\mathit{Adj}$ or
$\mathit{Ne}$, we would have to generate a new set of Monte Carlo
events because particle masses affect the kinematics of events.  But
all OSET parameters are not equally expensive to vary---the rates
$\sigma_i$ and $\mathrm{Br}_i$ are freebies.  This is possible in
principle because rate parameters have no effect on kinematics; it
is possible in practice because we have chosen to organize Monte
Carlo one process at a time. This can be of great computational
advantage, especially given the high cost of realistic detector
simulation.  Reweighting after Monte Carlo generation is
accomplished by editing the default rate values in
Figure~\ref{fig:pairOSETprocess}.

\subsection{CP Conjugation and Recycling Monte Carlo}

There are two other features of MARMOSET that make it practical for
real world applications.  The first is automatic handling of CP
conjugate processes.  Unlike the case of $u \bar{u} \rightarrow Z'$
vs. $Z' \rightarrow u \bar{u}$ from Section~\ref{sec:simpleOSET}
where we argued that it was impractical to always include both the
production and decay channel, in most situations one does want to
include CP conjugate processes.  This is because it is difficult to
observe CP-violation at low luminosity at the LHC and CP-conjugate
processes often populate similar final states.  For the same reason,
MARMOSET \emph{does not} assume any kind of flavor universality or
isospin conservation as not only are those symmetries  violated by
most models, but experimental resolutions are very different for,
say, electrons, taus, and neutrinos. Of course, these operating
modes and assumptions can be easily altered.

\begin{figure}
\begin{center}
\begin{minipage}{5.5in}\begin{verbatim}
# New Particles
Adj      : charge=0 color=8 mass=992.0
QU QUbar : charge=2 color=3 mass=700.0
QD       : charge=-1 color=3 mass=700.0
Ne2      : charge=0 color=0 mass=197.0
Ne       : pdg=100022 charge=0 color=0 mass=89.0

# Production Modes
g g    > Adj Adj : matrix=1
u ubar > Adj Ne2 : matrix=1
u ubar > Adj Ne2 : matrix=2

# Adj Decays
Adj > QU ubar
Adj > QD dbar

# Quark Partner Decays
QU > u Ne
QD > d Ne

# Neutral Decays
Ne2 > Ne Z0
\end{verbatim}\end{minipage}
\caption{\label{fig:cpOSET}The OSET input file for the diagrams in Figure~\ref{fig:ex1and2}.
Both \texttt{QU} and \texttt{QD} and complex particles whose CP-conjugate modes are
\texttt{QUbar} and (by default) \texttt{QD\~}.  Conjugate decay modes like
\texttt{QD\~{} > dbar Ne} are assumed.  MARMOSET knows how to efficiently handle
Monte Carlo files, such that if a new decay mode like \texttt{QU > u Ne2} were added,
only the relevant new processes would be created and the pre-existing processes would be reused.}
\end{center}
\end{figure}

Consider the OSET in Figure~\ref{fig:cpOSET}, which generates the
diagrams from Figure~\ref{fig:ex1and2}.  To begin, based on the
study in Section \ref{sec2}, we concluded that only $\Msq \propto 1$
was required to reproduce \verb=Adj= pair-production, but that both
$\Msq \propto 1,(1-1/X)$ were required to reproduce kinematics of
the associated channel, and we have chosen the \verb=matrix=
variables accordingly.

Here, we have also introduced the notation for particles in complex
representations. There are two different ways to specify complex
particles.  We have given \verb=QU= an explicit CP-conjugate state
\verb=QUbar=.   Considering its quantum numbers, \verb=QD= also must
have a conjugate state which has the default name \verb=QD~=.
Because they are in real representations of $U(1)_{\rm EM}$ and
$SU(3)_C$, \verb=Adj= and \verb=Ne= are assumed to be
self-conjugate.  If a conjugate were specified explicitly, it would
be treated in the same way as the conjugates previously mentioned.

Though we have not specified decay modes for \verb=QUbar= or
\verb=QD~=, conjugate decay modes (e.g. \verb=QD~ > dbar Ne=) are
assumed to proceed with the same rate as their conjugate processes.
Likewise, we assume \verb=Adj > QUbar u= proceeds with a rate equal
to \verb=Adj > QU ubar=.  For production modes, both CP-conjugate
and beam reversal symmetries are assumed. For example, for $W'$
production the user need only specify
\begin{equation}\begin{minipage}{5.5in}\begin{verbatim}
u dbar > Wprime+
\end{verbatim}\end{minipage}\end{equation}
and MARMOSET will automatically add in the contributions from
\texttt{dbar u > Wprime+}, \texttt{ubar d > Wprime-}, and \texttt{d
ubar > Wprime-}.  Note that the reported cross section or branching
ratio associated with a production or a decay mode includes the
contribution from the CP-conjugate process.

The second feature of MARMOSET is the automatic reuse of Monte Carlo
files when appropriate. We have already mentioned the computational
efficiency gained by separately generating and simulating Monte
Carlo for each process.  Because these processes are independent,
they need not be explicitly associated with a particular OSET!
Suppose that create a new OSET input file where we vary the mass of
\verb=Ne2=, but not of \verb=Adj=, \verb=QU=, or \verb=QD=.  This
will change the kinematics for the processes involving \verb=Ne2=,
but does not affect the pair-production processes, so the associated
channels must be remade but the pair production channels will be
reused.  In this way, the same process can receive different weights
in different OSETs!  Here, too, we have gained in computational cost
by keeping track of the minimal information required to achieve the
appropriate kinematics for that process.

We could also add new decays (e.g. \verb=QU > u Ne2=) to this OSET,
maintain all Monte Carlo associated with the old decays, and
generate new Monte Carlo only for the newly added processes.  It is
clear that in a complicated model, the number of processes to
generate will grow quite large. But the total number of events that
one must generate does not grow so quickly---the more branching
possibilities there are, the fewer events one expects to populate
each one, and so the total number of events one should generate
grows far more modestly with model complexity.   Finally, for the
fickle user, the names of particles (e.g. \verb=Adj=, \verb=gluino=,
\verb=KKgluon=) in the OSET input file can change, but the old Monte
Carlo files will still be valid.

\subsection{Details of Pythia Interface and Implementation}

The user interface to MARMOSET hides many of the the details of
Monte Carlo generation, which occurs in \Pythia\ \cite{Sjostrand:2006za}.  From the user
point of view, each OSET input file is parsed to generate a list of
processes, weight functions, and rate parameters.  The user can then
decide whether to generate Monte Carlo for one, some, or all
processes, and whether the output should be left in \StdHEP\ \cite{STDHEP} format
or passed through a detector simulation like PGS4 \cite{PGS}.  The user has
full control over how many Monte Carlo events should be generated
for each process.  The resulting Monte Carlo files can be analyzed
individually, and the process weight and rate information can be
used for inclusive analyses.  Details on specific MARMOSET
executables is available at \texttt{http://marmoset-mc.net}.

The important behind-the-scenes work occurs within each process.
Associated with each process number (e.g.  \verb=p001=) is a string
that fully specifies the topology and dynamics of the process
(\verb=pXXX.strg=) and the minimal OSET fragment required to
reproduce the process in question (\verb=pXXX.model=).  From this
information, three files are generated for use by Pythia
executables:
\begin{enumerate}
\item \texttt{pXXX.card} contains a list of Pythia variable settings
necessary for MARMOSET execution.  These include radiation settings,
initialization of SUSY (for the validity of particle code
\texttt{1000022}), and selection of one of the three new
MARMOSET-specific processes: \verb!MSUB(481)=1! ($2 \rightarrow 1$
production), \verb!MSUB(482)=1! ($2 \rightarrow 2$ production), or
\verb!MSUB(483)=1! ($2 \rightarrow 3$ production).
\item \texttt{pXXX.proc} contains MARMOSET-specific information for controlling
 processes \texttt{481}, \texttt{482}, and \texttt{483}.  This includes the beam
  types, hard scattering final states, and selected matrix element parametrization.
\item \texttt{pXXX.brtb} is a decay table in \Pythia-native format which gives
the decay topology after the hard scattering.  In order to avoid conflicts with
existing Monte Carlo tools, PDG numbers between 6000000 and 6009999 are used to
label the decaying particles.
\end{enumerate}

Processes \texttt{481}, \texttt{482}, and \texttt{483} are
implemented in a new \Pythia\ function \verb!PYSGGE!, which is
called by extended \verb!PYSCAT! and \verb!PYSIGH! subroutines
modified for MARMOSET.  Compared to existing \Pythia\ processes,
\texttt{48X} are unique because the beam types, final state
particles, color flow information, and cross sections are all
decided at runtime.  As long as the \texttt{pXXX.proc} information
is provided, though, \texttt{48X} could also be used in ordinary
\Pythia-based analyses independent of MARMOSET.

Decays are handled through the \Pythia-native subroutine
\verb!PYUPDA!, which reads in decay tables and allows definitions of
arbitrary new particles.  To specify different \emph{unique} decay
modes for each particle at different vertices, every decay node is
treated as a distinct particle.  For example, in \verb=p004= from
Figure~\ref{fig:pairOSETprocess}, the two \verb=Adj= particles are
treated as two different particles, one with a 100\% branching ratio
to $c \bar{c}\,\mathit{Ne}$ and the other with a 100\% branching
ratio to $b \bar{b}\, \mathit{Ne}$.

The current \Pythia\ implementation of MARMOSET is fast, flexible
and simple to use, but generalization to generators like \MadGraph\
would allow MARMOSET to leverage full spin and vertex information
with OSETs, providing the possibility of using variable detail in
Monte Carlo generation. Just as all \Pythia\ input files are
generated by MARMOSET's C++ executables, the input files for
\MadGraph\ can be similarly constructed with minimal modification of
the MARMOSET input format to include extra topology and spin
information.  Work in this direction with Johan Alwall is ongoing.

\subsection{\Bard: A Related Tool and Analysis Strategy} \label{sec:UGfuture}

Solving the ``LHC Inverse Problem'' will undoubtedly be a challenging
task, requiring a variety of tools for initial characterization of
data.  These include both event generation and signature analysis from
the top-down, based on a full, theoretically motivated model at the
TeV scale, and bottom-up approaches that seek to piece
together partial descriptions of a model based on observed
discrepancies in data, and build models with these descriptions in mind.

MARMOSET is focused on topology and rate characterizations, trading 
model realism for model flexibility.  
Alternatively, one can start from an effective
Lagrangian based description and systematically search through
Lagrangian realizations of different classes of topologies in an
effort to {\it construct} and constrain pieces of the underlying
Lagrangian. This is the organizing principle of \Bard\ \cite{Knuteson:2005ev,Knuteson:2006ha}.

Before focussing on differences, it is important to stress the
similarities in the two approaches.   They are both bottom-up, meant
to describe significant anomalies in data, and thus are
signature-based. They are not geared towards a certain type of
beyond the standard model physics. They both separate rates from
mass scales and kinematic features, which is particularly useful at
hadron colliders where K-factors, ``invisible'' decay modes, or
detector cuts can skew the naive cross section estimate. Both tools aim
 to explain the data with a certain level of coarseness,
 focusing on a simple characterization of the most significant excesses.

The main difference is in the approach toward explicitly dealing with many channels of
signal. \Bard\ is final-state based, which means that \Bard\ attacks discrepancies in a serial mode.
The final state with the most significant discrepancy in $\sum
p_T$ from the Standard Model expectation is treated first, 
followed by the next, and so on, using information
gained from treating the previous discrepancy.
MARMOSET is vertex (blob)-based, explicitly keeping track of branching 
fractions.  This means that MARMOSET is 
most effective in understand correlated excesses, 
because a given set of vertices will populate many final states.

There is also a difference in the way the two programs interact with the user.
In \Bard, stories are derived algorithmically with
a computer tool.  In MARMOSET, the user selects which vertex structure are of most interest, relying on intuition and experience to guide the creation of candidate OSETs.
These are complementary approaches which can be viewed as
optimizations of two extremes.  On one hand, there may be too many
signals and ideas to explain at once.  In this case, an
inspired guess with MARMOSET may lead to the first conceptual
understanding of the data.  When there are few channels, an
exhaustive search of possible explanations through \Bard\ is likely
to find the right answer first.

\Bard\ is currently optimized to solve the Inverse Problem in
scenarios with fairly simple event topologies, most relevant to the
Tevatron.  The \Sleuth\ \cite{Abbott:2000fb,Abbott:2000gx,Knuteson:2001dq,Knuteson:2000xg,Knuteson:2001qs,Aktas:2004pz} algorithm isolates statistically
signficant anomalies in a particular final state, for example on the
tail of the distribution of the total summed scalar transverse
momentum in the event.  Then, \Bard\ begins by exhaustively listing reasonable possibilities of
particles and interactions that fit the final state, using
\MadGraph~\cite{Stelzer:1994ta} to systematically generate
all renormalizable diagrams entailed by these new terms. The resulting diagrams are
partitioned into stories, collections of diagrams in which the
existence of any single diagram in the story implies the existence
of the others.  Depending on the final state, \Bard\ will generate
between a few and a hundred stories as potential explanations
for the observed discrepancy. Each story introduces several new
parameters. These parameters are the masses and widths of the
introduced particles, and the couplings at each vertex.

\Bard\ passes the new Lagrangian terms ${\cal L}_{\text{new}}$ to
\Quaero\ \cite{Abazov:2001ny,Knuteson:2003dn,Knuteson:2003rq,Caron:2006fg}, which has been prepared with the interesting subset of the
data highlighted by \Vista\ \cite{Knuteson:2004nj,Knuteson:2005ev,Knuteson:2006ua} or \Sleuth. \Quaero\ uses \MadEvent\ to
integrate the squared amplitude over the available phase space and
to generate representative events, and uses
\Pythia~\cite{Sjostrand:2006za} for the showering and
fragmentation of these events. \Quaero\ performs the analysis,
numerically integrating over systematic errors, returning as output
$\log_{10}{\cal L}$, where ${\cal L} = p({\cal D}|{\cal H})/p({\cal
D}|\text{SM})$ is a likelihood ratio, representing the probability
of observing the data ${\cal D}$ assuming the hypothesis ${\cal H}$
divided by the probability of observing the data ${\cal D}$ assuming
the Standard Model alone.  The region in the parameter space of the
story that maximizes $\log_{10}{\cal L}$ is determined, providing
also an error estimate on the parameter values.  Repeating this
process in parallel for each story enables an ordering of the
stories according to decreasing goodness of  fit to the data.

While the ultimate goal is to construct Lagrangian terms, the
approach of first searching through the space of Lagrangians may not
be numerically practical for the possible situation where many new
physics channels open up simultaneously. This will likely be the operating environment at the LHC, where the drastic increase in energy and luminosity compared to the Tevatron means that 100--10,000 clean new physics events in the first year is not an unreasonable expectation.  Speed and efficiency are therefore important
considerations, as the number of event topologies consistent with the observed final states multiplied by the number of parameter choices consistent with those topologies could overwhelm computational resources.  When the new physics signal is clearly visible over background, MARMOSET can help find the basin of attraction for a new physics model by allowing the user to quickly validate OSET hypotheses without any more computational overhead than traditional analyses.

%% file: MARMinPractice.tex
\section{Building OSETs from Data --- An Example}
\label{sec:MARMinPractice} In Section~\ref{sec2}, we have shown that
the OSET parameterization for new physics---in terms of particle
masses, cross-sections, and branching ratios, with a highly
simplified, approximate structure at each vertex---is sufficient to
determine reconstructed object multiplicities and many kinematic
distributions.  The success of this formulation motivates the
MARMOSET Monte Carlo tool of Section~\ref{sec:tool}. We claim that
the circle can be completed---that this simple kinematic and
multiplicity data points to a
unique consistent OSET or small collection of them.  
OSETs are therefore a useful intermediate characterization of discrepancies in
collider data.

The techniques illustrated in this section can be viewed as a very
simple-minded use of the template method. A template approach is
naturally compatible with the additive structure of OSETs. Moreover,
to keep our discussion simple, we restrict attention to a few
kinematic and object multiplicity distributions. More modern
techniques can and should also be used for data analysis with
OSETs, but the effectiveness of Mass and Rate Modeling (MARM)
despite its apparent simplicity underscores a crucial point: the
challenge of interpreting LHC data is not only in finding better
metrics for distinguishing very similar hypotheses but also in
generating and evaluating a large number of very different
hypotheses.

To illustrate how this approach could work in practice, we return to
the ``Michigan Black Box'' discussed in the introduction, defined in
Figure~\ref{fig:MichSpectrum}.  Our goal is to find an accurate OSET
characterization of the signal from this black box--because it
\emph{is} a new-physics signal, we should be able to. We will then
ask: how unique is this characterization? We will consider several
variations and argue that a combination of experimental and
theoretical arguments favor the OSET corresponding to the ``true''
physics.  We will also illustrate how OSET-level data analysis sheds
light on the qualitative structure and mass hierarchy of the
underlying SUSY model.

In this study, we have made necessary compromises between realism
and practicality.  We have used a detector simulator---PGS3 \cite{PGS}, created
by John Conway, with local modifications---that mimics some of the
limitations of real detection (finite energy resolution, object
mis-identification, and $b$-tagging efficiency, for example) but
does not fully represent any LHC detector, and omits certain
important effects.  Similarly, we have not generated Monte Carlo for
the Standard Model, but we will impose cuts on the signal (based on
a typical analysis for low-mass SUSY in the CMS Physics TDR \cite{CMSTDR2})
estimated to reduce the background to be much smaller than the
signal.  This is discussed in Section~\ref{sec:sigisolation}.
Further general study with full treatment of Standard Model
background, a representative detector simulation, and inclusion of
important systematic effects is warranted.

In Section \ref{sec:goodoset}, we show that a seemingly complicated
model can be represented effectively by a simple OSET with a small
number of production and decay channels.   With 1 fb$^{-1}$ of
signal Monte Carlo, we find that the OSET equivalent to the correct
Lagrangian is consistent with the object multiplicities and simple
kinematics of the data. This motivates us to consider several SUSY
models that are consistent with the OSET and makes clear what
classes of models are inconsistent.  For purposes of presentation,
the choice of initial OSET is presented as a ``lucky guess,'' though
originally the OSET considered was deduced through blind analysis.

Through the next two sections, we backtrack and consider plausible
alternative OSETs.  In Section~\ref{sec:OSETmasses}, we demonstrate
that for OSETs with the correct topology, the masses of the new
species are resolvable by eye to within $\sim 100$ GeV.  We then
consider a number of alternative OSETs with different spectra in
Section~\ref{sec:falsestarts}, and in which decays proceed through
different topologies. Though these new-physics events have different
kinematics, in some cases they can look approximately consistent
with the data in the crude kinematic variables we consider.  We
emphasize that these ambiguities are real, and that it is essential
to approach them both by seeking new experimental discriminators and
by considering indirect, theoretical consequences.  We contrast the
problem of ``degeneracies'' we encounter when working with OSETs to
that present working within a specific model framework like the
MSSM.

Finally, in Section~\ref{sec:tailprocess} we note a discrepancy at
high $H_T$ that is not accounted for by the OSET we have developed
so far at 1 fb$^{-1}$.  Working with 5 fb$^{-1}$ of signal Monte
Carlo, we outline an argument within the OSET framework, that this
new physics cannot be attributed simply to an error in the matrix
element parameterization for the dominant process. Instead, it must
be described by a genuinely new production process, which in turn
sheds light on the underlying physics model.

\subsection{Signal Isolation}
\label{sec:sigisolation}
In the analysis that follows, we will work with new-physics signal
Monte Carlo only, corresponding to 1 fb$^{-1}$ of integrated
luminosity.  We have not generated Monte Carlo for the relevant
Standard Model backgrounds, and this omission will artificially
improve the precision of our results.  To offer some assurance that
proper treatment of Standard Model backgrounds would not be
catastrophic to our approach, we must exhibit a set of cuts that
achieves signal purity $S/B \gg 1$, and does not sculpt the signal
beyond hope of interpretation.  We will work with only the signal
events that pass these cuts.  We approach this task conservatively,
using a benchmark set of cuts modeled on the CMS Physics TDR \cite{CMSTDR2} for which the Standard Model background has been
estimated.

We have simulated the signal events for the ``Michigan Black Box''
in PGS 3.  We apply triggers to fully reconstructed PGS events that
are modeled closely on a subset of CMS high-level triggers (HLT) for
low-luminosity running found in the CMS Physics TDR Table E.12
\cite{CMSTDR2}.  We have used hadronic, jet$+E_T^{\rm miss}$,
and single-lepton trigger streams---our trigger levels are
summarized in Table \ref{table:triggers}.  We apply cuts (again
based on the CMS Physics TDR, Table 4.2 and optimized for QCD
rejection in a low-mass SUSY scenario \cite{CMSTDR2}) on $H_T$,
jet $E_T$'s, and $E_T^{\rm miss}$, and vetoed events with $E_T^{\rm
miss}$ collinear with a jet.  These cuts are summarized in Table
\ref{table:susylmcuts}.  We have \emph{not} imposed the cuts used in
the CMS analysis to reduce electroweak background, but assume an
$\mathcal{O}(1)$ acceptance.
\begin{table}
\begin{center}
\begin{tabular}{|c|c|}
\hline
\textbf{Trigger} & \textbf{Reco. Obj. $p_T$ Thres. (GeV)} \\
\hline
Inclusive $e/\mu$ & 30 \\
\hline
Double-jet & 350 \\
\hline
Triple-jet & 195 \\
\hline
Four-jet & 80 \\
\hline
Jet $+ E_T^{\rm miss}$ & 180, 80 \\
\hline
Two-jet $+ E_T^{\rm miss}$ & 155, 80 \\
\hline
Three-jet $+ E_T^{\rm miss}$ & 85, 80 \\
\hline
Four-jet $+ E_T^{\rm miss}$ & 35, 80 \\
\hline
\end{tabular}
\caption{A simplified trigger table based on CMS Physics TDR Table
E.12, a high-level trigger menu for $\mathcal{L}=2 \times 10^{33}
\mbox{cm}^{-2} \mbox{s}^{-1}$.  Thresholds are evaluated on fully
reconstructed objects in PGS3, and $E_T^{\rm miss}$ is simply the
vector sum of transverse momenta of reconstructed objects. The 30
GeV $e/\mu$ trigger is used in place of separate $e$ (26 GeV) and
$\mu$ (19 GeV) triggers in the CMS table; in $n$-jet + $E_T^{\rm
miss}$ states, the first threshold applies to the $n$'th jet and the
second to $E_T^{\rm miss}$. Events that pass \emph{any one} of these
triggers are considered in the analysis.}\label{table:triggers}
\end{center}
\end{table}
\begin{table}
\begin{center}
\begin{tabular}{|l|l|}
\hline
\textbf{Requirement} & \textbf{Purpose} \\
\hline
Pass Trigger in Table \ref{table:triggers}& \\
$E_T^{\rm miss}>200$ GeV & Signal Signature\\
$N_{jet} \geq 3$, $|\eta_d(j_1)| \leq 1.7$ & Signal Signature\\
\hline
$p_T(j_1) > 180$ GeV / $p_T(j_2) > 110$ GeV & Signal/Background Optimization$^*$\\
$p_T(j_3) > 20$ GeV & \\
$H_T^* \equiv p_T(j_2) + p_T(j_3) + p_T(j_4) + E_T^{\rm miss} > 500$ GeV & \\
\hline
$\delta\phi_i \equiv \delta\phi(j_i, E_T^{\rm miss}) < 0.3$ rad ($i=1, 3$)
& QCD Rejection\\
$\delta\phi_2 < 20^\circ$ & \\ $R_1 \equiv
\sqrt{\delta\phi_2^2 + (\pi - \delta\phi_1)^2} < 0.5$ & \\
$R_2 \equiv \sqrt{\delta\phi_1^2 + (\pi - \delta\phi_2)^2} < 0.5$ & \\
\hline
\end{tabular}
\caption{Cuts (a subset of the cuts listed in CMS Physics TDR for
low-mass SUSY analysis, Table 4.2) used to reduce background
contamination.  We implement these cuts on PGS 3 reconstructed
events, and compare signal rates to background rate estimates from
the CMS Physics TDR.  We do not include the cuts intended to reduce
electroweak backgrounds, but assume that these preserve an
$\mathcal{O}(1)$ fraction of the signal.  Note that
signal/background optimization was performed for a  benchmark
scenario quite different from the signal we consider; with
appropriately optimized cuts, signal efficiency and purity could
probably be significantly improved.}\label{table:susylmcuts}
\end{center}
\end{table}
These cuts retain the ``Black Box'' signal with $\sim 7\%$ efficiency
(1600 events in 1 fb$^{-1}$).  The TDR estimates a background, after
cuts, of roughly 250 events (107 QCD, 137 from $Z+\mbox{jets}$, $W+\mbox{jets}$, and
$t\bar t$).  Though this estimate is not directly comparable with our
PGS simulation because of differences in detector modeling and
reconstruction, we will take it as evidence that the signal is not
hopelessly lost.  We emphasize that these cuts are not optimized for
the topology and low mass scale that dominates this signal; most
events have $\geq 5$ jets, so it is likely that allowing events with
more jets but lower $E_T^{\rm miss}, H_T^*$ than the TDR analysis could
significantly increase signal efficiency and/or purity.

In the analysis that follows, we will work with only the new-physics
signal that passes the cuts of Table \ref{table:susylmcuts}. The use
of signal only will unrealistically improve the precision of the
results that follow, but the qualitative features that the analysis
relies on appear easily resolvable even with order-1 uncertainties
on a background $\sim 200$ events.  Further study with background
uncertainties included and a representative detector simulation is
warranted.
\begin{figure}[tbp]
\begin{center}
\includegraphics[width=3in]{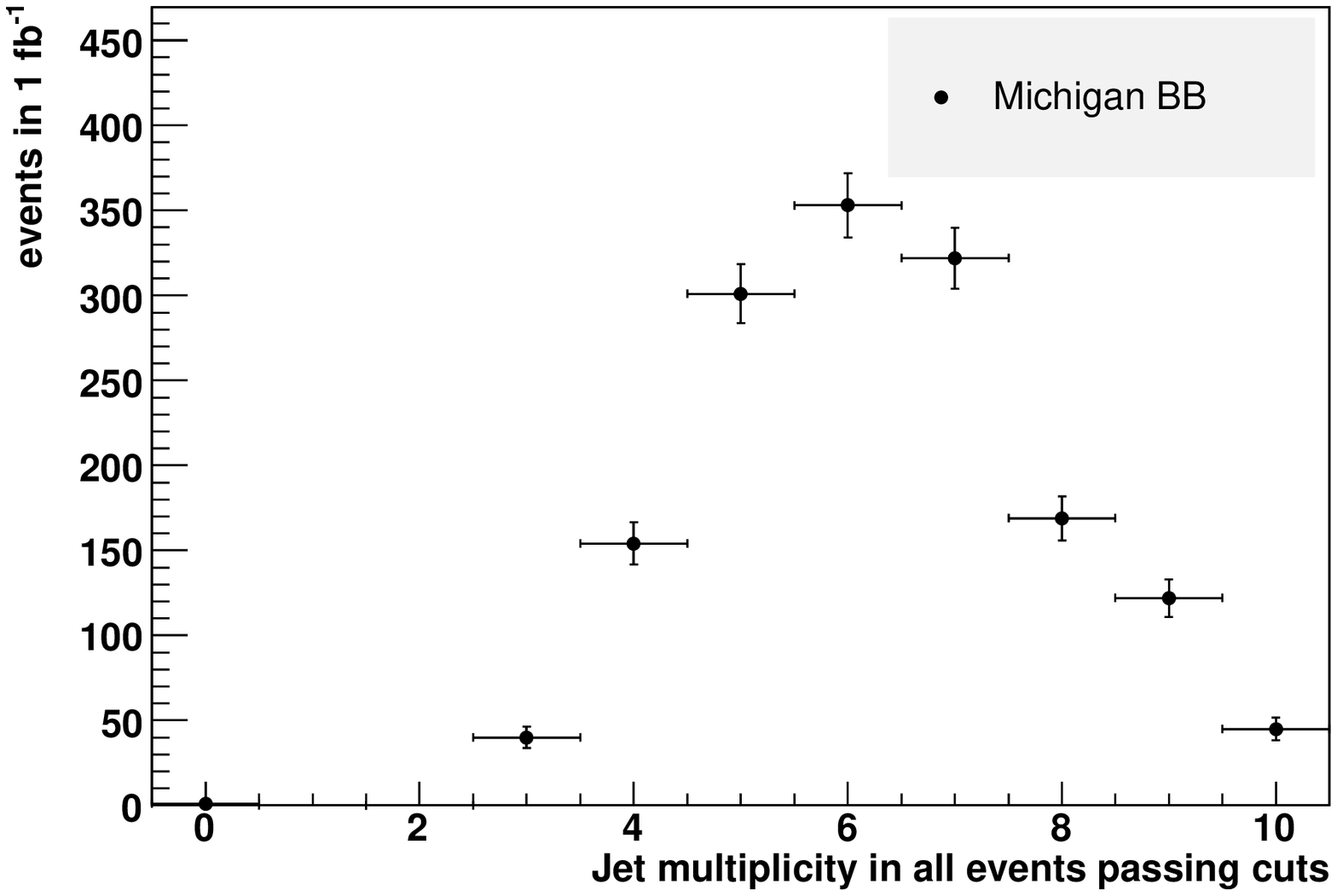}
\includegraphics[width=3in]{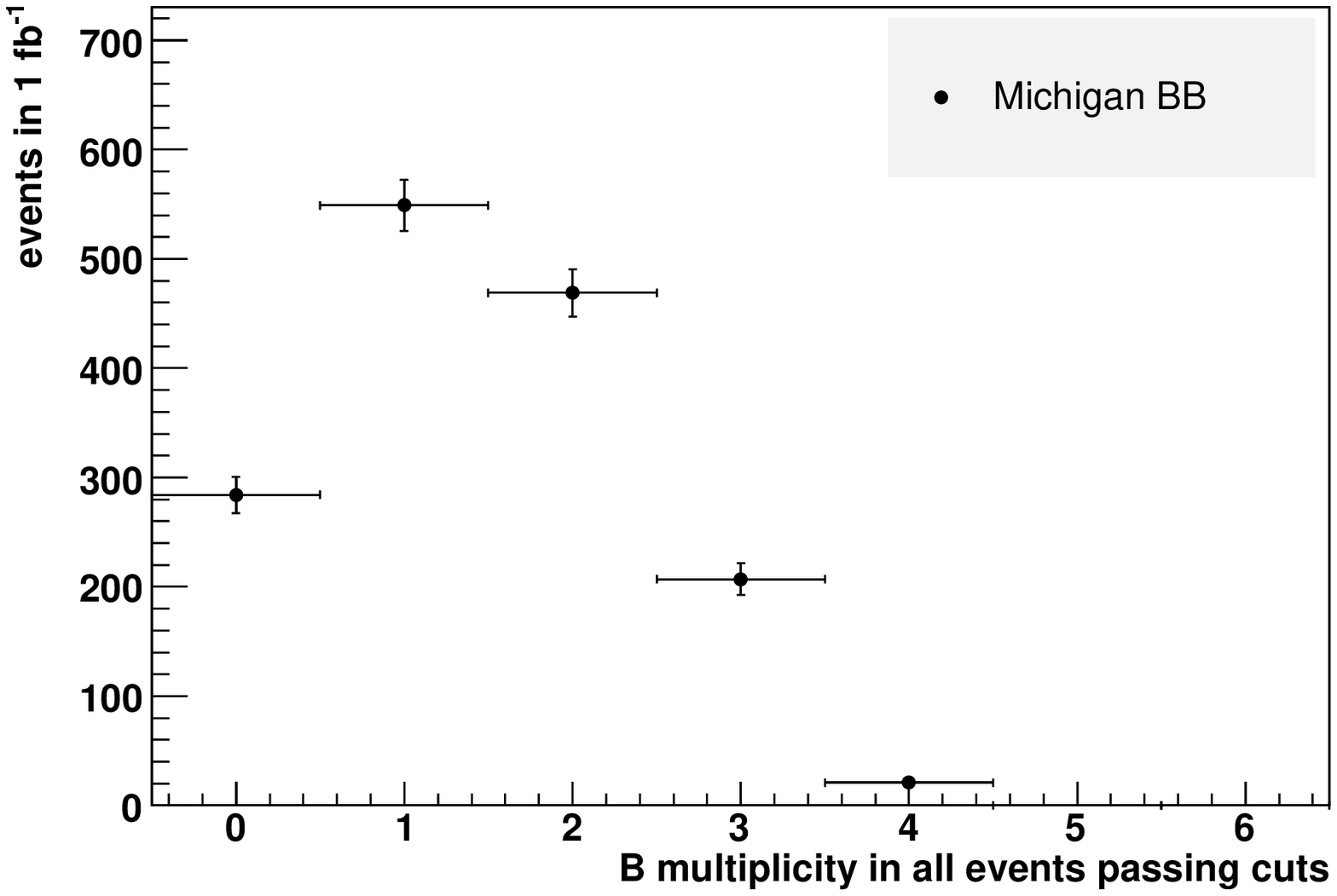}
\caption{\label{fig:michCounts}Jet (left) and $b$-tagged jet (right)
multiplicities in ``Michigan Black Box'' signal after analysis cuts
have been applied. Jets with reconstructed $p_T > 50$ GeV are counted.}
\end{center}
\end{figure}
\begin{figure}[tbp]
\begin{center}
\includegraphics[width=3in]{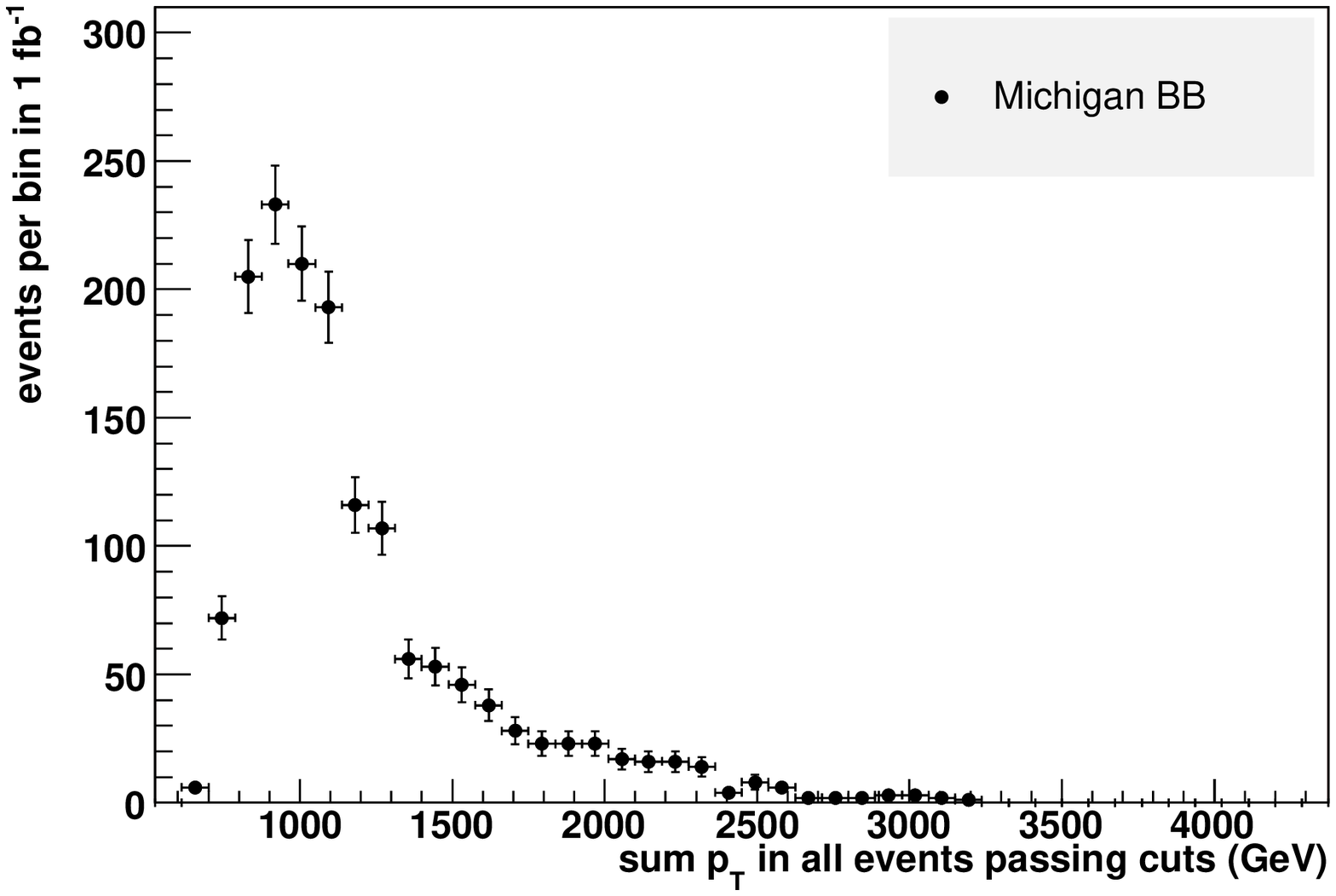}
\includegraphics[width=3in]{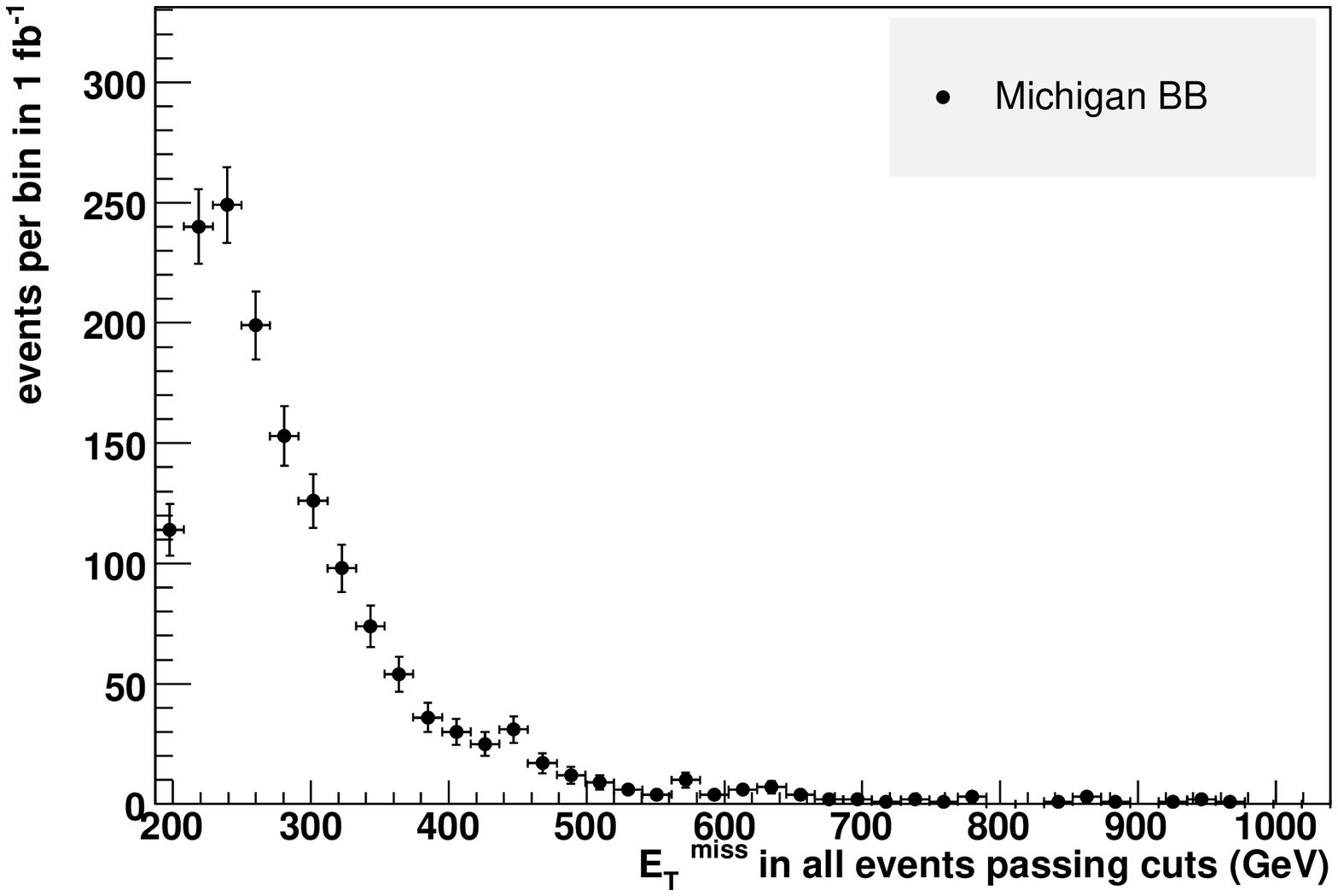}

\includegraphics[width=3in]{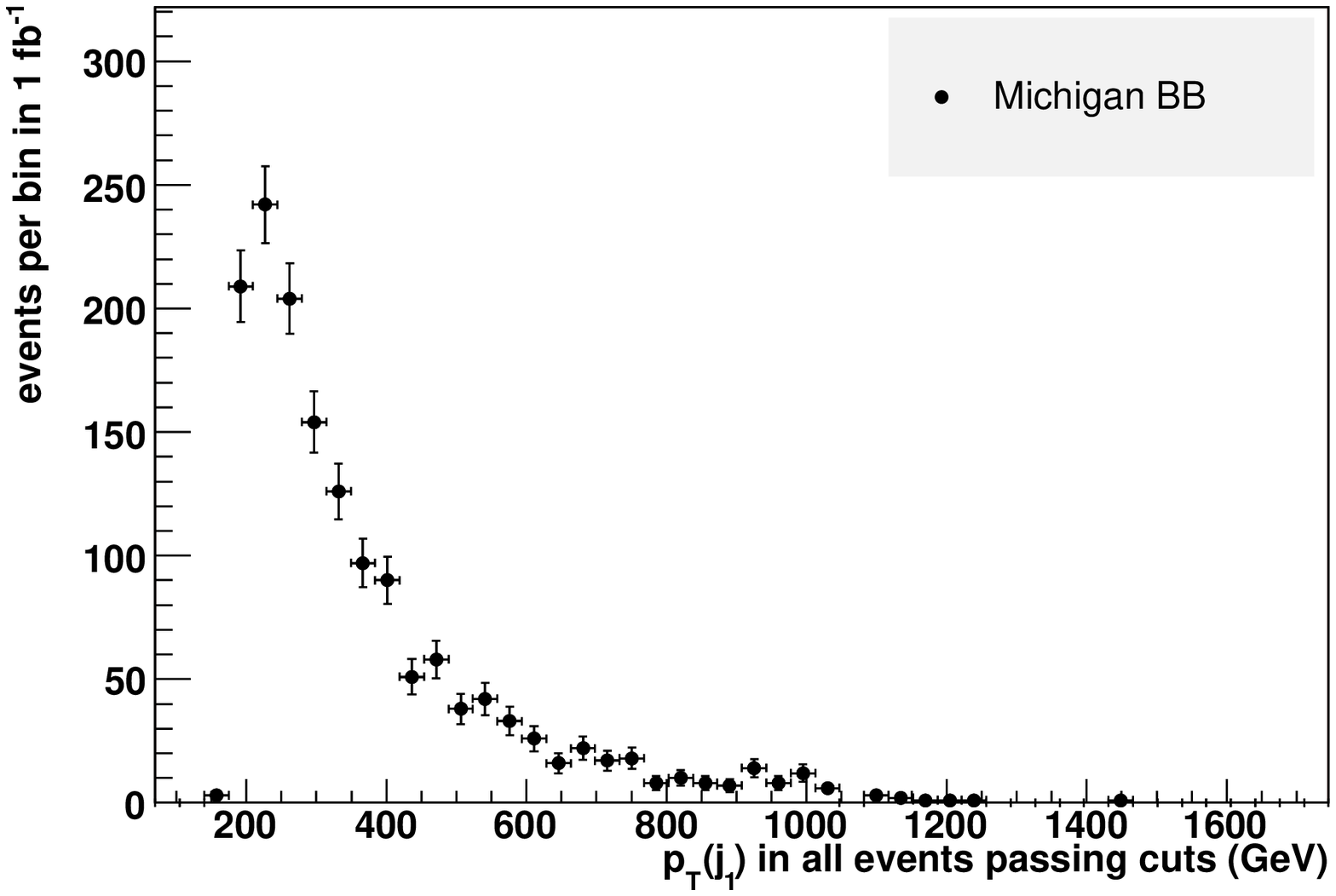}
\includegraphics[width=3in]{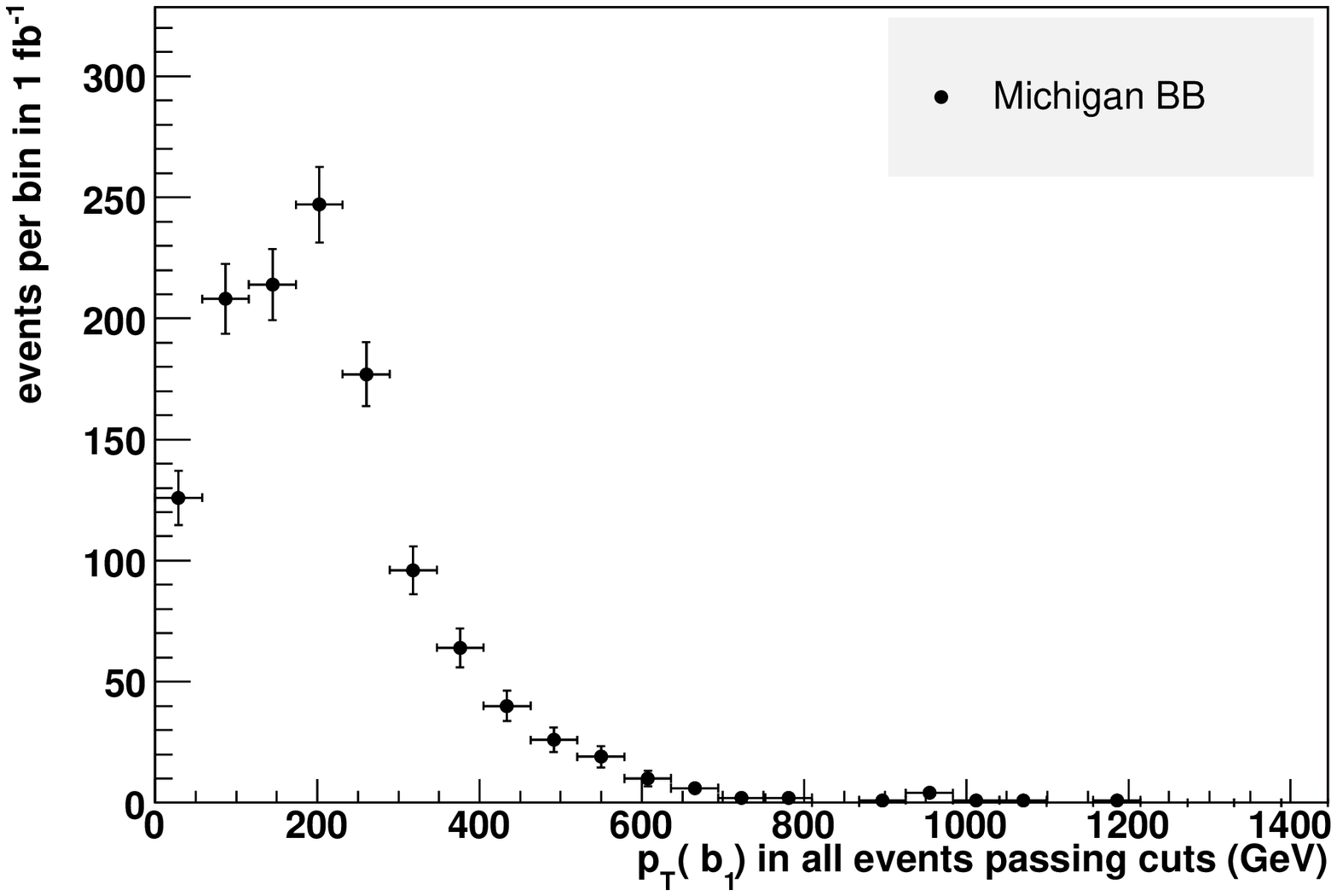}
\caption{\label{fig:michPlots}Kinematic distribution in the ``Michigan
Black Box'' signal after analysis cuts have been applied.  Top: scalar
sum of all object $p_T$'s (left), $E_T^{\rm miss}$ (right).  Bottom:
$p_T$ of the hardest jet (left) and hardest $b$-tagged jet (right).
For comparison, CMS Physics TDR \cite{CMSTDR2} estimates suggest $\sim
250$ Standard Model background events pass cuts analogous to those of
Table \ref{table:susylmcuts} and additional cuts to reduce $W/Z+\mbox{jets}$
and $t \bar{t}$ backgrounds with leptonically decaying bosons.}
\end{center}
\end{figure}

\subsection{Building an OSET for the Signal}
\label{sec:goodoset}
Signal-only distributions in $H_T \equiv \sum p_T$, $E_T^{\rm miss}$,
and $p_T(j_1)$ are shown in Figure \ref{fig:michPlots} below, as are
histograms of reconstructed jet and $b$-tagged jet counts (for a cone
size $\Delta R = 0.7$) in Figure \ref{fig:michCounts}.  These
distributions and all that follow are after application of the cuts of
Table \ref{table:susylmcuts}.  The signal we have found is dominantly
hadronic with appreciable missing energy; high $b$-jet multiplicities
suggest that third-generation partners play a special role.  We can
try to make more precise claims about the partonic final state of the
signal events, but must do so with care---multiple processes are
involved, and tagging efficiencies are momentum-dependent.  We will
restrict our attention here to new-physics interpretations with
SUSY-like decay topologies, where $E_T^{\rm miss}$ arises from a pair
of stable, neutral decay products.  It is also prudent to evaluate the
consistency of the observed excess with topologies with no new
invisible products, where $E_T^{\rm miss}$ is the result of jet
mis-measurement or comes from neutrinos.  Note that our definition of
$H_T$ as the sum of $p_T$'s of all reconstructed objects and missing
energies differs from the definition (here $H_T^*$) used in the CMS
Physics TDR and in our cuts.

\begin{figure}[tbp]
\begin{center}
\includegraphics[width=4.5in]{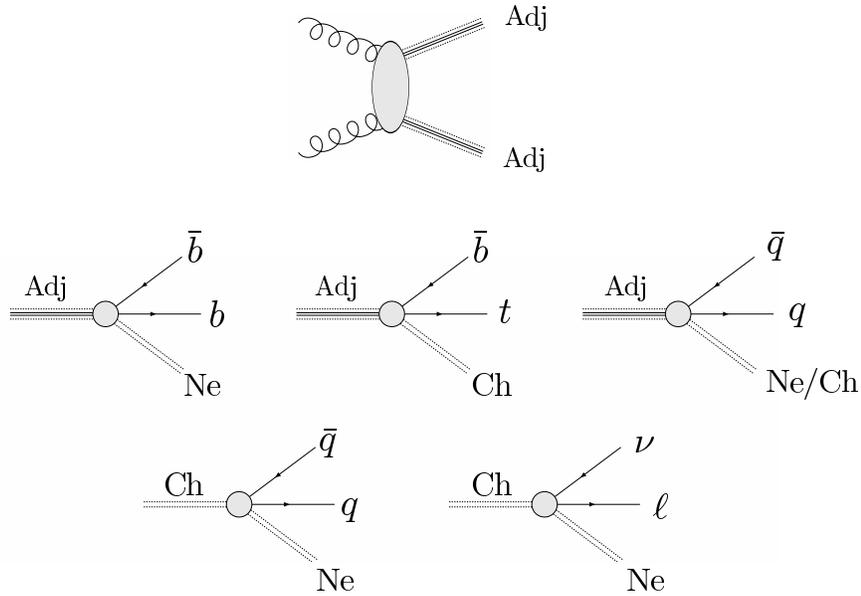}
\caption{\label{fig:OSETDiagrams}Diagrams in an OSET containing three
  new particles: a heavy SU(3) adjoint {\it Adj}, and lighter charged
  and neutral states {\it Ch} and {\it Ne}.  We have included only the
  dominant visible production mode, i.e. adjoint pair-production.
  These processes are generated by the MARMOSET input file is given in
  Figure \ref{fig:OSETInput}.}
\end{center}
\end{figure}

\begin{figure}
\begin{center}
\includegraphics[width=1.5in]{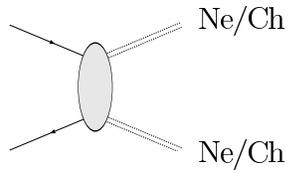}
\caption{This electroweak production process is quite reasonable in
  the context of the OSET of Figure \ref{fig:OSETDiagrams}.  We do not
  include it here because the trigger rate in the trigger streams we
  consider is negligible; therefore, our analysis does not constrain this
  production mode.  Also, even if we included the relevant trigger streams, Standard Model electroweak backgrounds would make it difficult to observe this channel.}
\label{fig:michEW}
\end{center}
\end{figure}

Let us begin by checking that the OSET we would guess from the correct
Lagrangian is indeed successful.  We assume the correct
spectrum---a new adjoint ($m_{\it Adj}=450 \mbox{ GeV}$) a stable neutral particle
($m_{\it Ne}=124 \mbox{ GeV}$) to which it decays, and a charged state nearly degenerate
with the neutral state (the hypothesis of near-degeneracy could be
disproved by the presence of extra hard leptons in the decays). We
will focus on the graphs of Figure \ref{fig:OSETDiagrams}, ignoring the electroweak processes in Figure \ref{fig:michEW} because of their low trigger rate and high background contamination.  A MARMOSET
input file for this OSET is shown in Figure \ref{fig:OSETInput}.  For
these masses, $t \bar t {\it Ne}$ decays are kinematically forbidden, and we
will not include them in our analysis.

\begin{figure}[tbp]
\begin{center}\begin{minipage}{5.5in}\begin{verbatim}
# Standard Model Particles
d dbar : pdg=1 charge=-1 color=3 mass=0.33
u ubar : pdg=2 charge=2 color=3 mass=0.33
b bbar : pdg=5 charge=-1 color=3 mass=4.8
t tbar : pdg=6 charge=2  color=3 mass=175.0
g      : pdg=21 charge=0 color=8 mass=0
e-   e+      : pdg=11 charge=-3 color=0 mass=0.00051
nu_e nu_ebar : pdg=12 charge=0  color=0 mass=0.0

# New Particles
Adj    : charge=0 color=8 mass=450.0
Ne     : pdg=1000022 charge=0 color=0 mass=124.0
Ch Ch~ : charge=3 color=0 mass=130.0

# Decay Modes
Adj > b bbar Ne
Adj > t bbar Ch~
Adj > u ubar Ne
Adj > u dbar Ch~
Ch  > u dbar Ne
Ch  > e+ nu_e Ne

# Production Modes
g g > Adj Adj : matrix=1

# Invisible Production Modes (not included)
# u ubar > Ch Ch~ : matrix=1
# u dbar > Ch Ne  : matrix=1
\end{verbatim}\end{minipage}\end{center}
\caption{\label{fig:OSETInput} MARMOSET input file for the OSET in
Figure \ref{fig:OSETDiagrams}.  For a first pass, we have chosen flat
matrix elements for all production modes.  Because leptons from
\texttt{Ch} decays will be soft, we have only included electrons in
the \texttt{Ch} decays to avoid a proliferation of non-essential
processes.  Note that in the context of this OSET, the decay
\texttt{Adj > t tbar Ne} is kinematically forbidden and therefore not
included.  From this file, MARMOSET creates 10 \Pythia\ processes.  At
the bottom of the file, the associated chargino and neutralino
production modes have been commented out; these processes (see Figure
\ref{fig:michEW}) do not trigger at an appreciable rate in the trigger
streams we consider.}
\end{figure}
\begin{figure}[tbp]
\begin{center}
\includegraphics[width=3in]{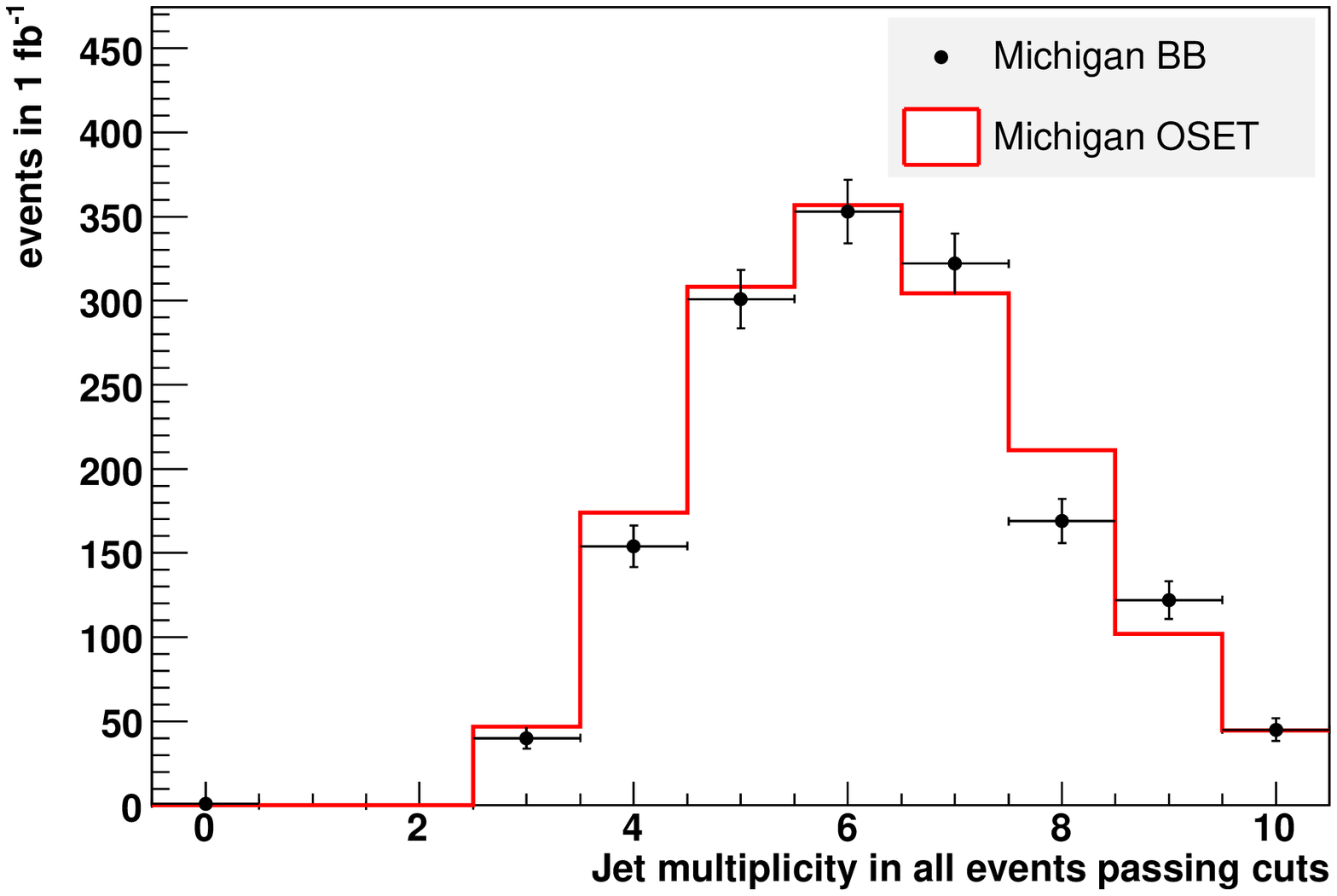}
\includegraphics[width=3in]{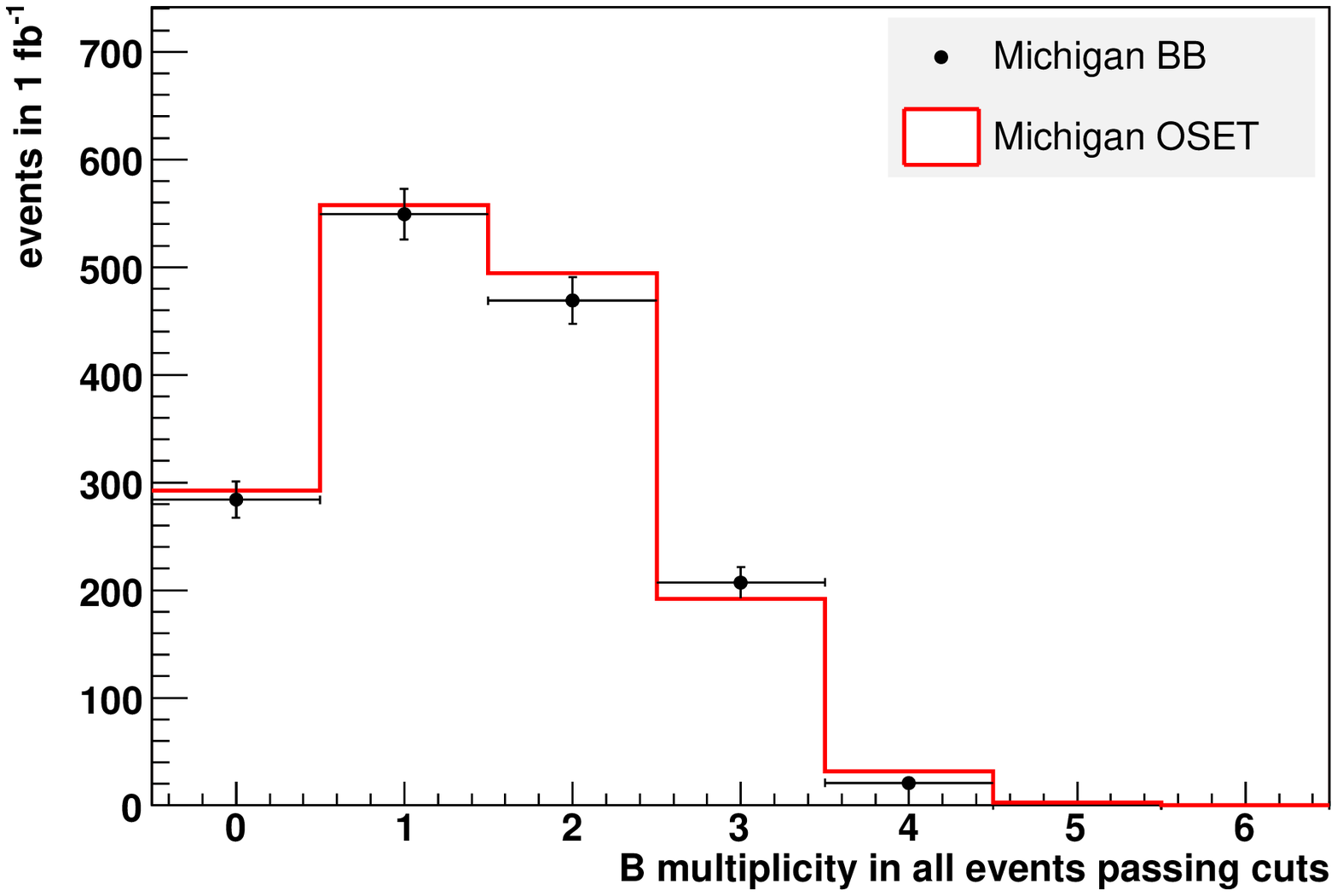}

\includegraphics[width=3in]{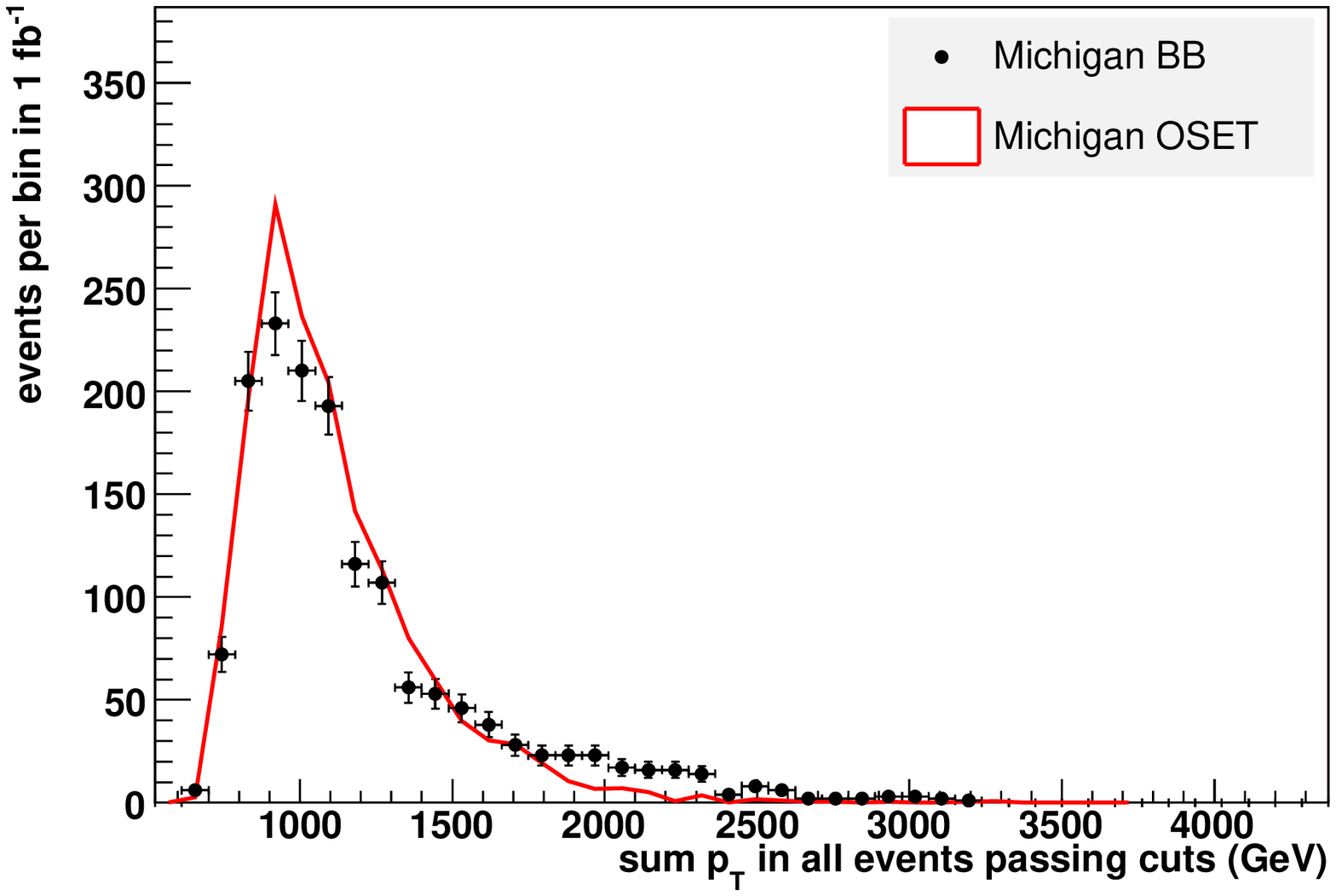}
\includegraphics[width=3in]{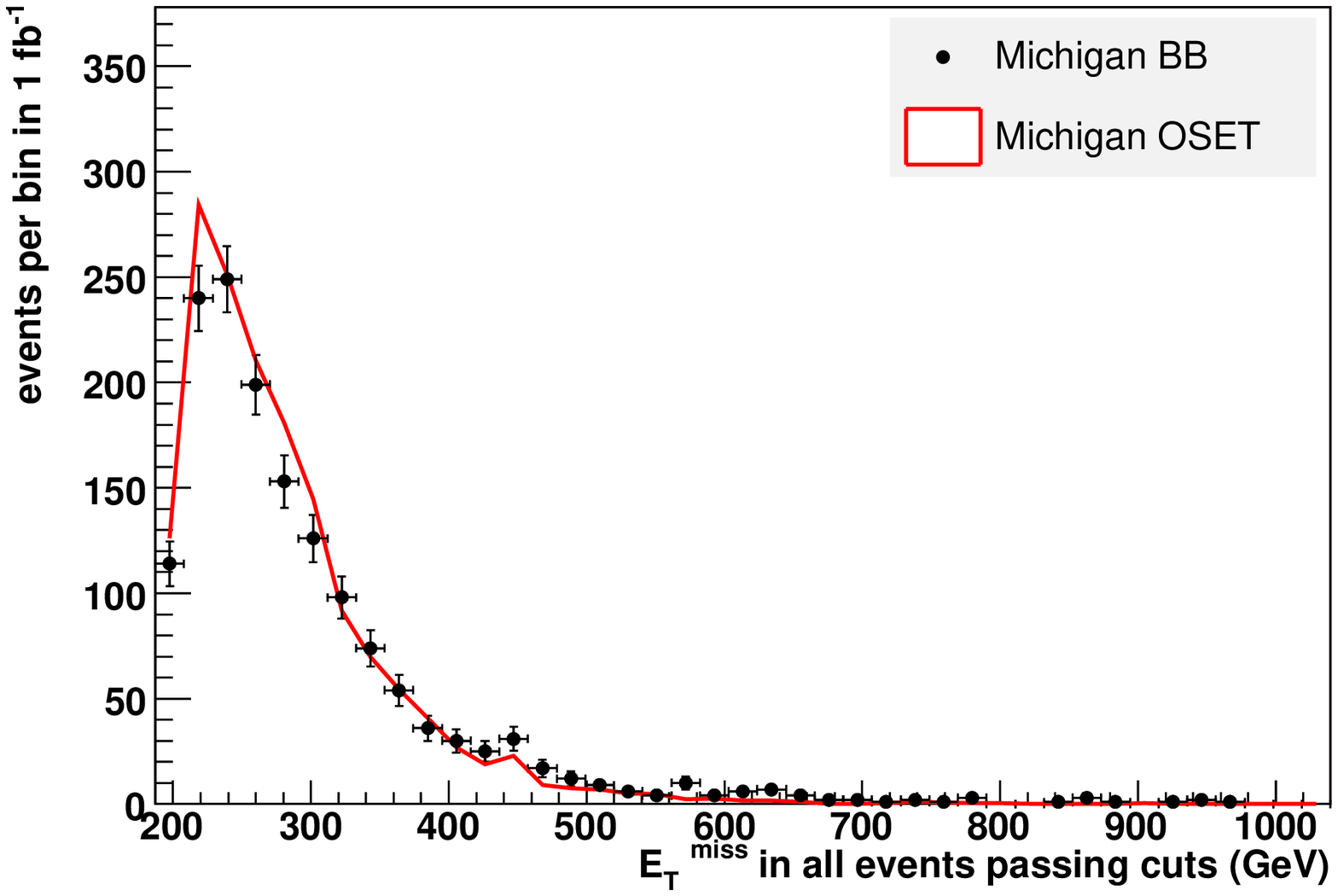}

\includegraphics[width=3in]{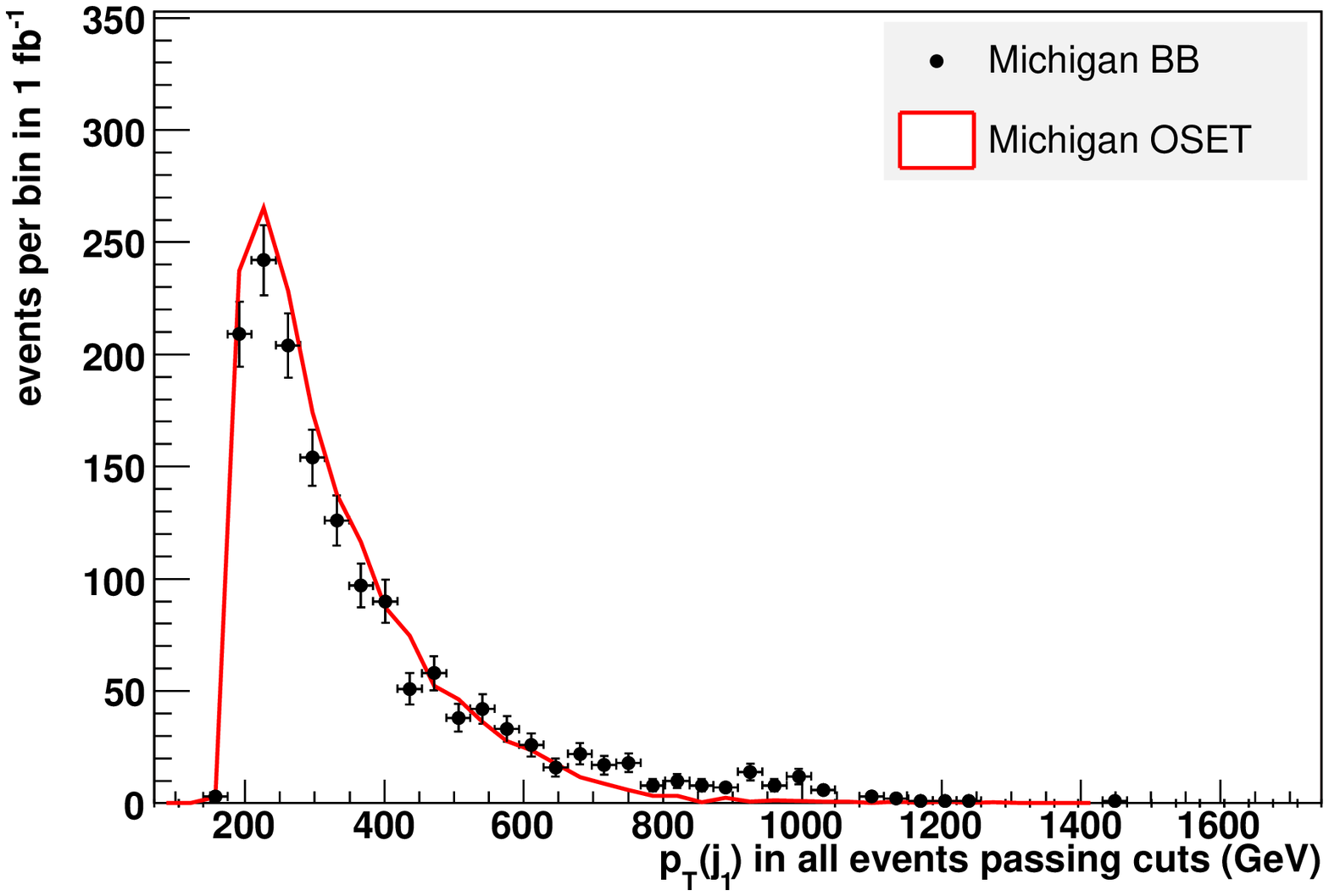}
\includegraphics[width=3in]{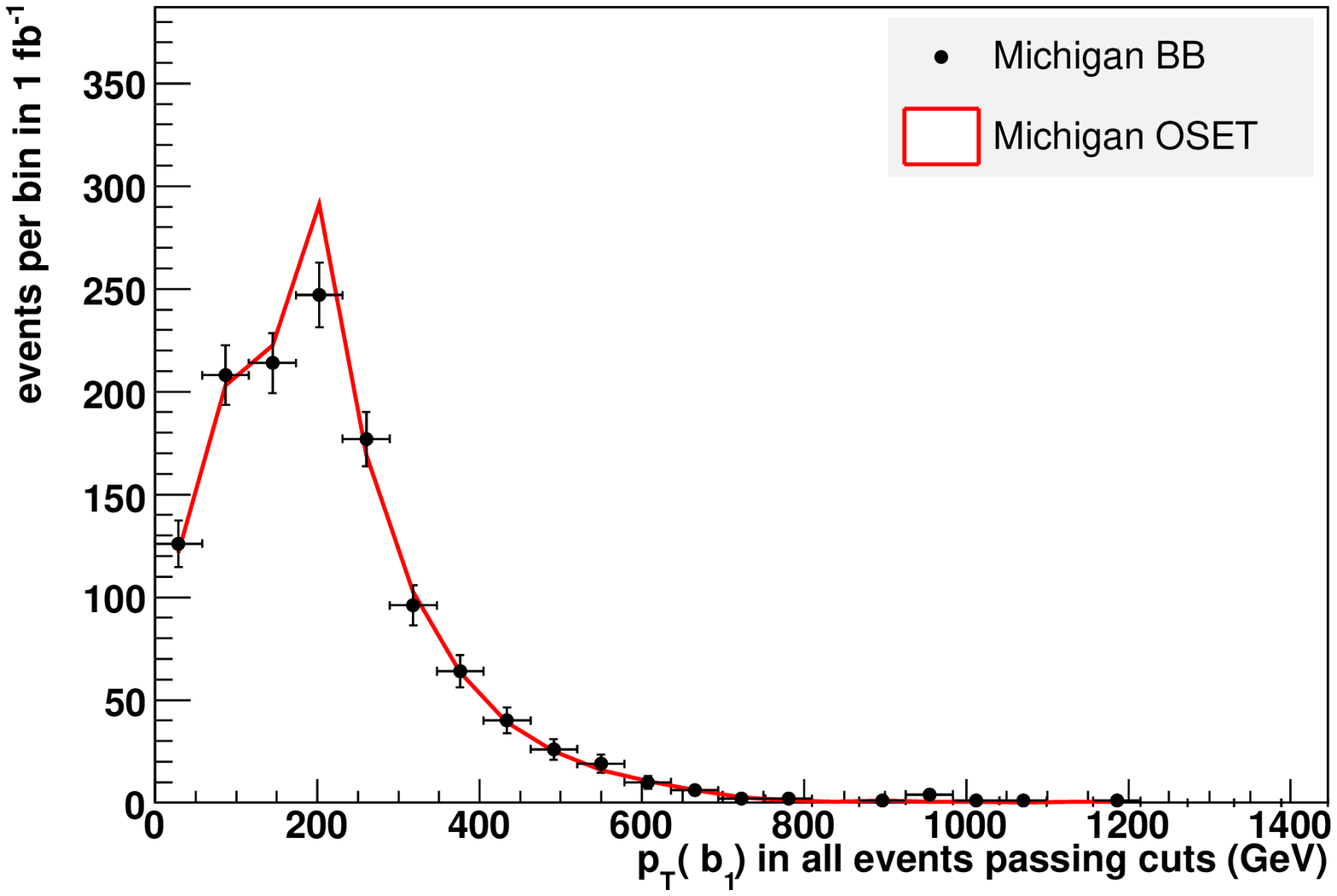}
\caption{\label{fig:fitAA}Comparisons of the Michigan Black Box data
  against an OSET fit, obtained by optimizing cross-sections and
  branching ratios in the OSET specified in Figures
  \ref{fig:OSETDiagrams} and \ref{fig:OSETInput}.  In this case,
  optimization is done by a simple $\chi^2$ fit to correlated object
  multiplicities; the branching ratios obtained by our fit are
  summarized in Table \ref{table:rates}.  Top: Histograms of jet
  (left) and $b$-tagged jet (right) multiplicities.  Middle: $H_T$ and
  $E_T^{\rm miss}$ distributions.  Bottom: distributions of $p_T$ for
  the hardest jet (left) and hardest $b$-tagged jet (right).  The
  diagreement at high $H_T$ and high jet $p_T$ will be discussed in Section \ref{sec:tailprocess}.}
\end{center}
\end{figure}

These three decay mode of \textit{Adj} populate final states
differently: $b b$ and $t b$ modes produce four $b$-partons in each
$Adj$ pair-production event, and hence have more $b$-tagged jets
than $j j$; $t b$ differs from $b b$ in having more but softer jets,
and sometimes producing leptons.  A naive fitting procedure can be
used to distinguish these modes.  We count events in the Black Box
signal with $j$ reconstructed jets ($p_T>50$), $b$ of them b-tagged,
and $\ell$ leptons ($p_T > 10$) for triples $(j, b, \ell)$ covering
the signal region.  These bins are further subdivided into bins
according to $H_T = \sum |p_T| + E_T^{\rm miss}$, and we have used
divisions at 700, 1200, and 1500 GeV. 10000 Monte Carlo events are
also generated for each of the 16 processes in the OSET; these are
run through the PGS detector simulator, and events are counted in
the same way.  A least-squares fit of the detector-simulated count
data produces accurate fits to both the cross-section and $t b {\it
Ch}$ and $b b {\it Ne}$ branching fractions as shown in Table
\ref{table:rates}.   At $2 \sigma$ the di-jet decay rate is
consistent with zero.  Comparisons of signal and OSET Monte Carlo
are shown in Figure \ref{fig:fitAA}. Note that at this level, the
well-motivated OSET used here provides a model-independent
characterization of the data.
\begin{table}
\begin{center}
\begin{tabular}{|r|c|c|}
\hline
\textbf{Process} & \textbf{Fit Rate} & \textbf{Actual Rate} \\
\hline
$\sigma(g g \rightarrow {\it Adj} \, {\it Adj})$ & 43.1 $\pm$ 2.9 fb&  28.0 fb\\
\hline
$\mbox{Br}(Adj \rightarrow \bar{t} \bar b {\it Ch}^+ \mbox{ or } c.c.)$ & 0.74 $\pm$ 0.04 & 0.77 \\
$\mbox{Br}(Adj \rightarrow b \bar b {\it Ne})$ & 0.25 $\pm$ 0.03 & 0.22 \\
$\mbox{Br}(Adj \rightarrow q \bar q {\it Ne})$ & 0.01 $\pm$ 0.01 & 0.01 \\
\hline
$\mbox{Br}({\it Ch} \rightarrow q \bar q' {\it Ne})$ & 0.88 $\pm$ 0.11 & 0.60 \\
$\mbox{Br}({\it Ch} \rightarrow e/\mu \bar \nu {\it Ne})$ & 0.12 $\pm$ 0.11 & 0.40 \\
\hline
\end{tabular}
\end{center}
\caption{\label{table:rates}Fit results at $1 \mbox{ fb}^{-1}$. Error bars quoted are for
  uncorrelated modification of parameters subject to constraints of
  the form $\sum_X \mbox{Br}(Adj \rightarrow X)=1$; correlated errors are
  comparable.  Note that statistical and systematic uncertainties due
  to Standard Model background would decrease the precision of the fit
  if they were properly included.  Another systematic effect---the
  pull on parameters by a channel we have not included---could play a
  role, but we have attempted to reduce that effect put choosing $H_T$
   bin divisions that isolate the missing channel.  We have fit
  to $Ch$ branching ratios by treating  $e$ and $\mu$ identically in
  the fit, and including only the $e$ decay mode in an OSET.  As we will see in Table \ref{table:ratesat5}, with increased statistics, the overestimate of the total cross section is reduced.}
\end{table}

\begin{figure}[tbp]
\begin{center}
\includegraphics[width=2in]{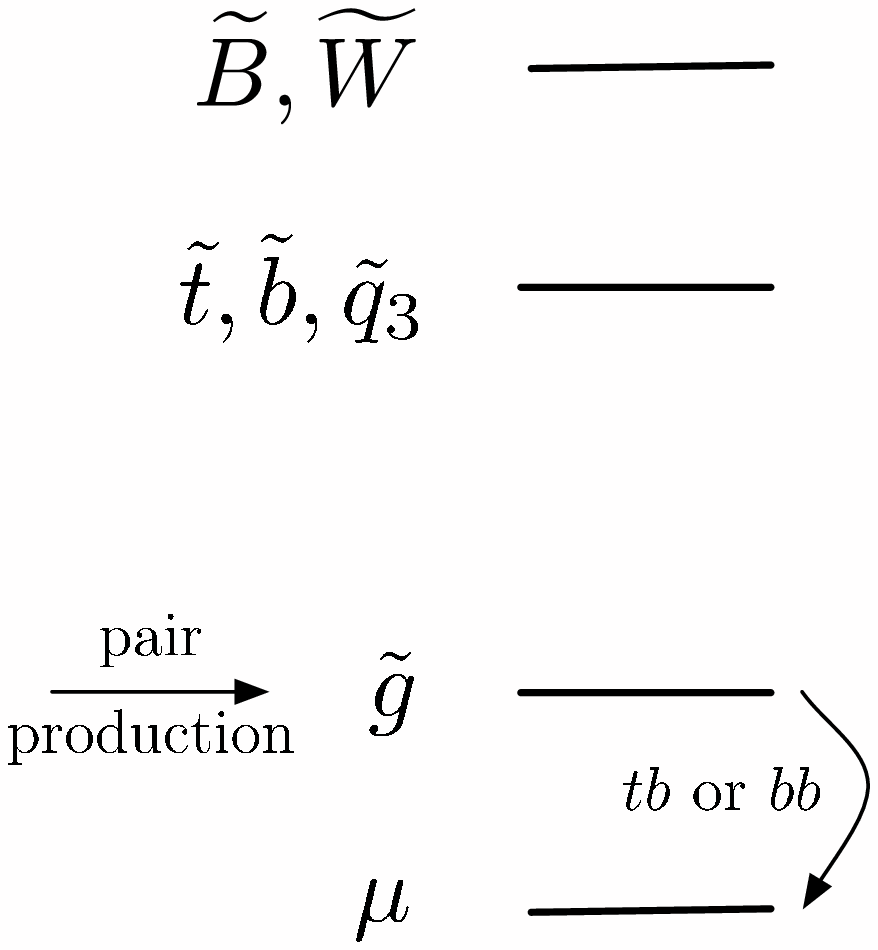}
\includegraphics[width=2in]{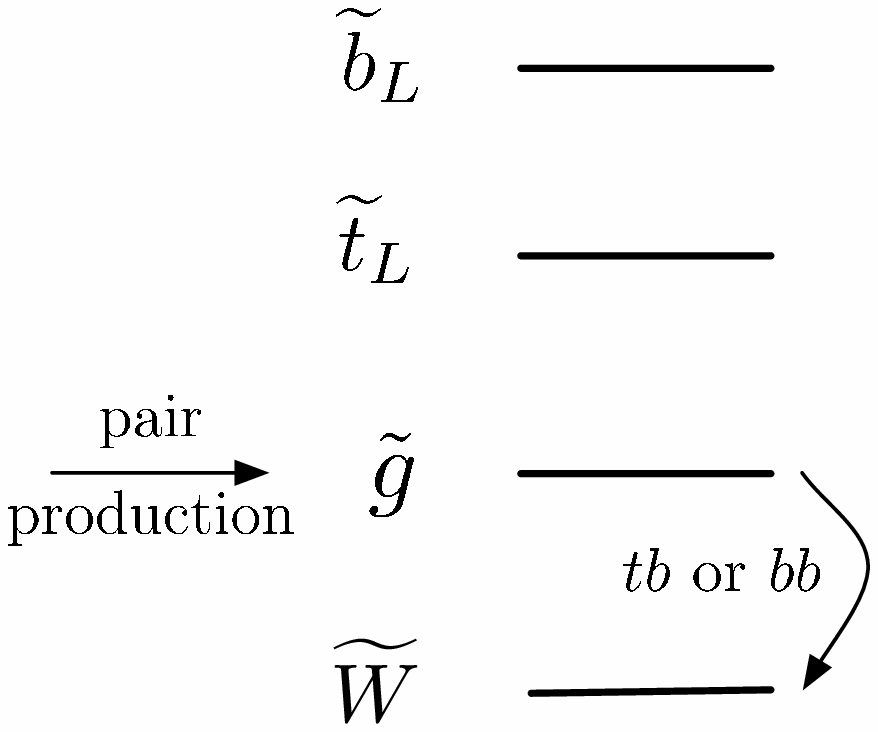}
\caption{The prevalence of third-generation
   decays for {\it Adj} in our OSET fit suggests two patterns of
   masses in the MSSM that could be consistent with the signal.  The
   3rd-generation richness can be explained by either a Higgsino LSP
   (stronger couplings to the third generation) or a large splitting
   between 3rd-generation squarks and those associated with the lighter
   generations, in which case first-generation $\tilde g$ decays are
   suppressed by the large squark mass in an off-shell propagator.  A
   Bino LSP cannot reproduce the nearly degenerate neutral and charged
   states of our OSET.
}
\label{fig:susyPictures}
\end{center}
\end{figure}

Is this picture consistent with supersymmetry?  And if so, what are
the SUSY parameters?  Two notable aspects of this story are a
nearly-degenerate ``chargino'' (\textit{Ch}) near the ``LSP'' (\textit{Ne}) and the
preponderance of 3rd-generation quarks among decay products.  Already
these facts, readily apparent from the OSET, significantly constrain
SUSY explanations---a pure Bino LSP is ruled out, and our attention
must be focused on less standard parameter choices.  Both properties
are accounted for if the LSP is mostly Higgsino, but are also
consistent with a Wino or nearly degenerate Wino and
Bino---in the latter two cases, the preponderance of
3rd-generation decays could be explained by 3rd-generation squarks
much lighter than the other squarks.  At this point, we cannot readily
distinguish the two scenarios in Figure \ref{fig:susyPictures}.  In either case, study of soft leptons
from ${\it Ch} \rightarrow {\it Ne}$ leptonic decays could constrain the ${\it Ch}$--${\it Ne}$
mass-splitting, shedding light on the electroweak-ino mixing.

This OSET has served its two purposes well.  It suggests regions of
theory space consistent with the data.  At the same time, it offers a
more efficient and transparent characterization of the data than the
full model would; the cross-section and branching ratios we have
found are determined in the SUSY models not only by the on-shell
masses but also by many neutralino-mixing parameters and squark
masses. But could we have arrived at it without knowing the model in
advance?

\subsection{Mass Determination}\label{sec:OSETmasses}

As is well-known, accurately determining absolute mass scales is
difficult in cases where full event reconstruction is not possible.
We do not add any new techniques for mass determination here, and only
comment on how mass scales can be inferred in an OSET context. We
have chosen the masses in this OSET with foresight, but could they
have been determined from the data?  At low luminosities and with
the smearing of jet energies, we cannot easily reconstruct any sharp
kinematic features (edges and endpoints at the boundaries of phase
space) that permit precise mass measurement.  Nonetheless, if the
detector response to jets is well-calibrated on Standard Model
processes, we can resolve various particle masses from the shapes of
$p_T$ distributions.

\begin{figure}[tbp]
\begin{center}
\includegraphics[width=3in]{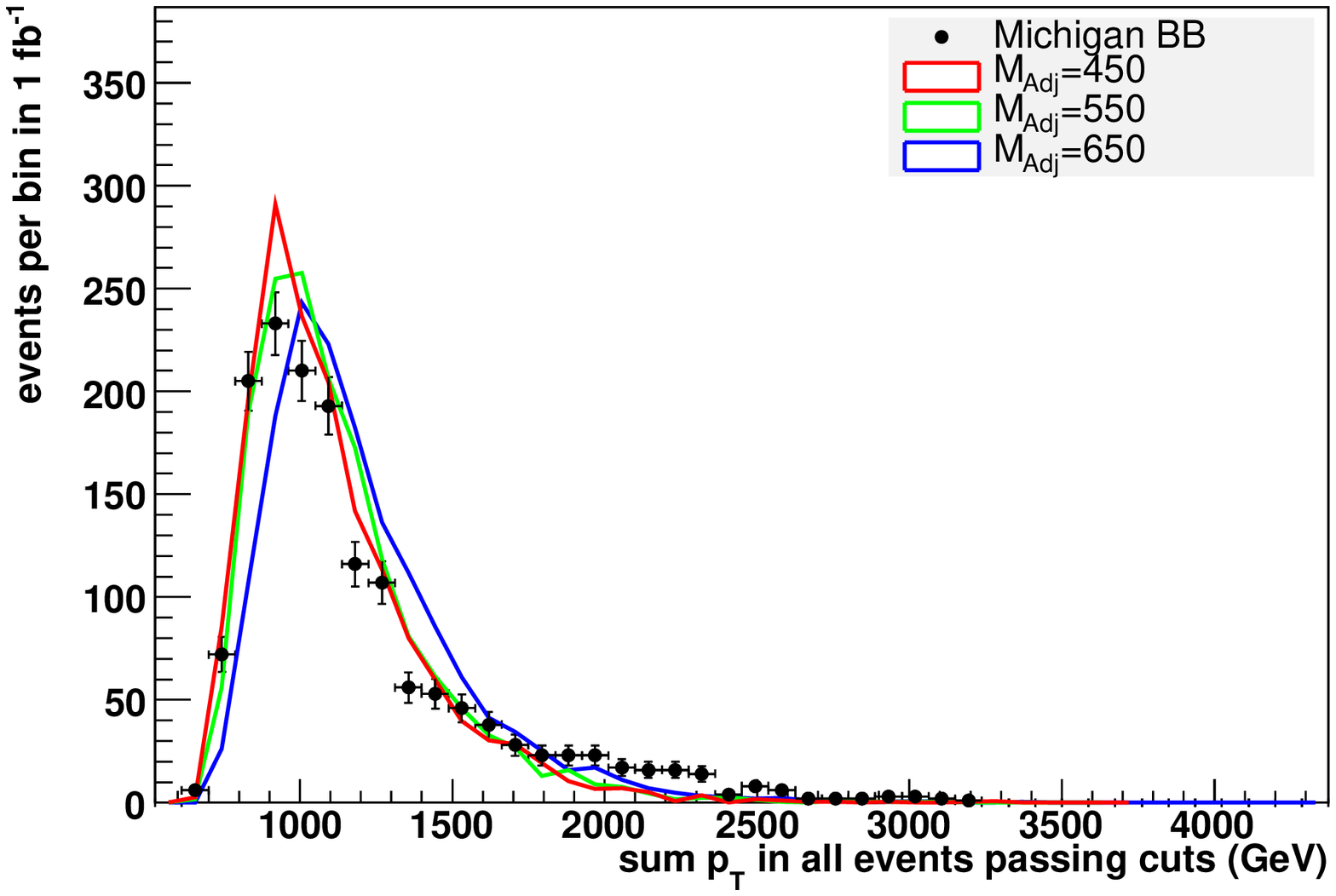}
\includegraphics[width=3in]{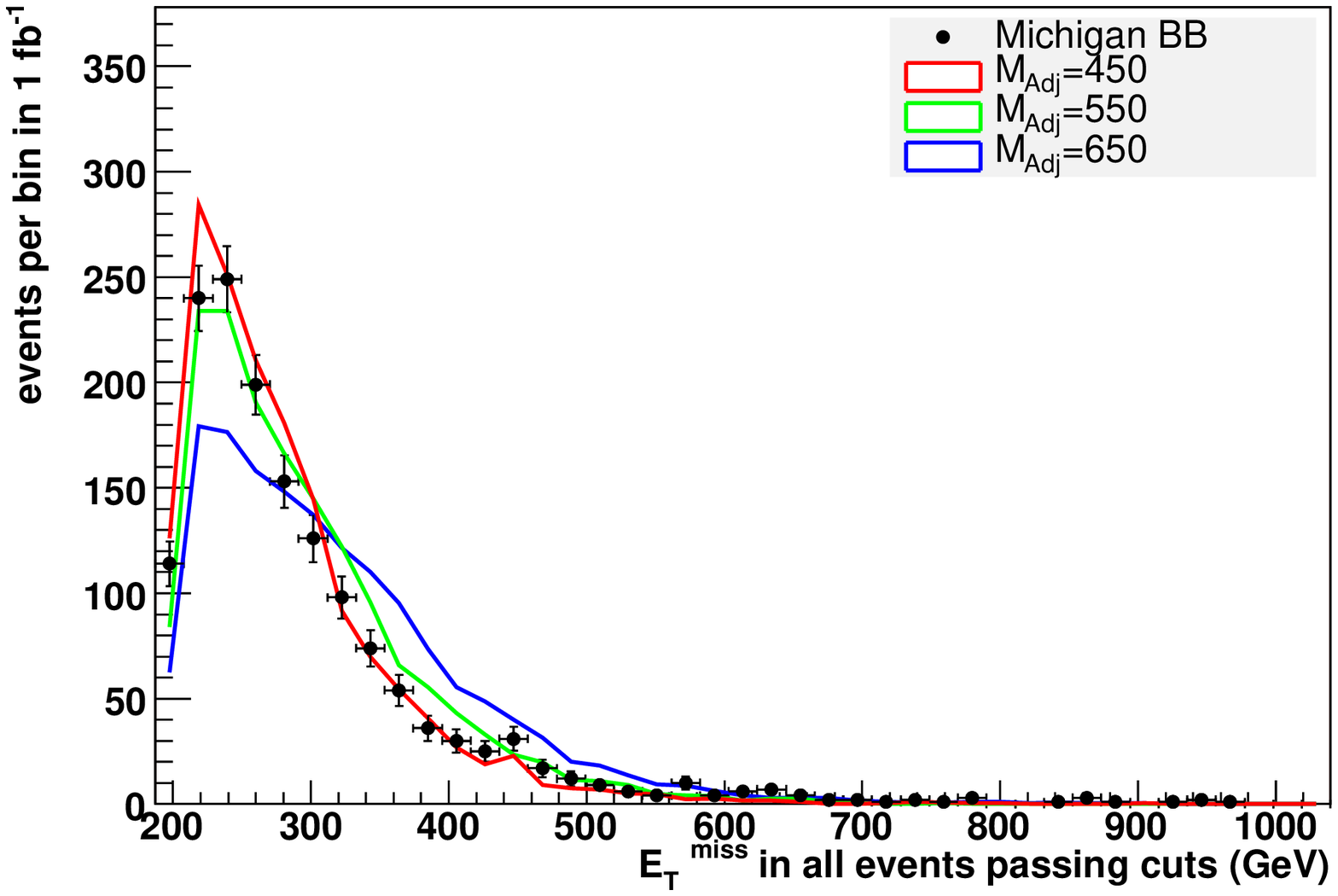}
\caption{\label{fig:adjointMassShift}Neither $H_T$ nor $E_T^{\rm miss}$
distributions can be matched accurately when the mass of the produced
adjoint particle is varied by 200 GeV.}
\end{center}
\end{figure}

For example, raising the mass of the $Adj$ particle that is
pair-produced significantly affects the kinematics of decays---all
products carry more momenta, so that $H_T$ and $E_T^{\rm miss}$ (shown in
Figure \ref{fig:adjointMassShift}), as well as individual object
$p_T$'s, increase correspondingly.  A shift of 100 GeV begins to be
resolvable.

\begin{figure}[tbp]
\begin{center}
\includegraphics[width=3in]{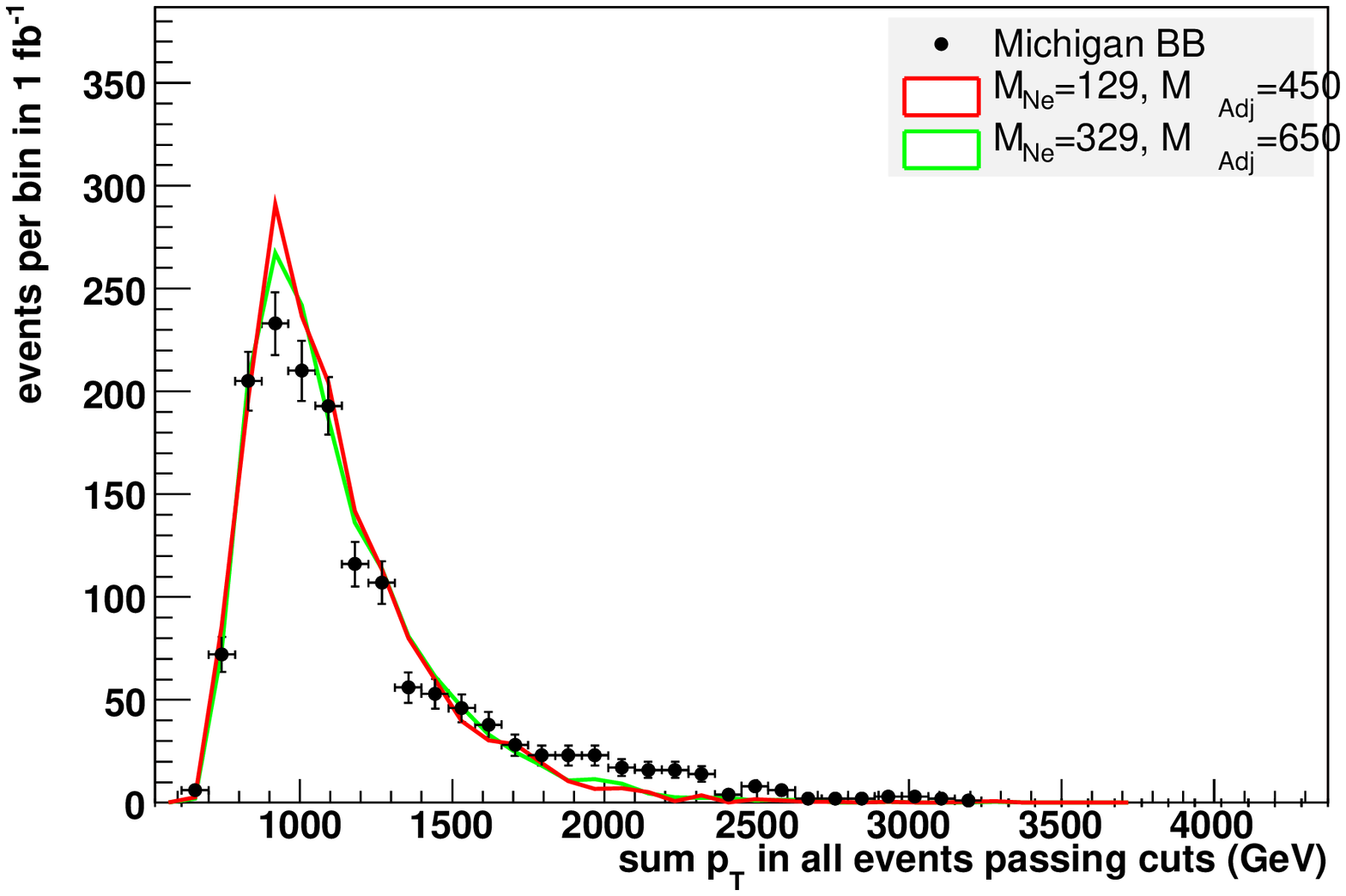}
\includegraphics[width=3in]{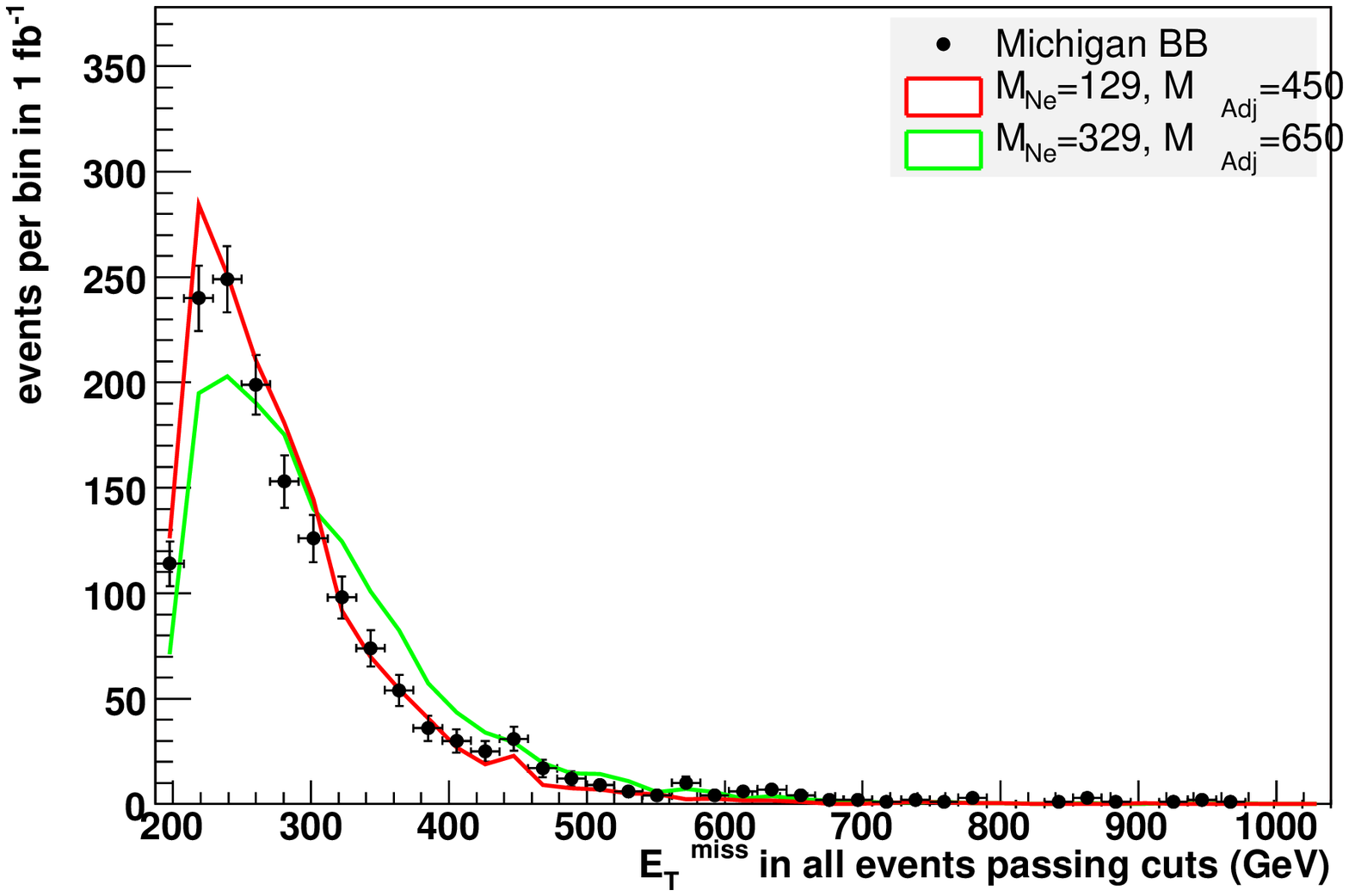}
\caption{\label{fig:bothMassShift} Here, we have increase the adjoint
mass by 200 GeV in the OSET. A consistent fit to counts binned over $H_T$
is possible when the mass of the produced adjoint is
coupled to a mass shift in the final-state invisible particle, but is
irreconcilable with the $E_T^{\rm miss}$ distribution.  The ability to
resolve this difference requires clear understanding of Standard Model
backgrounds and calibration of missing energy distributions, and may
be impossible in early LHC running.}
\end{center}
\end{figure}

The leading effect of $m_{\it Adj}$ on the available decay phase space can
be reduced by increasing the mass of ${\it Ne}$, the stable final-state
particle, in lockstep.  This shift, illustrated in Figure
\ref{fig:bothMassShift}, keeps visible object $p_T$'s fixed, but
we see that some difference in $E_T^{\rm miss}$ is still visible.
It should be noted that the mass scale determined by these means
\emph{assuming} a flat production matrix element will be inaccurate
if a threshold-suppressed contribution to $|\mathcal{M}|^2$ is
important.  Therefore, it would be important to test the systematic
effects of different matrix element choices on mass determination.

One other element that can be used to constrain masses indirectly is
the production cross section; within a model with no free couplings,
this is determined by particle masses.  In some regions of parameter
space, the rate depends only on the mass of the particles being
produced, while in others, a number of intermediate-state masses and
couplings play an important role in determining the production cross
section.  This is one reason that leaving $\sigma$ as a free
parameter, as is done in an OSET, is desirable as there are often
\emph{theoretical} uncertainties in assumed mass/rate correlations.
Another difficulty in tying production rates to masses is that the
predictions differ among different models, so a mass measurement
assuming SUSY will be different from the mass measurement assuming a
different underlying model.  By using an OSET parametrization, these
theory systematics can be understood by varying matrix elements in a
model-agnostic way.

\subsection{False Starts}
\label{sec:falsestarts}
To achieve confidence in an OSET, it is essential not only to obtain
reasonable agreement between the data and OSET predictions, but also
to rule out reasonable alternative hypotheses.  The ease of generating
Monte Carlo for a \emph{class} of candidate OSETs all at once
--accounting for branching ratios by re-weighting events instead of
re-simulating them--makes this computationally difficult task far more
approachable. The interplay of quick comparisons of OSET-level event
generation to data and theoretical inferences is also essential.
Here, we will not rule out alternatives thoroughly, nor even to our
own satisfaction; we merely seek to outline how an analysis could
proceed, and highlight this interplay.

\begin{figure}
\begin{center}
\includegraphics[width=5in]{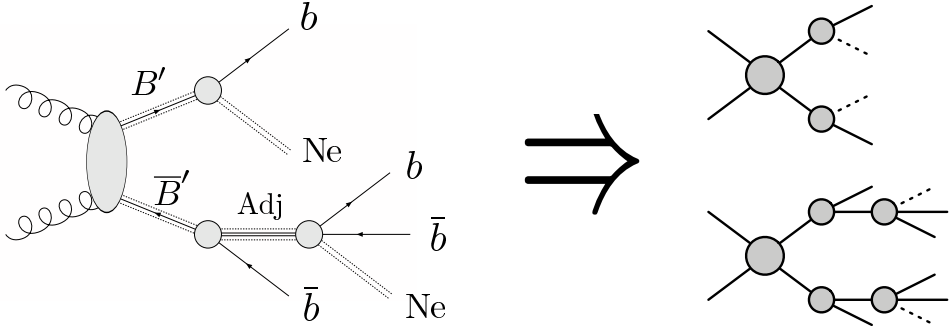}
\end{center}
\caption{\label{fig:false0}A triplet pair production diagram
consistent with the same mass hierarchy as the OSET of Figure
\ref{fig:OSETDiagrams}.  It is implausible that this topology
dominates the observed signal for three reasons: (1) it should be
dwarfed by production of the lower-mass adjoint intermediate state,
(2) it is necessarily accompanied by 2-quark and 6-quark topologies
(small diagrams above), and (3) the resulting events would be highly
asymmetric, with one jet much harder than the others, whereas the
signal displayed excellent agreement with a more symmetric OSET.}
\end{figure}

The distributions and counts we have seen and the agreement of the
data with the OSET of Section \ref{sec:goodoset} suggest production
of $b$ quarks and $W$'s; $b$-jet counts are consistent with four $b$'s
per event.  We will assume that deviations from this counting are not
dramatic.  Though the process shown in Figure
\ref{fig:false0} could produce the desired final state, we reject it
for several reasons: it is likely sub-dominant relative to the $Adj$
pair-production process of
Figure \ref{fig:OSETDiagrams}; it must be accompanied (if not dwarfed) by
the 2- and 6-quark symmetric decay chains that it implies, producing a
larger spread in jet multiplicity; and these decays would be quite
asymmetric, with one jet typically much harder than the others.

\begin{figure}[tbp]
\begin{center}
\includegraphics[width=5in]{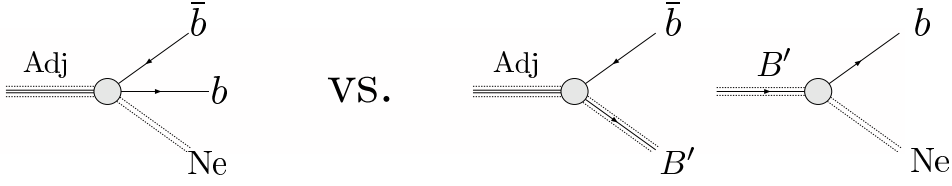}
\caption{\label{fig:falseA}A modification to the OSET of Figure
  \ref{fig:OSETDiagrams} that would yield the same partonic
  multiplicities in final states.  Here, the $Adj$ decays through a
  cascade of two-body processes rather than a three-body process.  If
  $B'$ were taken off-shell, this would reduce to Figure
  \ref{fig:OSETDiagrams}.  For the choice of masses in Table
  \ref{spectrumfalseA}, the kinematics of this process is difficult to
  distinguish from the off-shell process using the simple observables
  considered here.  An OSET containing this process could be
  disfavored by distinguishing these processes through refined
  kinematic observables, or by bounding the direct production of the
  light state $B'$.}
\end{center}
\end{figure}

Two more subtle modifications to the OSET of Figure
\ref{fig:OSETDiagrams} merit further scrutiny.  One possibility is
that the new triplets assumed to mediate the 3-body $Adj$ decay
could be on-shell rather than off-shell; another is that some or all
of the $W$'s could arise from electroweak cascades rather than $t$
quarks. An OSET for the first alternative is shown in Figure
\ref{fig:falseA} and Table \ref{spectrumfalseA}.

\begin{figure}[tbp]
\begin{center}
\includegraphics[width=3in]{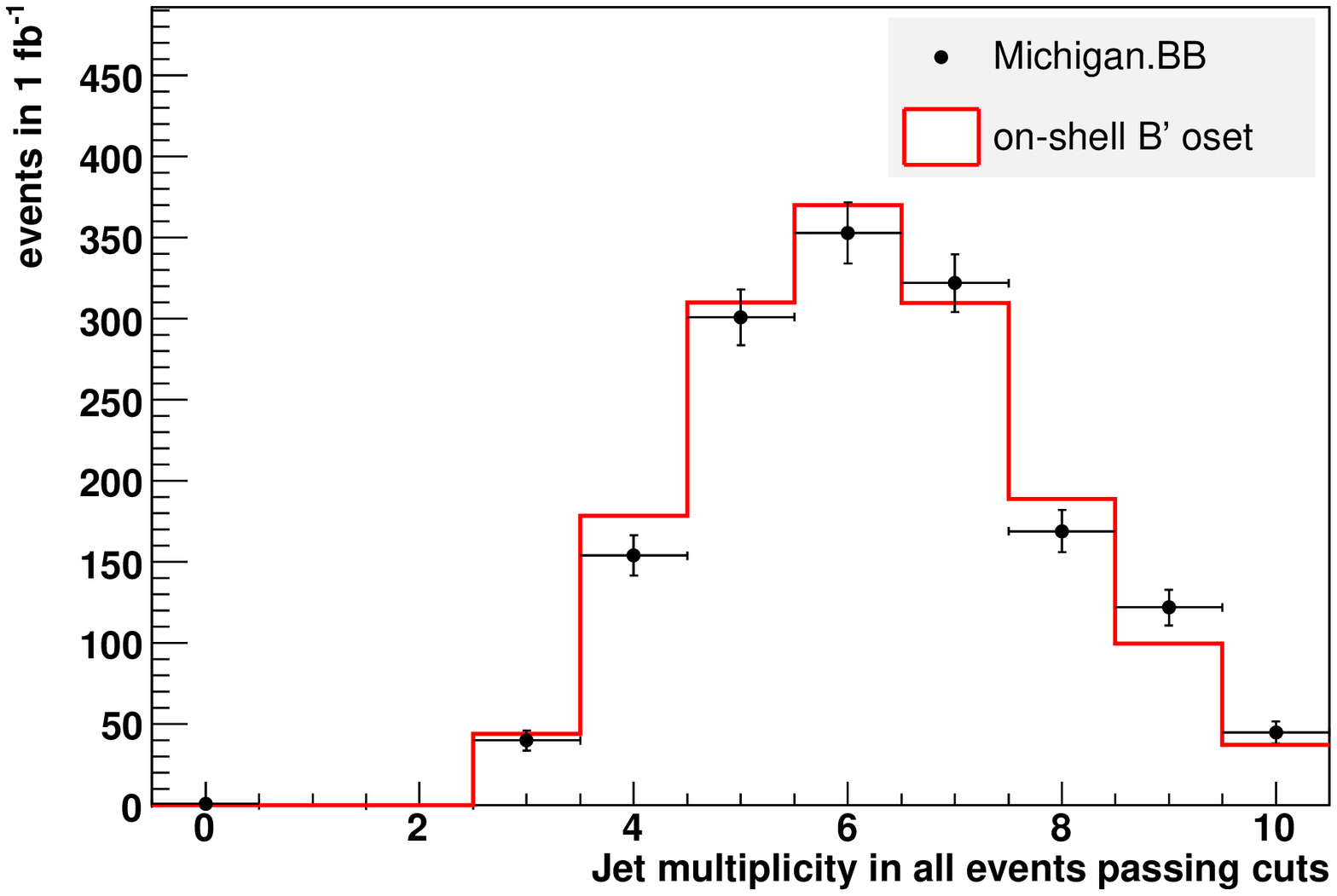}
\includegraphics[width=3in]{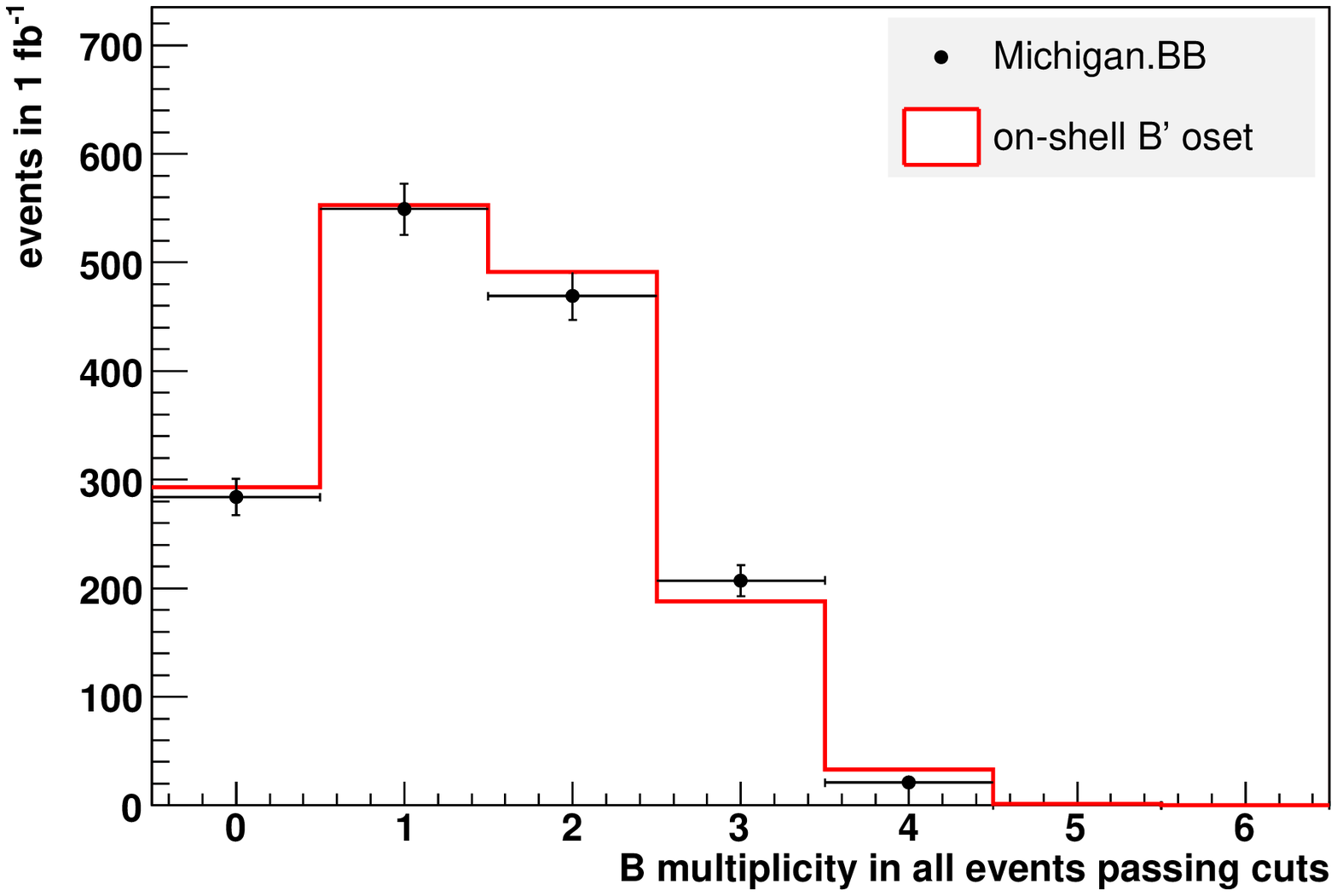}

\includegraphics[width=3in]{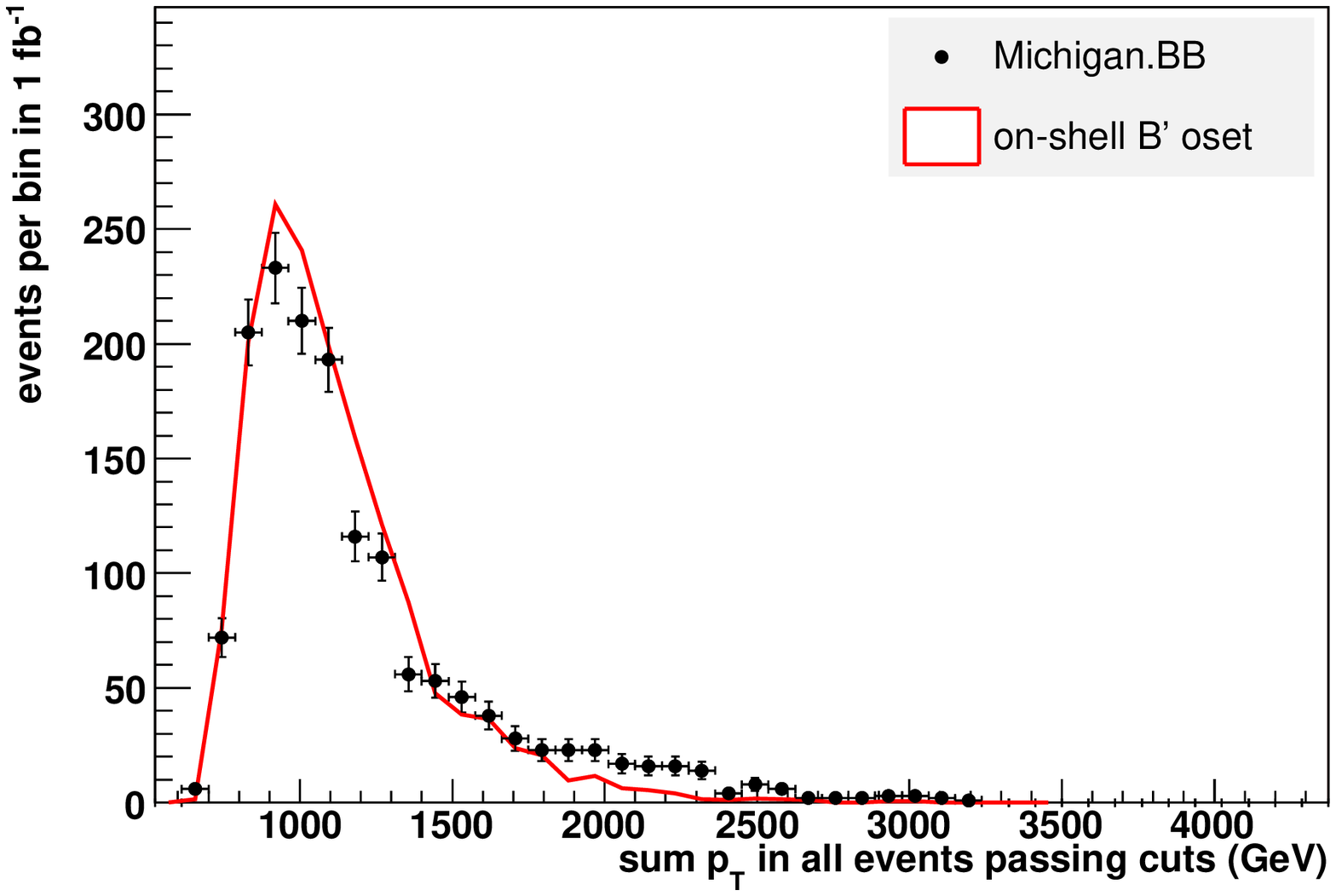}
\includegraphics[width=3in]{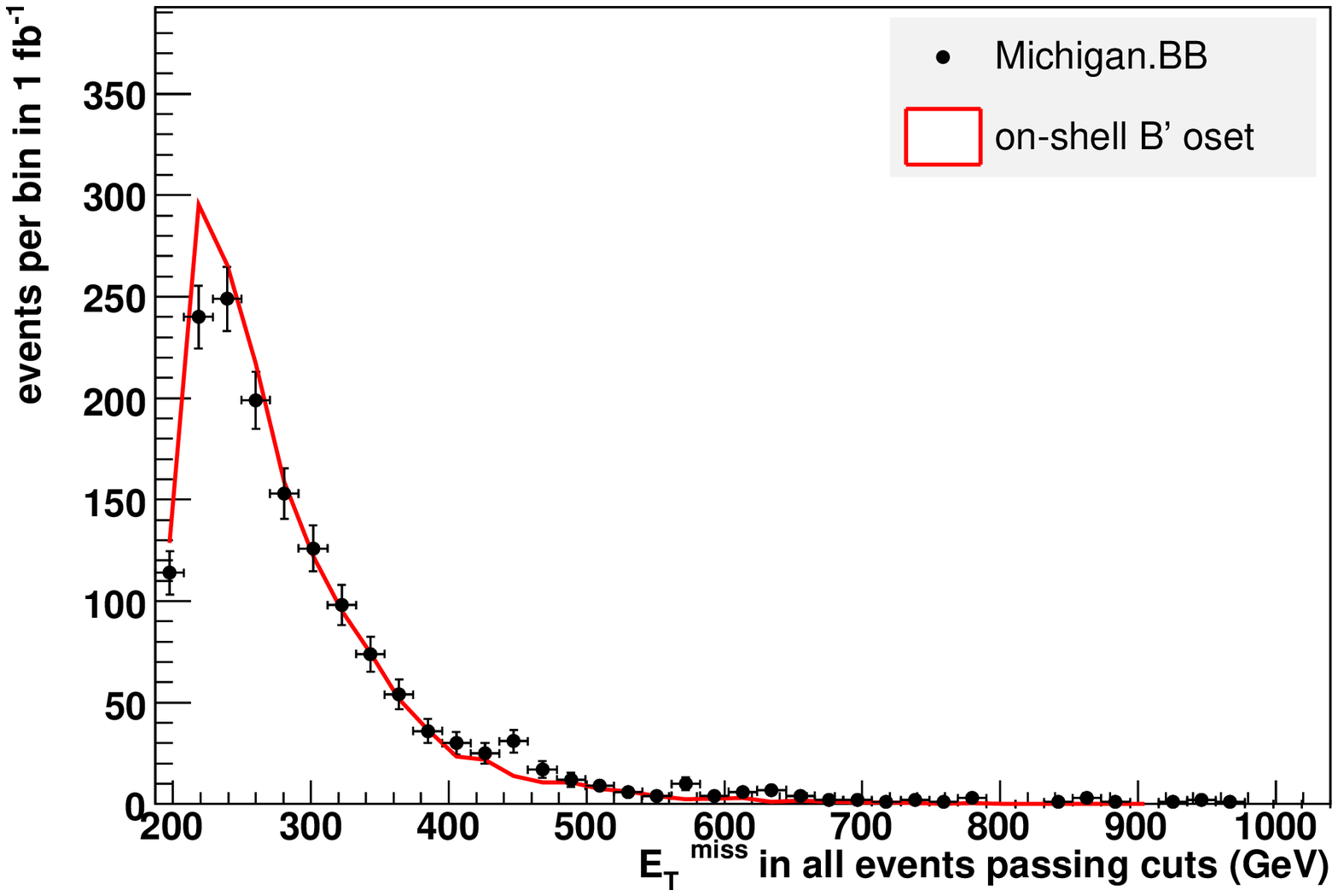}

\includegraphics[width=3in]{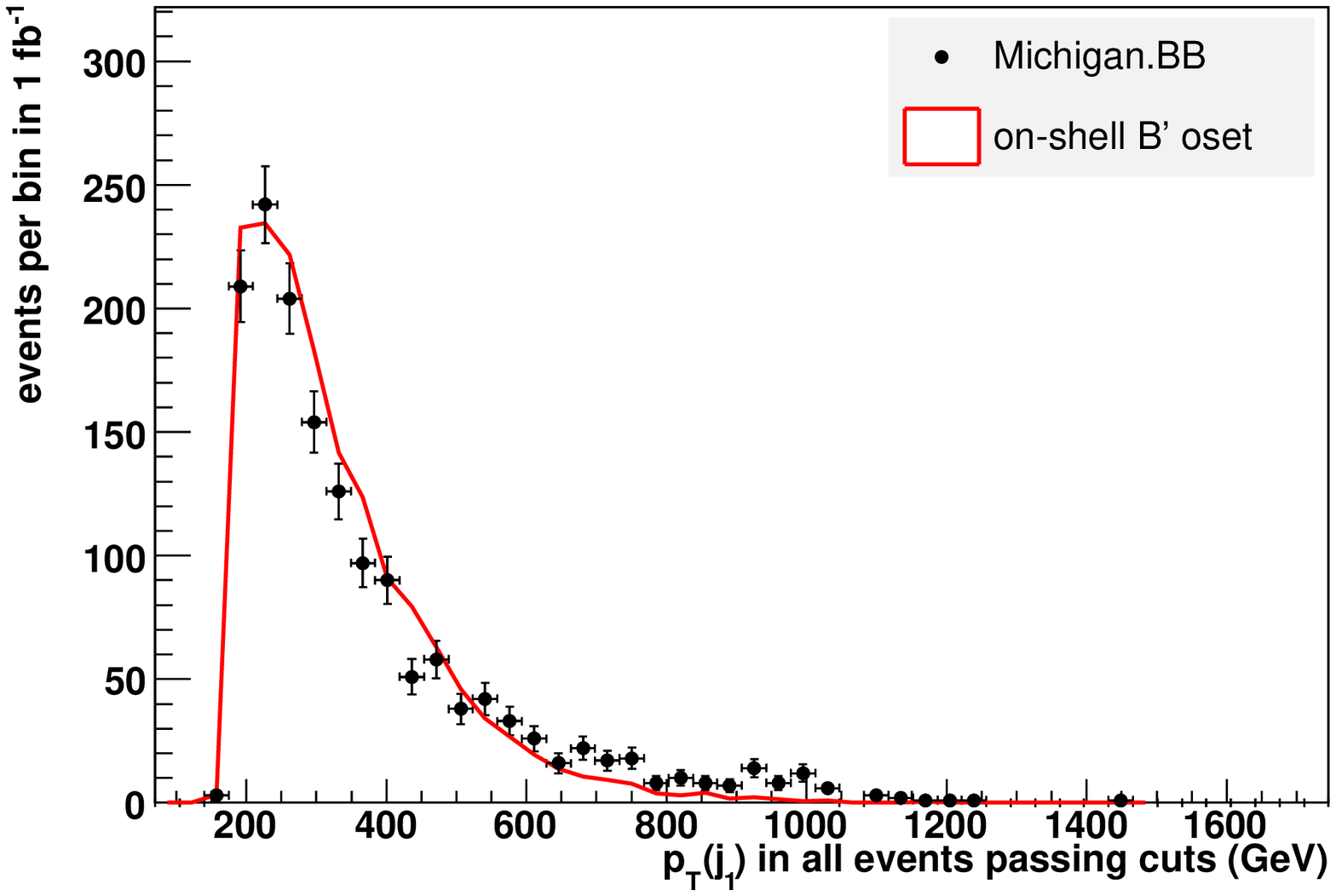}
\includegraphics[width=3in]{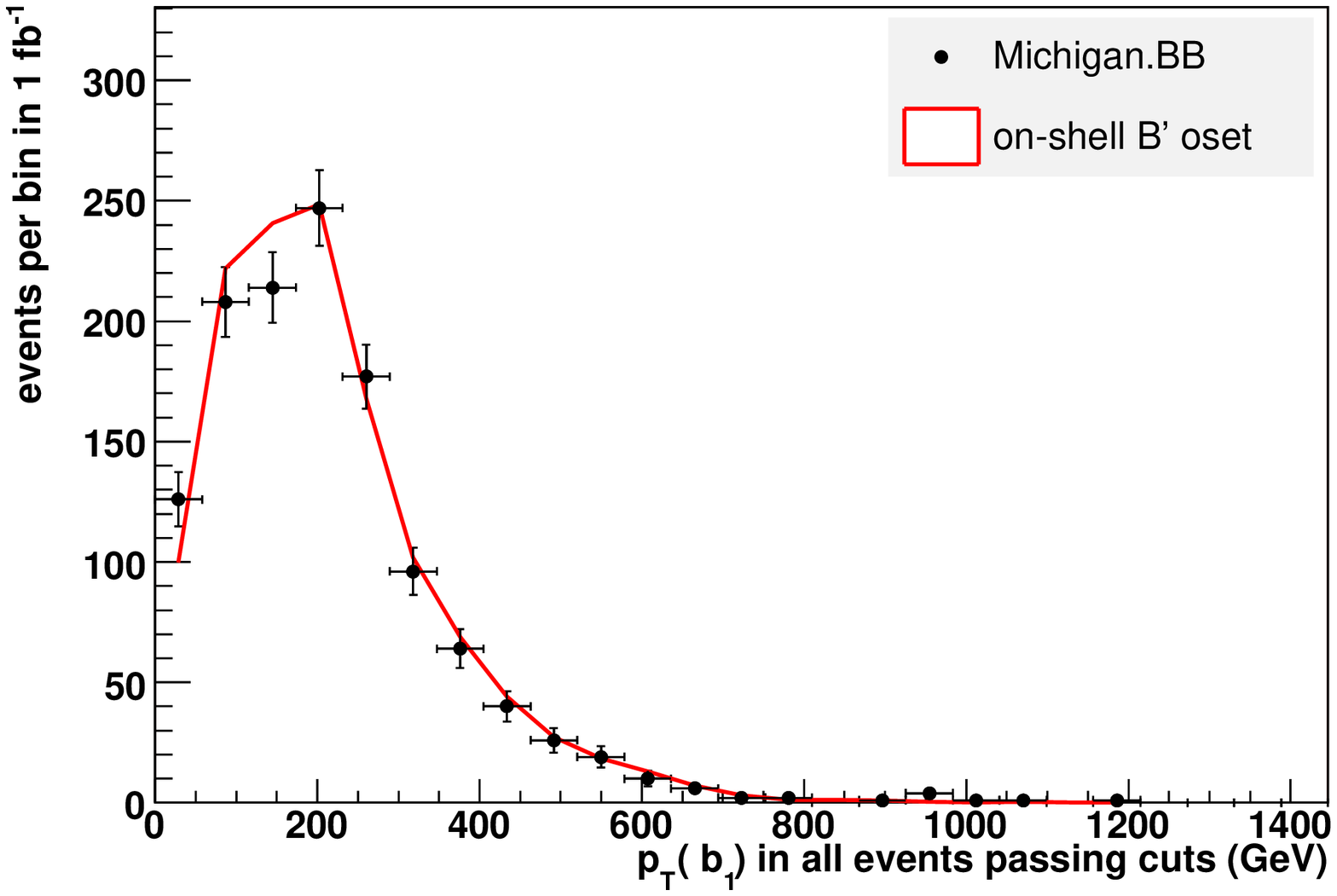}

\caption{\label{fig:falseAdists}Top: jet (left) and $b$-tagged jet
(right) multiplicity histograms for the OSET with an on-shell $B'$
described in Figure \ref{fig:falseA} and Table \ref{spectrumfalseA}.
Middle: $H_T$ and $E_T^{\rm miss}$ distributions.  Bottom:
distributions of $p_T$ for the hardest jet (left) and hardest
$b$-tagged jet (right).  With this choice of masses, these simple
observables do not distinguish effectively between three-body
${\it Adj}$ decays as in the correct OSET and the cascades of this
OSET.  More sophisticated discriminators or indirect arguments
(e.g. constraints on $B'$ direct production) are required to
distinguish these two scenarios.}
\end{center}
\end{figure}

The mass of the $B'$ has been {\it tuned} to minimize the difference
among transverse distributions used in this analysis. The minimal
differences with a tuned $B'$ mass are illustrated in Figure
\ref{fig:falseAdists}. Given the combinatorial background, it is
difficult to resolve a difference between on- and off-shell $B'$ by
using di-jet invariant mass distributions, looking for an edge
versus and endpoint. More sophisticated analysis should be employed
to resolve these cases. In this case, we resort to somewhat indirect
logic: if the $B'$ is strongly interacting, then it should be
minimally produced through QCD.
\begin{table}[tbp]
\begin{center}
\begin{tabular}{|c|c|}
\hline
\multicolumn{2}{|c|}{OSET Spectrum}\\
\hline
$Adj$ & 450 GeV \\
$B'$ & 360 GeV \\
$Ch$ & 128 GeV \\
$Ne_1$ & 124 GeV \\ \hline
\end{tabular}
\end{center}
\caption{A candidate spectrum for the OSET modification in Figure \ref{fig:falseA}.}\label{spectrumfalseA}
\end{table}

A sbottom search in the channel with a $b \bar{b} E_T^{\rm miss}$ final state (or potentially with a $b \bar{t} E_T^{\rm miss}$ or $t \bar{t} E_T^{\rm miss}$ final state with additional soft charged particles)
could be used to constrain the production of this $B'$.
If results are presented as limits on $\sigma\times\mbox{Br}(B'
\rightarrow b {\it Ne})^2$ for this topology, as functions of
$m_{B'}$ and $m_{Ne}$, they could be readily interpreted for
application to any model or to an OSET analysis.  Some mass range
will be excluded when one assumes only QCD production, with
different exclusions depending on the $B'$ spin, but the $\sigma
\times \mbox{Br}^2$ limit is robust. If minimal $B'$ production can
be excluded at the $B'$ mass for which other distributions do not
show a difference, then we can rule out this scenario. Arguments
along these lines could help restrict this class of model, with
$m_{B'} < m_{Adj}$.

\begin{figure}[tbp]
\begin{center}
\includegraphics[width=5in]{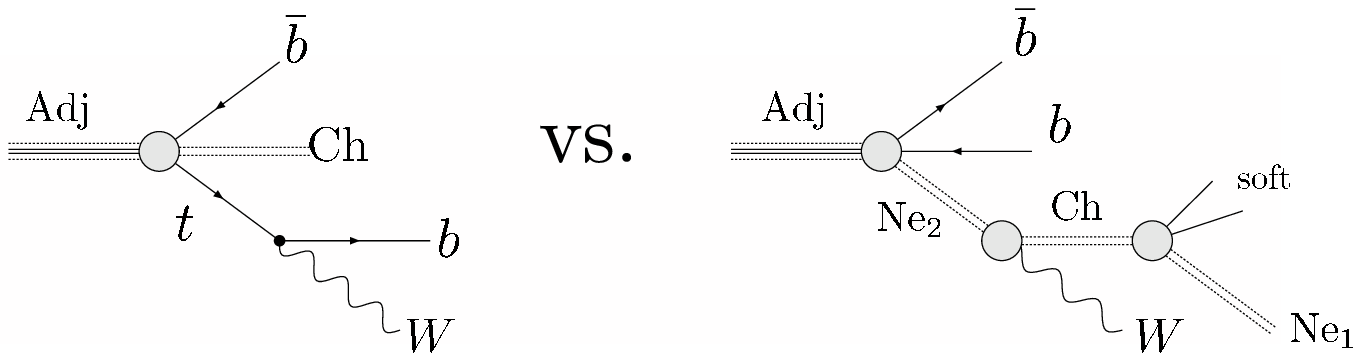}
\caption{\label{fig:falseB}Another OSET modification with the same final states as
  the one in Figure \ref{fig:OSETDiagrams}, but different internal
  kinematics: instead of $b$ and $W$ coming from a $t$, two $b$'s are produced
  directly and $W$ comes from an electroweak cascade.  A candidate spectrum for this scenario is
  given in Table \ref{falseBspectrum}.  }
\end{center}
\end{figure}
\begin{table}[tbp]
\begin{center}
\begin{tabular}{|c|c|}
\hline
\multicolumn{2}{|c|}{OSET Spectrum}\\
\hline
${\it Adj}$ & 450 GeV \\
${\it Ne}_2$ & 300 GeV \\
${\it Ch}$ & 128 GeV \\
${\it Ne}_1$ & 124 GeV \\ \hline
\end{tabular}
\caption{Spectrum for the OSET defined pictorially in Figure
\ref{fig:falseB}, where the decay of ${\it Ne}_2$ to ${\it Ch}$ yields enough $W$s such that tops need not appear in decay chains.}\label{falseBspectrum}
\end{center}
\end{table}

\begin{figure}[tbp]
\begin{center}
\includegraphics[width=3in]{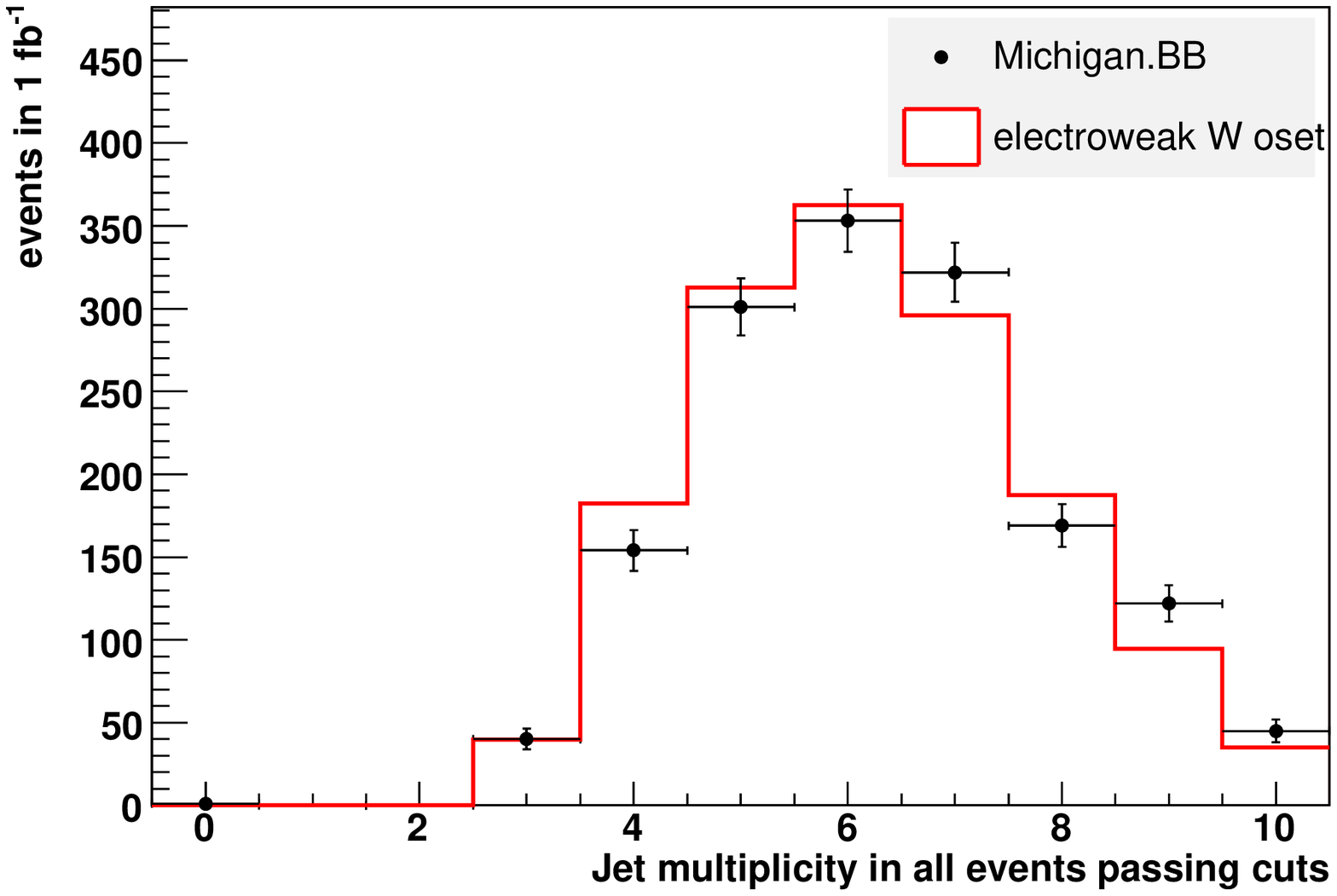}
\includegraphics[width=3in]{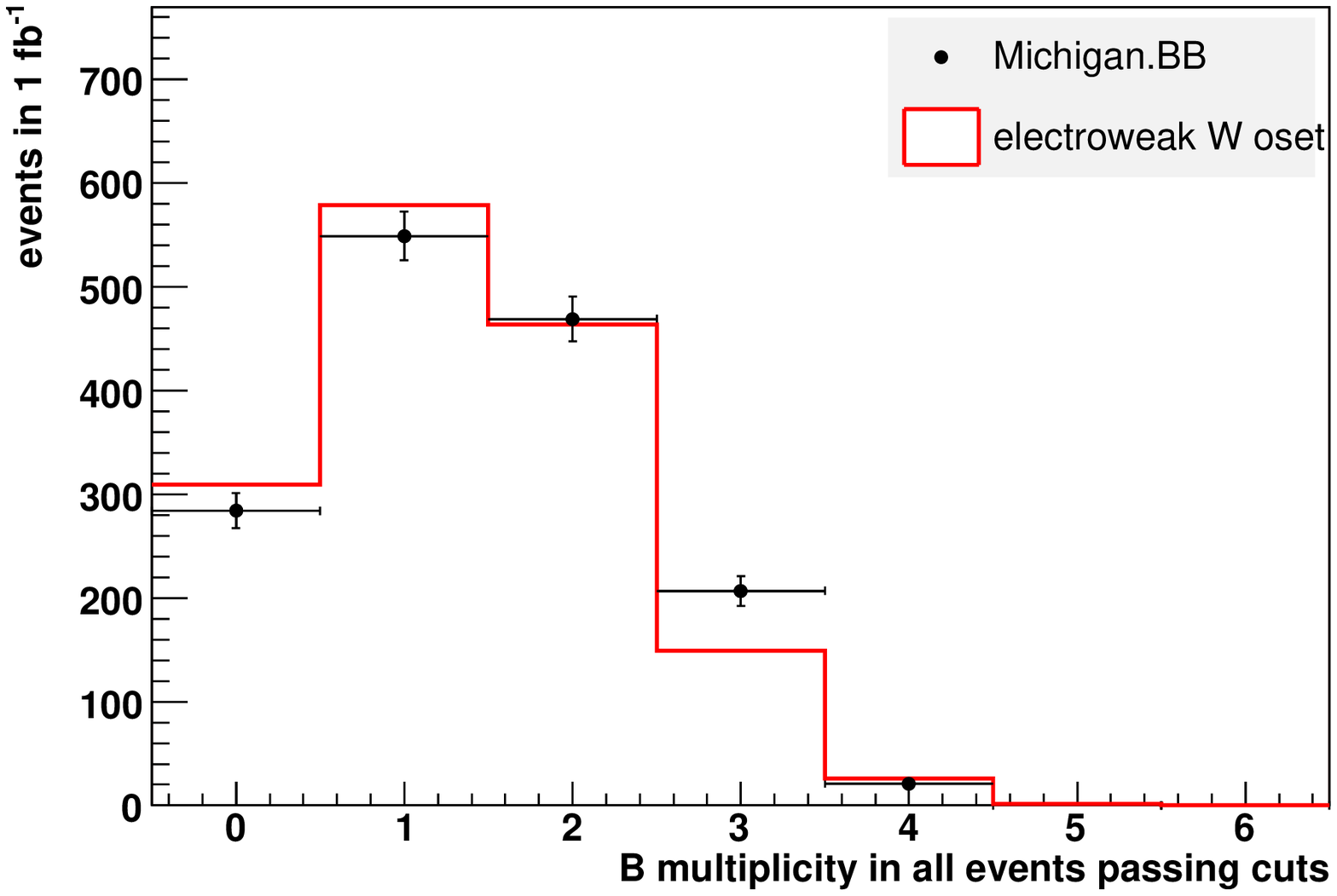}
\caption{\label{fig:falseBcounts}Histogram of jet (left) and
$b$-tagged jet (right) multiplicities in the OSET with the decay shown
  in Figure \ref{fig:falseB}. Without an explicit handle on the number of
  tops in the sample, it is consistent for the $b$s and $W$s to come from cascade decays.}
\end{center}
\end{figure}

\begin{figure}[tbp]
\begin{center}
\includegraphics[width=3in]{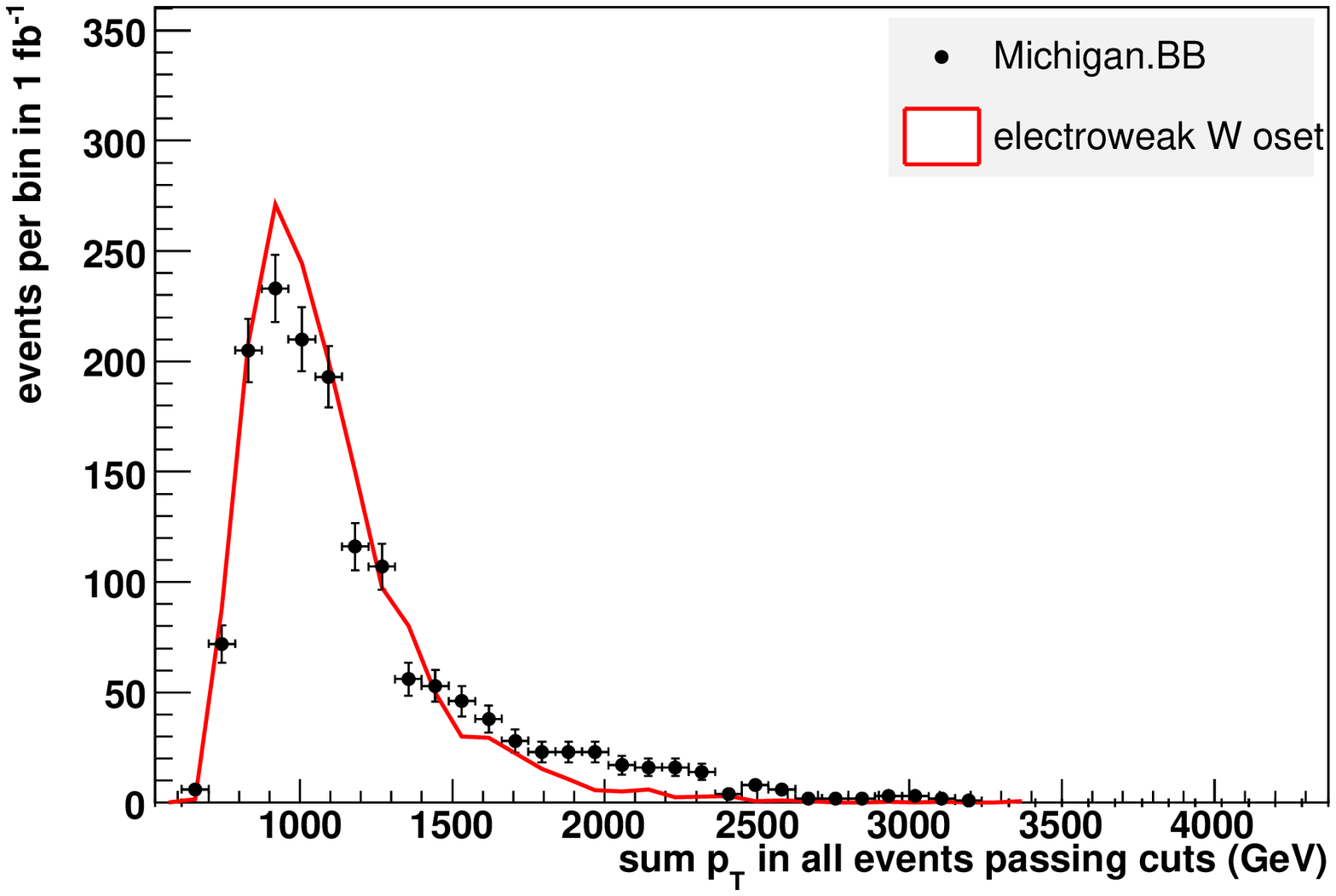}
\includegraphics[width=3in]{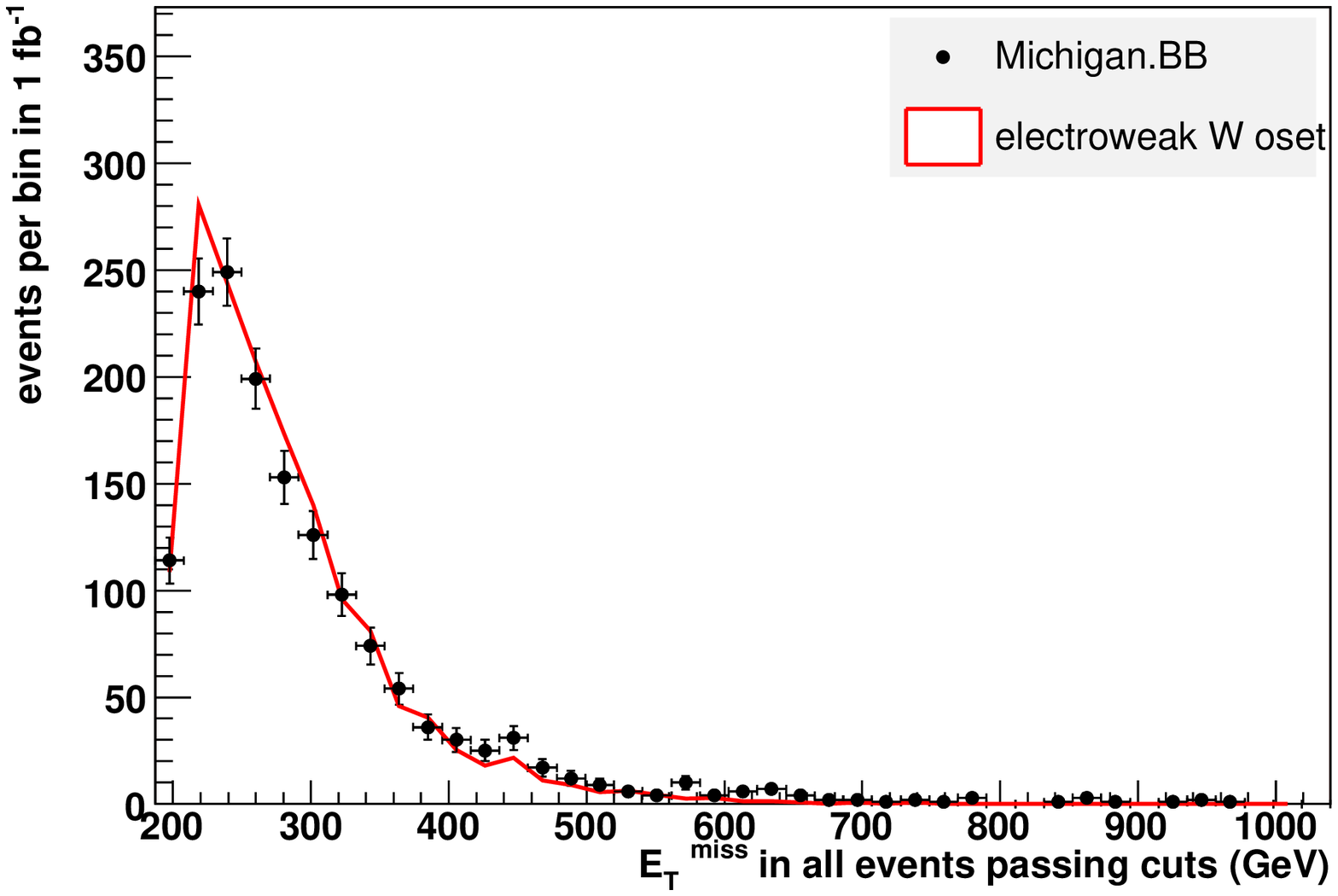}

\includegraphics[width=3in]{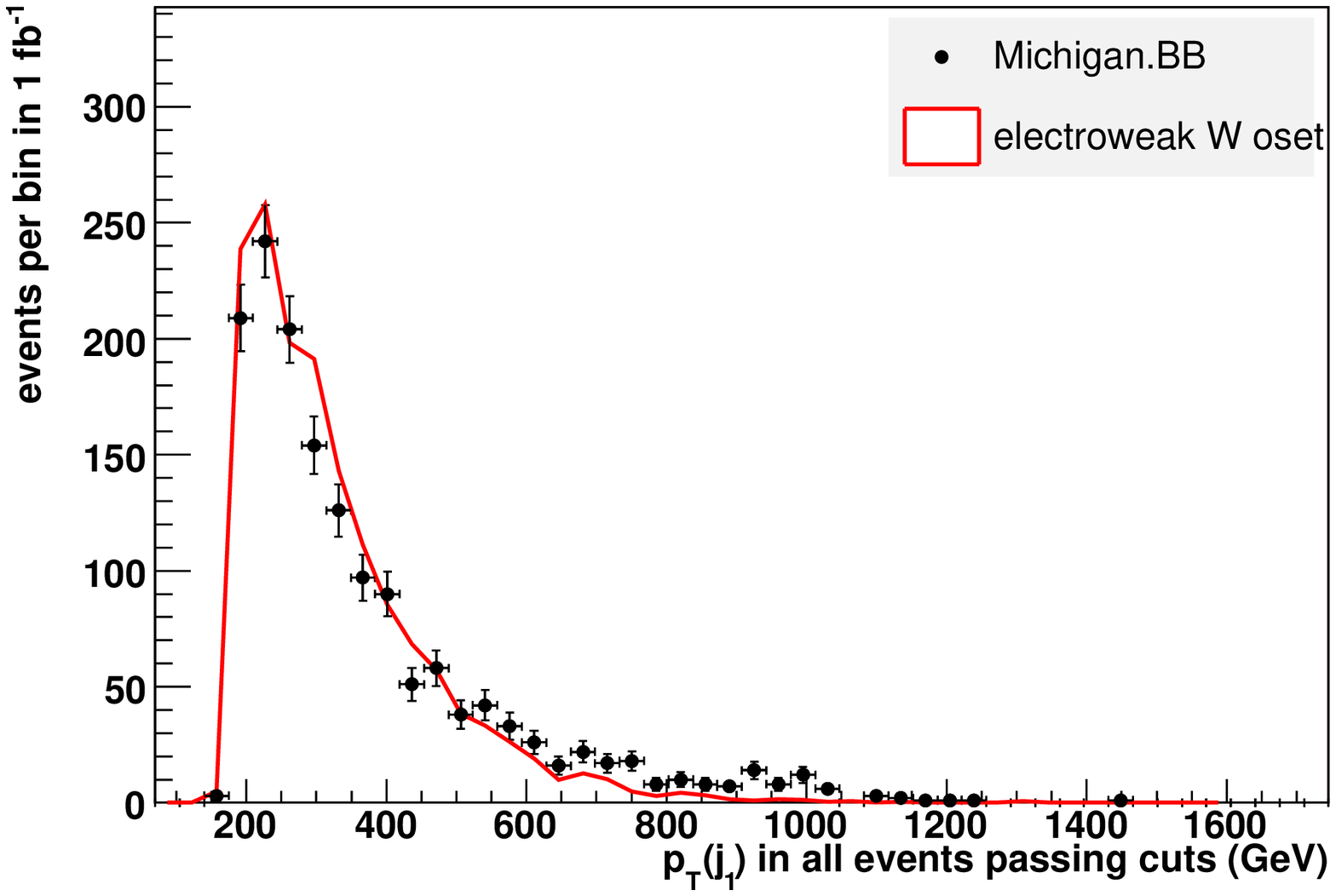}
\includegraphics[width=3in]{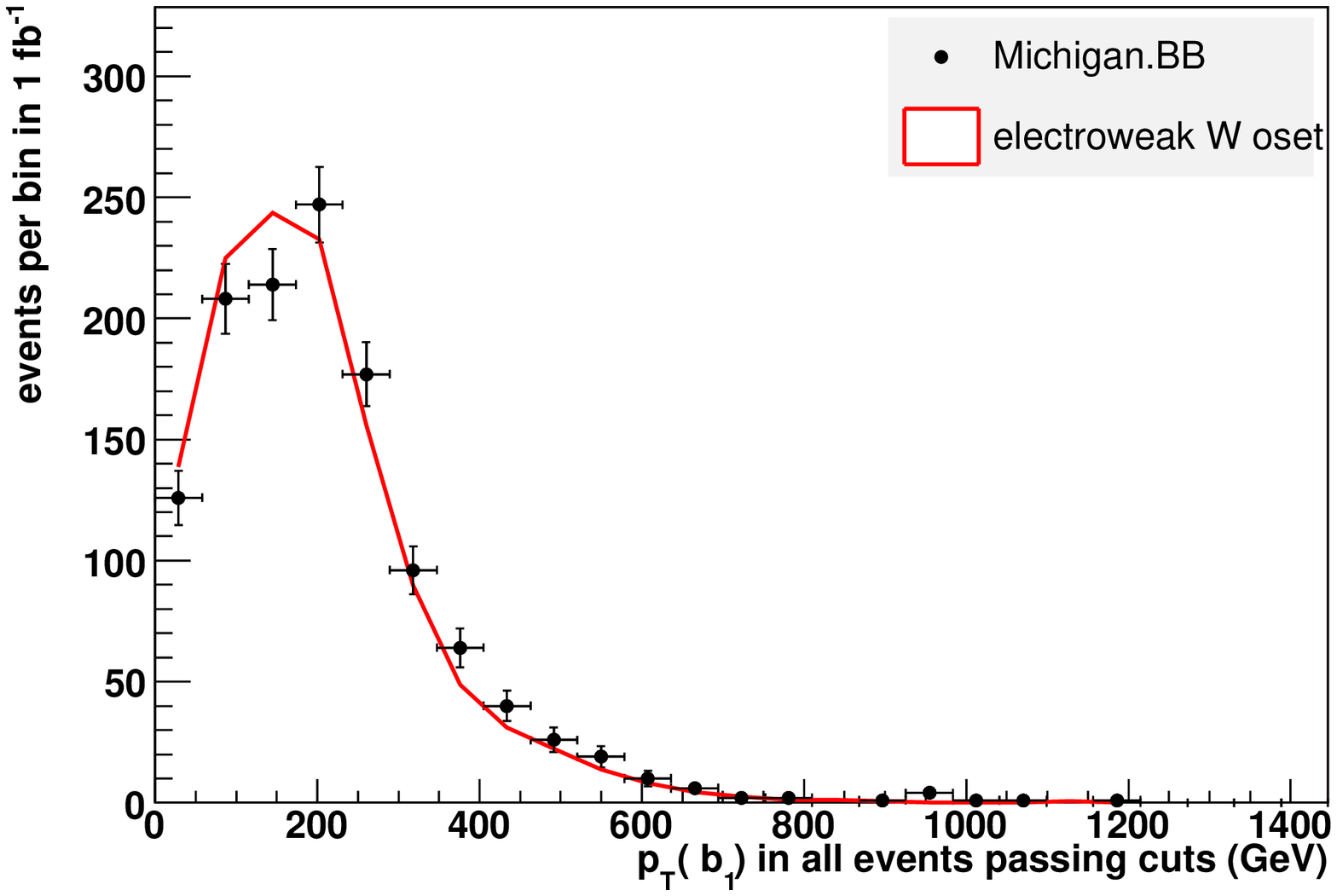}
\caption{\label{fig:falseBdist}Several distributions for the OSET with
  the decay of Figure \ref{fig:falseB} and spectrum of Table
  \ref{falseBspectrum}.  Top: sum of all object $p_T$'s (left) and
  $E_T^{\rm miss}$ (right).  Bottom: Hardest jet (left) and $b$-jet (right) $p_T$'s.
  The distributions are not  quantitatively off, indicating that ${\it Adj} \rightarrow b t {\it Ch}$ decays could be replaced by ${\it Adj} \rightarrow b t {\it Ne}_2$ decays, as long as extra $W$s come from the decay of ${\it Ne}_2$.}
\end{center}
\end{figure}

The third possibility ($W$'s arising from electroweak cascades, as in Figure
\ref{fig:falseB}) is more interesting: with this choice of ${\it Ne}_2$
mass, no clear differences exist in single-particle momenta, $H_T$, or
$E_T^{\rm miss}$ distributions, or counts (Figures \ref{fig:falseBdist}
and \ref{fig:falseBcounts}).  So in fact the $\sim 75\%$ branching
ratio $Adj \rightarrow b \bar b W^\pm$ could consist of any
combination of $t b {\it Ch} $ (Figure \ref{fig:OSETDiagrams}) and $b b {\it Ne}_2$
(Figure \ref{fig:falseB}) as far as this analysis could tell.


Given this ambiguity, two directions should be explored
simultaneously; one is experimental, the other theoretical. With the
complicated detector environment, we have made no attempt to
explicitly reconstruct tops, but a thoughtful look at more
correlated observables could show evidence of top quarks in the
decays. One can also look at additional variables to try and
discriminate among these event shapes. On the other hand,
theoretical considerations suggest other signals to look for: if the
decay ${\it Ne}_2 \rightarrow W^\pm {\it Ch}^\mp$ is present, and
${\it Ch}^\mp$ and ${\it Ne}_1$ form an $SU(2)_L$ doublet, then
${\it Ne}_2 \rightarrow Z^0 {\it Ne}_1$ should also appear with a
comparable rate.  By searching for leptonically decaying $Z$'s in
these events, we can indirectly limit the decay rate of ${\it Ne}_2
\rightarrow W^\pm {\rm Ch}^\mp$.

\subsection{Sub-dominant Process or Poorly Modeled Tail?}
\label{sec:tailprocess}
The $H_T$ plot for the best-fit OSET has captured the bulk of the
signal events, but the overall count with $H_T > 1500 \mbox{ GeV}$ is inconsistent
with the OSET at $\sim 4 \sigma$.  The actual significance of this
excess at the LHC at $1 \mbox{ fb}^{-1}$ depends on the Standard Model
contribution and its uncertainty, and on the precise cuts used in the
analysis.  But assuming that it would eventually be identifiable as
new physics, we wish to consider its interpretation \emph{within}
this class of possibilities: is it ``old'' new physics (a poorly
modeled tail of the underlying model's $Adj$ pair-production process)
or a new process altogether?

\begin{figure}[tbp]
\begin{center}
\includegraphics[width=4in]{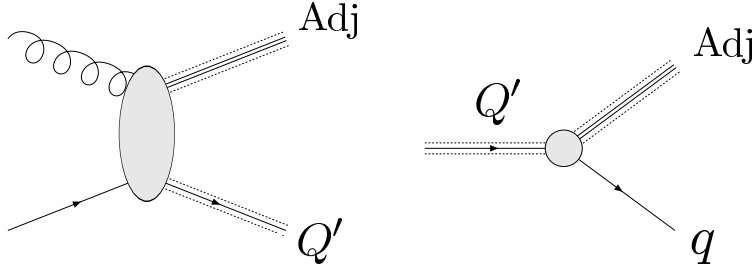}
\caption{A new production and decay mode that we could add to the
Michigan OSET of Figure \ref{fig:OSETDiagrams} to account for the high $H_T$ tail.  In a SUSY scenario where gluino decays are mediated by off-shell squarks, squark-gluino associated production is indeed expected at some level.  If this mode is the explanation for the high $H_T$ tail, the Wino LSP hypothesis from Figure \ref{fig:susyPictures} would likely be ruled out, as it would be difficult to explain why first- and second-generation squark-gluino associated production was visible, even though the gluino decays favor the third-generation.}
\label{fig:MichiganMore}
\end{center}
\end{figure}

Among new processes, two well-motivated
ones deserve particular attention.  The three-body $Adj$ decays must
be mediated by new color triplets.  Assuming flavor isn't badly broken,
these are bottom- and/or top-type $B'/T'$; they can be pair-produced,
but associated production with $Adj$ is probably quite suppressed.
Approximate flavor universality would again suggest light-quark
partners $Q'$, and these could be produced in association with
quarks as in Figure \ref{fig:MichiganMore}.

The former possibility can of course be evaluated by comparing the
high-$H_T$ data to model Monte Carlo, but the matrix element
parameterization of the previous section allows a similar analysis
within an OSET.  For clarity, we will study this smaller effect in $5
\mbox{ fb}^{-1}$ of data.  This is not a realistic discovery threshold, but
the point at which the differences between the models we consider are
apparent, in the absence of backgrounds, through our simple
observables.

\begin{figure}[tbp]
\begin{center}
\includegraphics[width=3in]{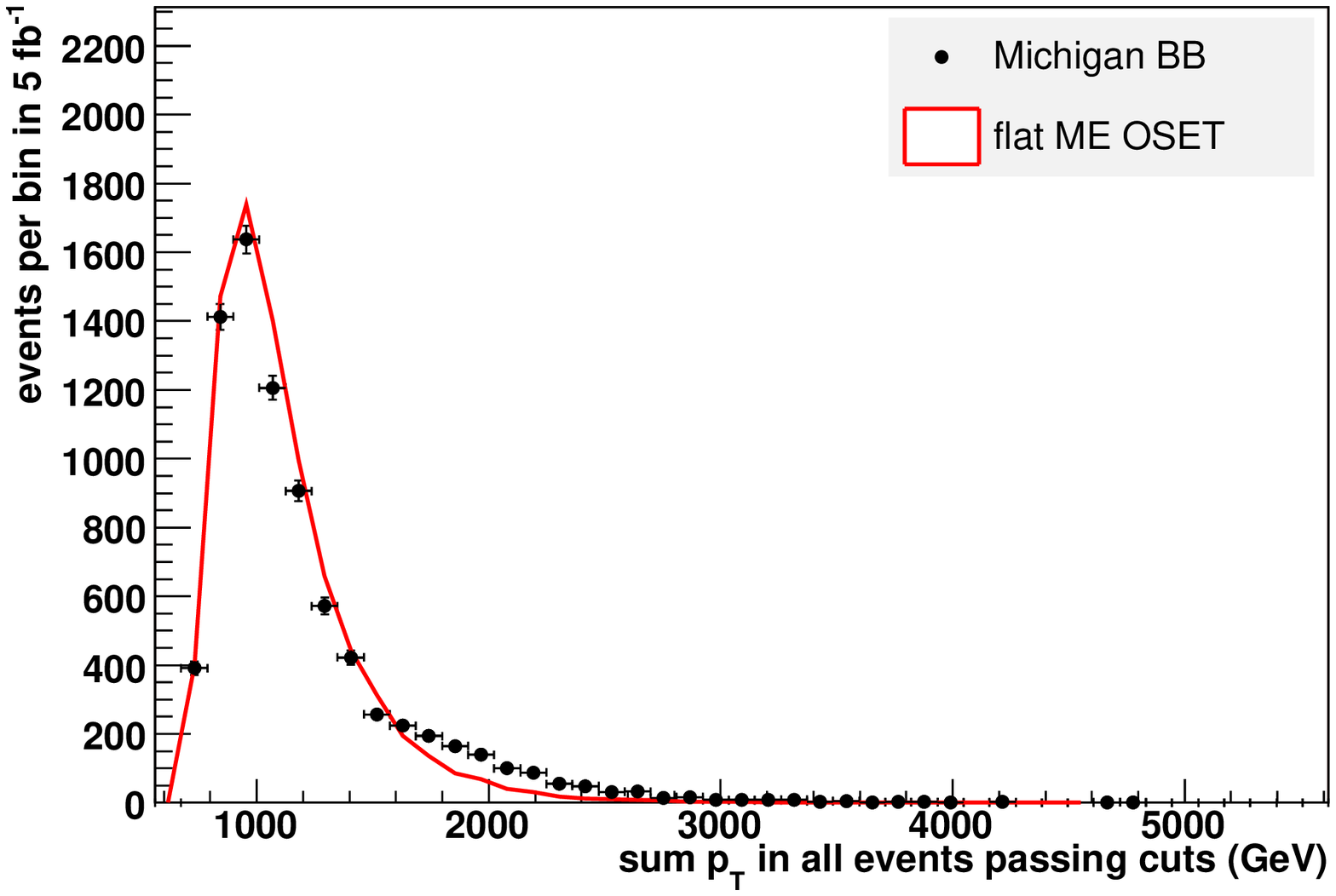}
\includegraphics[width=3in]{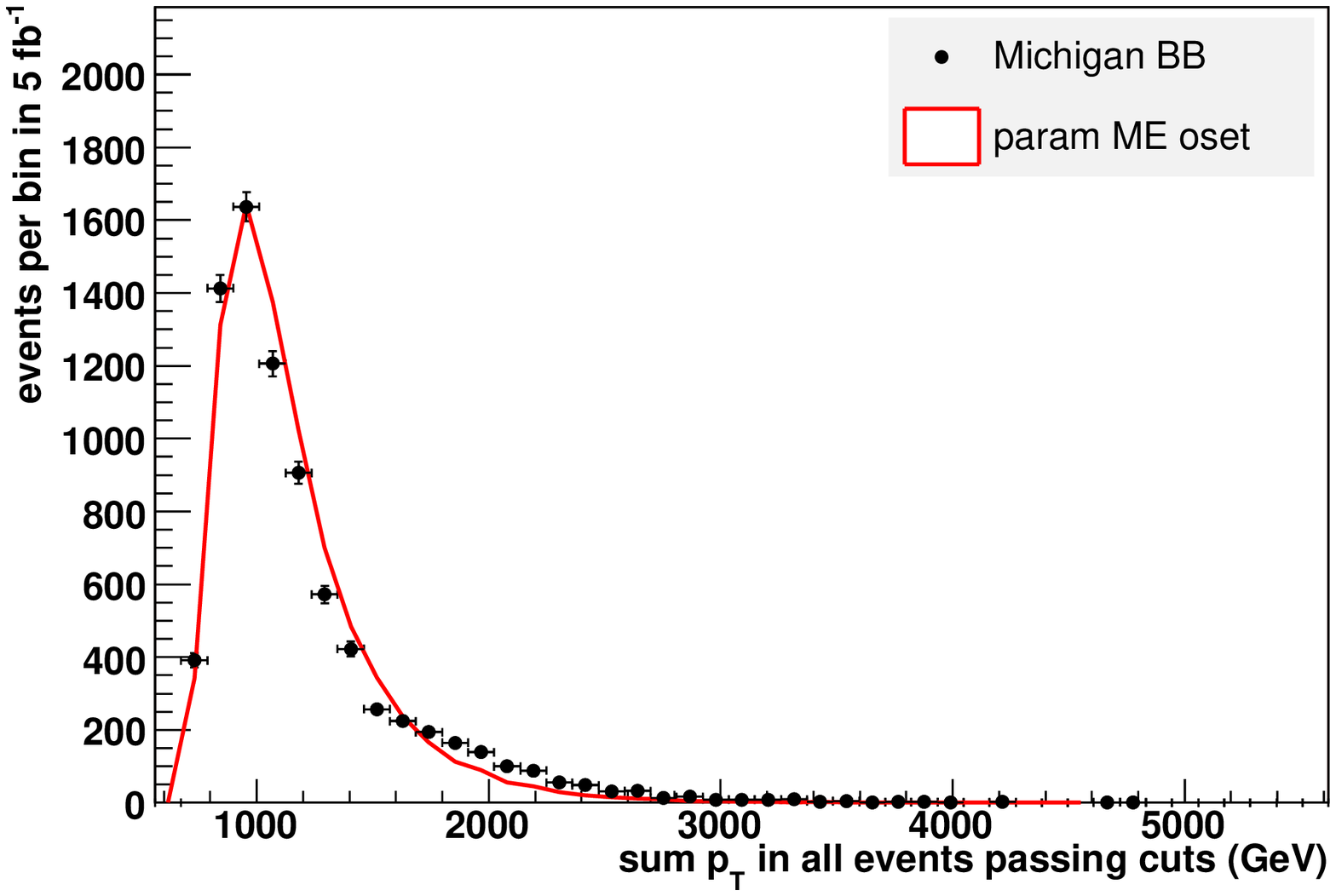}
\caption{\label{fig:tail}Results of a fit to $H_T$ and lepton/jet/btag
  counts in the original OSET of Figure \ref{fig:OSETDiagrams} (left)
  and in an OSET extended to include higher-dimension contact terms
  $|\mathcal{M}|^2 \sim X, X^2$ in the $Adj$ pair-production process (right).  The contact term hypothesis does not seem to adequately address the $p_T$ spectrum.
  }
\end{center}
\end{figure}

\begin{figure}[tbp]
\begin{center}
\includegraphics[width=3in]{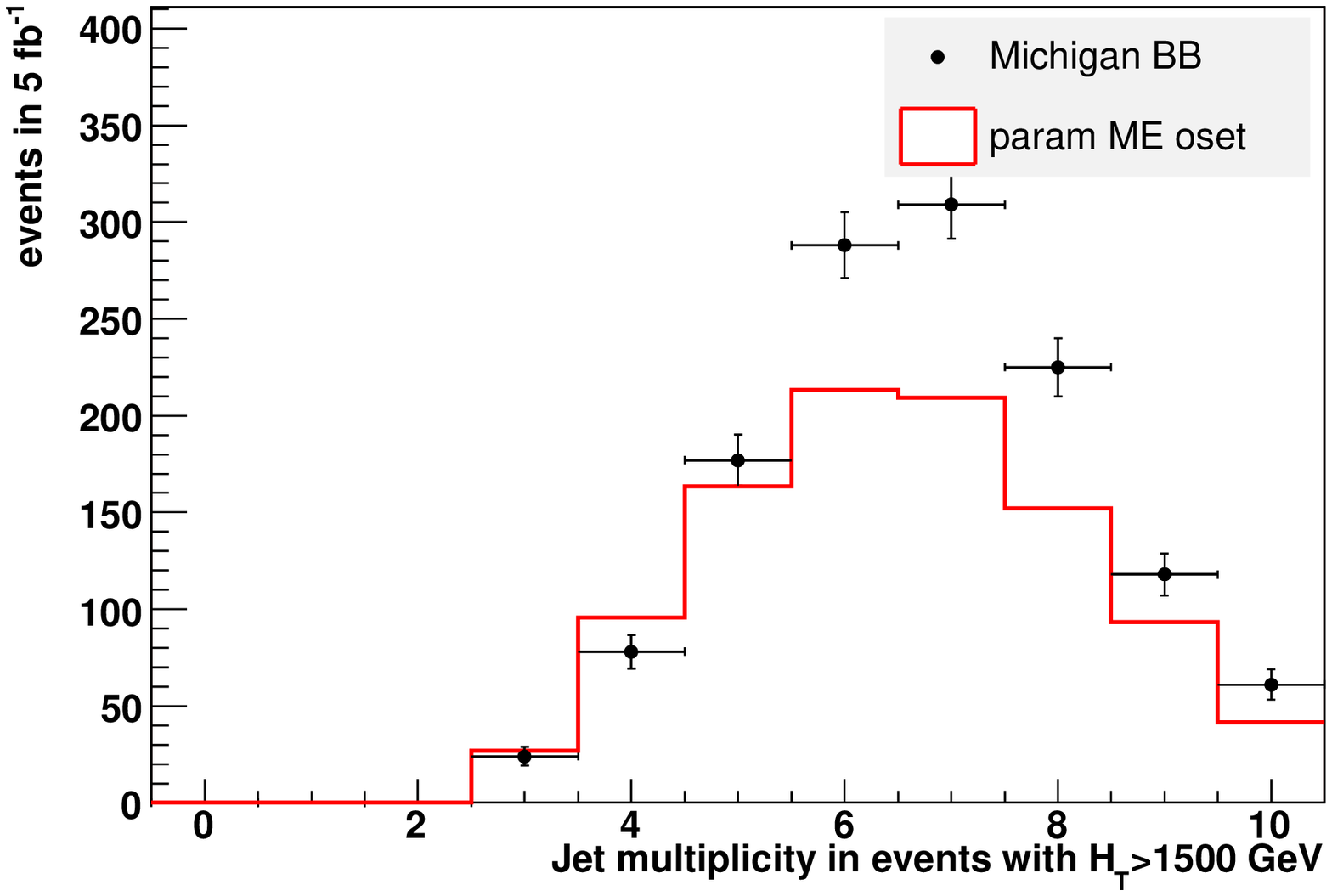}
\includegraphics[width=3in]{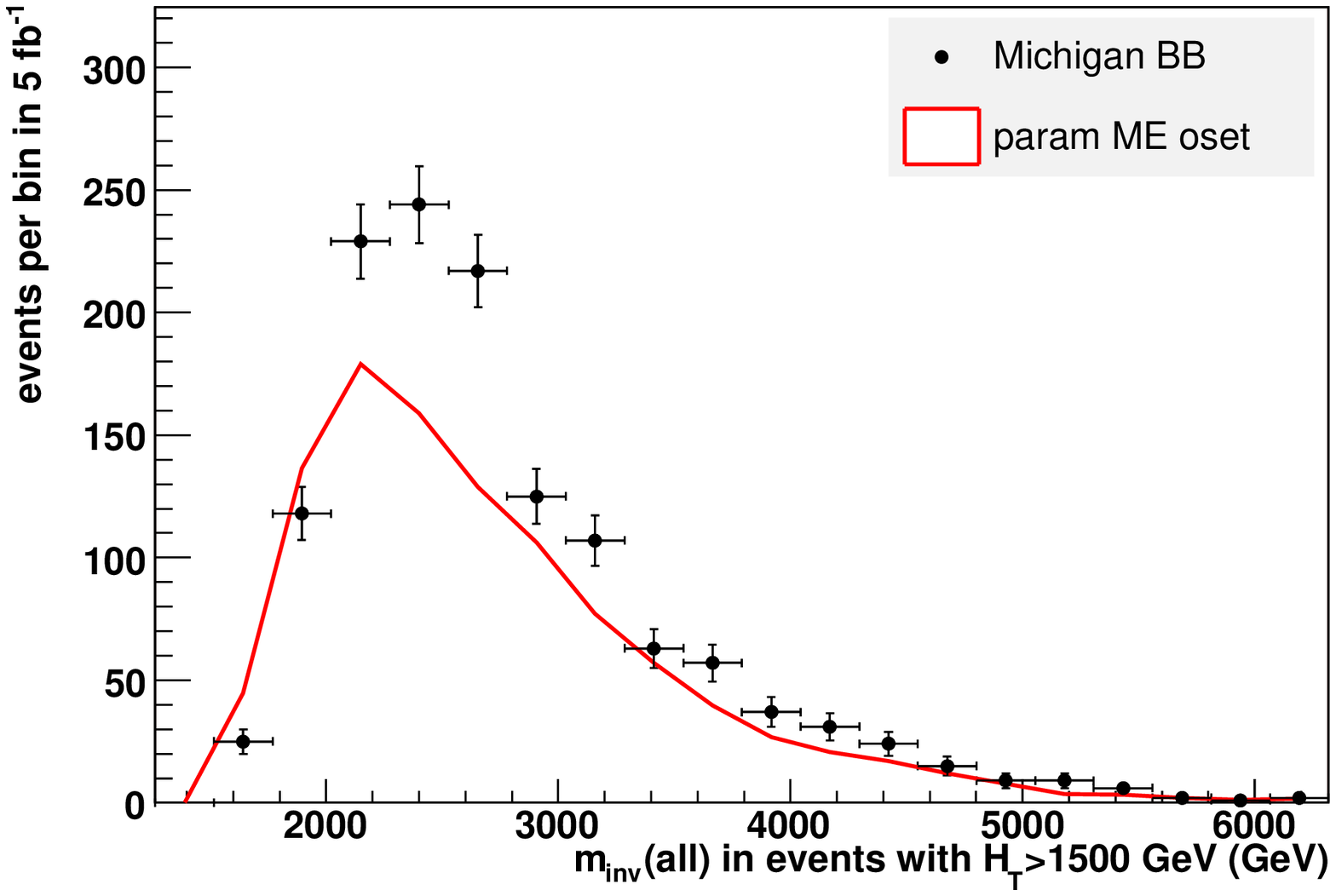}

\includegraphics[width=3in]{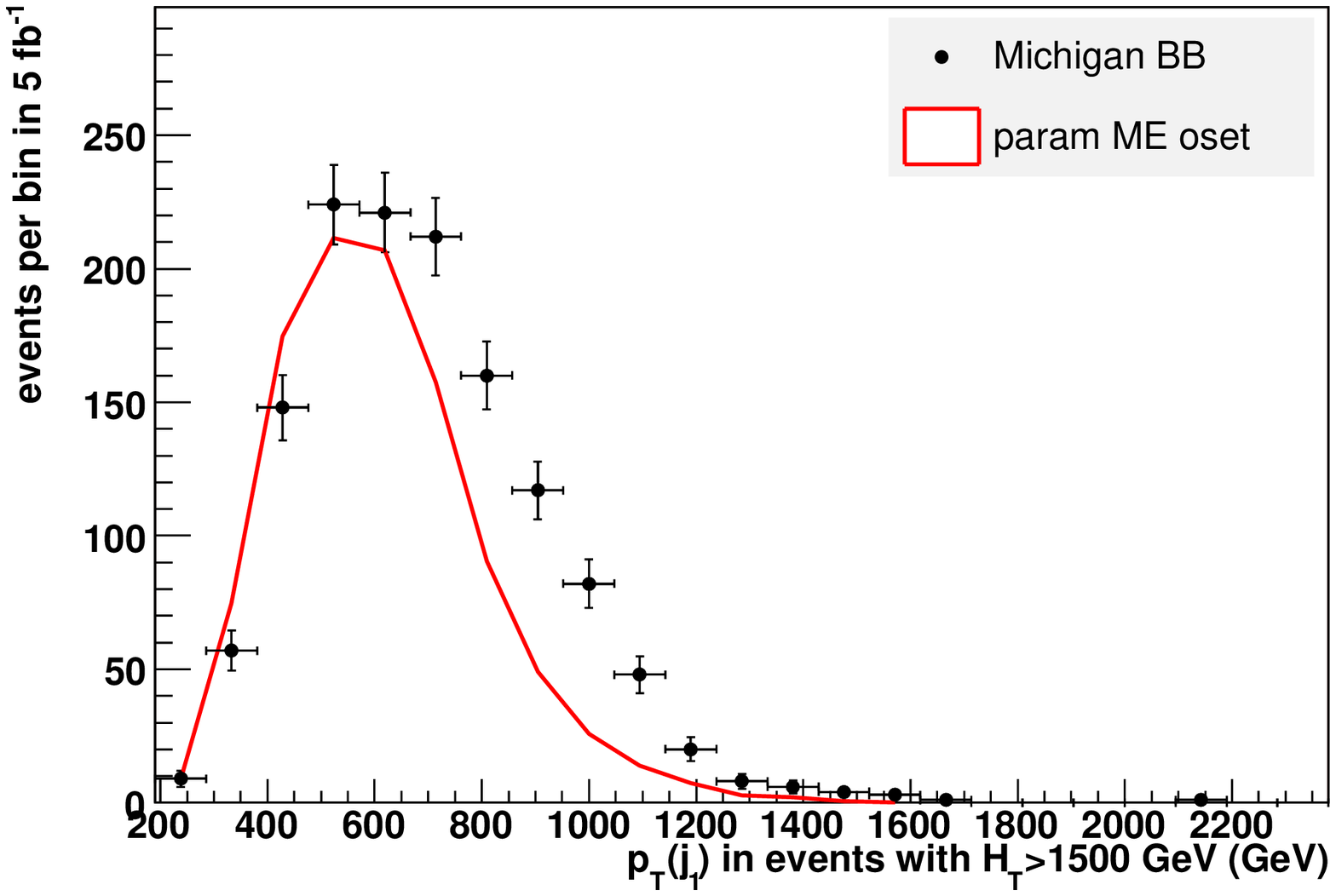}
\includegraphics[width=3in]{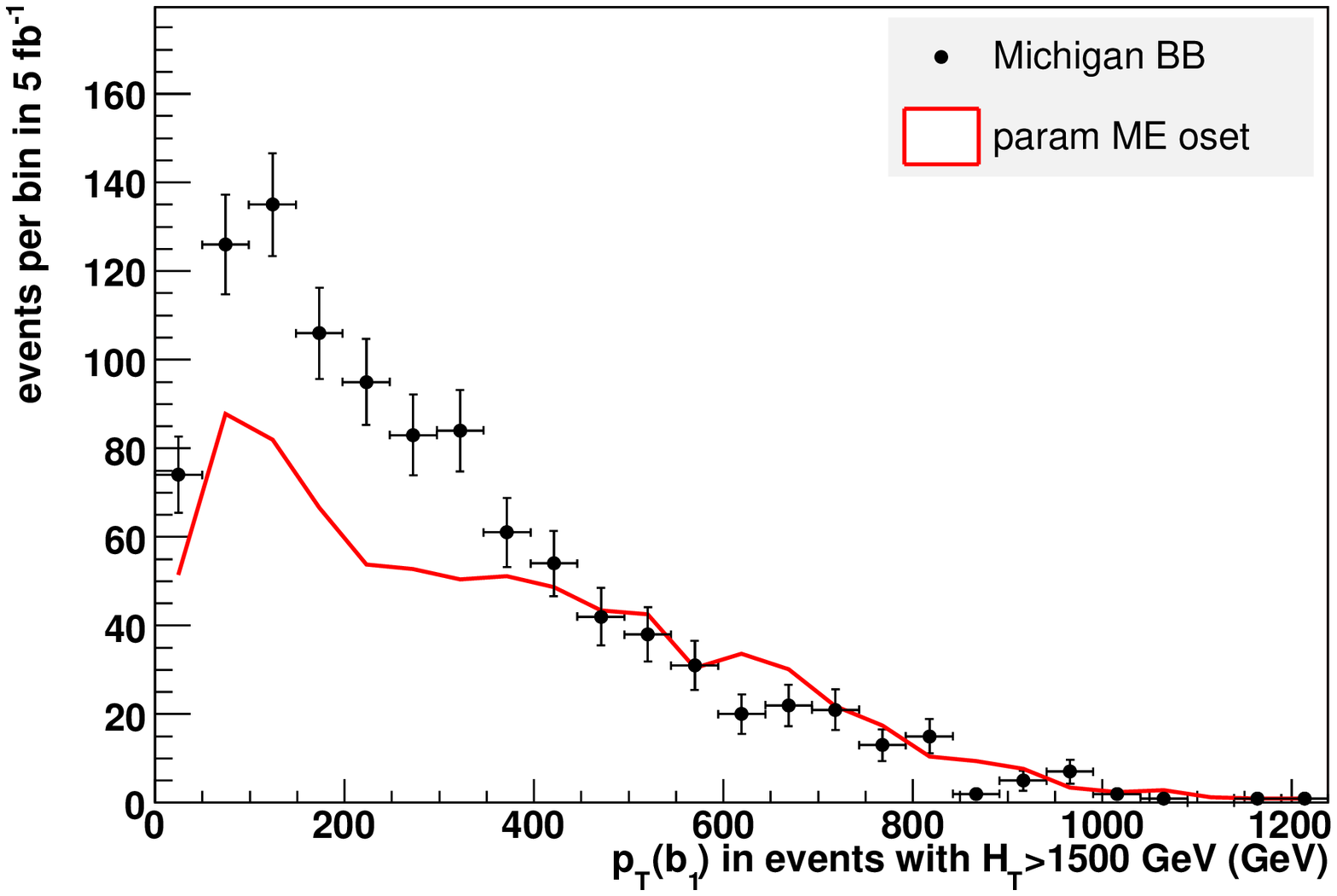}
\caption{\label{fig:de_tail}Results of a fit to $H_T$ and lepton/jet/btag
  counts in the improved OSET with higher-dimension contact terms
  $|\mathcal{M}|^2 \sim X, X^2$ in the $Adj$ pair-production process.
  We have focused on events in the region of the apparent excess,
  $H_T > 1500 \mbox{ GeV}$.  Top: jet multiplicities (left) and invariant mass of all reconstructed objects (right).  On bottom, $p_T(j_1)$ (left) and 
  $p_T(b_1)$ (right).  Neither multiplicity information nor $p_T$ spectra favor the contact term hypothesis.}
\end{center}
\end{figure}

Figure \ref{fig:tail} illustrates the discrepancy between our OSET fit
with flat matrix elements and the Black Box ``data'' signal, as well
as the distribution in an OSET where we have attempted to absorb the
tail into a more general matrix element.  We have added terms $\sim X,
X^2$ (in the parameterization of Section \ref{sec2:Production}) to the
production matrix element, as these are the terms that fall slowest,
even well above threshold.  We also examine the high-$p_T$ tails of
the improved OSET in more detail in Figure \ref{fig:de_tail}.  In fact this is
not enough---because we know that modifying the matrix element shifts
the peaks of $p_T$ distributions (and hence of $H_T$) as well as the
tail.  Therefore, we also search for OSETs that compensate for this effect by lowering
$m_{\it Adj}$.

\begin{figure}
\begin{center}
\includegraphics[width=3in]{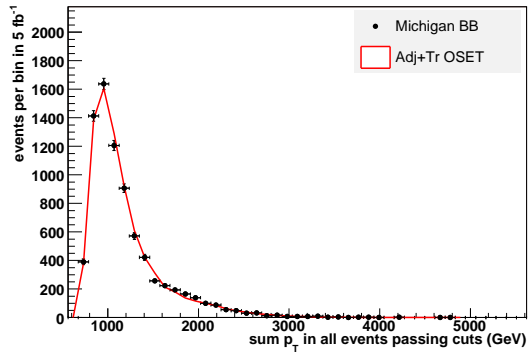}
\caption{\label{fig:aaat} Once the $Q'$--${\it Adj}$ associated channel from Figure \ref{fig:MichiganMore} is added, the $H_T$ distribution is in very good agreement with the Black Box data.}
\end{center}
\end{figure}

\begin{figure}[tbp]
\begin{center}
\includegraphics[width=3in]{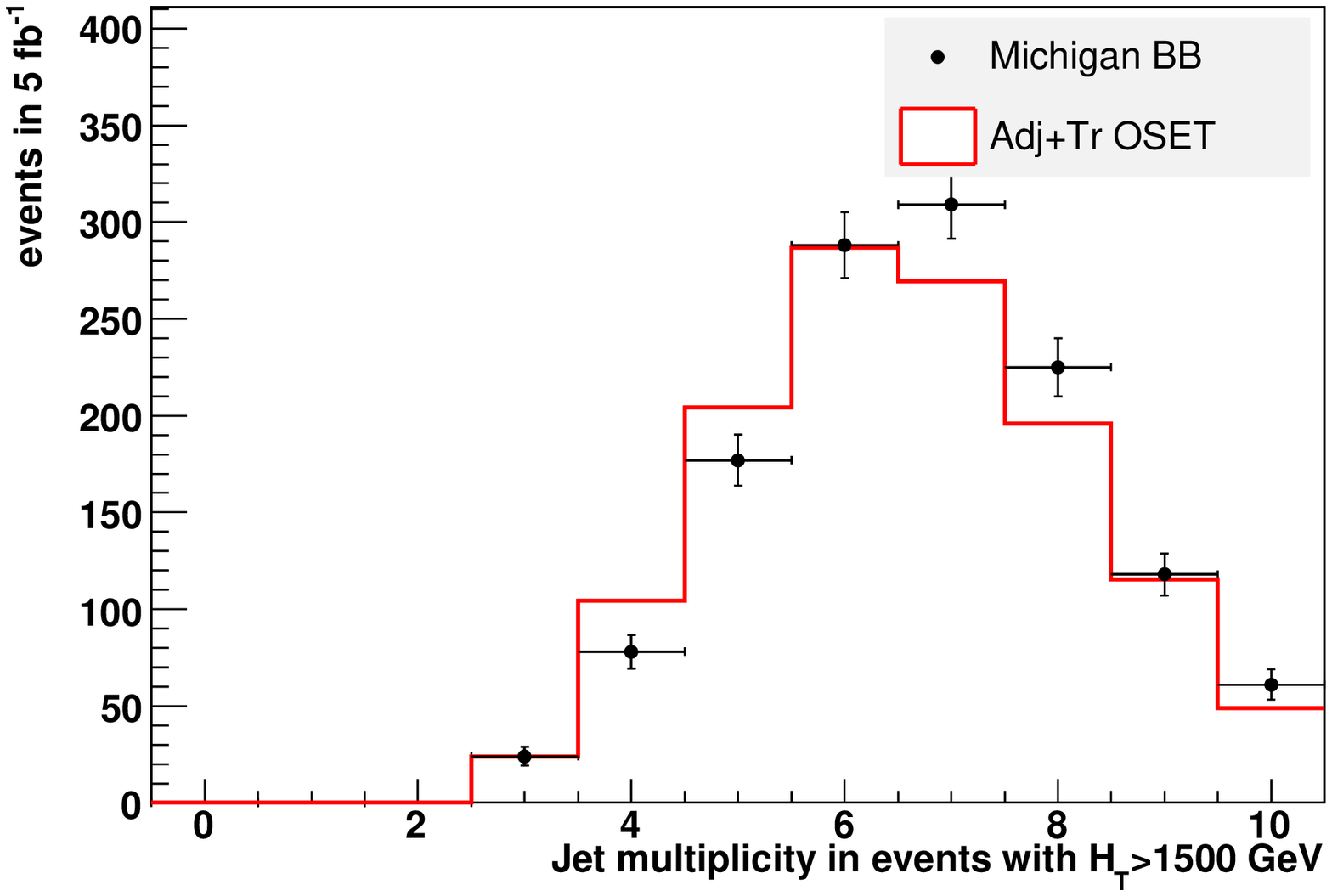}
\includegraphics[width=3in]{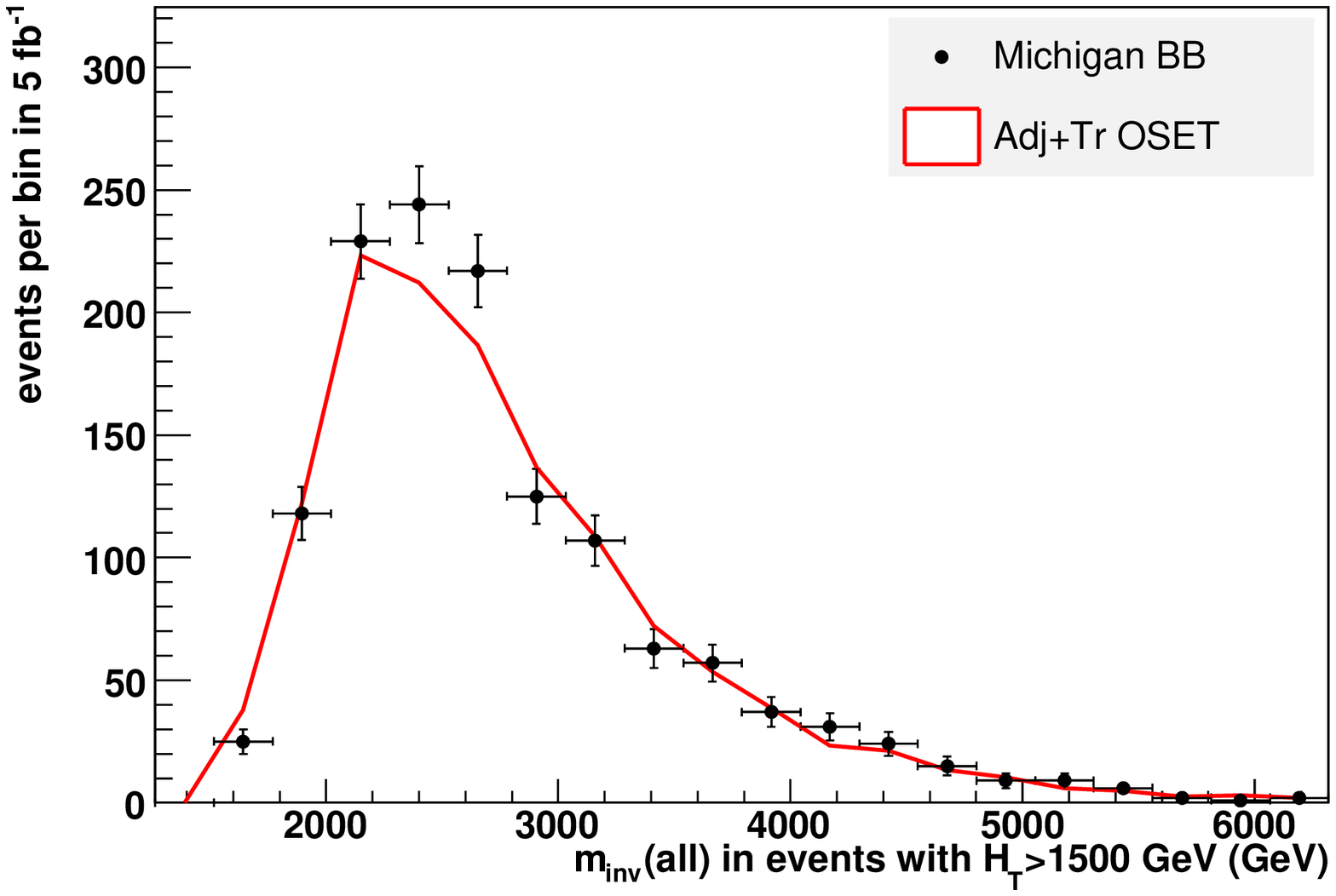}

\includegraphics[width=3in]{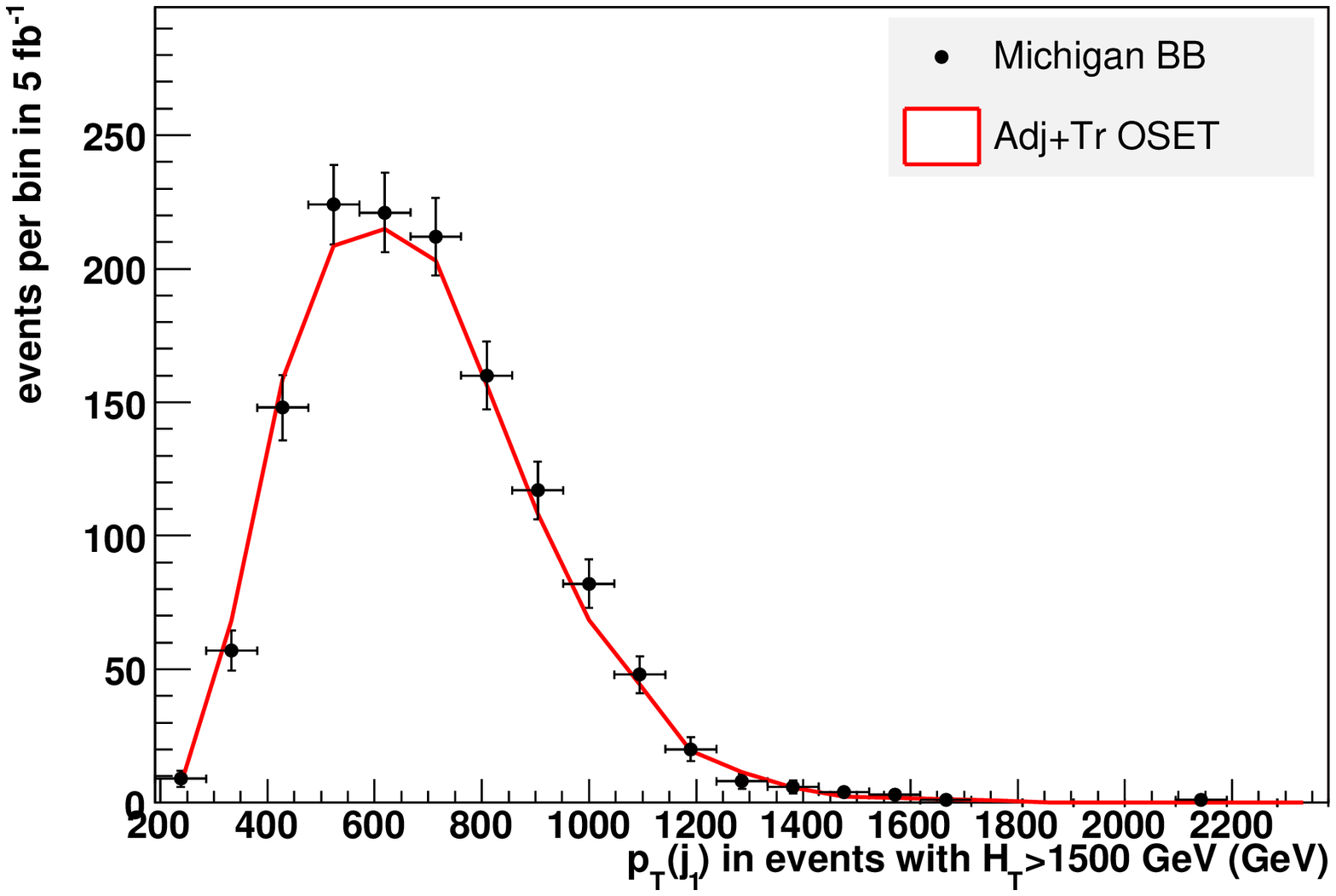}
\includegraphics[width=3in]{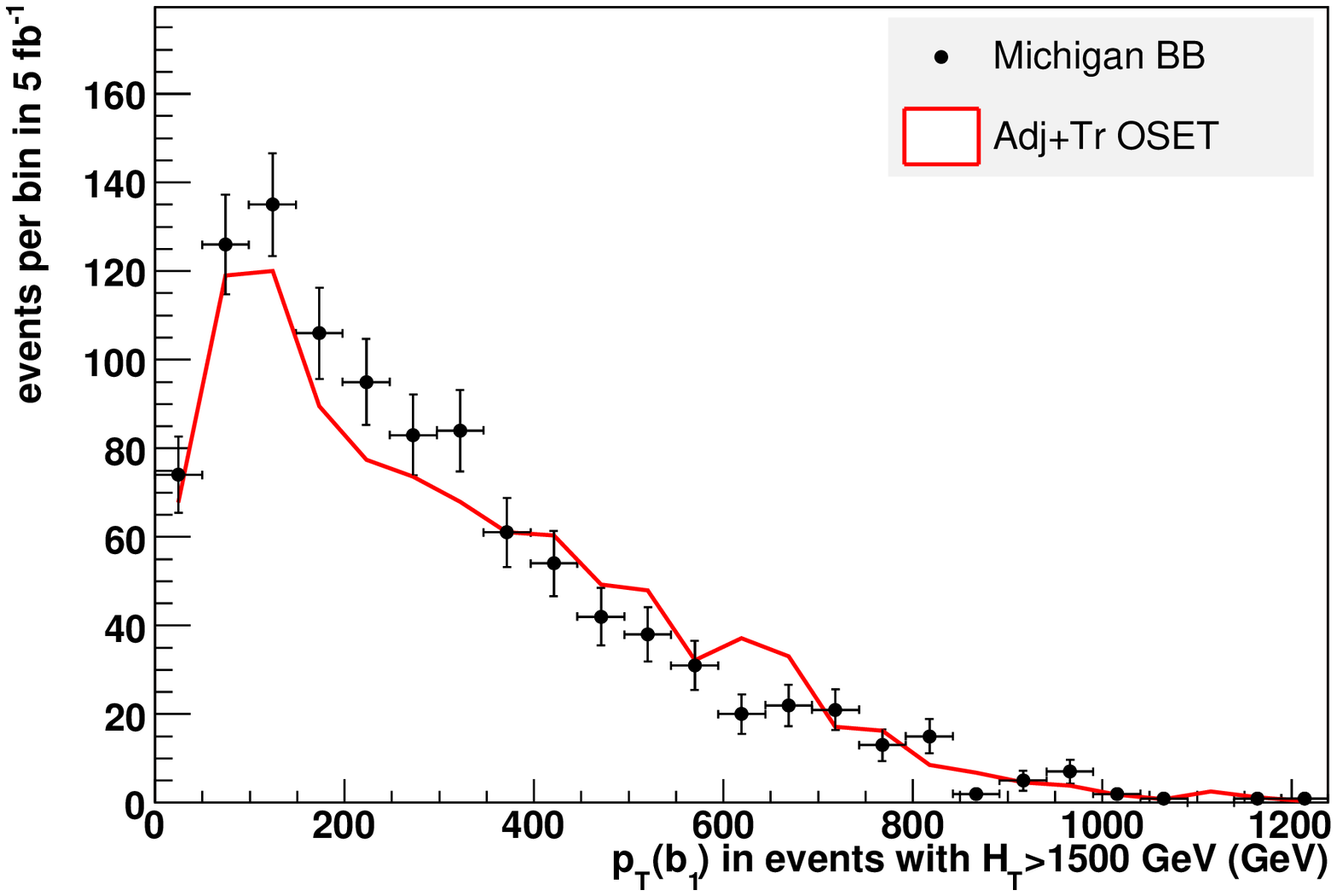}
\caption{\label{fig:aaat_tail}Results of a fit to $H_T$ and
  lepton/jet/btag counts in an OSET with associated production of a
  heavy triplet added.  We have focused on the region of the apparent
  excess, $H_T > 1500 \mbox{ GeV}$.  On top, inclusive jet
  counts and invariant mass of all reconstructed objects.  On bottom, $p_T(j_1)$ (left) and $p_T(b_1)$ (right).  The $Q'$--${\it Adj}$ channel qualitatively accounts for the high $H_T$ excess.}
\end{center}
\end{figure}

We note a few features of the events on this tail that are not
consistent with a matrix element correction: these events have higher
jet counts, but comparable $b$-tagged counts to the OSET.  The leading
jet is harder than in the OSET, while the leading $b$-jet is softer.
Such an asymmetry can only come from a new production process.
Both of these facts suggest that the events on the tail come from a process
that produces an extra light-flavor jet---given our previous theory
prejudice, a heavy triplet in associated production with $Adj$ is an
excellent guess.  We have added the 1870 GeV triplet $Q'$ to the
model, which decays to $u + {\it Adj}$ (here $u$ is a stand-in for any
light-generation quark).  The overall $H_T$ distribution for a fit to
this OSET is shown in Figure \ref{fig:aaat}.  The plots restricted to the
high-$H_T$ tail are shown in Figure \ref{fig:aaat_tail}.  Comparing
Figures \ref{fig:de_tail} and \ref{fig:aaat}, we conclude that the high
$H_T$ excess is more consistent with the additional processes suggested
in Figure \ref{fig:MichiganMore} than with a mismodeling of the adjoint
pair production matrix element.  A final rate fit with the new channel added is given in Table \ref{table:ratesat5}.
\begin{table}[tbp]
\begin{center}
\begin{tabular}{|c|c|}
\hline
\multicolumn{2}{|c|}{OSET Spectrum}\\
\hline
$Q'$ & 1870 GeV \\
${\it Adj}$ & 450 GeV \\
${\it Ch}$ & 128 GeV \\
${\it Ne}$ & 124 GeV \\ \hline
\end{tabular}
\caption{A candidate spectrum for the OSET modification in Figure \ref{fig:MichiganMore}.}
\end{center}
\end{table}

\begin{figure}[tbp]
\includegraphics[width=3in]{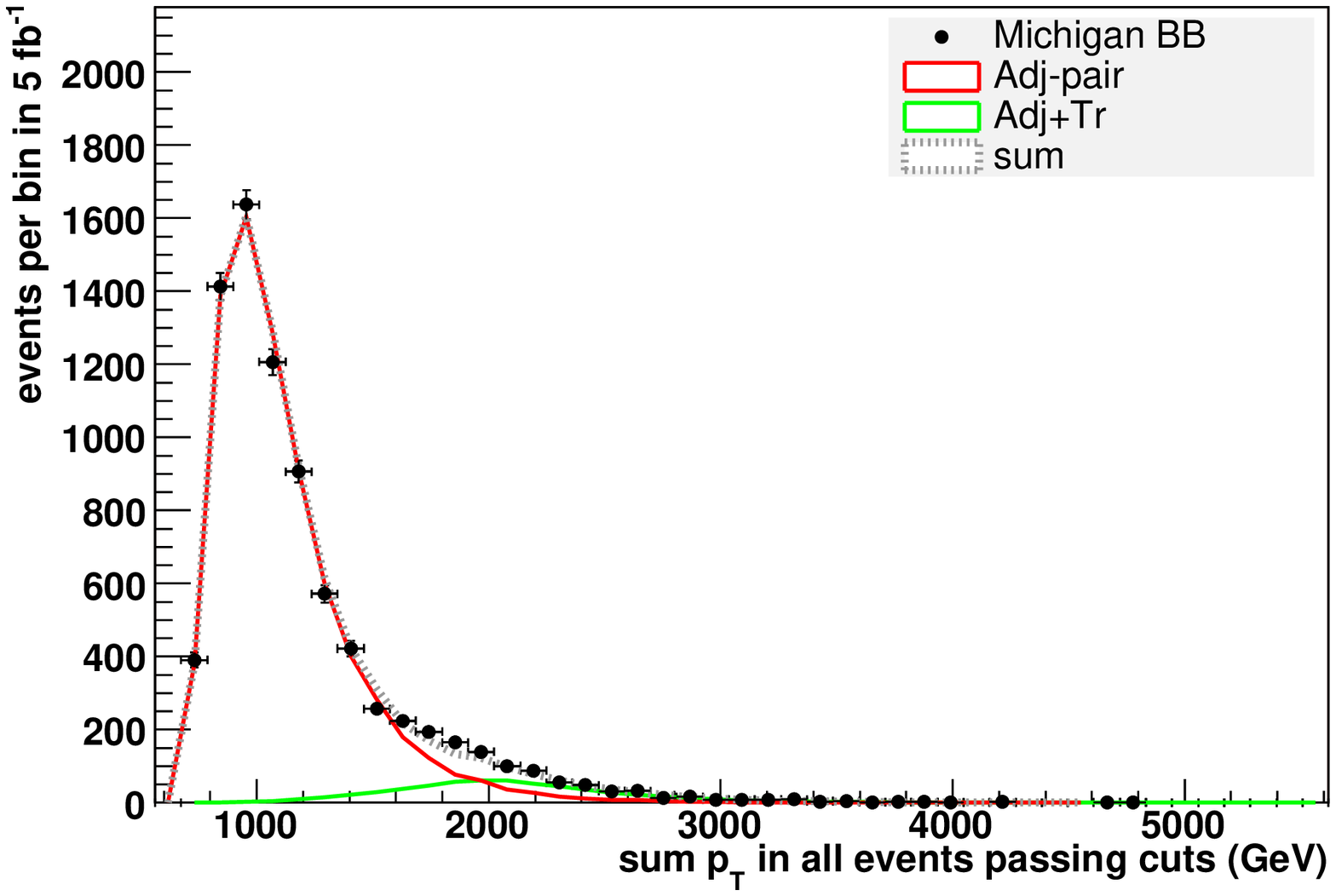}
\includegraphics[width=3in]{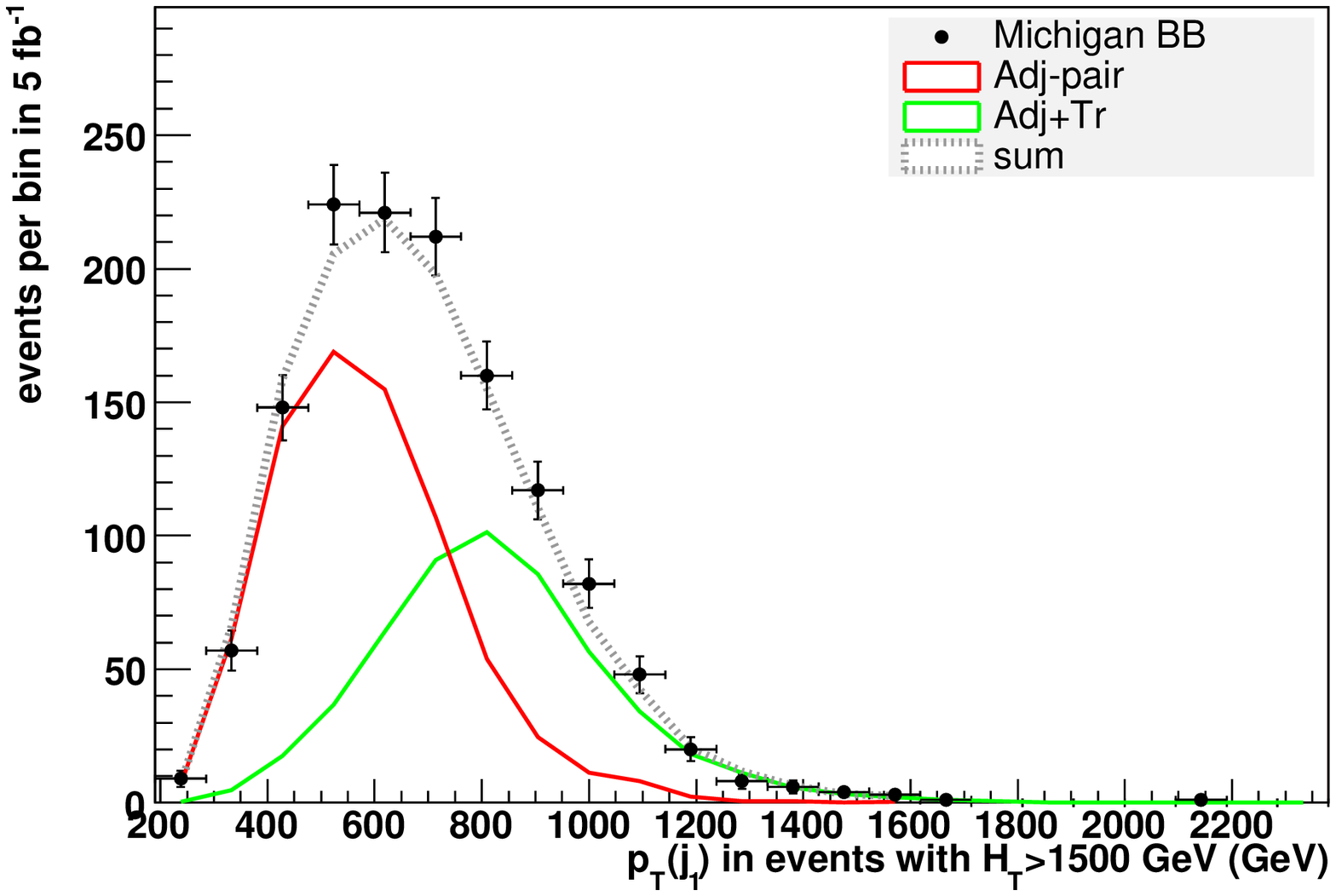}
\caption{\label{fig:newprocess}Error bars: $H_T$ distribution for
  Michigan Black Box data.  The colored lines are different production
  modes in an OSET fit to the data: $Adj$-pair production (red) and
  ${\it Adj}$--$Q'$ associated production (green).  The sum contribution (dashed) gives a good overall fit to the data.}
\end{figure}

\begin{table}
\begin{center}
\begin{tabular}{|r|c|c|}
\hline
\textbf{Process} & \textbf{Fit Rate} & \textbf{Actual Rate} \\
\hline
$\sigma(g g \rightarrow {\it Adj} \, {\it Adj})$ & 30.1 $\pm$ 0.9 fb&  28.0 fb\\
$\sigma(g u \rightarrow {\it Adj} \, Q')$ & 0.31 $\pm$ 0.04 fb&  0.41 fb\\
\hline
$\mbox{Br}(Q' \rightarrow u {\it Adj})$ & 1.0 & 1.0 \\
\hline
$\mbox{Br}(Adj \rightarrow \bar{t} \bar b {\it Ch}^+ \mbox{ or } c.c.)$ & 0.82 $\pm$ 0.03 & 0.77 \\
$\mbox{Br}(Adj \rightarrow b \bar b {\it Ne})$ & 0.17 $\pm$ 0.02 & 0.22 \\
$\mbox{Br}(Adj \rightarrow q \bar q {\it Ne})$ & 0.01 $\pm$ 0.01 & 0.01 \\
\hline
$\mbox{Br}({\it Ch} \rightarrow q \bar q' {\it Ne})$ & 0.56 $\pm$ 0.10 & 0.60 \\
$\mbox{Br}({\it Ch} \rightarrow e/\mu \bar \nu {\it Ne})$ & 0.43 $\pm$ 0.10 & 0.40 \\
\hline
\end{tabular}
\end{center}
\caption{\label{table:ratesat5}Fit results at $5 \mbox{ fb}^{-1}$, including the ${\it Adj}$--$Q'$ associated production channel. Error bars quoted are for
  uncorrelated modification of parameters subject to constraints of
  the form $\sum_X \mbox{Br}(Adj \rightarrow X)=1$.  While there appears to be a 5\% systematic error in the ${\it Adj}$ branching ratios, the qualitative agreement between the actual rates and the OSET parameterization is encouraging.  Note that with increased statistics, the best fit total cross section is closer to the correct value compared to Table \ref{table:rates}.}
\end{table}

\subsection{OSETs and Data Characterization}
\label{sec:subconclusion}
The process of zeroing in on the correct theoretical description of an
unexplained signal, whether it arises from new physics or from the
Standard Model, will not be easy.  No characterization scheme can remove the
fundamental difficulties associated with uncertainties in Standard
Model implementation, nor the loss of information inherent in hadronic
collisions and decays.  These questions are hard to answer, but the
OSET description has at least facilitated asking them.

\begin{figure}[tbp]
\includegraphics[width=6in]{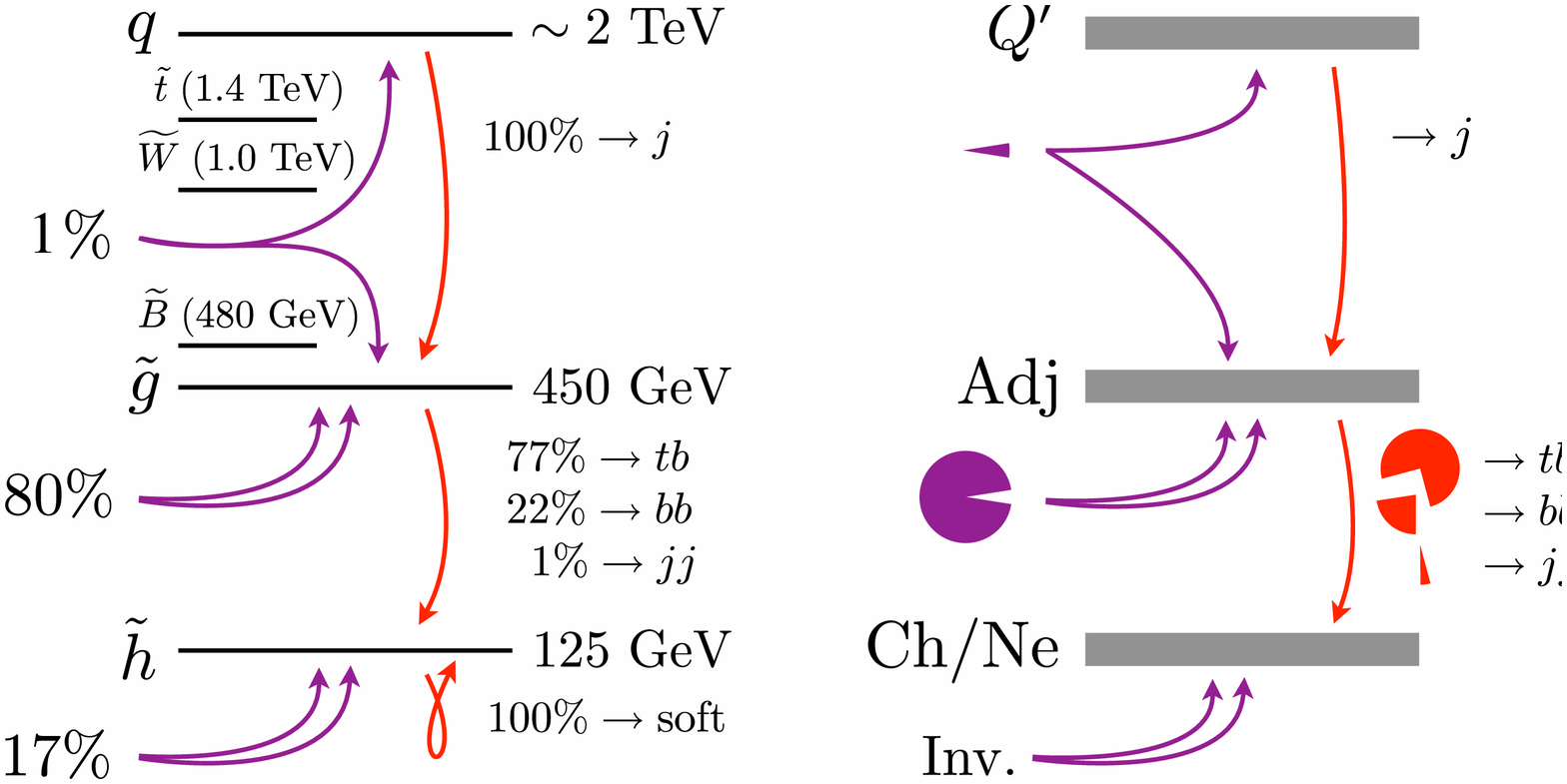}
\caption{\label{fig:qualitativeMich}Left: graphical depiction of the
  spectrum, production and decay modes for the Michigan Black Box.
  Right: OSET fits have given a qualitatively correct picture of the
  underlying physics.}
\end{figure}

The success of simple rate-fitting techniques in this example should
be viewed skeptically---the Standard Model background is absent, the
detector simulator agrees perfectly with the detector, and systematic
errors in this approach are difficult to quantify.  The primary
benefit of an OSET fit is in rapidly determining key
\emph{qualitative} features of new physics---the types of new
particles, their approximate masses, and which decay modes dominate,
as represented in Figure \ref{fig:qualitativeMich}. The agreement is striking.
Even the qualitative level of detail of this description is enough to sharply focus our model-building intuition. The first OSET fit
has guided us to two possibly consistent MSSM scenarios---a $\widetilde H$
or $\widetilde W$ LSP.  The second relies on a mass splitting $m_{\tilde q}
\gg m_{\tilde t_1}, m_{\tilde b_1}$ to explain the hierarchy of
branching ratios---disfavored by the presence of a channel highly
suggestive of squark-gluino associated production.

%% file: endmatter.tex
\section{Concluding Remarks}
\label{sec:conclusion}
We have shown that On-Shell Effective Theories are
an effective bridge between LHC data and the fundamental theory
at the TeV scale,  in a wide range of possible scenarios for new
physics. This is especially important for theories with a zoo of new particles,
complicated production mechanisms, and long decay chains. MARMOSET is
a simple Monte Carlo tool for directly simulating OSET descriptions
for new physics.  An
OSET motivated by new physics signals captures important
qualitative features of the underlying model
of new physics and strongly constrains its structure.
The examples in this paper illustrate the productive interplay
between physically motivated data characterization and model-building.

The first data from the LHC has the potential to revolutionize our
understanding of fundamental physics.
The collective effort of
experimentalists and theorists should guide us the new Standard
Model. A physically meaningful characterization of collider data
will play a critical role in making this happen. OSETs
and MARMOSET can act as clear conduits of information between
experimentalists and theorists in the era of discovery. In the hands
of experimentalists, MARMOSET can be used to characterize new
physics discoveries and exclusion limits in a transparent way---this sort of information will be invaluable and easily understood by theorists
outside the collaborations. In turn,
theorists can cast their favorite Lagrangians in
terms of an OSET which can be readily compared to data.

An OSET characterization of LHC data is not an end in itself but a
waystation on the path to a fundamental theory. The
OSET framework will only be successful if it leads to
a compelling Lagrangian
that obviates the need for an OSET description. Ultimately,
understanding the structure of this
effective Lagrangian will unveil the new principles of Nature long
anticipated at the weak scale.

\section*{Acknowledgments}

We are indebted to Johan Alwall, Matthew Baumgart, Liam Fitzpatrick,
Tom Hartman, Jared Kaplan, and Itay Yavin for furthering the
development of MARMOSET through their programming efforts.  We thank
Rikard Enberg, Paddy Fox, Roni Harnik, Michele Papucci, and John
Terning for testing MARMOSET in its infancy.

We benefitted greatly from conversations with Dante Amidei, John
Conway, Su Dong, Melissa Franklin, Henry Frisch, Fabiola Gianotti,
Eilam Gross, Joao Guimaraes da Costa, Valerie Halyo, Tao Han, Beate
Heinemann, John Huth, Gordy Kane, Paul Langacker, Patrick Meade,
Michael Peskin, Aaron Pierce, Leonardo Rastelli, Matt Reece, Albert
de Roeck, Maria Spriopulu, Matt Strassler, Scott Thomas, Chris
Tully, Herman Verlinde, Gordon Watts, and Charlie Young. Discussion
of the ``Michigan Black Box'' draws heavily from the work of all
involved in the LHC Olympics, particularly Matt Bowen and Matt
Strassler for their inspiring first assault on the problem, the
members of the Harvard Michigan Black Box team, and the creator of
the black box, Gordy Kane. Special thanks to the entire LHCO
community for stimulating and facilitating the development and study
of numerous data challenges.

The work of N.A.-H. is supported by the DOE under contract
DE-FG02-91ER40654. P.S. and N.T. are each supported by National
Defence Science and Engineering Fellowships. J.T. is supported by a
fellowship from the Miller Institute for Basic Research in Science.
The work of L.-T.W. is supported by the National Science Foundation
under Grant No. 0243680 and the Department of Energy under grant
DE-FG02-90ER40542.  Any opinions, findings, and conclusions or
recommendations expressed in this material are those of the
author(s) and do not necessarily reflect the views of the National
Science Foundation. S.M. is supported by Fermi Research Alliance,
LLC under Contract No. DE-AC02-07CH11359 with the United States
Department of Energy.

%% file: AppendixProduction.tex
\section{Structure of Non-Resonant $2 \rightarrow 2$ Scattering}
\label{sec2:APP}

In this appendix, we address the physics of $2\rightarrow 2$
production processes. The main point of an OSET parametrization
scheme is to capture the possible kinematic shapes of variables that
are useful for untangling and interpreting new physics. For now, we
will focus on inclusive single object rapidity and $p_T$
distributions. As we'll see and as the reader might
find intuitive, we can tremendously improve our scheme over the
constant approximation by including the leading order near-threshold
behavior of matrix elements. Less intuitive and more useful to
exploit is that for rapidity and $p_T$ variables, there is a large
shape degeneracy among the possible matrix element behaviors. When
we exploit these shape degeneracies, we will find a simple
parametrization scheme that captures both rapidity and $p_T$
shapes.  Empirical results appear in Appendix~\ref{sec:empirical2to2} which
are more fully justified in Appendix~\ref{sec2:ShapeInv}.

\subsection{Parametrizing $2\rightarrow 2$ Particle Production}
\label{sec2:2to2}

Let's start by introducing some notation and basic formulas to
facilitate our analysis. For a general $a+b\rightarrow c+d$ process
(we assume $a$ and $b$ are massless), the kinematics can be
completely described in terms of the familiar Mandelstam variables
$\sh=(p_a+p_b)^2$, $\that=(p_a-p_c)^2$, and $\uh=(p_a-p_d)^2$, where
$p_{a,b,c,d}$ are the four-vectors for particles $a$, $b$, $c$, and $d$. If we are
working in the center-of-mass frame, it is convenient to express
both $\that$ and $\uh$ in terms of $\sh$ and $\xi\equiv \beta
\cos(\theta)$ as,
\begin{equation}
\label{eq:def:thatuhat}
  \that = -\f{1}{2}\left((\sh-m_c^2-m_d^2)-\sh\xi \right),
  \qquad \uh = -\f{1}{2}\left((\sh-m_c^2-m_d^2)+\sh\xi\right) ,
\end{equation}
where $\theta$ is the center-of-mass scattering angle, $m_{c,d}$ are the final
state masses, and
\begin{equation}
\beta^2=\left(1-\f{m_c^2}{\sh}-\f{m_d^2}{\sh}\right)^2-4\f{m_c^2}{\sh}\f{m_d^2}{\sh}.
\end{equation}


In the lab frame, it is convenient to express the final state
kinematics in term of the transverse momentum squared
\begin{equation}
p_T^2=\f{\that\uh-m_c^2m_d^2}{\sh},
\end{equation}
as well as the rapidity $y$ of one of the final states (say
particle $c$):
\begin{equation}
y=\f{1}{2}\log \left(\f{E_c+p_{z,c}}{E_c-p_{z,c}}\right),
\end{equation}
where $E,p_z$ are evaluated in the lab frame. As is well known,
rapidity is additive under longitudinal boosts, so it is often
easier to work in terms of the center of mass rapidity $\hat y$
and then boost by the initial state rapidity $\bar y$ to get back to
the lab frame.

In what follows, we will be interested in computing differential
distributions for the transverse momentum $\f{d\sigma}{d p_T}$ as
well as the rapidity $\f{d\sigma}{dy}$. Following the usual parton
model for hadronic collisions, we start with the full differential
cross section
\begin{equation}\label{eq:TotalXSection}
    \sigma(a+b\rightarrow c+d) = \int \! d\bar y  d\tau  d\that \;
    \f{f_a(x_a,Q^2)f_b(x_b,Q^2)}{\sh^2}
\left(\sh^2\f{d\hat{\sigma}}{d\that}\right),
\end{equation}
where the $f_{a,b}(x_{a,b},Q^2)$ are the parton distribution
functions, $Q^2$ is the momentum transfer squared of the process,
$x_{a,b}$ are the longitudinal momentum fractions of incoming
particle $a$ and $b$ respectively, $\tau=x_a x_b$, and $\bar
y=\f{1}{2}\log \left(\f{x_a}{x_b}\right)$. The squared matrix element
for any $2\rightarrow 2$ process is proportional to the differential
cross section for scattering, $|\mathcal{M}|^2 \sim
\sh^2\f{d\hat{\sigma}}{d\that}$, hence we've isolated this piece in
Eq.~(\ref{eq:TotalXSection}).

For convenience, we adopt a set of dimensionless variables with which to
express our results. For the process $a+b\rightarrow c+d$, let's
define dimensionless $\sh$ and $p_T^2$ variables as
\begin{equation}
X \equiv \f{\sh}{s_0}, \qquad x_T^2 \equiv \f{4p_T^2}{s_0},
\end{equation}
where $s_0=(m_c+m_d)^2$ is the threshold center-of-mass energy squared.
We also introduce final state mass asymmetry parameters $\Delta$ and $\Delta^{\prime}$ as
\begin{equation}
  \Delta \equiv 1-\f{4m_cm_d}{s_0},\qquad \Delta^{\prime} \equiv \f{m_c^2-m_d^2}{s_0} ,
\end{equation}
so that $\Delta,\Delta^{\prime}\rightarrow 0$ as $m_c\rightarrow
m_d$. In terms of our dimensionless variables, we can express the
kinematic quantities $\xi^2$ and $\beta^2$:
\begin{eqnarray}\label{eq:XiBetaSqRap}
  \xi^2 &=& 1-\f{1+x_T^2+\Delta}{X}+\f{\Delta}{X^2} \; =\;
  \left(1+\f{\Delta^{\prime}}{X}\right)^2\tanh(\hat y)^2, \nonumber \\
  \beta^2 &=& \xi^2+\f{x_T^2}{X}.
\end{eqnarray}
The lower boundary $X_{\rm min}(x_T,\Delta)$ of phase space occurs
when $\xi = 0$, and the upper boundary $X_{\rm max}$ depends on the
 beam center-of-mass energy squared $s_b$, yielding
\begin{equation}
X_{\rm min}(x_T,\Delta) = \f{(1+\Delta+x_T^2)+\sqrt{(1+\Delta+x_T^2)^2-4\Delta}}{2} ,
\qquad X_{\rm max}    = \f{s_b}{s_0}.
\end{equation}
Note that $1\leq X_{\rm min}(x_T,\Delta)$.  The longitudinal
momentum fractions can be expressed as
\begin{equation}
x_a=e^{\bar{y}} \sqrt{\f{s_0}{s_b}X}, \qquad x_b=e^{-\bar{y}} \sqrt{\f{s_0}{s_b}X},
\end{equation}
and we see that the initial state rapidity is constrained between
$-\bar{y}_*(X) \le \bar{y} \le \bar{y}_*(X)$, where
\begin{equation}
\bar{y}_*(X)=\f{1}{2}\log \left(\f{s_b}{s_0 X}\right).
\end{equation}

\begin{figure}[tbp]
\begin{center}
\includegraphics[width=3.5in]{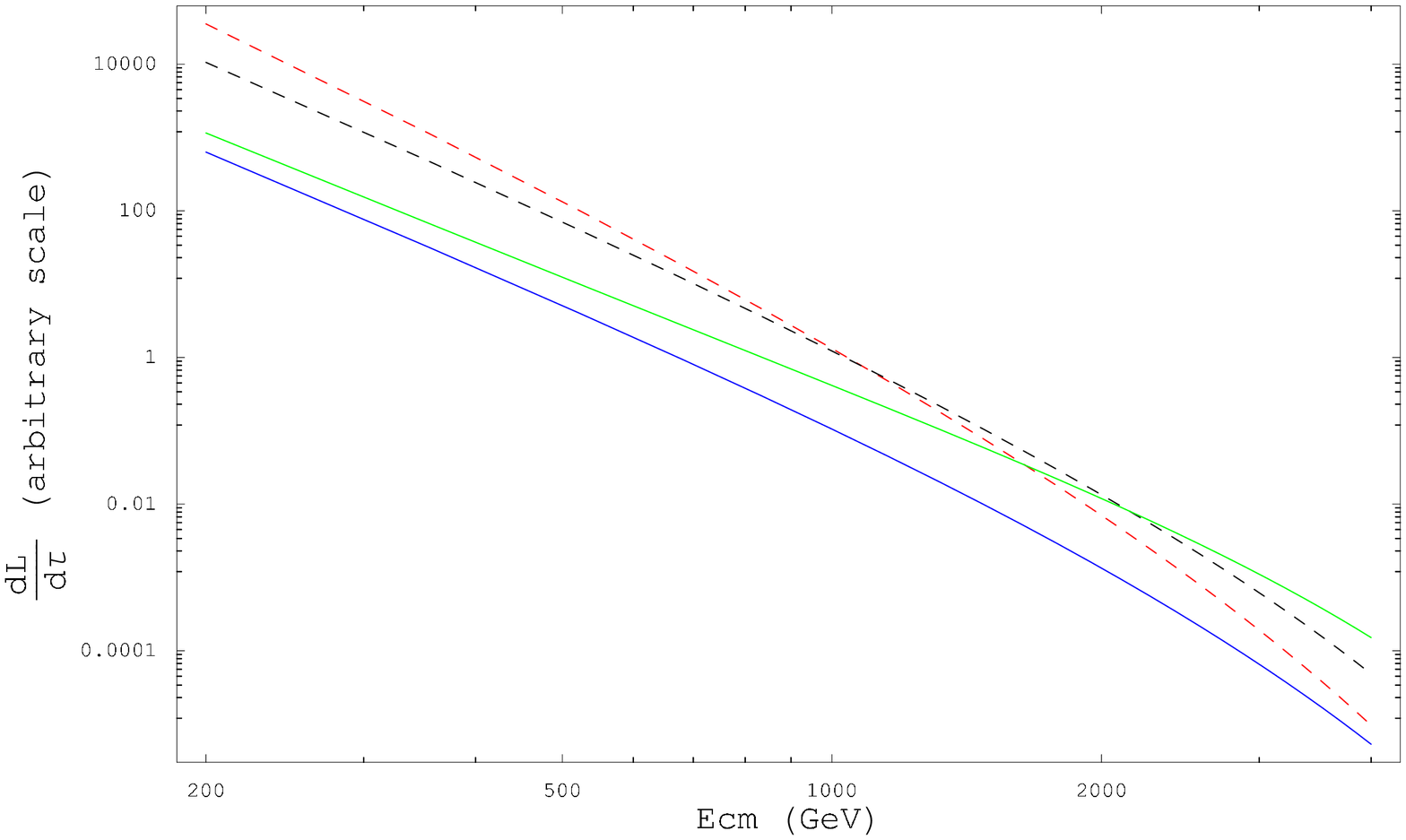}
\caption{Dimensionless ``parton luminosities'' integrated over center-of-mass rapidity at
$Q^2=(500 \mbox{ GeV})^2$.  We use the MRST PDFs \cite{Martin:2002aw}. The overall scale is arbitrary.  From top to bottom starting on the left of the graph are, $gg$, $ug$, $uu$, and
$u\bar{u}$ respectively. Note that locally, the parton luminosities
behave as homogenous functions of $E_{cm}$, especially below $\sim
2$ TeV.} \label{fig:partonLum}
\end{center}
\end{figure}

Before calculating differential distributions, it is convenient to define
dimensionless ``parton luminosity'' functions
\begin{equation}
\rho_{ab}(\bar{y},X,s_0)\equiv\f{x_af_a(x_a)x_bf_b(x_b)}{X^2}.
\end{equation}
Figure \ref{fig:partonLum} displays the integrated ``parton luminosities'' $\rho_{ab}(X,s_0)=\int
d\bar{y}\; \rho_{ab}(\bar{y},X,s_0)$ for several different initial
states at $Q^2=(500 \mbox{ GeV})^2$. For simplicity, the
logarithmic running of the PDFs as a function $Q^2$ is ignored in performing this
integration.
As can be seen, for a given value of $s_0$, it is often a very good
approximation to take $\rho(X,s_0)\approx A X^q$, valid over a
region from $X=1$ to $X\sim 10$. Typical values of $q$ fall in the
range of $q\sim -3$.  Many of the properties of
$\f{d\sigma}{d x_T}$ that we will discuss
later will follow from this simple parametrization.  Another useful fact that we will exploit is
that the parton distribution function $f_a$ can be parametrized as
\begin{equation}
\label{eq:wellf}
x_af_a(x_a)= x_a^{\eta_a}(1-x_a)^{\gamma_a}P_a(x_a),
\end{equation}
where $\eta_a, \gamma_a >0$ and $P_a(x_a)=A_a+B_ax_a+C_ax_a^2+
\cdots$. We will ignore $P_a(x_a)$, as it is often negligible and
will not effect our basic results in later subsections. For
understanding $\f{d\sigma}{d y}$, we will  use the fact that $f_a$
is well-behaved in $x_a$.


Throughout the remainder of our discussion, we will ignore the
logarithmic running of the $f_a$ with $Q^2$ and the running of any
couplings.   Having made that approximation, which is valid to the
$\sim 1\%$ level for most of the quantities we will compute, we can
express the full differential cross sections in dimensionless form.
Taking either $(\bar y, X, \xi)$, $(\bar y, X, x_T)$, or $(\bar y,
X, \hat y)$ as independent variables, we can rewrite Eq.~(\ref{eq:TotalXSection}) as
\begin{eqnarray}\label{DimlessXSections}
  s_0d^3\sigma(a+b\rightarrow c+d) &=& \f{1}{2}d\bar{y}dXd\xi
  \; \rho_{ab}(\bar{y},X,s_0) \left(\sh^2\f{d\hat{\sigma}}{d\that}\right) , \\
  &=& \f{1}{2}d\bar{y}dXdx_T \left(\f{x_T}{X\xi}\right)
  \; \rho_{ab}(\bar{y},X,s_0) \left(\sh^2\f{d\hat{\sigma}}{d\that}\right)
  ,\nonumber \\
  &=& \f{1}{2}d\bar{y}dXd\hat{y} \left(1+\f{\Delta^{\prime}}{X}\right)\left(1-\tanh(\hat y)^2\right)
  \; \rho_{ab}(\bar{y},X,s_0) \left(\sh^2\f{d\hat{\sigma}}{d\that}\right).
  \nonumber
\end{eqnarray}
Ultimately, we are interested in differential distributions for
$x_T$ and $y=\hat{y}+\bar{y}$. We can integrate
Eq.~(\ref{DimlessXSections}) to calculate $\f{d\sigma}{d x_T}$ and
$\f{d\sigma}{d y}$ as
\begin{eqnarray}\label{eq:DsigPt}
  s_0\f{d\sigma}{d x_T} &=& \int_{X_{\rm min}(x_T,\Delta)}^{X_{\rm max}}dX
  \int_{-\bar{y_*}}^{\bar{y_*}}d\bar{y}
  \left(s_0\f{d^3\sigma}{d\bar{y} dX dx_{T} }\right) ,\\
  \label{eq:DsigRap}
  s_0\f{d\sigma}{d y} &=& \int_{1}^{X_{\rm max}}dX\int_{-\bar{y_*}}^{\bar{y_*}}d\bar{y}
  \left(s_0\f{d^3\sigma}{d\bar{y} dX dy}\right).
\end{eqnarray}
In the next subsections, we explore the shapes of $\f{d\sigma}{dx_T}$ and
$\f{d\sigma}{d y}$ for different values of $|\mathcal{M}|^2\sim \sh^2\f{d\hat{\sigma}}{d\that}$.

\subsection{Rules of Thumb}
\label{sec:empirical2to2}

\begin{figure}[tbp]
\begin{center}
\includegraphics[width=3.1in,angle=0]{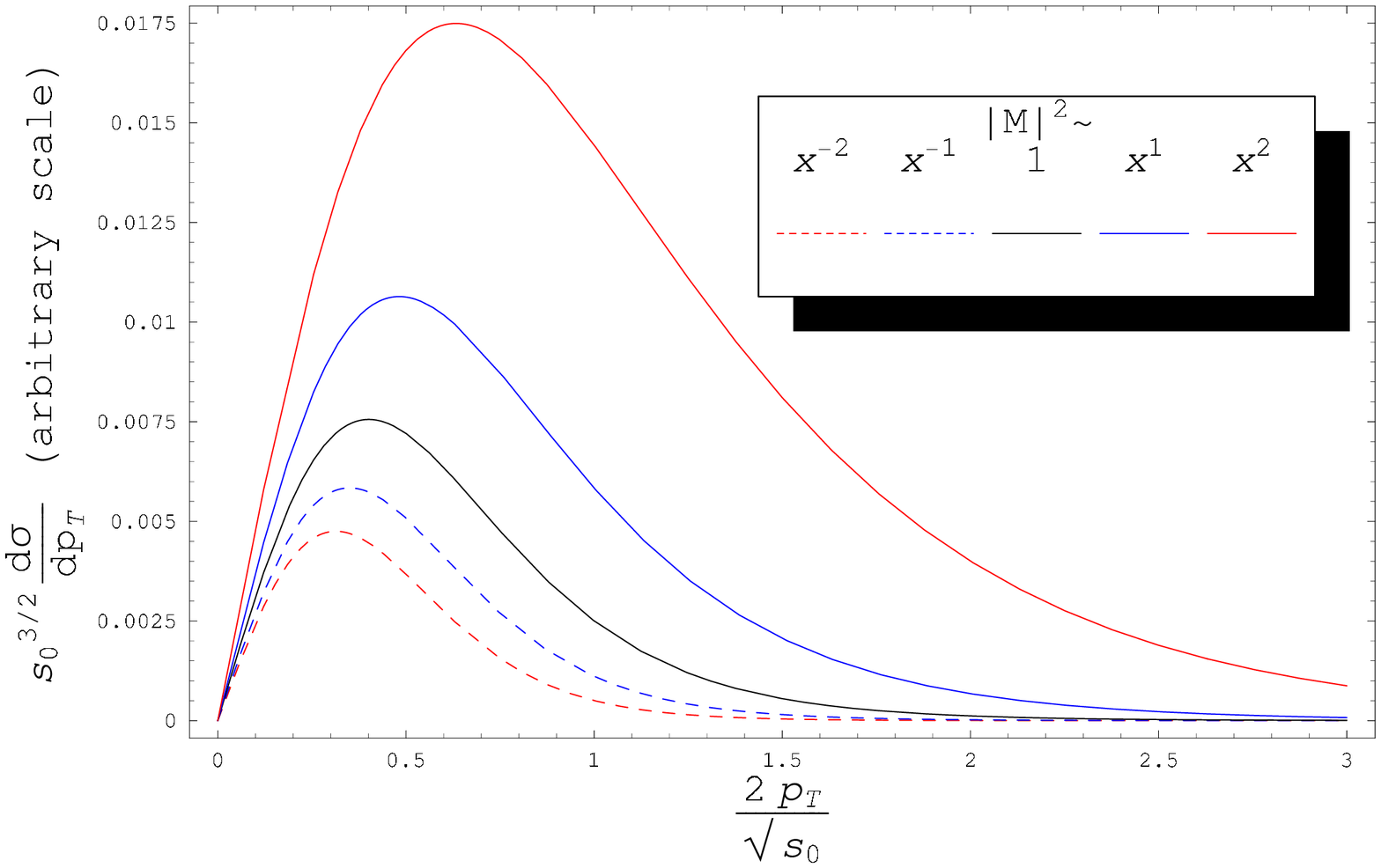}
\includegraphics[width=3.1in,angle=0]{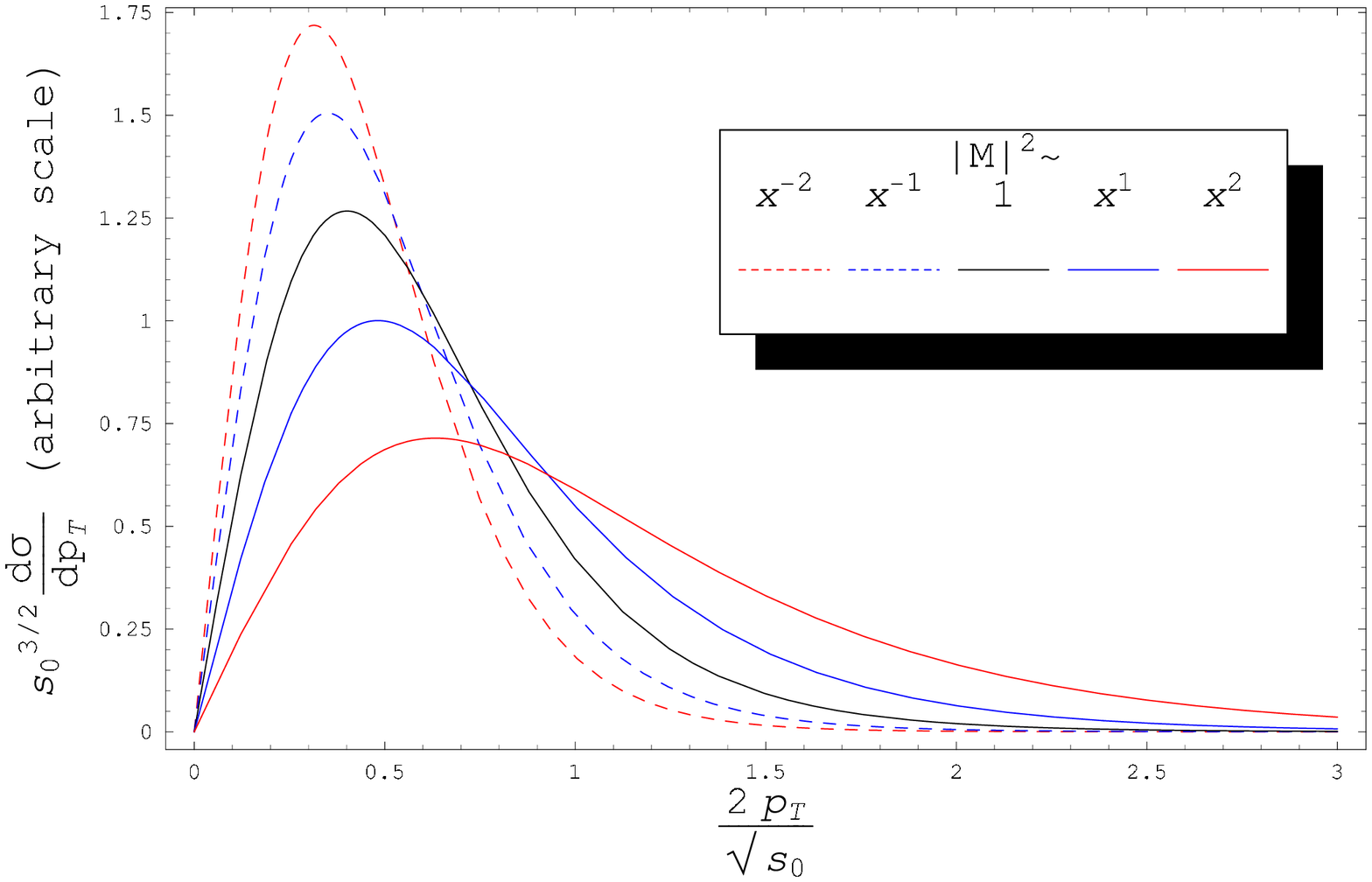}
\includegraphics[width=3.1in,angle=0]{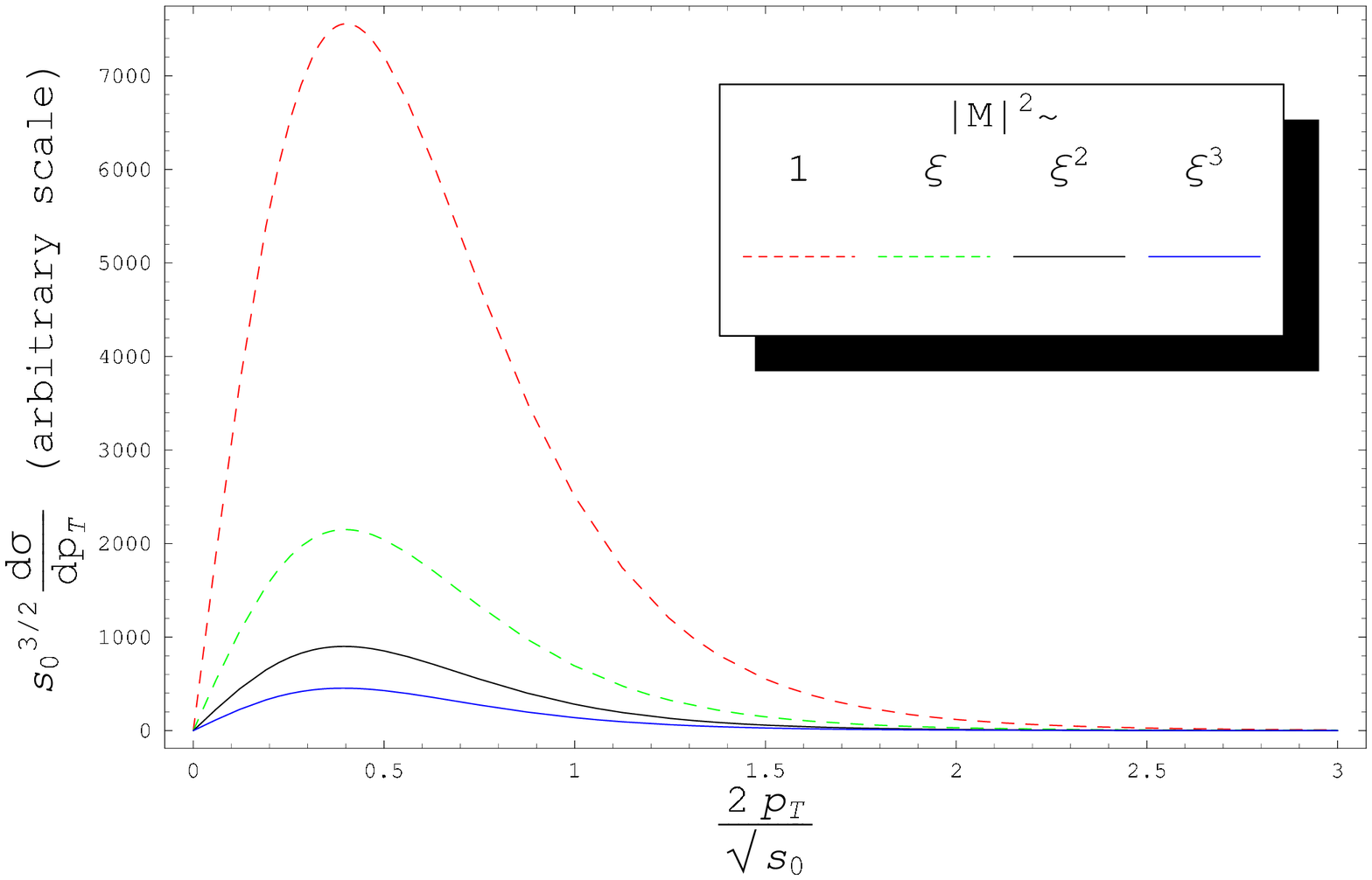}
\includegraphics[width=3.1in,angle=0]{sec2plots/CompXINorm.eps}
\caption{Inclusive $x_T$ distributions for different choices of
$|\mathcal{M}|^2$. Final state masses are taken as equal with $s_0=(1 \mbox{ TeV})^2$
and $Q^2=(500 \mbox{ GeV})^2$. The top left illustrates the relative
normalization differences for $|\mathcal{M}|^2\sim X^m$ behavior. The top
right illustrates shape differences by displaying the different
$|\mathcal{M}|^2$ with the same overall normalization. The bottom left and
right similarly illustrate normalization and shape difference
respectively, comparing $|\mathcal{M}|^2\sim \xi^n$ behavior instead.
We see that the transverse structure is largely independent of $\xi$.}
\label{fig:MEParaCompareXT}
\end{center}
\end{figure}

Perhaps the simplest way to parametrize $2\rightarrow 2$ scattering
is to approximate the behavior of the matrix element $|\mathcal{M}|^2\sim
\sh^2\f{d\hat{\sigma}}{d\that}$ as the sum of rational polynomials
in $X$ and $\xi$,
\begin{equation}
|\mathcal{M}|^2\approx \sum_{mn}C_{mn}X^m\xi^n .
\end{equation}
As an empirical exercise, we can compare the $y$ and $x_T$
distributions for matrix element of the form $|\mathcal{M}|^2\sim X^n\xi^m$.
This is illustrated in Figure \ref{fig:MEParaCompareXT} for $x_T$
distributions.
\begin{figure}[tbp]
\begin{center}
\includegraphics[width=3.1in,angle=0]{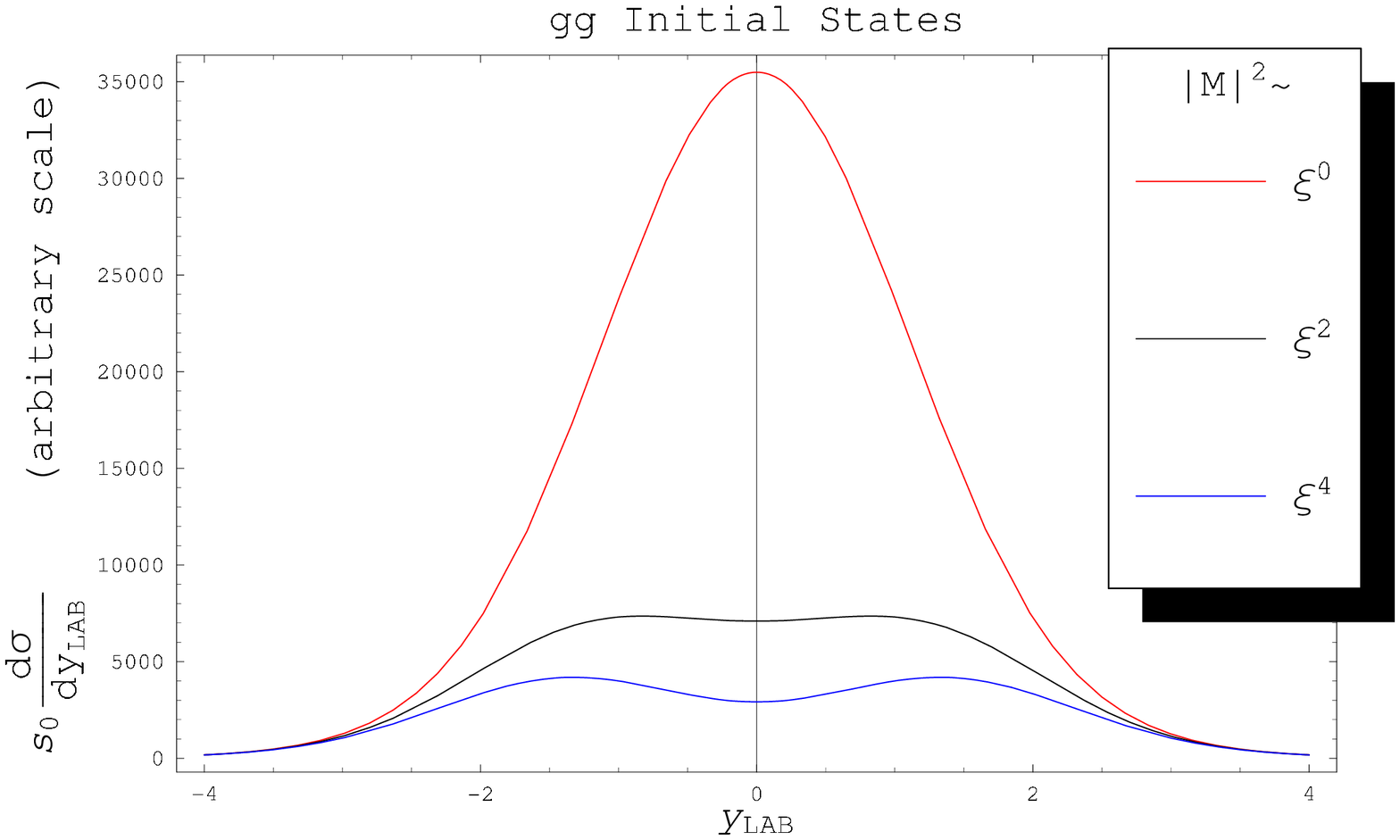}
\includegraphics[width=3.1in,angle=0]{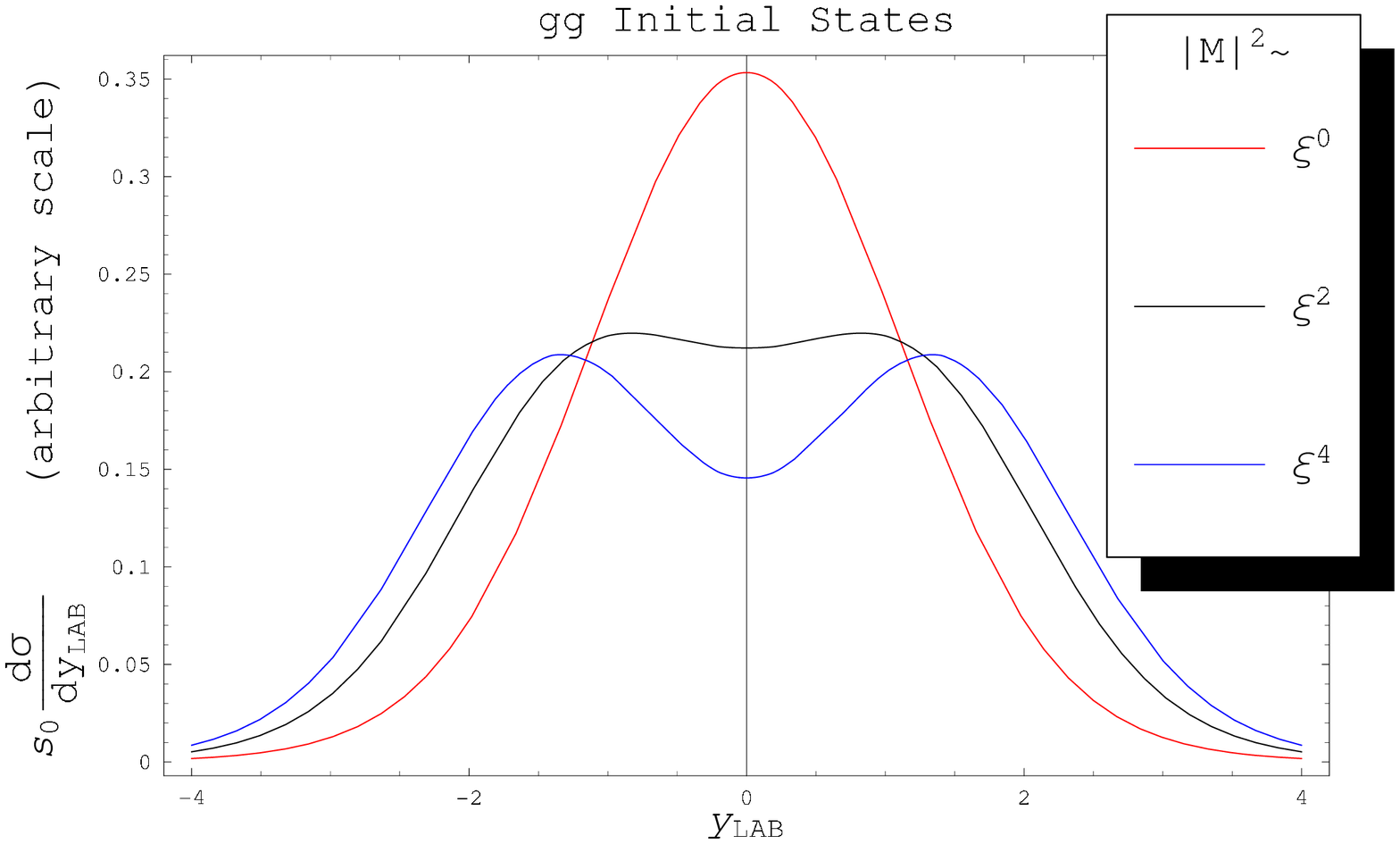}
\includegraphics[width=3.1in,angle=0]{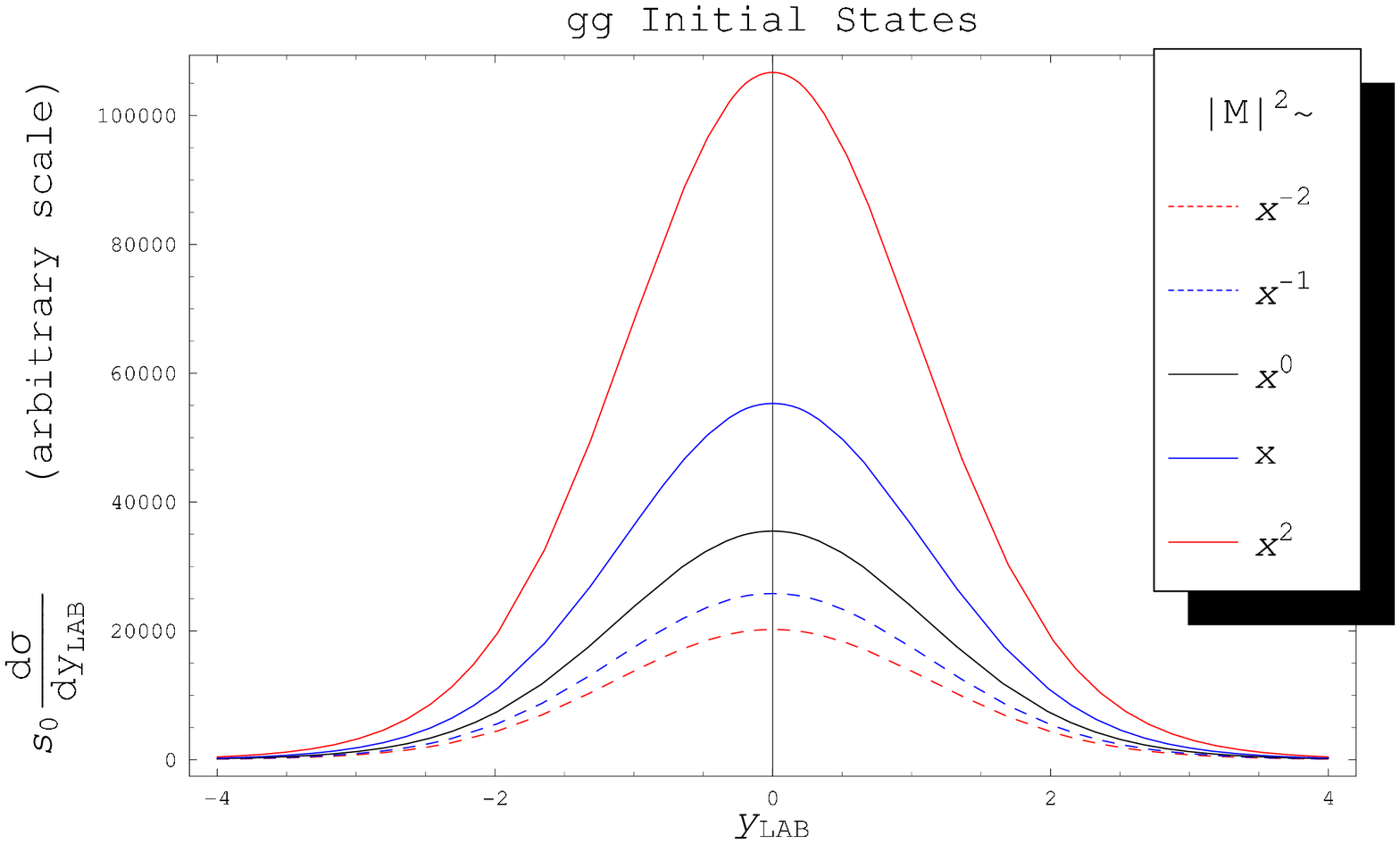}
\includegraphics[width=3.1in,angle=0]{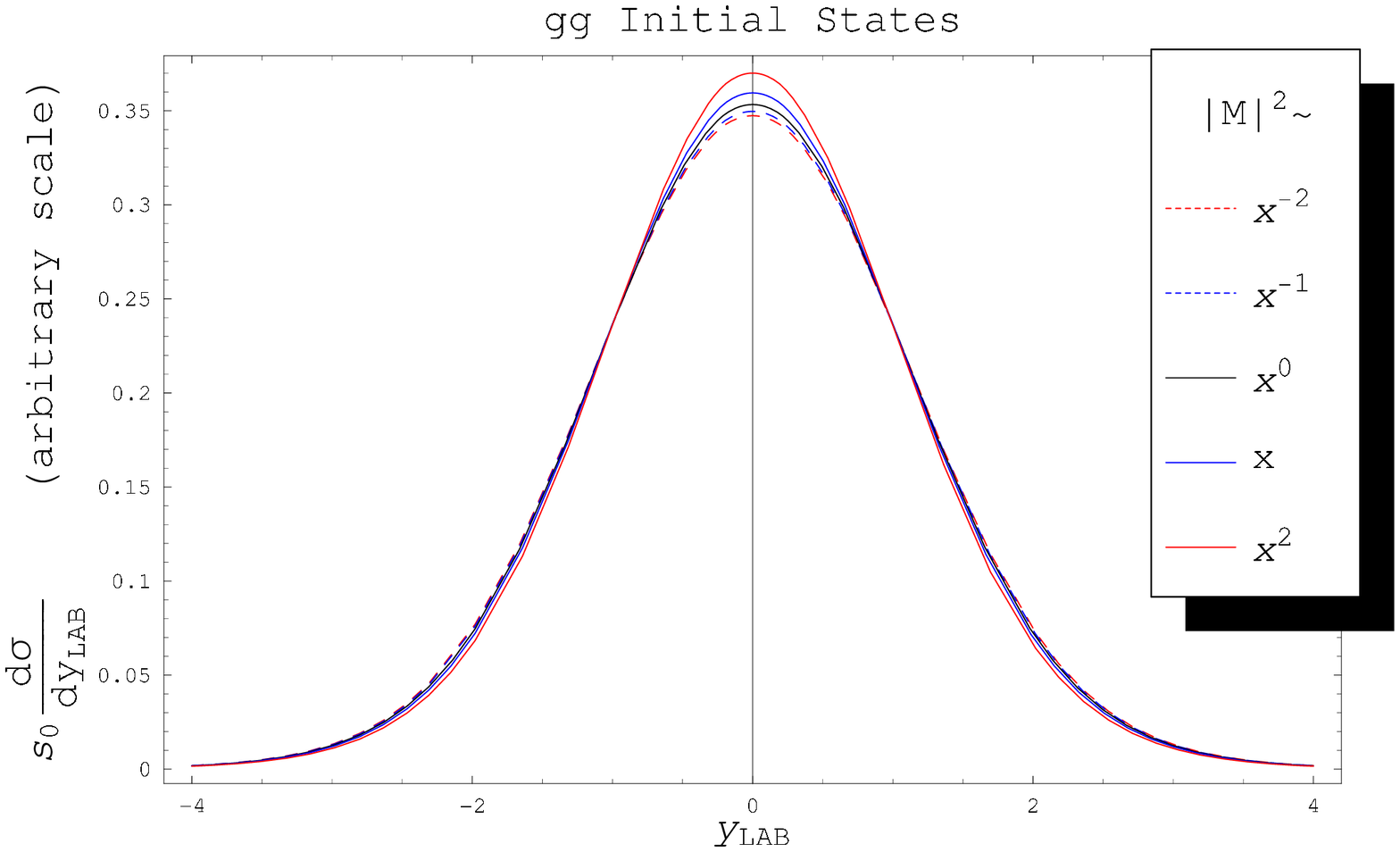}
\includegraphics[width=3.1in,angle=0]{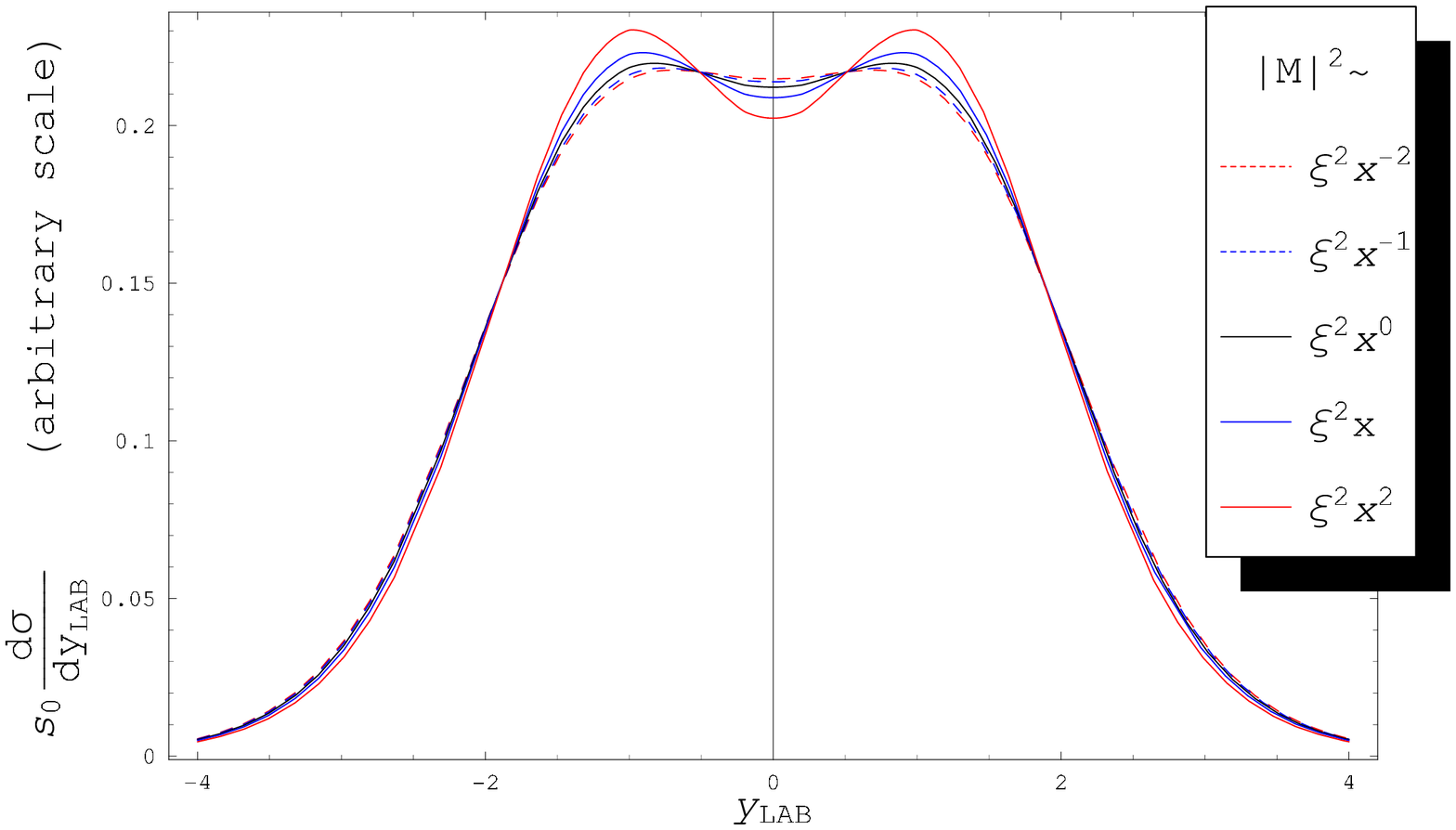}
\includegraphics[width=3.1in,angle=0]{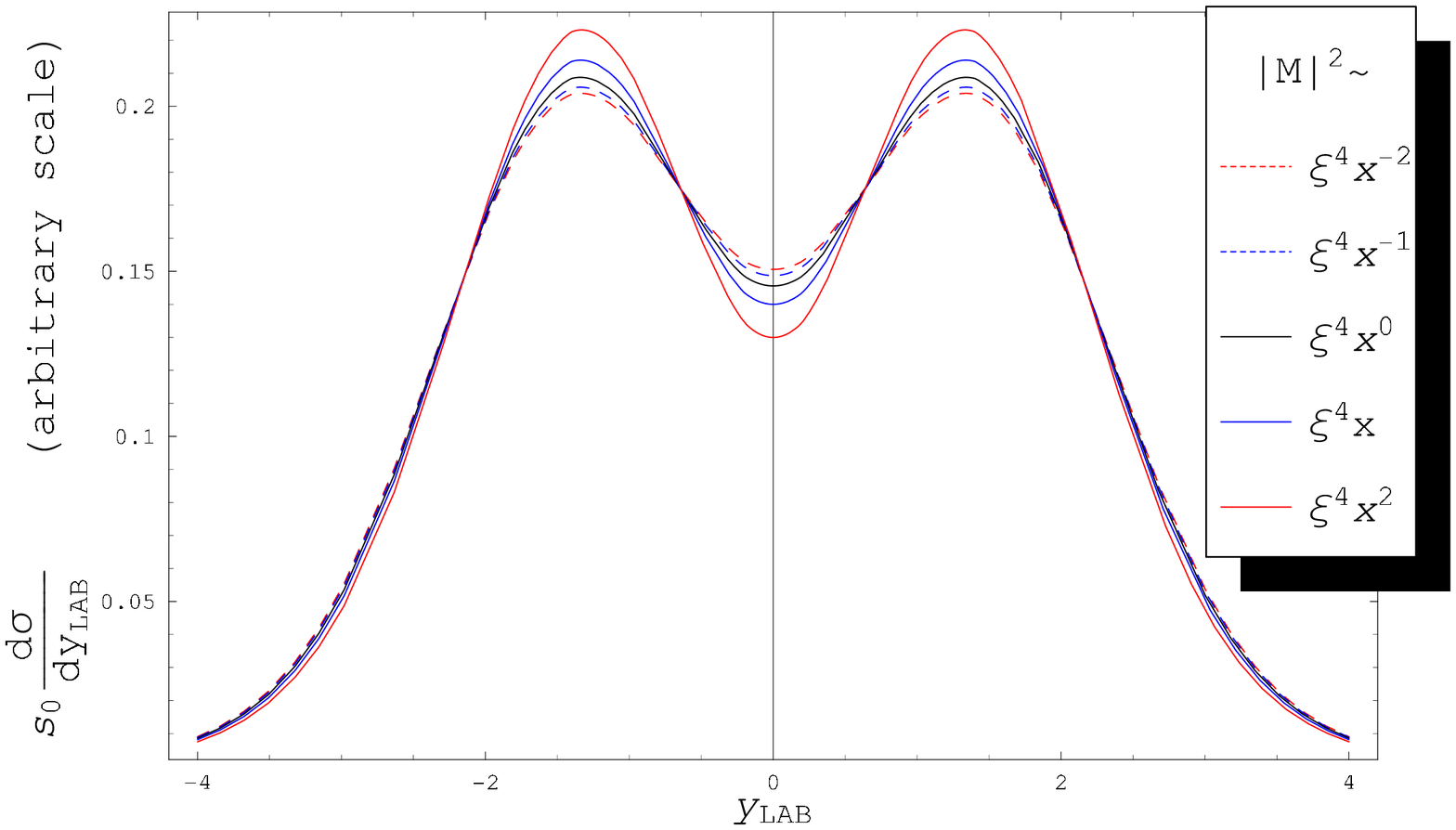}
\caption{Inclusive rapidity distributions shown for various choices
of $|\mathcal{M}|^2\sim X^m \xi^n$ with $gg$ initial states. Final
state masses are taken as equal with $s_0=1$ TeV$^2$ and $Q^2=(500
\mbox{ GeV})^2$. Top left/right: Un-normalized/normalized
distributions for $|\mathcal{M}|^2\sim \xi^n$. Middle left/right:
Un-normalized/normalized distributions for $|\mathcal{M}|^2\sim
X^m$. Bottom left: Normalized distributions for $|\mathcal{M}|^2\sim
\xi^2X^m$. Bottom right: Normalized distributions for
$|\mathcal{M}|^2\sim \xi^4X^m$. We see that the rapidity structure
is largely independent of $X$.} \label{fig:ggRapShapePlots}
\end{center}
\end{figure}

As is clearly evident and hardly surprising, for $|\mathcal{M}|^2\sim X^m$
the moments of the $x_T$ distributions increase/decrease with
increasing/decreasing powers $m$. More surprising is that the shapes
of the $x_T$ distributions do not depend on $\xi$! Illustrated in
Figure \ref{fig:ggRapShapePlots} are the rapidity distributions for
varying choices of $|\mathcal{M}|^2$. Evident is the strong correlation in the
shape structure of the rapidity distribution with $\xi$ while there
is essentially no dependence on $X$.

Equipped with these observations, we are ready to state some
generally applicable and very useful results for modeling $x_T$ and
$y$ observables. We argue that for general non-resonant
$2\rightarrow 2$ processes:
\begin{itemize}
\item \emph{The shape of $\f{d\sigma}{dx_T}$ is well-parametrized
by leading order in $X$ behavior without regard to $\xi$
dependence.} This works because of an approximate shape invariance
of $\f{d\sigma}{dx_T}$ under $|M|\rightarrow |M|\xi^p$. In most
cases, including only leading order $X$ behavior suffices to capture
the transverse structure at the $\sim 5\%$ level.
\item \emph{An approximate shape invariance of
$\f{d\sigma}{dy}$ under $|M|\rightarrow |M|X^p$ allows one to model the
$x_T$ structure without effecting the shape modeling of
$\f{d\sigma}{dy}$}.
\item Once the transverse structure is well modeled, \emph{the
rapidity structure is usually fixed by the PDF structure with
sub-leading correction in $\xi$}. Moreover, inclusion of these
sub-leading $\xi$ corrections does not effect the $x_T$
distributions by virtue of the $\xi$ shape invariance of
$\f{d\sigma}{dx_T}$.
\end{itemize}
We will rigorously demonstrate these points below.
As these results will help define our leading order OSET
parametrization scheme for particle production, we will also
identify where and why the above properties fail. The application of
these results to the modeling of full decay topologies is the topic
of Secs.~\ref{sec2} and \ref{sec:MARMinPractice}.

\subsection{Shape Invariance for Transverse and Rapidity Structure}
\label{sec2:ShapeInv}

Let's first understand the shape invariance of
$s_0\f{d\sigma}{dx_T}$ under $|\mathcal{M}|^2\rightarrow |\mathcal{M}|^2\xi^p$. We'll
start by returning to Eq.~(\ref{eq:DsigPt}) and write
\begin{equation}\label{eq:dsigmadxT}
s_0\f{d\sigma}{dx_T}= \f{1}{2}\int_{X_{min}(x_T,\Delta)}^{X_{max}}
dX\int_{\bar{-y_*}(X)}^{\bar{y_*}(X)} d\bar{y}\left(\f{x_T}{X\xi}\right)
  \rho_{ab}(\bar{y},X,s_0) \left(\sh^2\f{d\hat{\sigma}}{d\that}\right) .
\end{equation}
Expanding a general $2\rightarrow 2$ matrix element as
\begin{equation}\label{eq:MEexpansion}
\left(\sh^2\f{d\hat{\sigma}}{d\that}\right)= \sum_{mn}C_{mn}X^m\xi^n ,
\end{equation}
and taking advantage of the property illustrated in Figure
\ref{fig:partonLum} that the the integrated ``parton luminosity'' $\rho_{ab}(X,s_0)\approx A X^q$,
we have
\begin{equation}\label{eq:dsigmadxT3}
  s_0\f{d\sigma}{dx_T} \approx A\f{x_T}{2}\int_{X_{\rm min}(x_T,\Delta)}^{X_{\rm max}}
  dX\sum_{mn}C_{mn}X^{m+q-1}\xi^{n-1}.
\end{equation}
It is straightforward to evaluate this expression analytically order
by order in an expansion in the mass asymmetry parameter $\Delta$. We will drop small
logarithmic corrections arising from Altarelli-Parisi running. Expanding to first
order in $\Delta$,
\begin{eqnarray}
  s_0\f{d\sigma}{dx_T} &\approx& A\f{x_T}{2}\int_{1+x_T^2}^{X_{\rm max}}
  dX\sum_{mn}C_{mn}X^{m+q-1}\left(1-\f{1+x_T^2}{X}\right)^{\f{n-1}{2}}+O(\Delta) , \nonumber \\
   &\approx& A\f{x_T}{2}\sum_{mn}C_{mn}(1+x_T^2)^{m+q}\int_{0}^{1}d\eta \;
   \eta^{-(m+q+1)}(1-\eta)^{\f{n-1}{2}}+O(\Delta) ,
\end{eqnarray}
where we have defined $\eta\equiv \f{1+x_T^2}{X}$ and taken
$X_{\rm max}^{-1} \approx 0$, a good approximation given that the
integrand is dominated away from $\eta = 0$ (recall that $m+q+1\le 0$ for any reasonably
behaved matrix element and PDF). Evaluating these integrals,
\begin{eqnarray}\label{eq:dsigmadxT4}
  s_0\f{d\sigma}{dx_T} \approx
  A\sum_{mn}C_{mn}\f{x_T(1+x_T)^{m+q}}{2}B\left(-m-q,\f{n+1}{2}\right)+O(\Delta),
\end{eqnarray}
where $B(\mu,\nu)=\f{\Gamma(\mu)\Gamma(\nu)}{\Gamma(\mu+\nu)}$ is
the Euler Beta function.

\begin{figure}[tbp]
\begin{center}
\includegraphics[width=3.1in,angle=0]{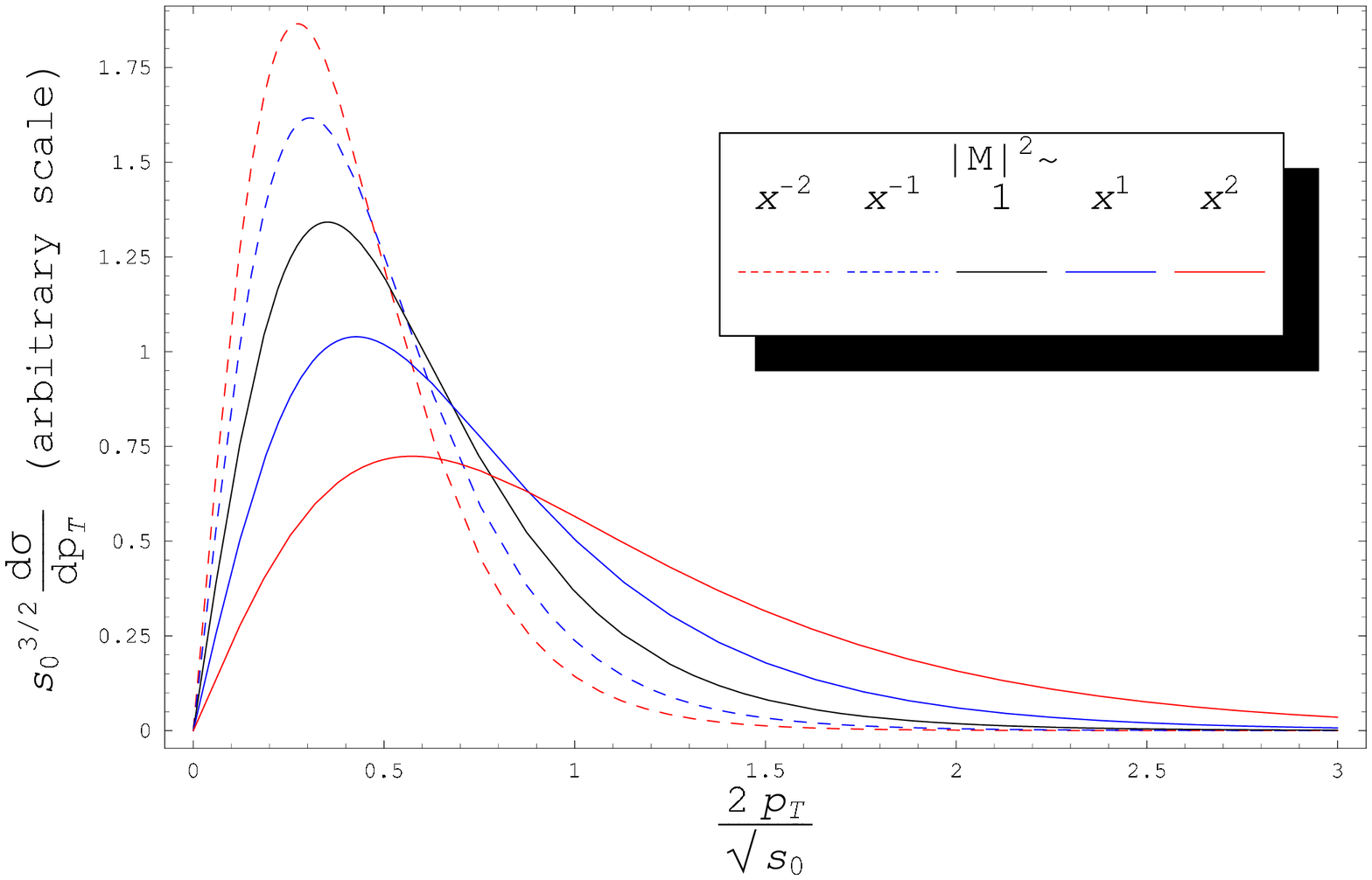}
\includegraphics[width=3.1in,angle=0]{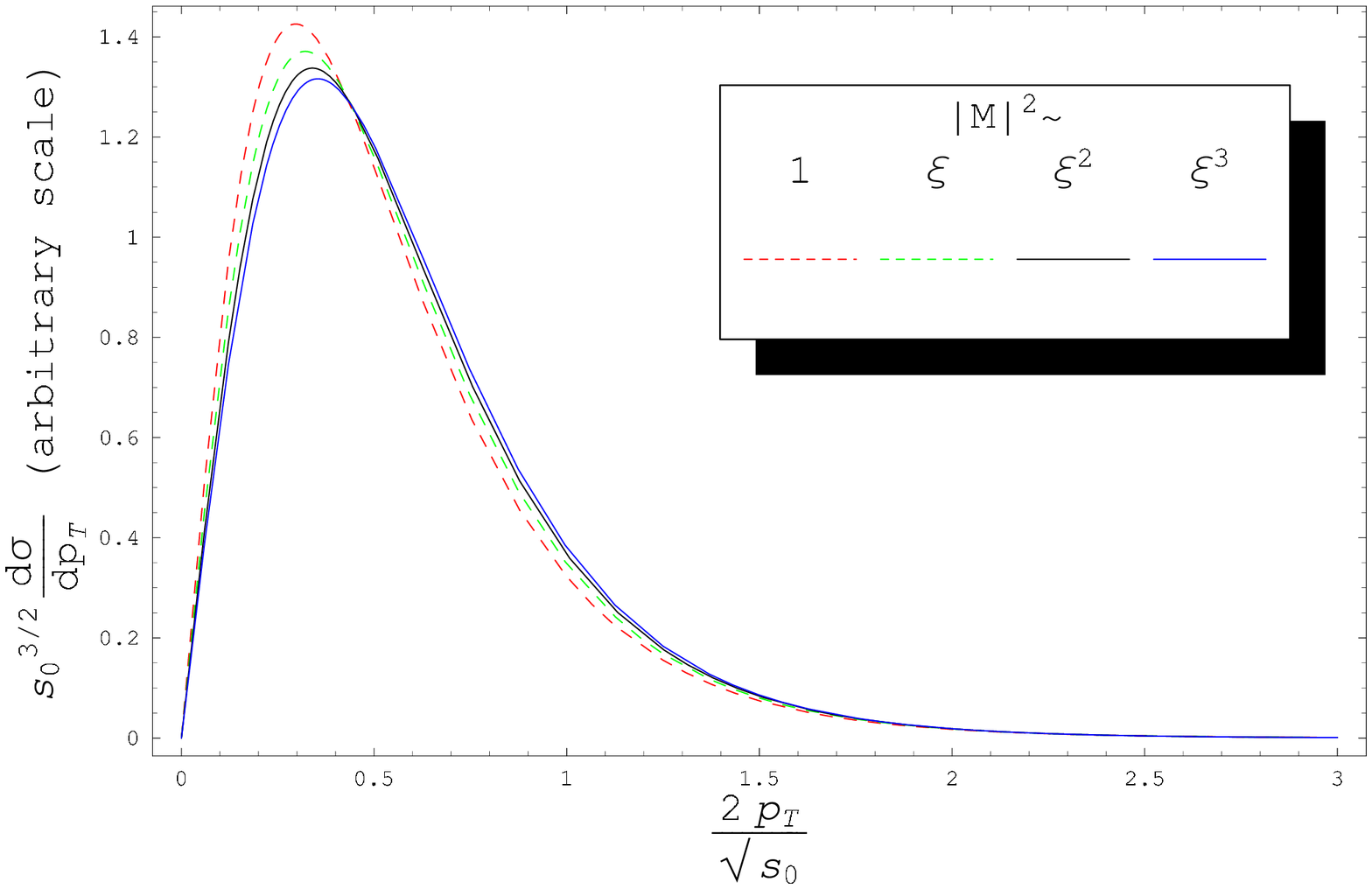}
\caption{Inclusive $p_T$ distributions (normalized to 1) shown for
$|\mathcal{M}|^2\sim X^m$ (left) and $|\mathcal{M}|^2\sim\xi^n$ (right) with a final state
mass asymmetry of $\Delta=0.5$ ($m_c\approx 6m_d$). In contrast to
the approximate shape invariance when $\Delta=0$, note that there is
an additional $\sim 5\%$ shift in the the peak and mean of the $p_T$
distribution as $n$ is varied. Fortunately, this effect is small in
all but the most extreme kinematic situations (i.e. $m_c\gg m_d$).
In practice, not including $\Delta$ effects could lead to a small
error in a measurement of $s_0$ using shape and scale information.}
\label{fig:DeltaEffectPt}
\end{center}
\end{figure}

From Eq.~(\ref{eq:dsigmadxT4}), it's clear that the shape invariance
is the result of the approximate homogeneity in $X$ of both the
parton luminosity functions and $\xi(X,x_T)$. Of course, there are
$O(\Delta)$ corrections that are not shape invariant, but these are
suppressed not only by $\Delta$, but by $(1+x_T^2)$ and Euler Beta
function factors with respect to the leading order behavior. An
illustration of the shape deformation in the partonic $p_T$
distribution is shown in Figure \ref{fig:DeltaEffectPt} with
$\Delta=0.5$ (which corresponds to $m_c \approx 6 m_d$). Another
important fact can be read off from Eq.~(\ref{eq:dsigmadxT4}), namely
the Euler Beta suppression with decreasing powers of $X$ explains
why the leading order behavior is usually sufficient to capture the
$s_0\f{d\sigma}{dx_T}$ shape.

The above observations make it clear that the leading order
transverse structure is controlled by the leading order $X$
dependence of $|\mathcal{M}|^2$. We also see that this should start to fail in
cases where the final state mass asymmetry is large (i.e.
$\Delta\approx 1$), in which case small shape corrections may be
needed. For small $\Delta$, note that we can actually get away with choosing the wrong
PDFs as long as we compensate for the transverse shape by also choosing the
wrong $X$ dependence of the matrix element.  If one is interested in extracting
the correct $X$ dependence, then we will see that rapidity information can be used
to constrain the PDFs.

We turn now to understanding the approximate shape invariance of
$s_0\f{d\sigma}{dy}$ under $|\mathcal{M}|^2\rightarrow |\mathcal{M}|^2X^p$. Let's start
with Eq.~(\ref{eq:DsigRap}) and write,
\begin{eqnarray}\label{eq:RapCalc}
  s_0\f{d\sigma}{d y} &=& \f{1}{2}\int_{1}^{X_{\rm max}}dX\int_{-\bar{y_*}}^{\bar{y_*}}d\bar{y}
  \left(1+\f{\Delta^{\prime}}{X}\right)\left(1-\tanh(y-\bar{y})^2\right)
  \rho_{ab}(\bar{y},X,s_0)\left(\sh^2\f{d\hat{\sigma}}{d\that}\right)
  ,\nonumber \\
  &=& \f{1}{2}\sum_{mn}C_{mn}\int_{1}^{X_{max}}dXX^m
  \left(1+\f{\Delta^{\prime}}{X}\right)^{1+n}g_n(y,X,s_0),
\end{eqnarray}
where
\begin{equation}
g_n(y,X,s_0)=\int_{-\bar{y}_*(X)}^{\bar{y}_*(X)}d\bar{y}
\left(1-\tanh(y-\bar{y})^2\right)\tanh(y-\bar{y})^n
  \rho_{ab}\left(\sqrt{\epsilon X}e^{\bar{y}},\sqrt{\epsilon X}e^{-\bar{y}}\right),
\end{equation}
and we've used the expansion in Eq.~(\ref{eq:MEexpansion}),
Eq.~(\ref{eq:XiBetaSqRap}), and defined a small expansion parameter
$\epsilon=\f{s_0}{s_b}\ll 1$ that controls the available initial
rapidity phase space.  We're also taking advantage of the fact that
$\rho_{ab}(\bar{y},X,s_0)$ only depends on $\sqrt{\epsilon
X}e^{\pm\bar{y}}$.

From Eq.~(\ref{eq:wellf}), we see that $\rho_{ab}$ has a pole in
$\epsilon$ near $\epsilon = 0$. Factoring out the pole, the
integrand of $g_n(y,X,s_0)$ has a well-behaved polynomial form if we
expand in $\sqrt{\epsilon}$. But $\epsilon$ and $X$ in $\rho_{ab}$
only appears in the combination $\sqrt{\epsilon X}$, so we know that
at leading order in $\sqrt{\epsilon}$, the $y$ and $X$ dependence of
the integrand must factorize.  After doing the integral, we are
guaranteed to find
\begin{equation}
g_n(y,X,s_0)=\hat{g}_n(y,s_0) \f{\left(\epsilon
X\right)^{\eta_a+\eta_b}}{X^2} \left(1 +
\mathcal{O}\left(\frac{1}{\log \epsilon} \right) \right),
\end{equation}
where $\hat{g}_n$ is a function of $y$, $n$, and $s_0$ alone. The
PDF information is largely contained in $\hat{g}_n$. To see why the
corrections occur at $\mathcal{O}\left(\frac{1}{\log
\epsilon}\right)$ instead of $\mathcal{O}(\sqrt{\epsilon})$, recall
that the boundaries of phase space are given by $\bar{y}_*= - \log
\sqrt{\epsilon X}$, so the integral over $\bar{y}$ spoils the
factorization before the higher order corrections from $\rho_{ab}$.

\begin{figure}[tbp]
\begin{center}
\includegraphics[width=3.1in,angle=0]{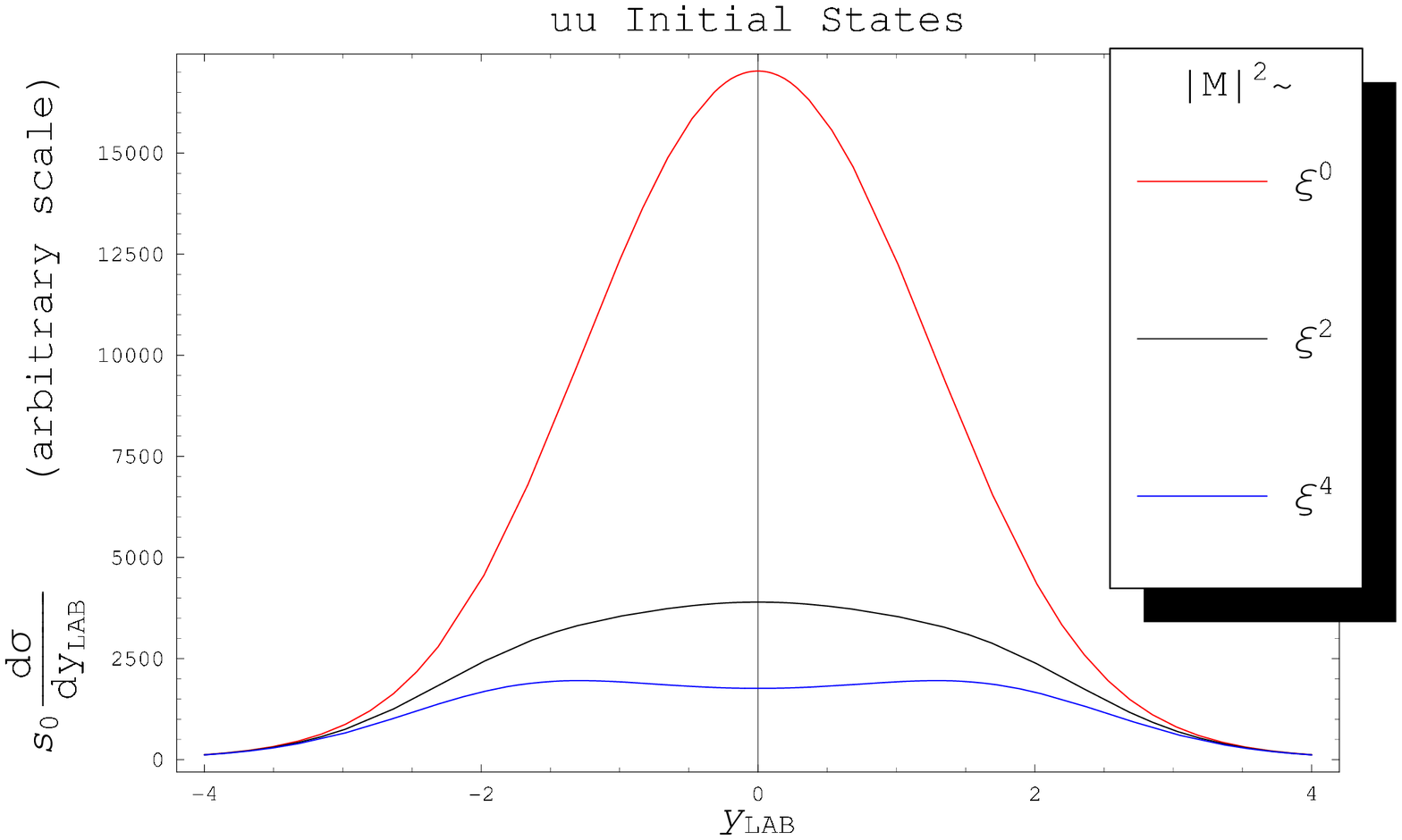}
\includegraphics[width=3.1in,angle=0]{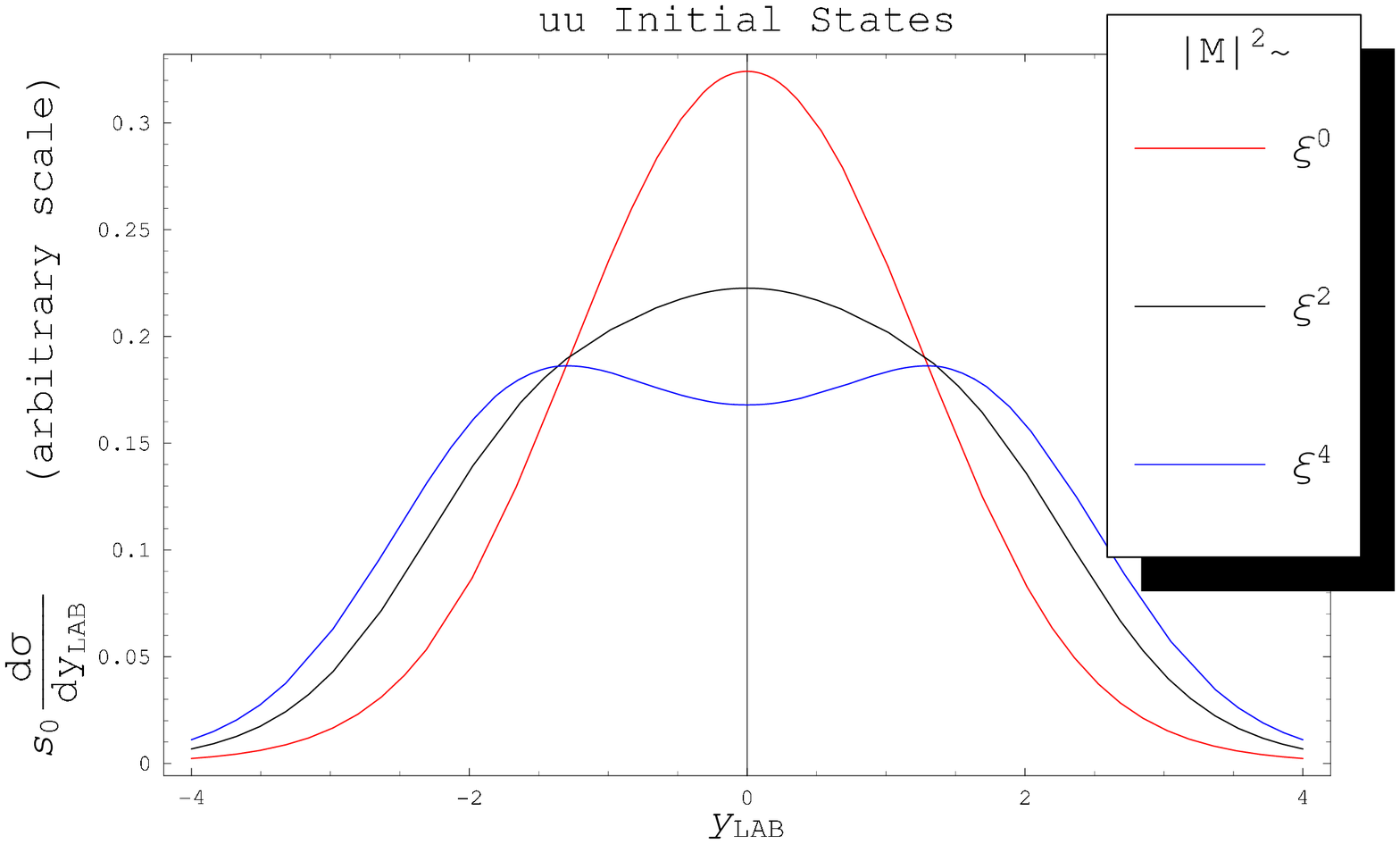}
\includegraphics[width=3.1in,angle=0]{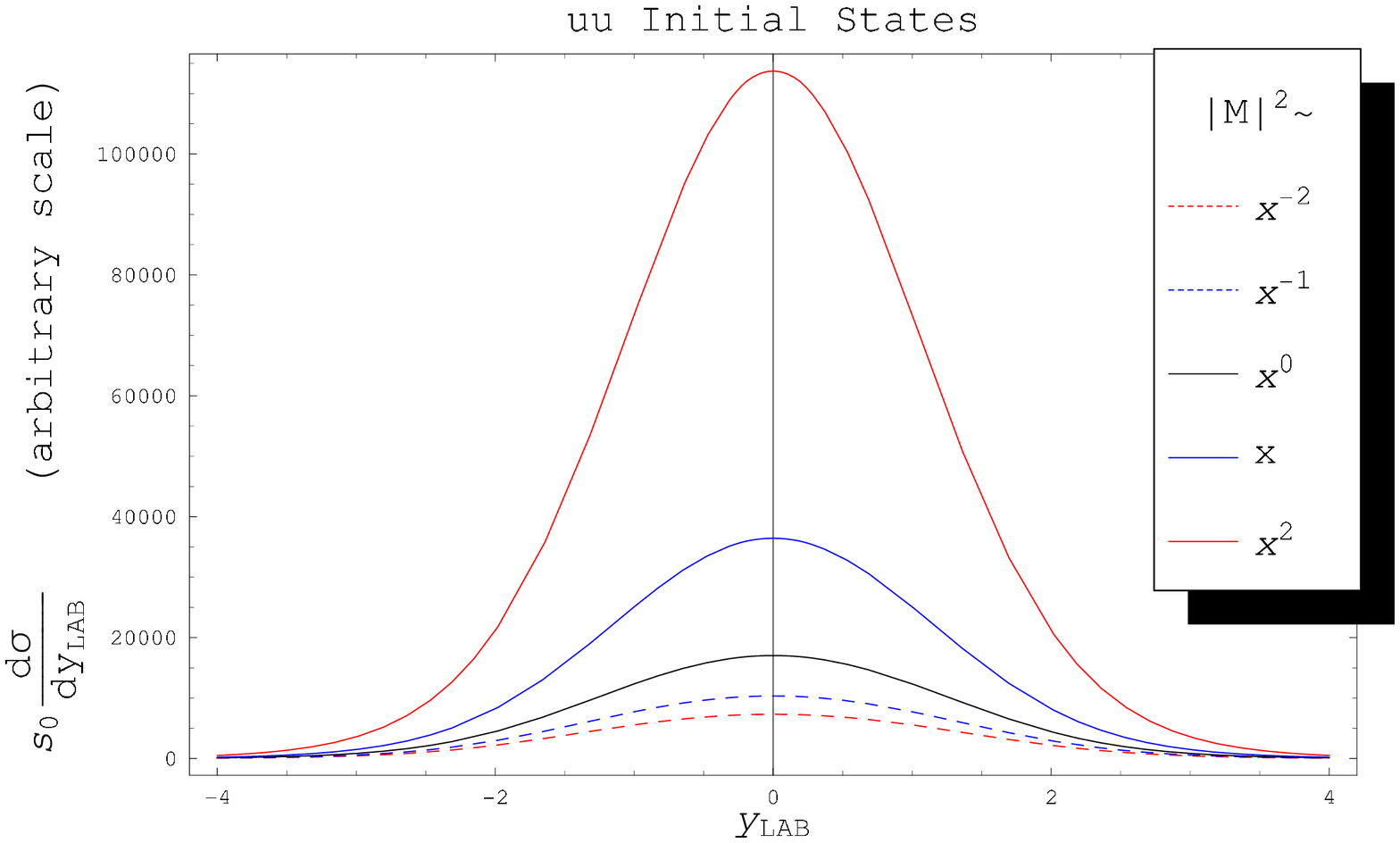}
\includegraphics[width=3.1in,angle=0]{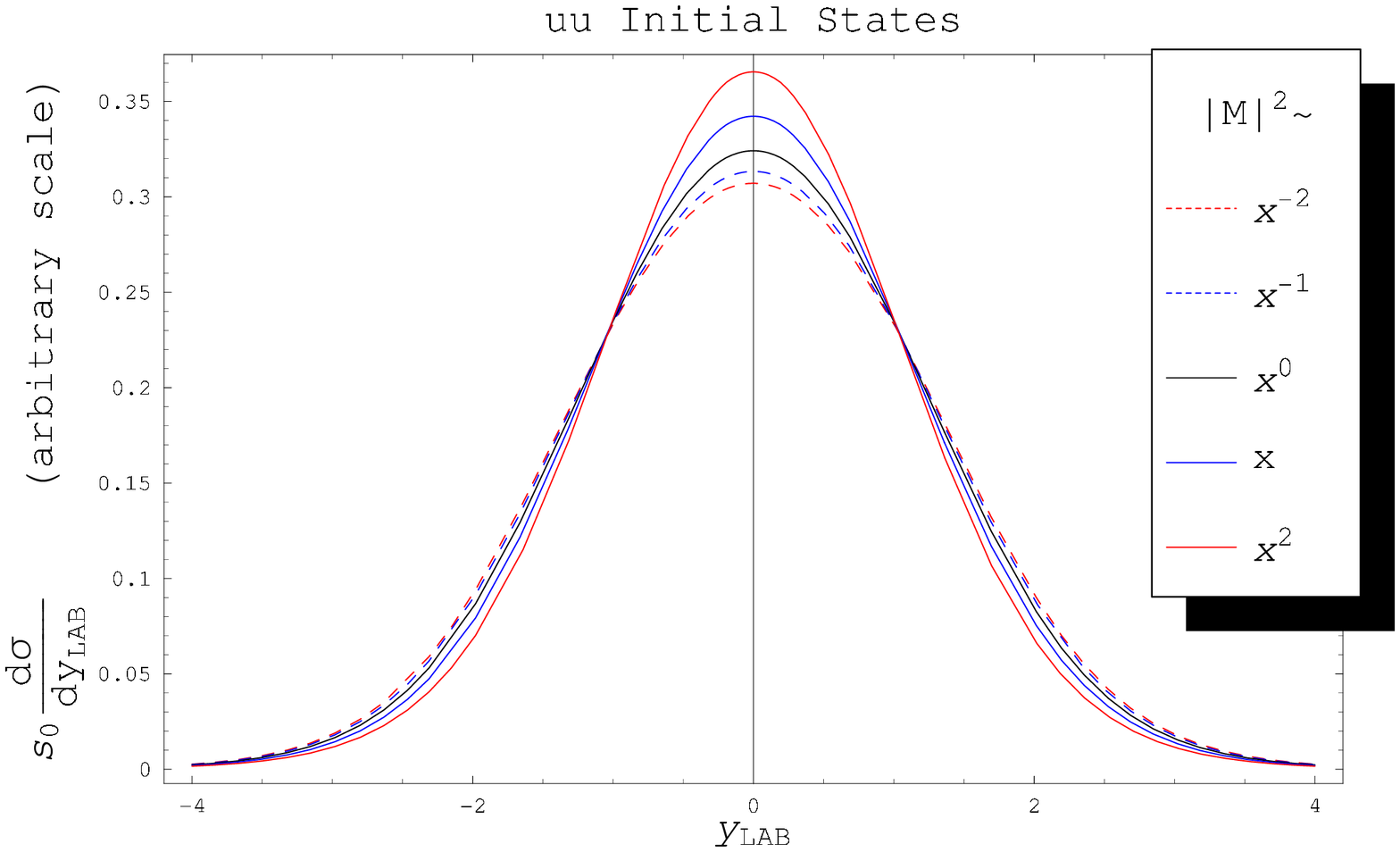}
\includegraphics[width=3.1in,angle=0]{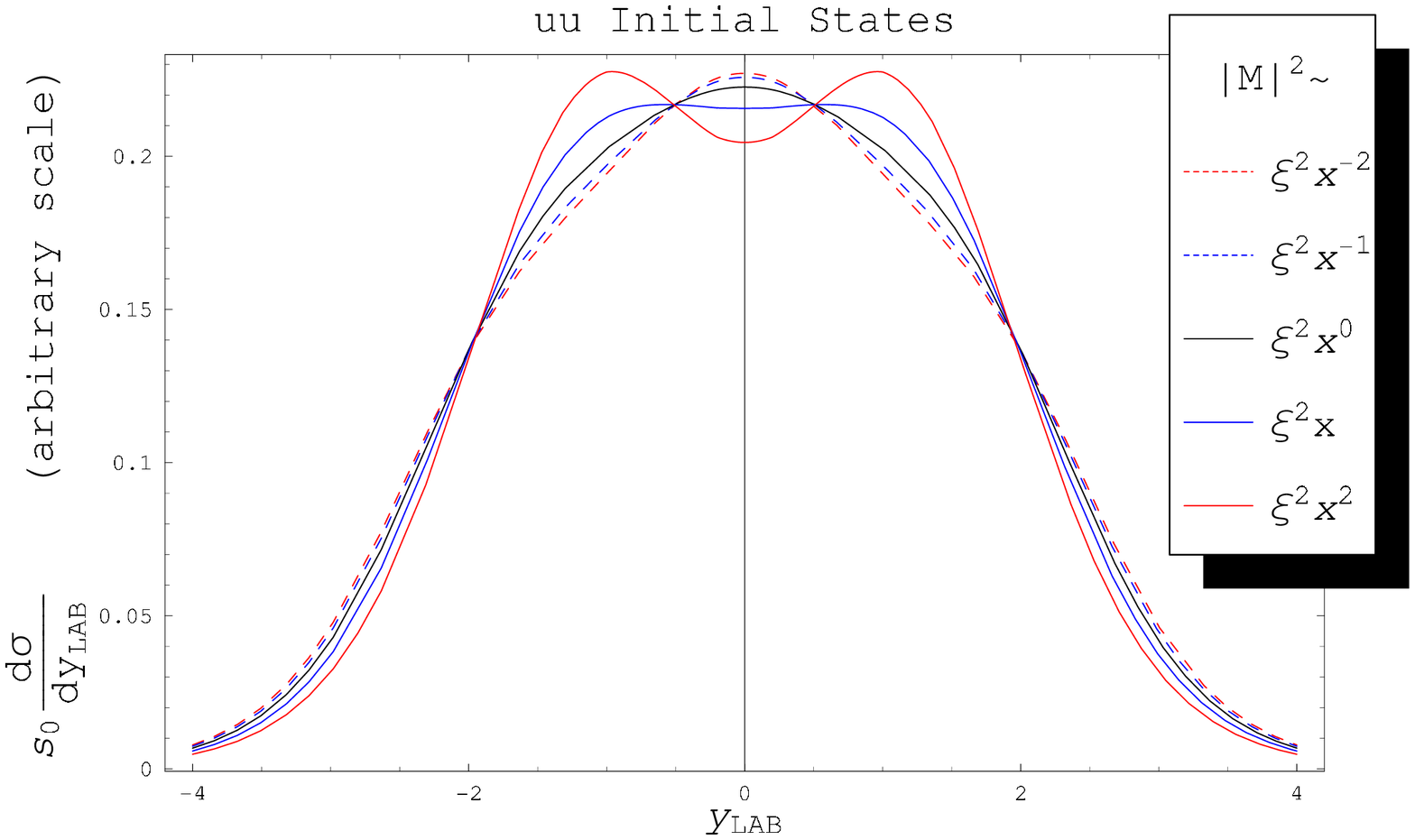}
\includegraphics[width=3.1in,angle=0]{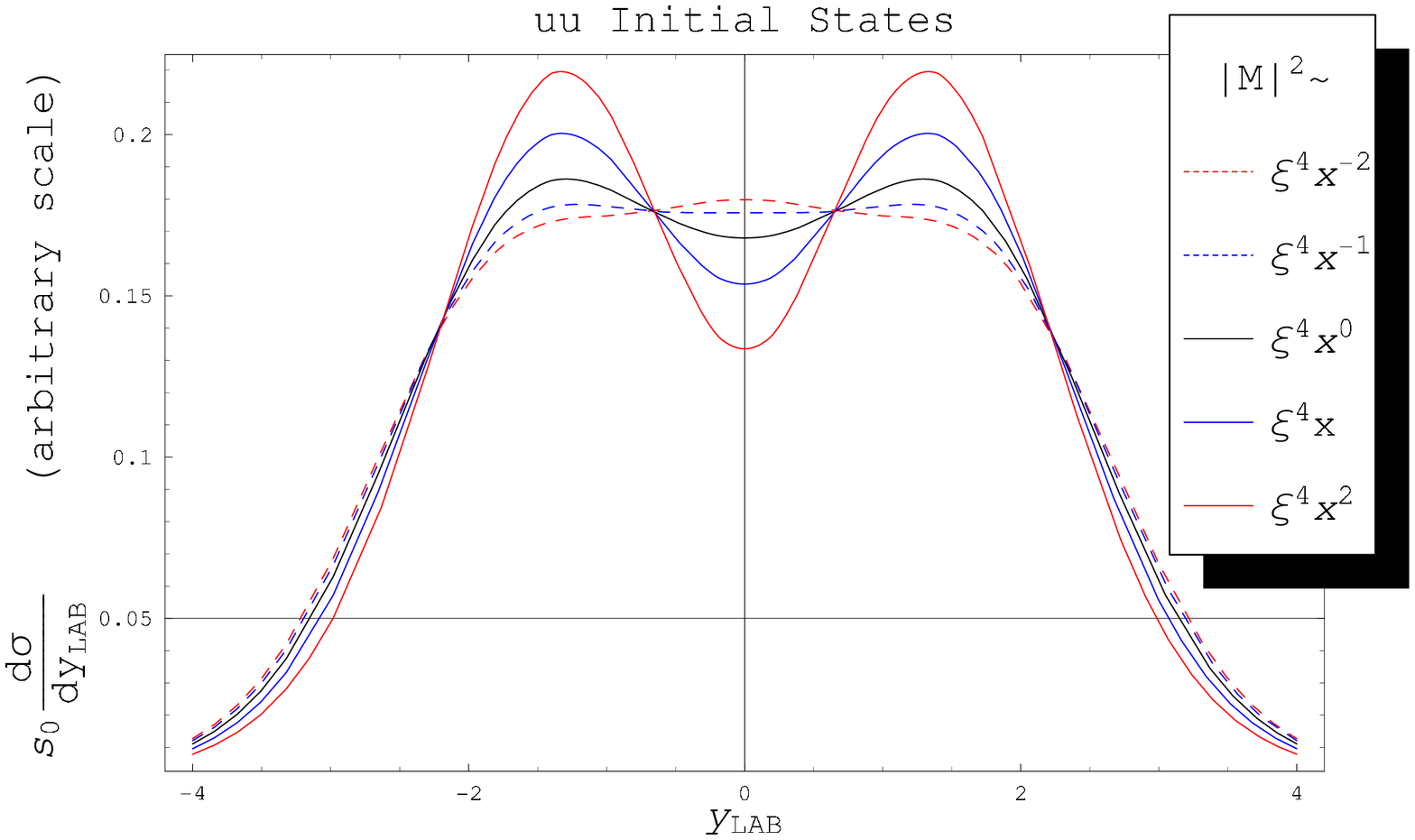}
\caption{Inclusive rapidity distributions shown for various choices
of $|\mathcal{M}|^2\sim X^m \xi^n$ with $uu$ initial states. Final
state masses are taken as equal with $s_0=1$ TeV$^2$ and $Q^2=(500
\mbox{ GeV})^2$. Top left/right: Un-normalized/normalized
distributions for $|\mathcal{M}|^2\sim \xi^n$. Middle left/right:
Un-normalized/normalized distributions for $|\mathcal{M}|^2\sim
X^m$. Bottom left: Normalized distributions for $|\mathcal{M}|^2\sim
\xi^2X^m$. Bottom right: Normalized distributions for
$|\mathcal{M}|^2\sim \xi^4X^m$. We see that the rapidity structure
is largely independent of $X$.} \label{fig:uuRapShapePlots}
\end{center}
\end{figure}
\begin{figure}[tbp]
\begin{center}
\includegraphics[width=3.1in,angle=0]{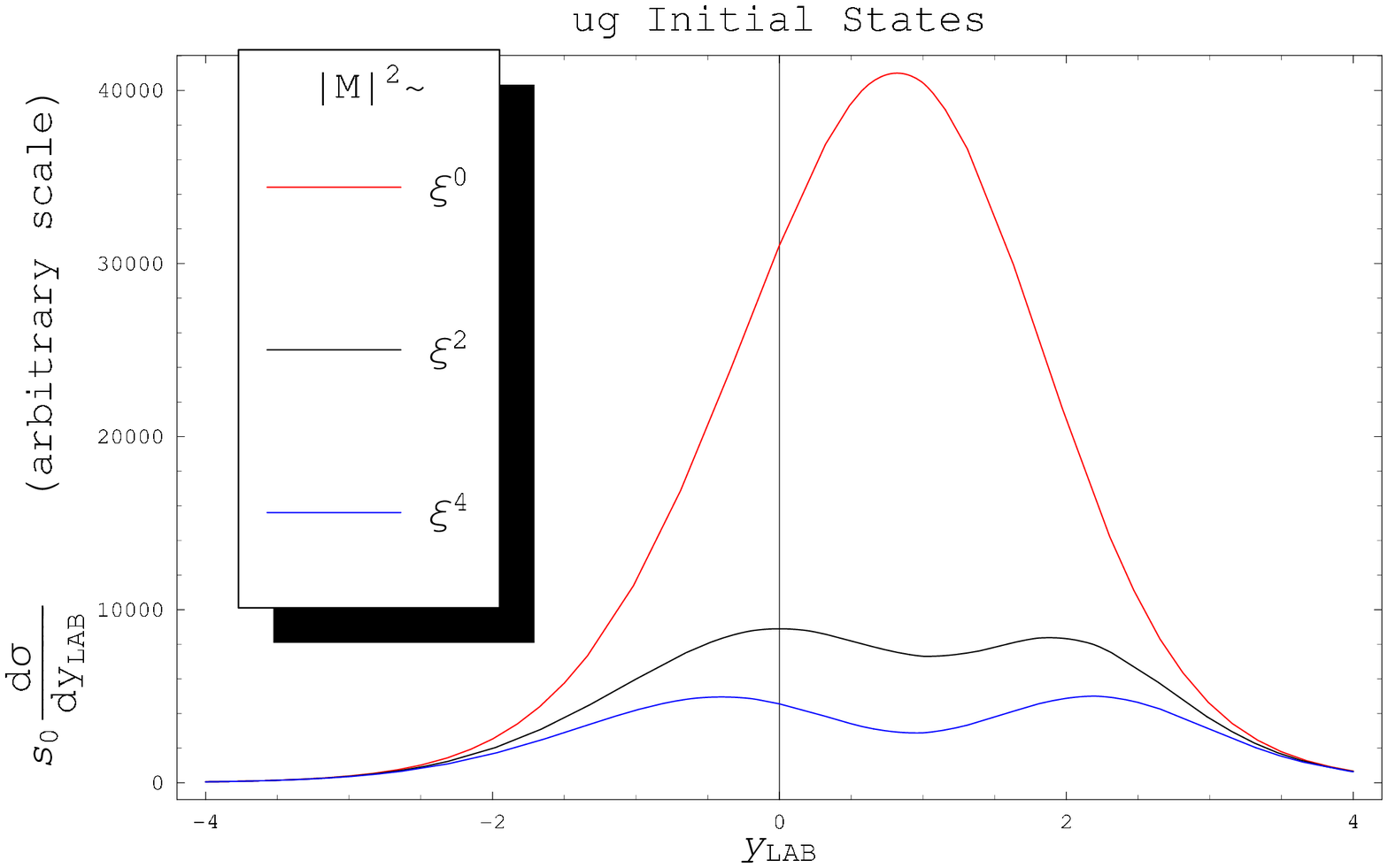}
\includegraphics[width=3.1in,angle=0]{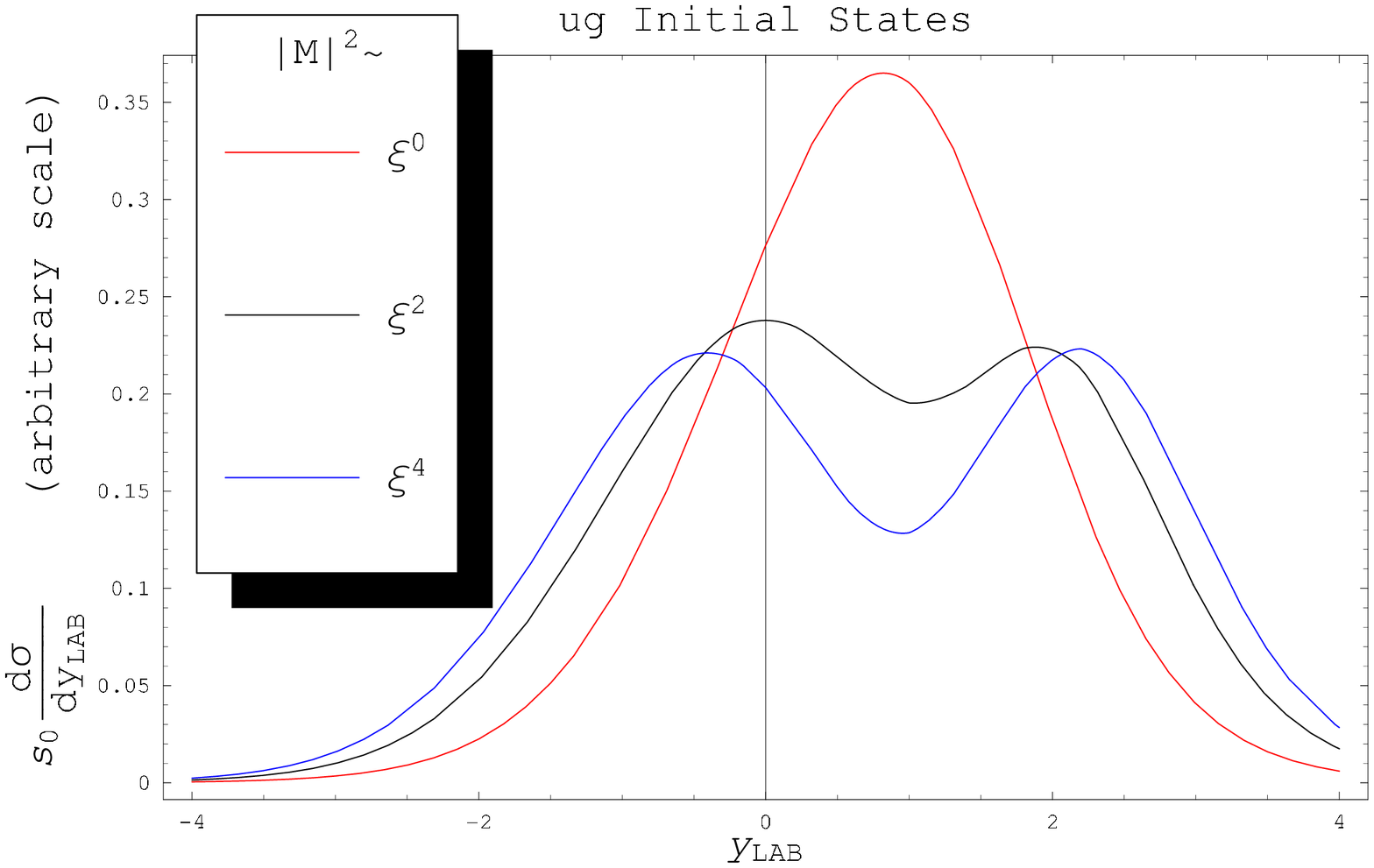}
\includegraphics[width=3.1in,angle=0]{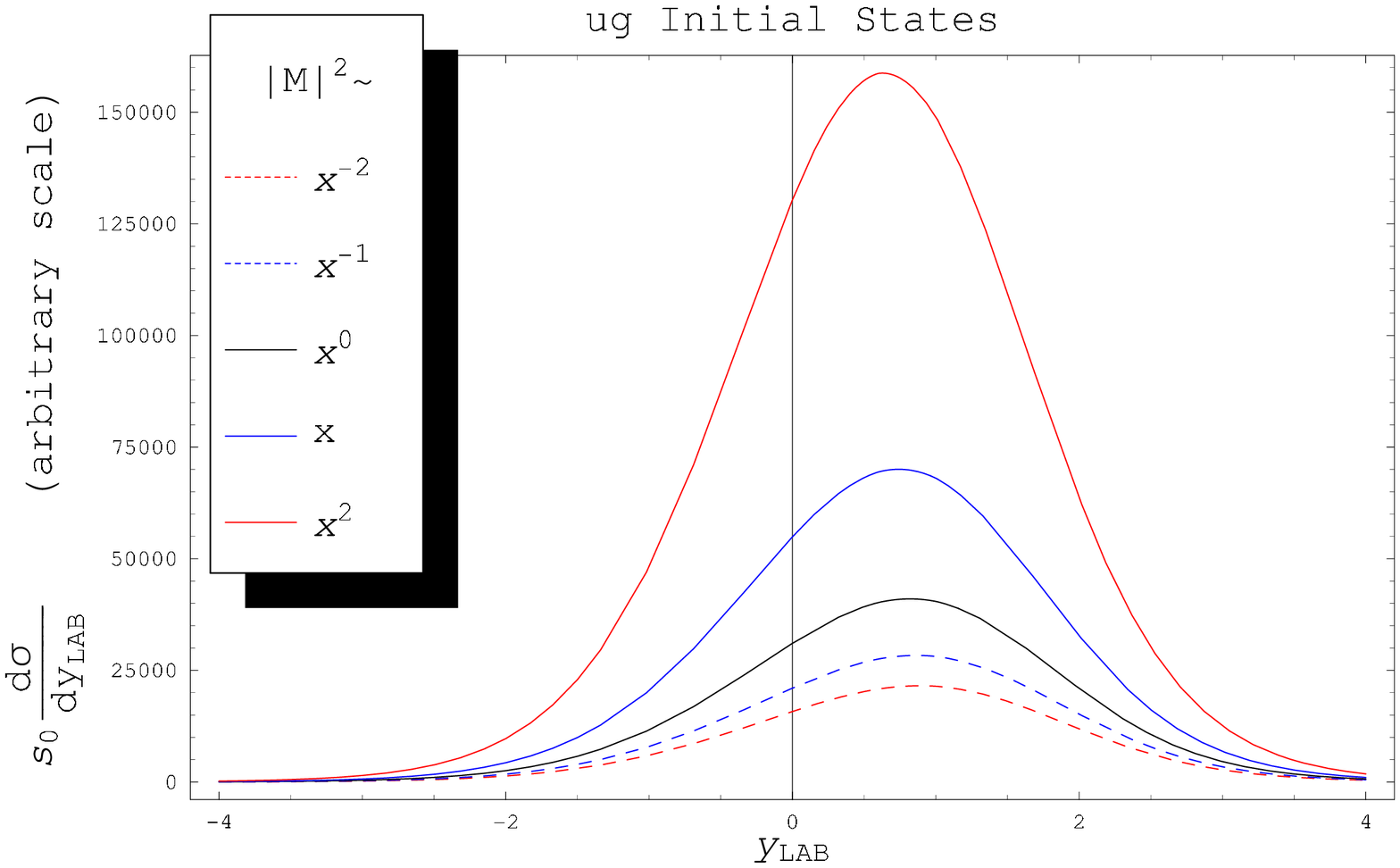}
\includegraphics[width=3.1in,angle=0]{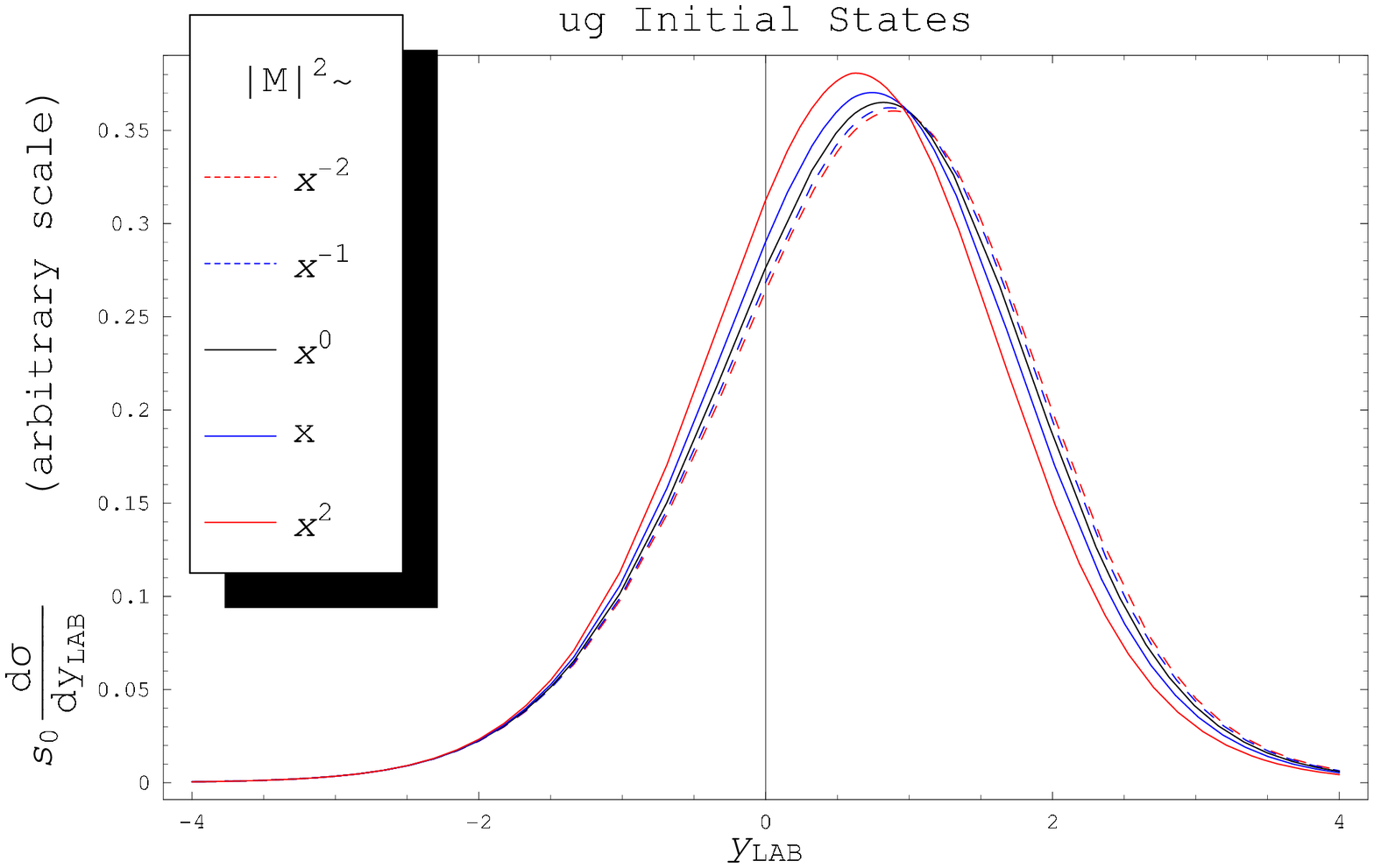}
\includegraphics[width=3.1in,angle=0]{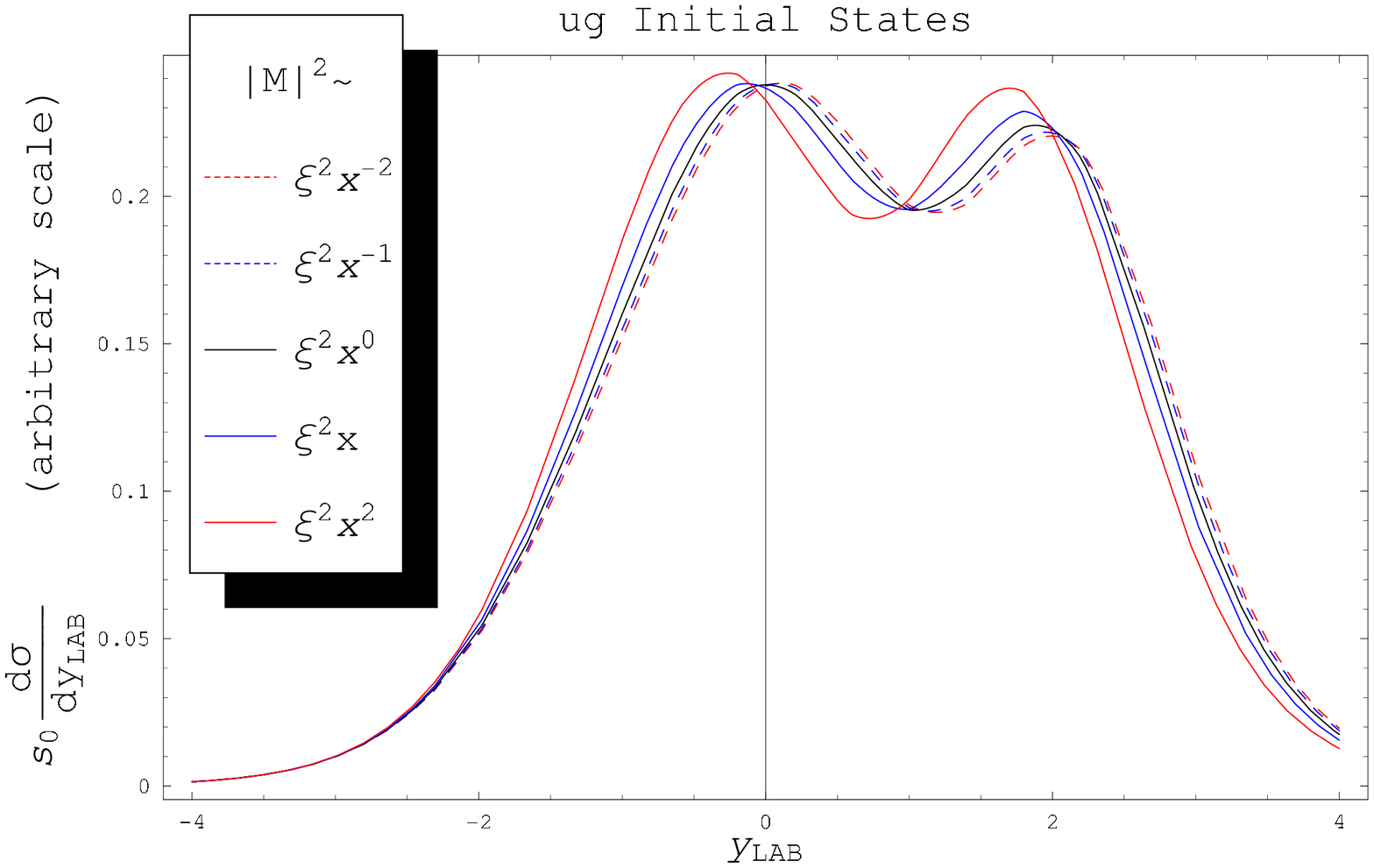}
\includegraphics[width=3.1in,angle=0]{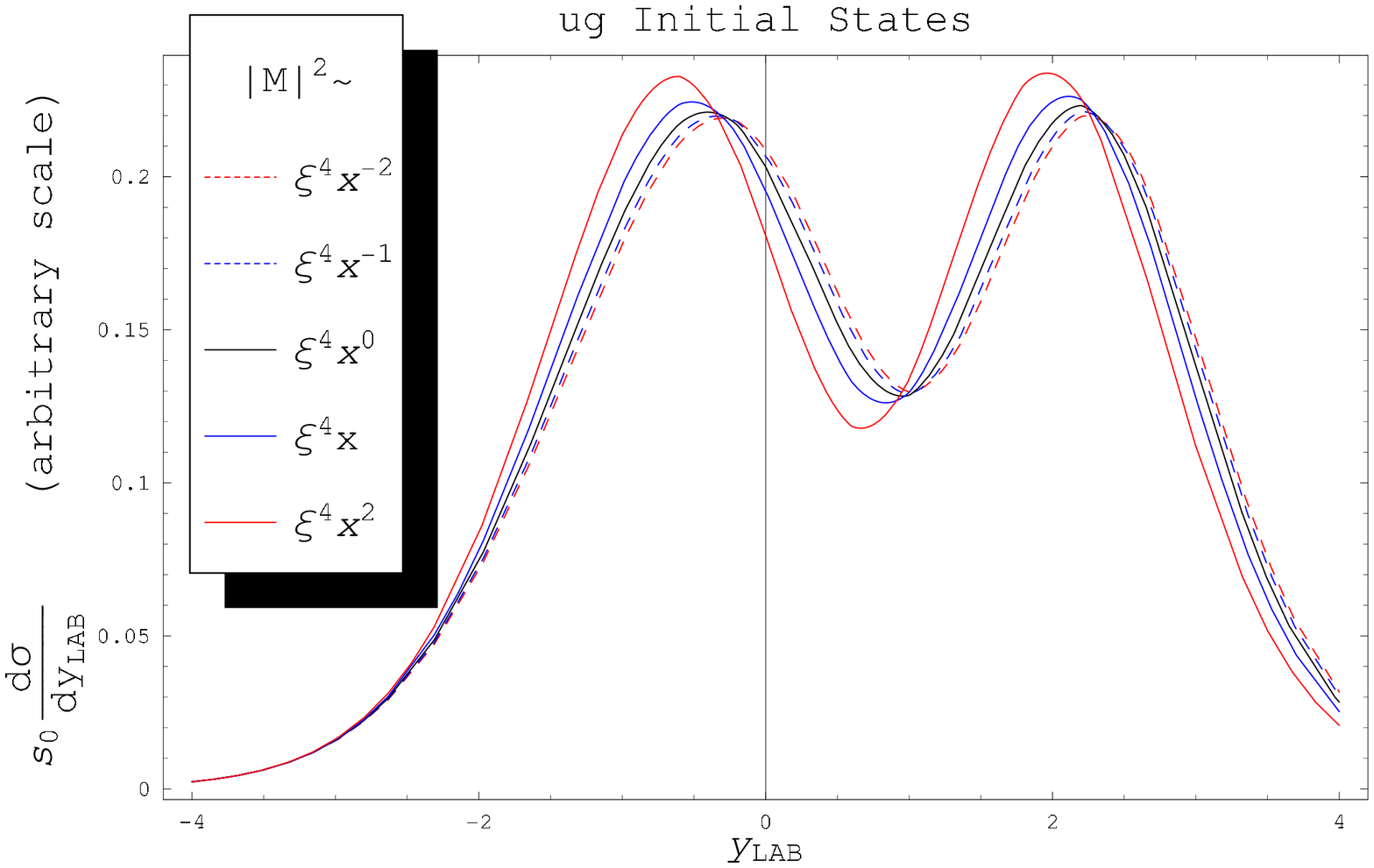}
\caption{Inclusive rapidity distributions shown for various choices
of $|\mathcal{M}|^2\sim X^m \xi^n$ with $ug$ initial states. Final
state masses are taken as equal with $s_0=1$ TeV$^2$ and $Q^2=(500
\mbox{ GeV})^2$. Top left/right: Un-normalized/normalized
distributions for $|\mathcal{M}|^2\sim \xi^n$. Middle left/right:
Un-normalized/normalized distributions for $|\mathcal{M}|^2\sim
X^m$. Bottom left: Normalized distributions for $|\mathcal{M}|^2\sim
\xi^2X^m$. Bottom right: Normalized distributions for
$|\mathcal{M}|^2\sim \xi^4X^m$. We see that the rapidity structure
is largely independent of $X$.} \label{fig:ugRapShapePlots}
\end{center}
\end{figure}

With this result, Eq.~(\ref{eq:RapCalc}) takes the form
\begin{equation}
s_0\f{d\sigma}{d y}\approx \f{1}{2}\sum_{mn}C_{mn} \,
\hat{g}_n(y,s_0) \int_{1}^{X_{max}}dX
X^{m-2}\left(1+\f{\Delta^{\prime}}{X}\right)^{1+n} \left(\epsilon
X\right)^{\eta_a+\eta_b} \left(1 + \mathcal{O}\left(\frac{1}{\log
\epsilon} \right) \right).
\end{equation}
From this, we see that at leading order in $1/\log \epsilon$, the integral over $X$ does not affect the
$y$ dependence of $s_0\f{d\sigma}{dy}$!
Consequently, we expect only the normalization of
$s_0\f{d\sigma}{dy}$ to change at leading order under
$|\mathcal{M}|^2\rightarrow |\mathcal{M}|^2X^p$. We also see that the leading order
rapidity shape structure is controlled by the PDFs quite explicitly!

The above argument tells us that the success of factorization
depends on whether or not we explore phase space boundaries.  For
$|\mathcal{M}|^2\sim X^m\xi^n$ with non-zero $n$, the matrix element
favors large rapidity so the $X$-dependent $\bar{y}_*$ boundary is
more pronounced, leading to largish violations of shape invariance.
Figures \ref{fig:uuRapShapePlots} and \ref{fig:ugRapShapePlots} show
rapidity shapes and variations for different choices of
$|\mathcal{M}|^2$ behavior.  The rapidity shape invariance under
$|\mathcal{M}|^2\rightarrow |\mathcal{M}|^2X^p$ turns out to be
numerically less accurate than the transverse shape invariance under
$|\mathcal{M}|^2\rightarrow |\mathcal{M}|^2\xi^p$, but because the
rapidity distributions are controlled mostly by PDFs anyway, getting
a robust matrix element parametrization is less important than
choosing the right initial state partons.


%% file: AppendixExampleFits.tex
\section{A Useful Parametrization With Examples}
\label{app:ExampleFits}

In Appendix \ref{sec2:APP}, we argued that once PDFs are folded into
partonic matrix elements, there exists useful shape invariance
relations.  These shape invariances allow us to capture the leading
LHC signatures through a small set of basis matrix elements.

\begin{figure}[tbp]
\begin{center}
\includegraphics[width=6in,angle=0]{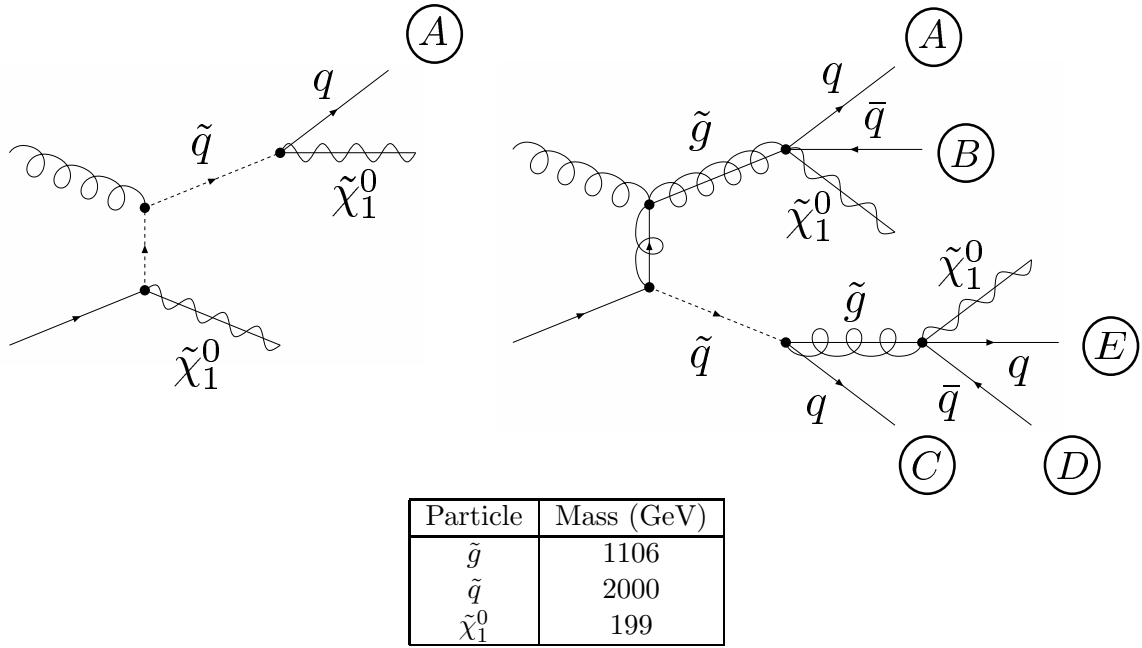}
\begin{tabular}{|c|c|c|}
\hline Particle & Mass (GeV)
\\ \hline
$\tilde g$ & 1106  \\
$\tilde q$ & 2000  \\
$\tilde{\chi}^0_1$   & 199   \\
\hline
\end{tabular}
\caption{The left figure illustrates the $\tilde \chi_1 \tilde q_L$
associated production example. The squarks are left handed, first
and second generation, and decay to first and second generation
quarks and the LSP. The right figure shows the topology and labeling
scheme for the $\tilde g\tilde q_L$ production example. In this
case, $\tilde g$ decays through an off-shell squark to light quarks
and an LSP. For simplicity, we force $\tilde g$ to decay to
$u\bar{u}\tilde \chi_1$. The spectrum for both of these examples is
given in the table, and the LSP $\tilde \chi_1$ is Bino-like.}
\label{fig:ex3and4}
\end{center}
\end{figure}

Here, we give several concrete examples where the matrix element
parametrization of Section \ref{sec2} is indeed effective.  In
Appendix \ref{app:twopara} we give two examples where a
two-parameter matrix element fit is sufficient to accurately
reproduce almost all final state kinematic distributions.  In
Appendix \ref{app:spinwash}, we show how spin information is washed
out both by PDF integration and cascade decays.  While we have made
no attempt to justify the OSET parametrization in every conceivable
topology, the examples in this appendix provide ample anecdotal
evidence to warrant a more complete study within the context of a
realistic detector simulation.  All examples are either based on the
topologies from Figure \ref{fig:ex1and2} or on new topologies
defined in Figure \ref{fig:ex3and4}.

Indeed, for an experimental collaboration, these kinds of case
studies can be used to optimize the discovery potential for
particular topologies, regardless of the dynamics of the underlying
physical model. It may be easier to search for certain pathological
matrix elements that violate the shape invariance relations we
found, or the persistence of shape invariance might influence the
types of experimental cuts used in general searches.  Once a
discovery is made, the shape of kinematic distributions may provide
hints as to the underlying model, so it will be important to know
which distributions are most sensitive to detailed dynamics. In the
plots that follow, we look only at single object observables;
undoubtedly, there exist correlated observables that can be analyzed
at high luminosity for which higher order correction must be
included in the OSET parametrization scheme.

\subsection{Two-Parameter Fits}
\label{app:twopara}

\begin{figure}[htbp]
\begin{center}
\includegraphics[width=2.2in,angle=-90]{sec2plots/chi2gl_glpt.eps}
\includegraphics[width=2.2in,angle=-90]{sec2plots/chi2gl_Apt.eps}
\includegraphics[width=2.2in,angle=-90]{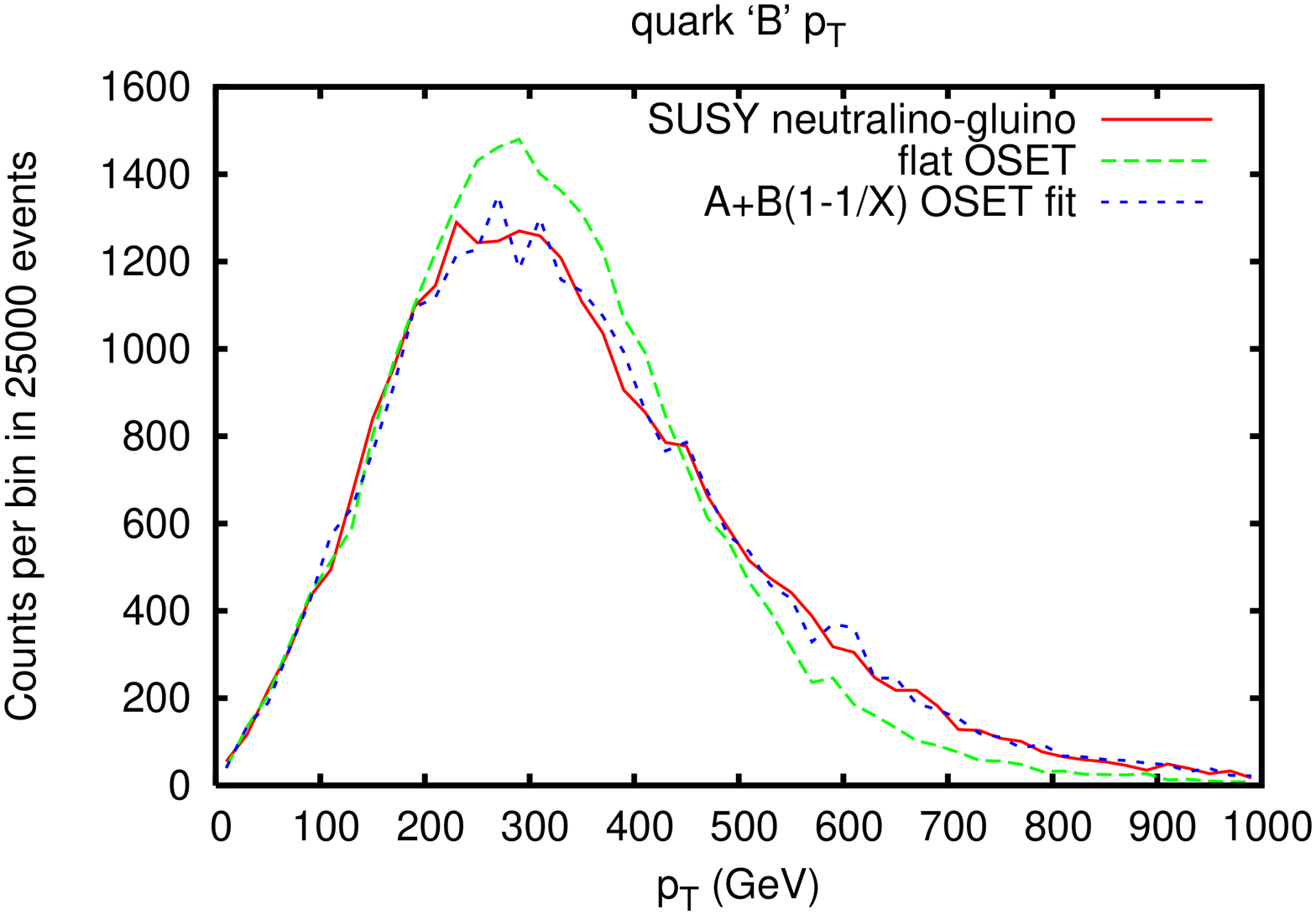}
\includegraphics[width=2.2in,angle=-90]{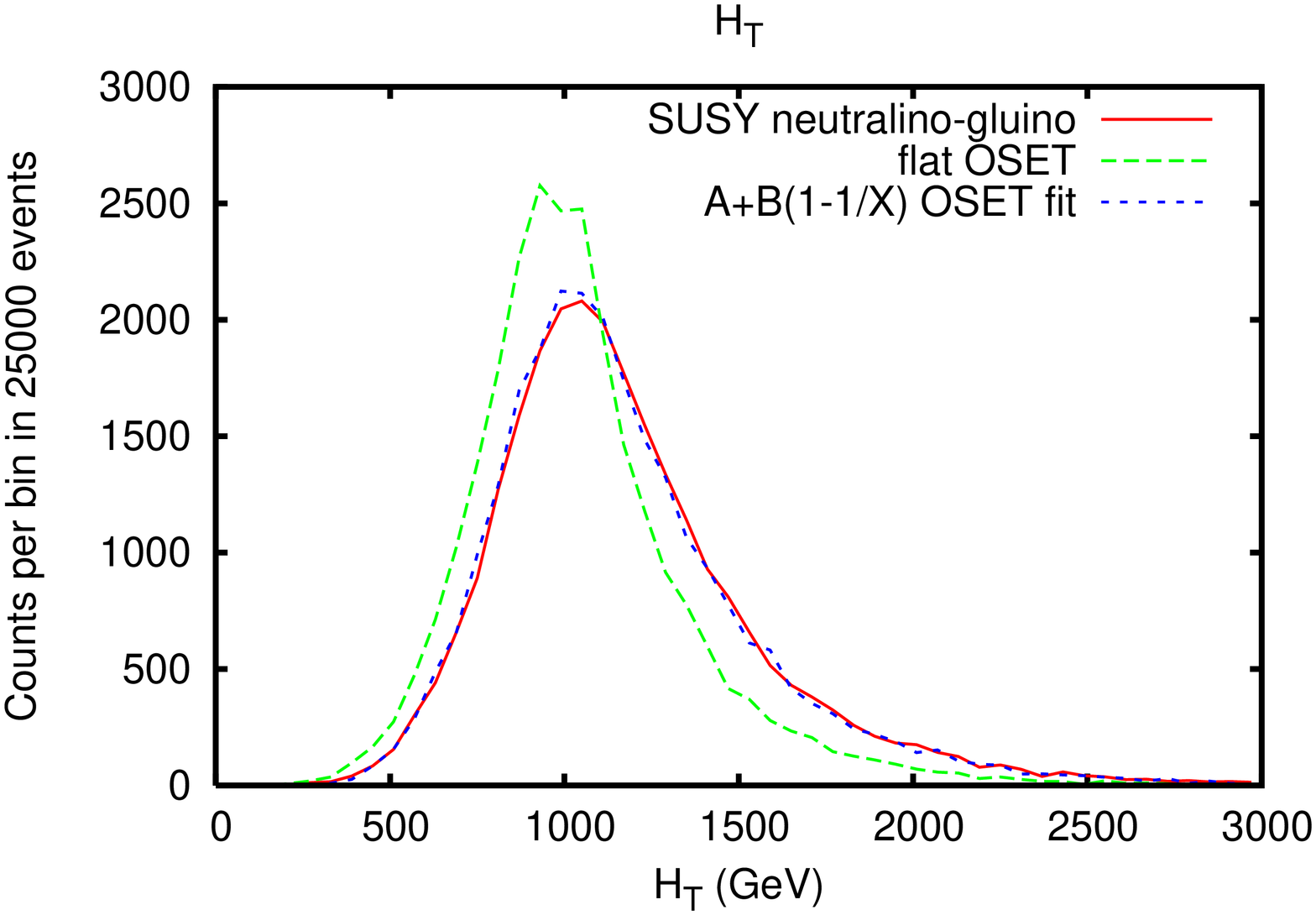}
\includegraphics[width=2.2in,angle=-90]{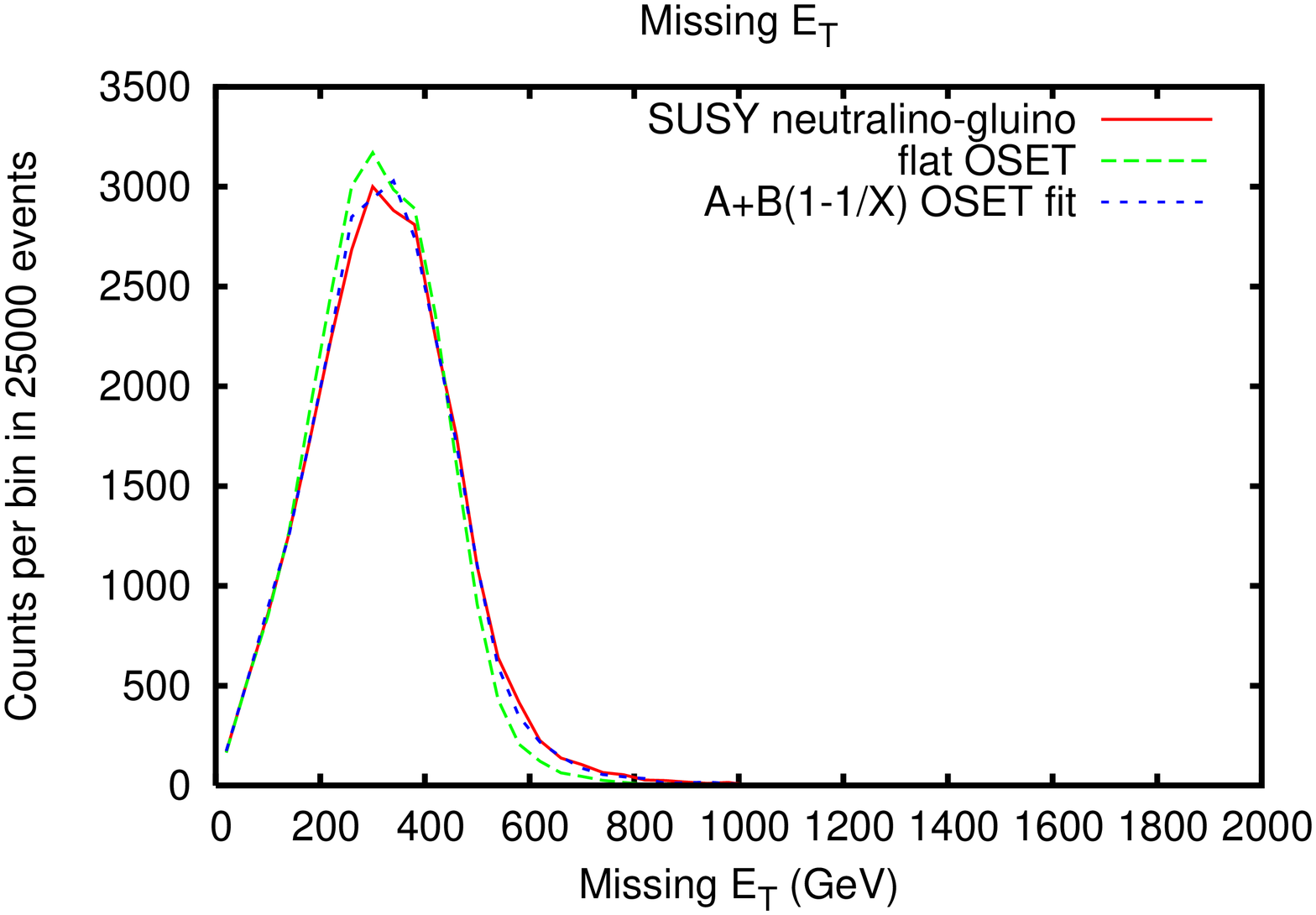}
\caption{Inclusive $p_T$ and $H_T$ distributions for the SUSY
$\tilde \chi_2 \tilde g$ associated production process in Figure
\ref{fig:ex1and2} compared with an OSET with a flat matrix element
$|\mathcal{M}|^2 \propto 1$ and an OSET fit using our leading order shape
parametrization (LOSP). In this case, the LOSP consists of
$\Msq=A+B(1-1/X)$. The fit is obtained by fitting to the shape of the
$Z$ $p_T$ distribution. In all cases, the correct $s_0$ and $\Delta$
are chosen. Top left: $p_T$ of partonic gluino. Top right: $p_T$ of
quark ``A''. Middle left: $p_T$ of quark ``B''. Middle right: Inclusive
$H_T$ distribution. Bottom: Missing $E_T$ distribution.}
\label{fig:FitChi2GlPt}
\end{center}
\end{figure}
\begin{figure}[htbp]
\begin{center}
\includegraphics[width=2.2in,angle=-90]{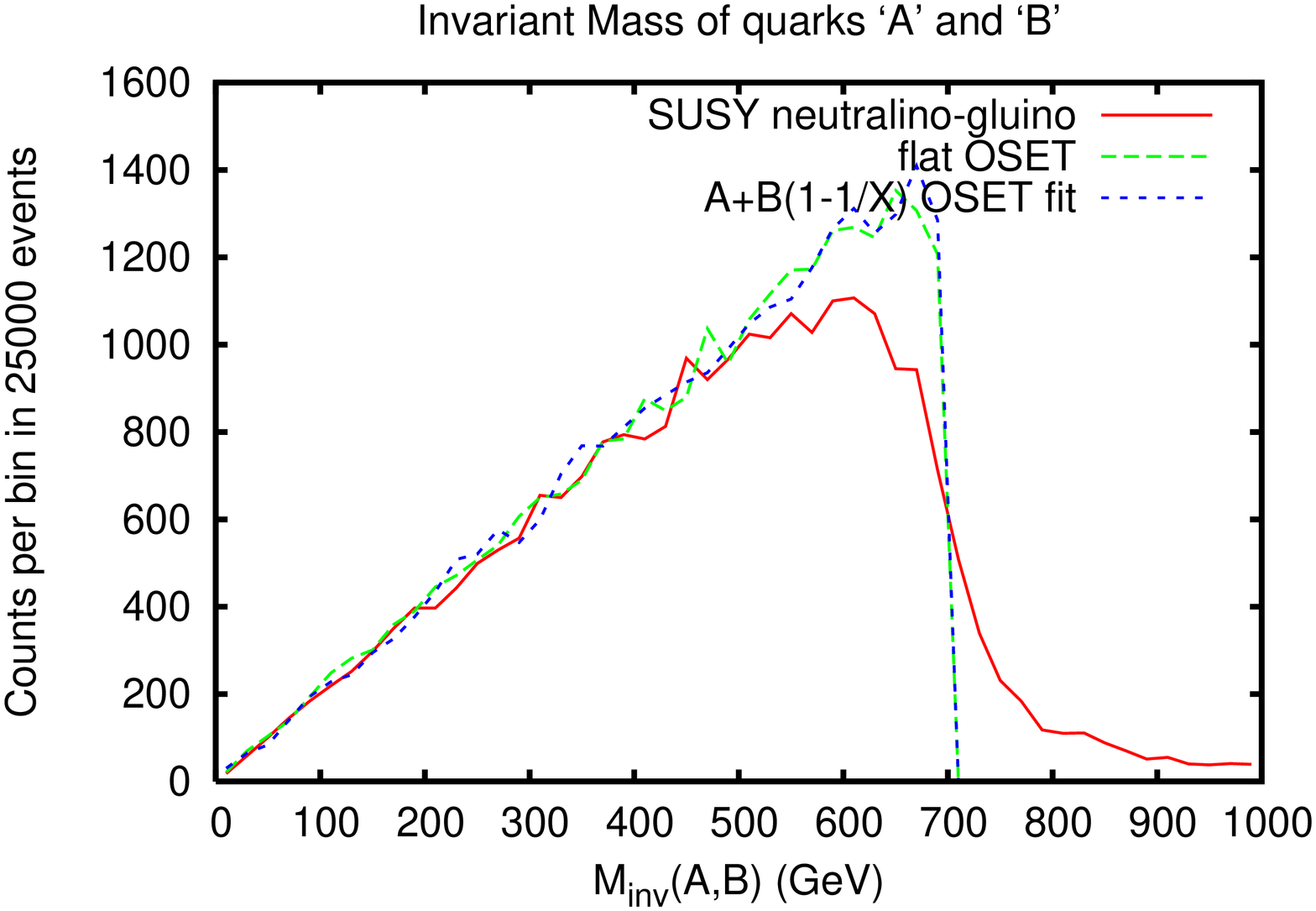}
\includegraphics[width=2.2in,angle=-90]{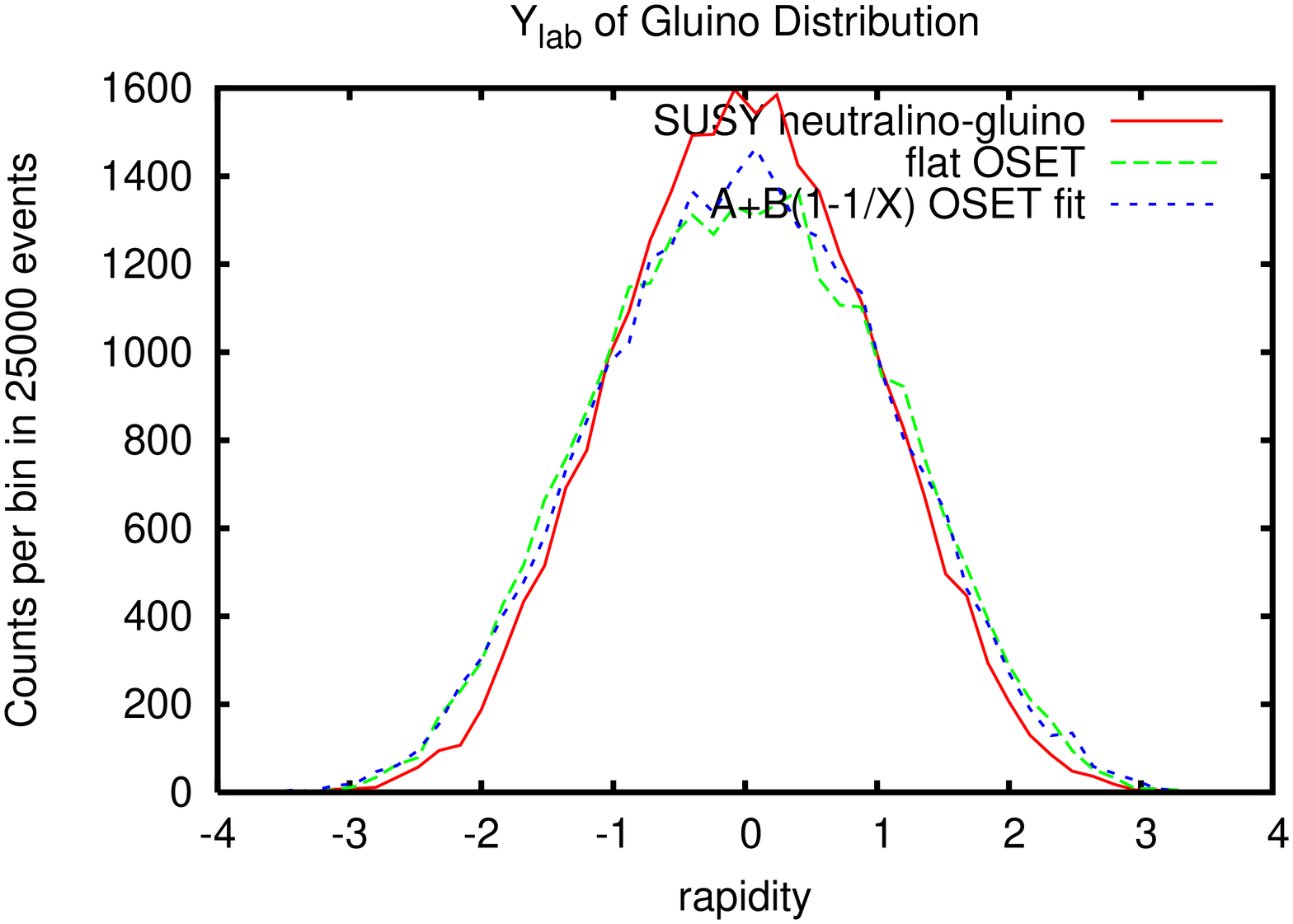}
\includegraphics[width=2.2in,angle=-90]{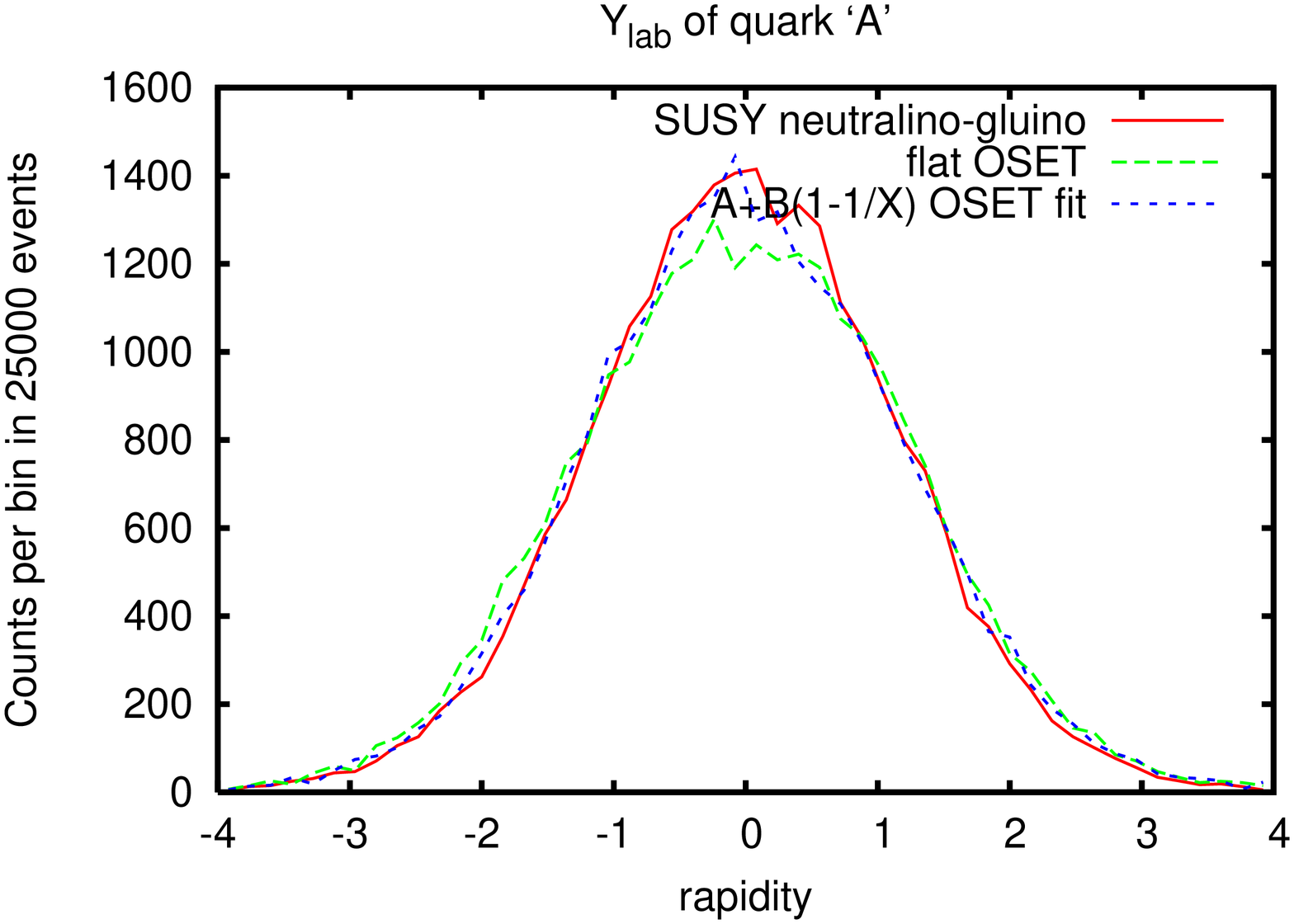}
\includegraphics[width=2.2in,angle=-90]{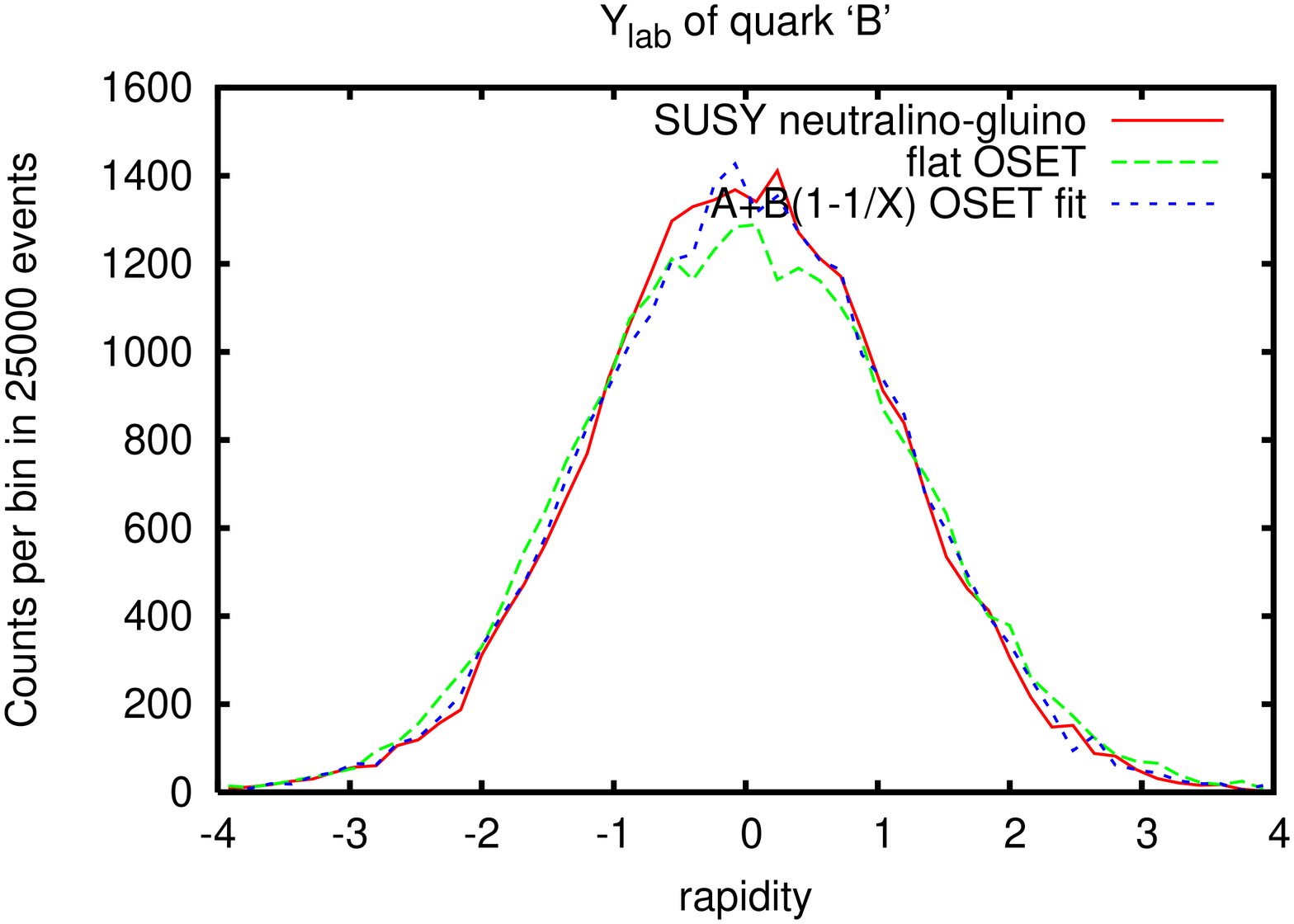}
\caption{Dijet invariant mass and inclusive rapidity distributions
for the SUSY $\tilde \chi_2 \tilde g$ associated production process
in Figure \ref{fig:ex1and2} compared with an OSET with a flat matrix
element $\Msq \propto 1$, and an OSET fit using our leading order shape
parametrization (LOSP). In this case, the LOSP consists of
$\Msq=A+B(1-1/X)$. The fit is obtained by fitting to the shape of the
$Z$ $p_T$ distribution. In all cases, the correct $s_0$ and $\Delta$
are chosen. Top left: $M_{AB}$ of ``A'' and ``B'' quarks. Top right:
rapidity of partonic gluino. Bottom left: rapidity of ``A'' quark.
Bottom right: rapidity of ``B'' quark.}\label{fig:FitChi2GlRap}
\end{center}
\end{figure}
\begin{figure}[htbp]
\begin{center}
\includegraphics[width=2.2in,angle=-90]{sec2plots/chi2gl2700_glpt.eps}
\includegraphics[width=2.2in,angle=-90]{sec2plots/chi2gl2700_Apt.eps}
\includegraphics[width=2.2in,angle=-90]{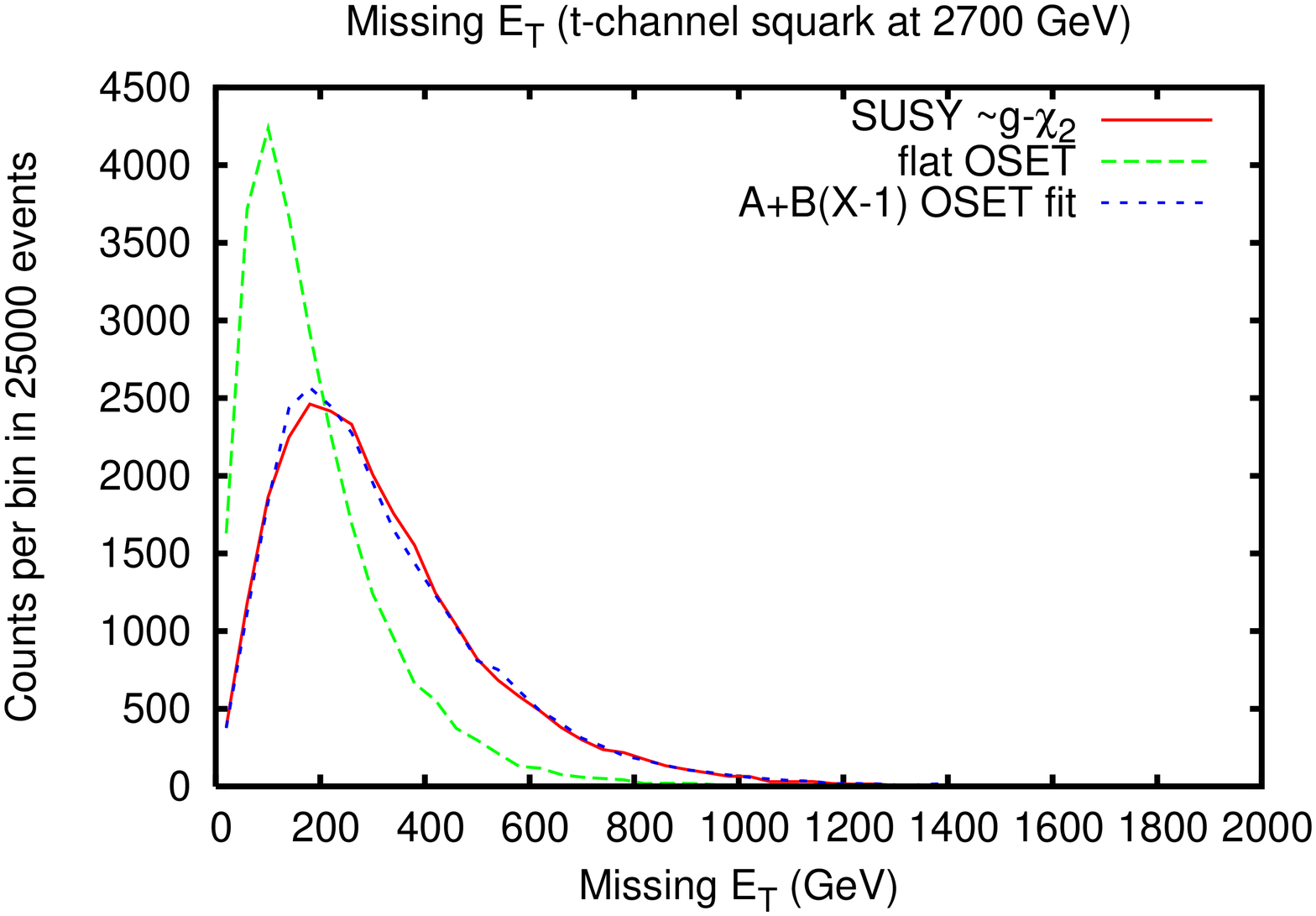}
\includegraphics[width=2.2in,angle=-90]{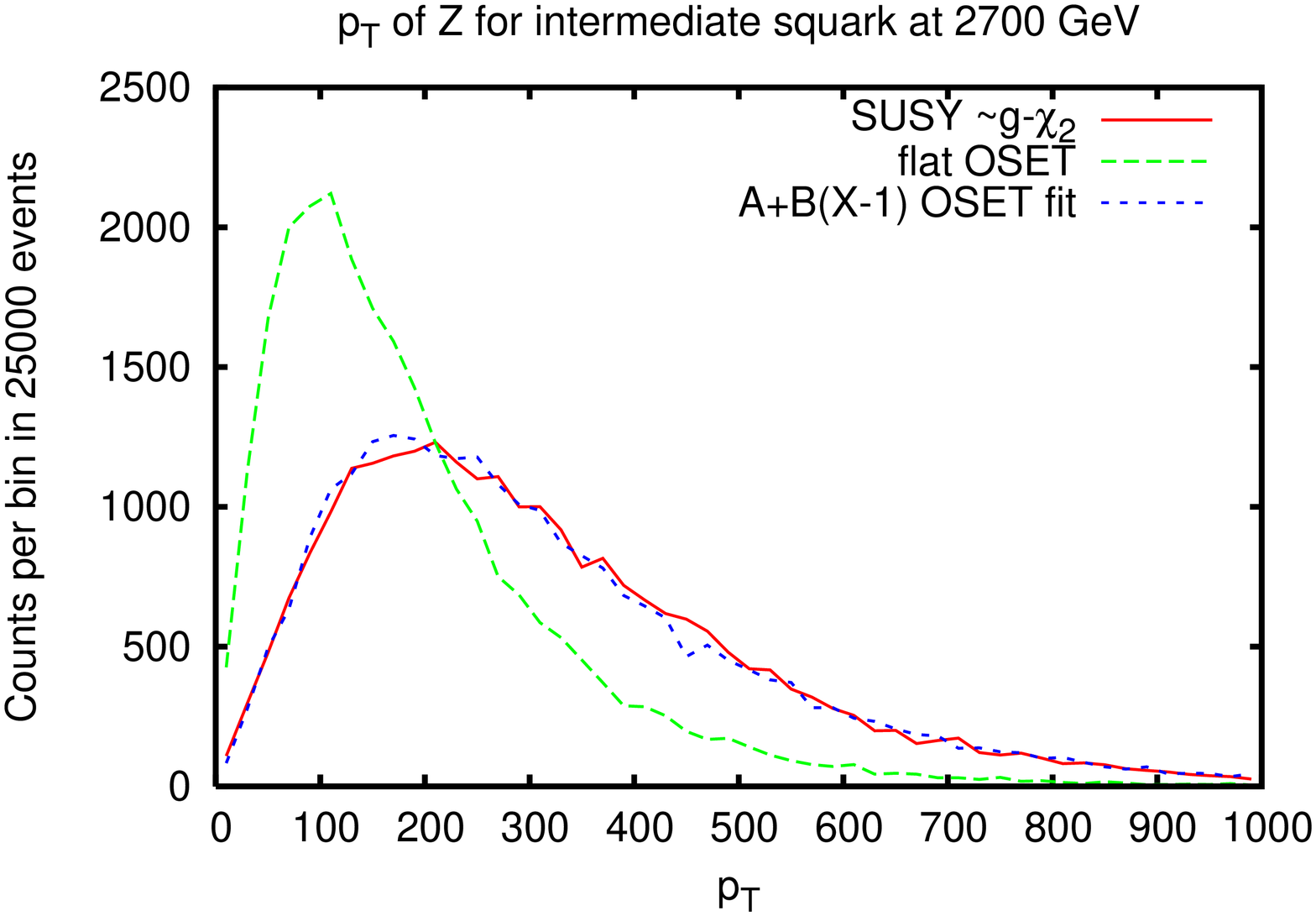}
\includegraphics[width=2.2in,angle=-90]{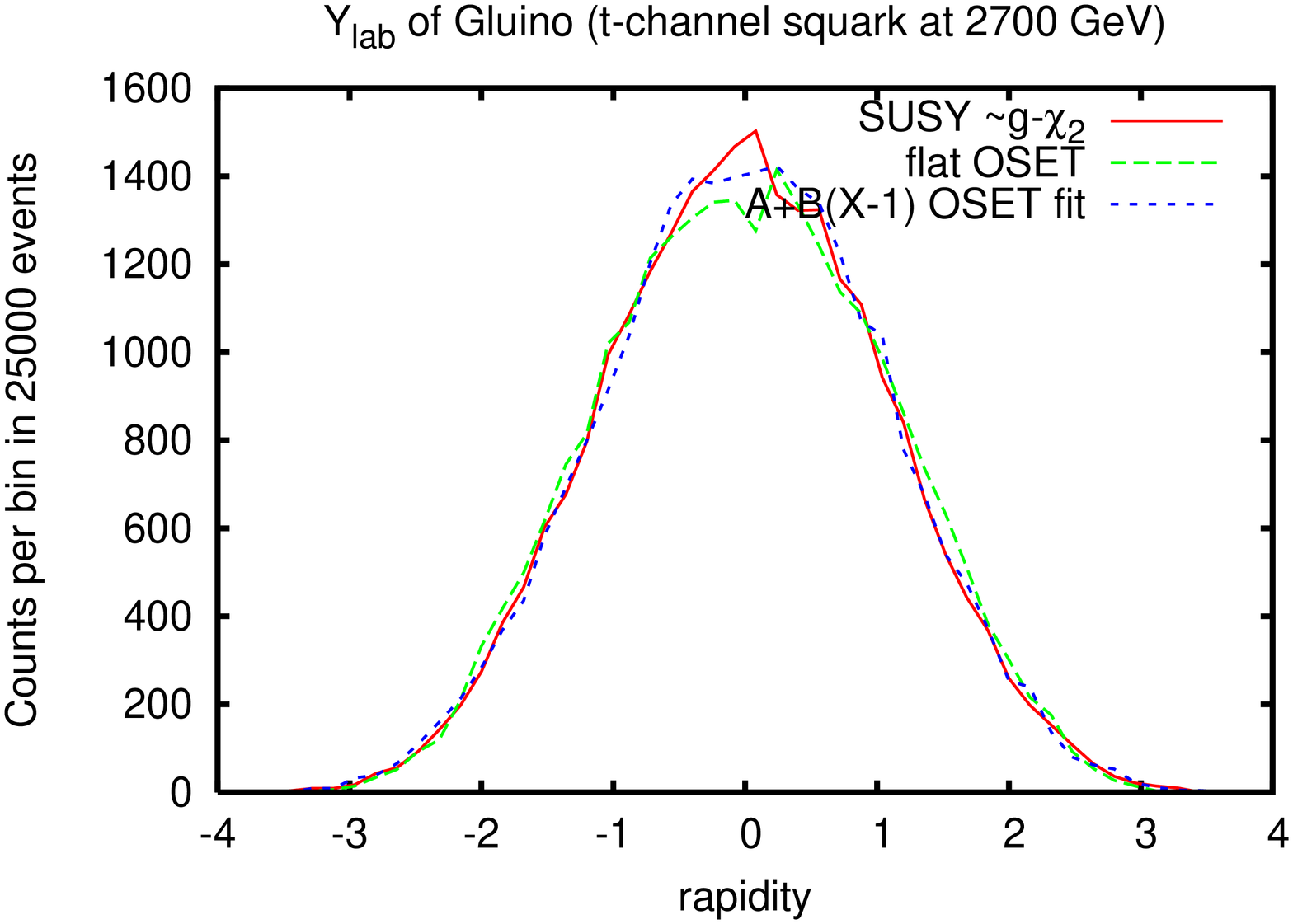}
\includegraphics[width=2.2in,angle=-90]{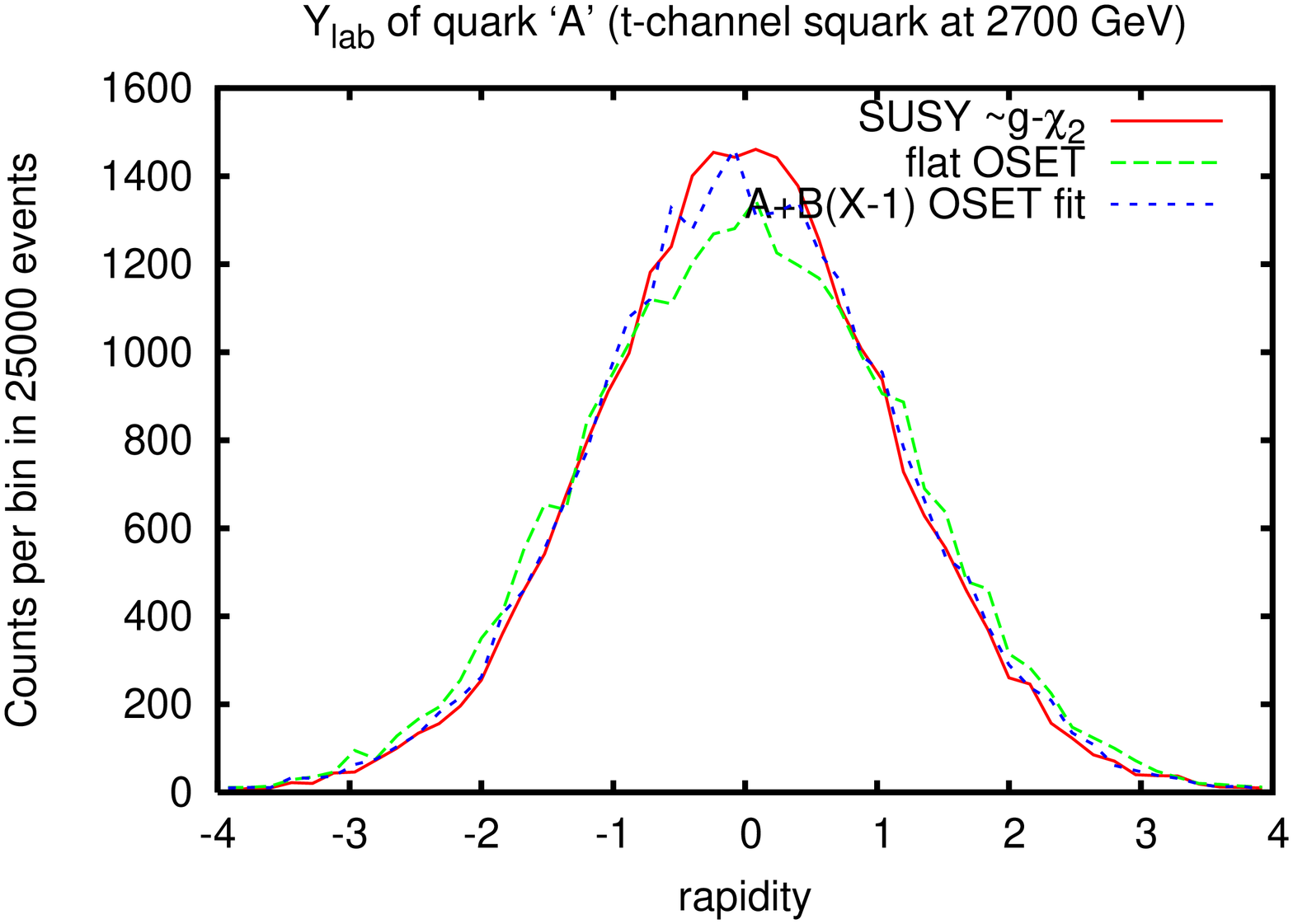}
\caption{Inclusive $p_T$ and rapidity distributions for SUSY $\tilde
\chi_2 \tilde g$ associated production with a heavy (2700 GeV)
$t$-channel propagator, compared with an OSET with a flat matrix
element $|\mathcal{M}|^2 \propto 1$ and an OSET fit using our leading
order shape parametrization (LOSP). In this case, the LOSP consists of
$\Msq=A+B(X-1)$, with $A$ negligible. The fit is obtained by fitting
to the shape of the gluino $p_T$ distribution. In all cases, the
correct $s_0$ and $\Delta$ are chosen. Top left: $p_T$ of partonic
gluino. Top right: $p_T$ of quark ``A'' (note unlike Figure
\ref{fig:ex1and2}, gluino decay is directly 3-body). Middle left:
Missing $E_T$. Middle right: $p_T$ of $Z$ boson.  Bottom: Rapidity
distributions for gluino (left) and quark from gluino decay (right).}
\label{fig:FitChi2GlPt2700}
\end{center}
\end{figure}
\begin{figure}[htbp]
\begin{center}
\includegraphics[width=2.2in,angle=-90]{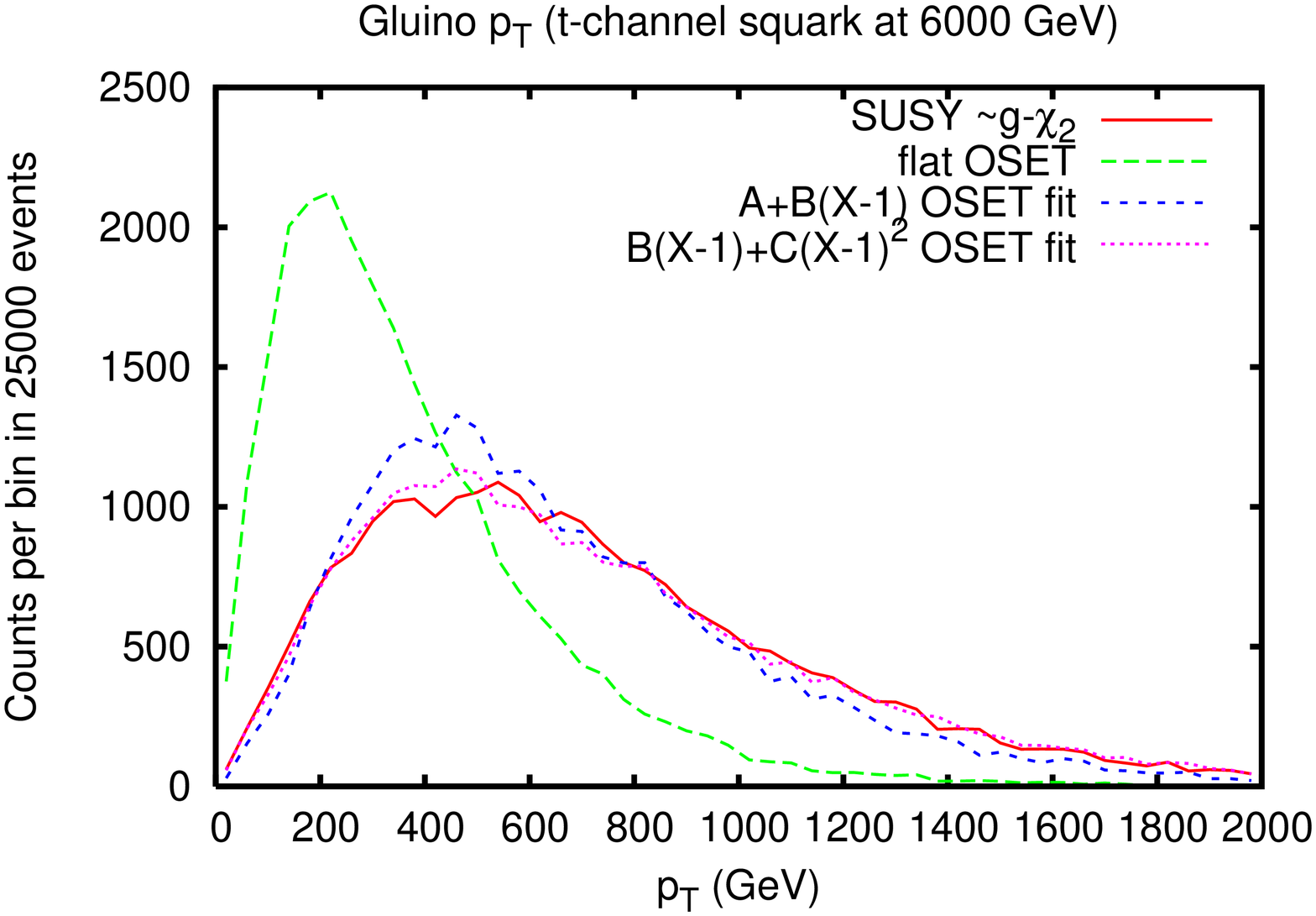}
\includegraphics[width=2.2in,angle=-90]{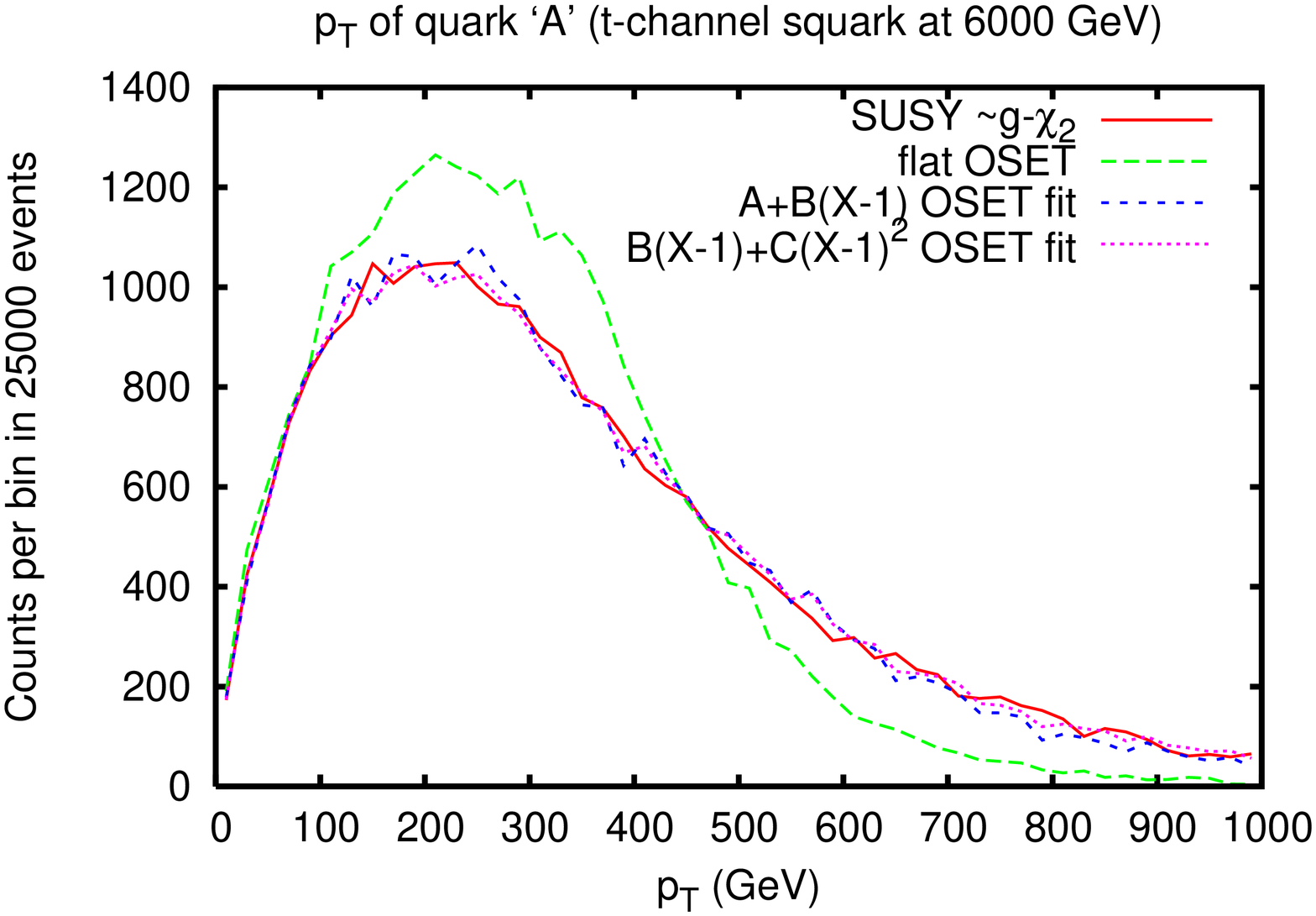}
\includegraphics[width=2.2in,angle=-90]{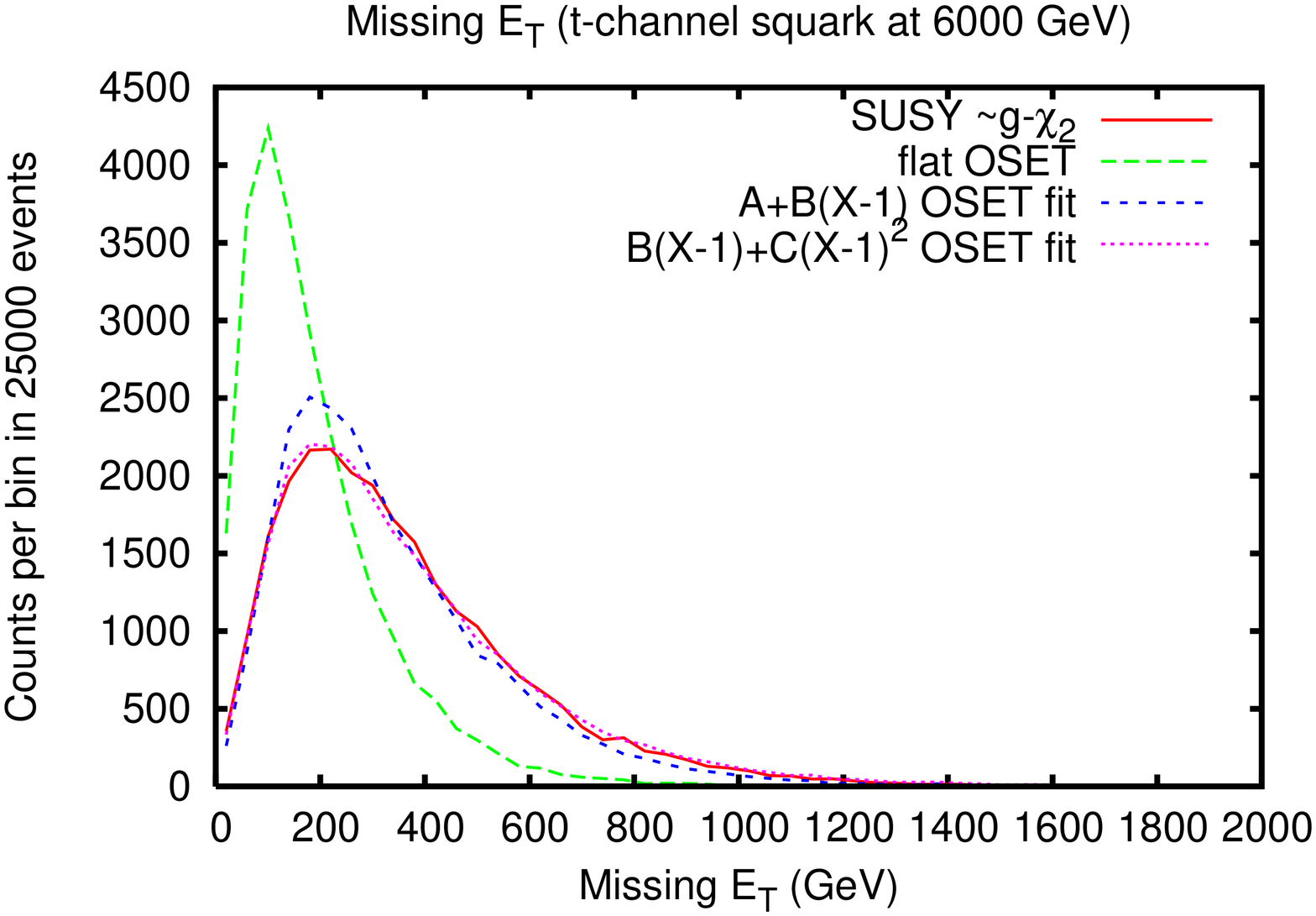}
\includegraphics[width=2.2in,angle=-90]{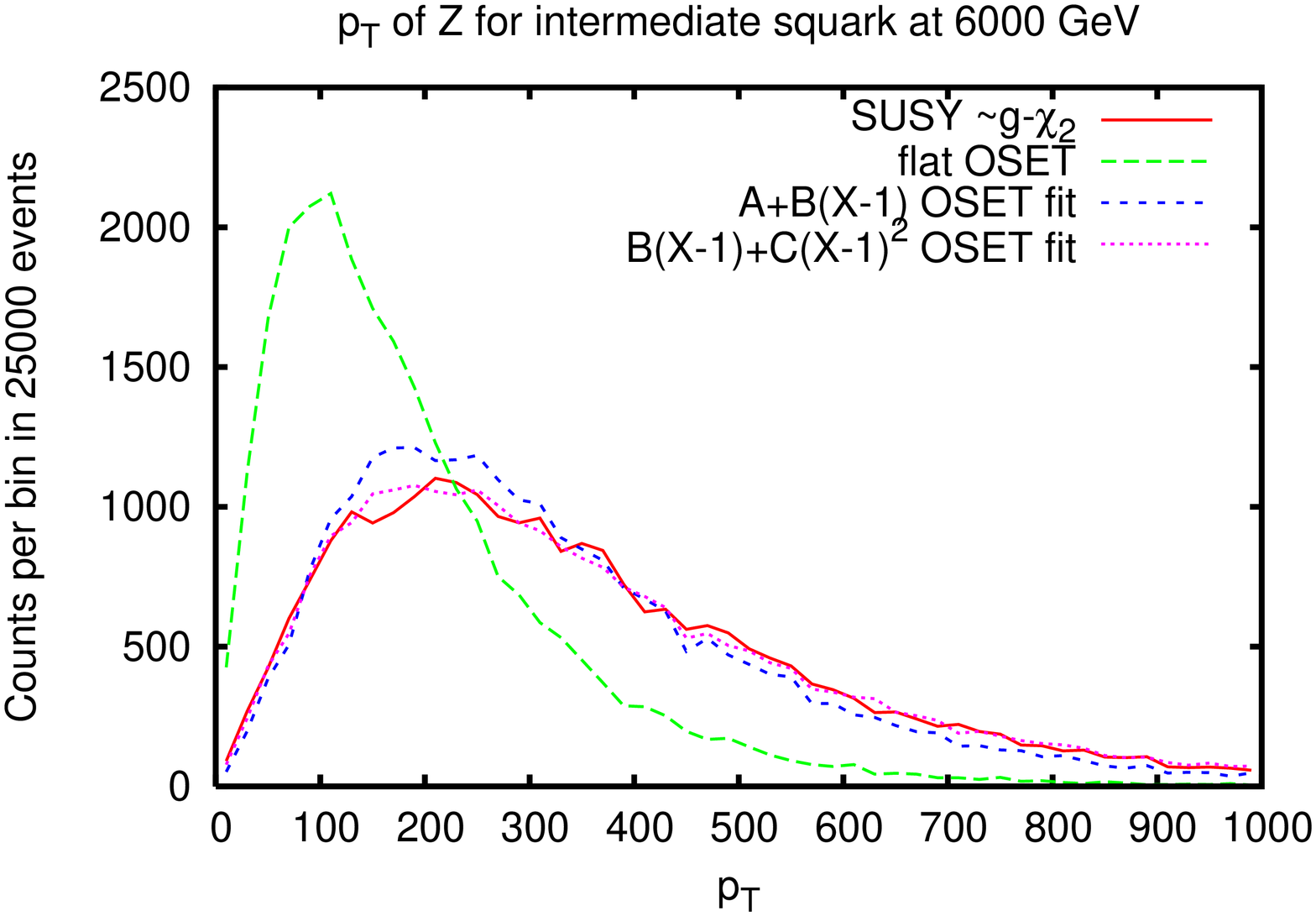}
\includegraphics[width=2.2in,angle=-90]{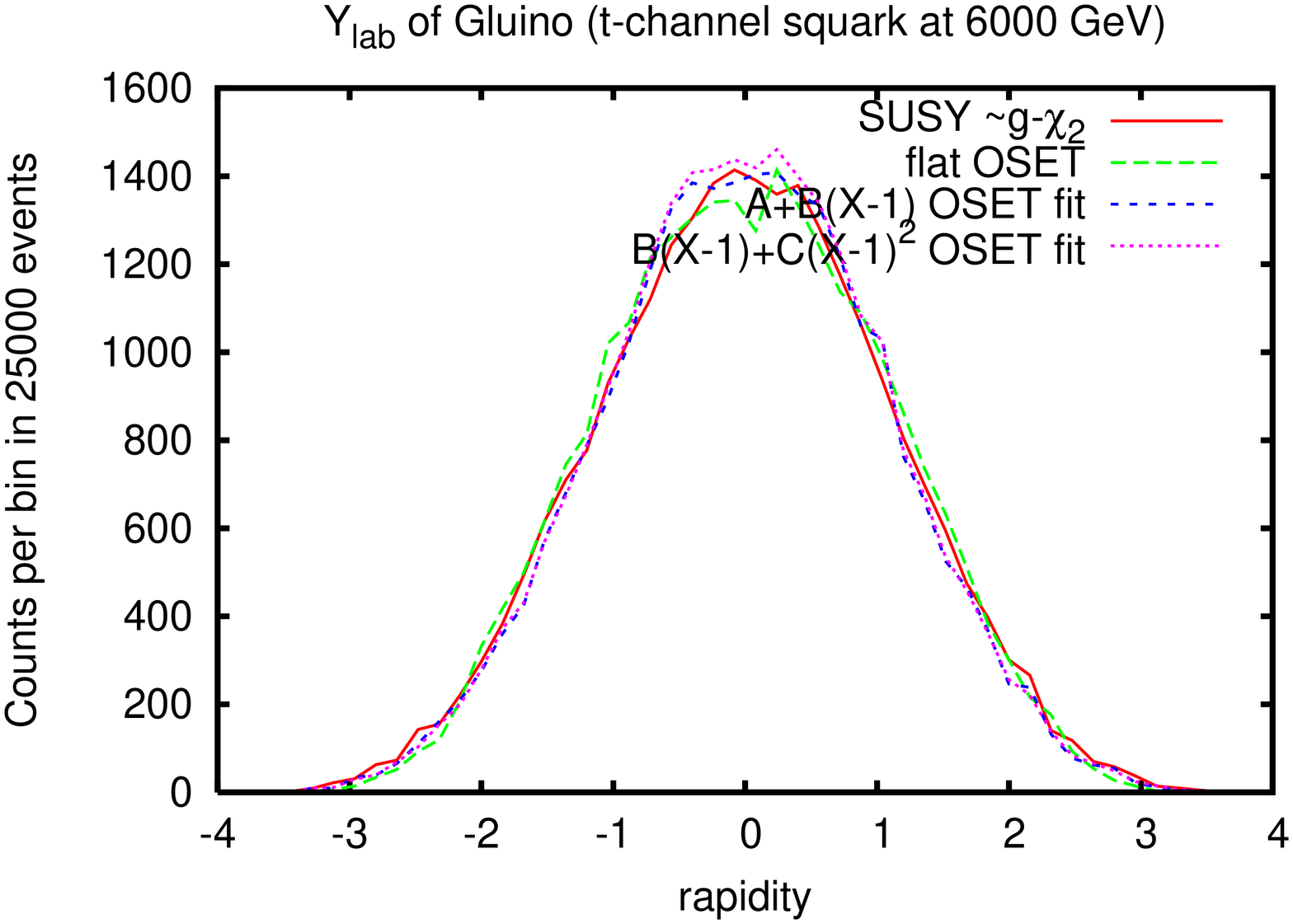}
\includegraphics[width=2.2in,angle=-90]{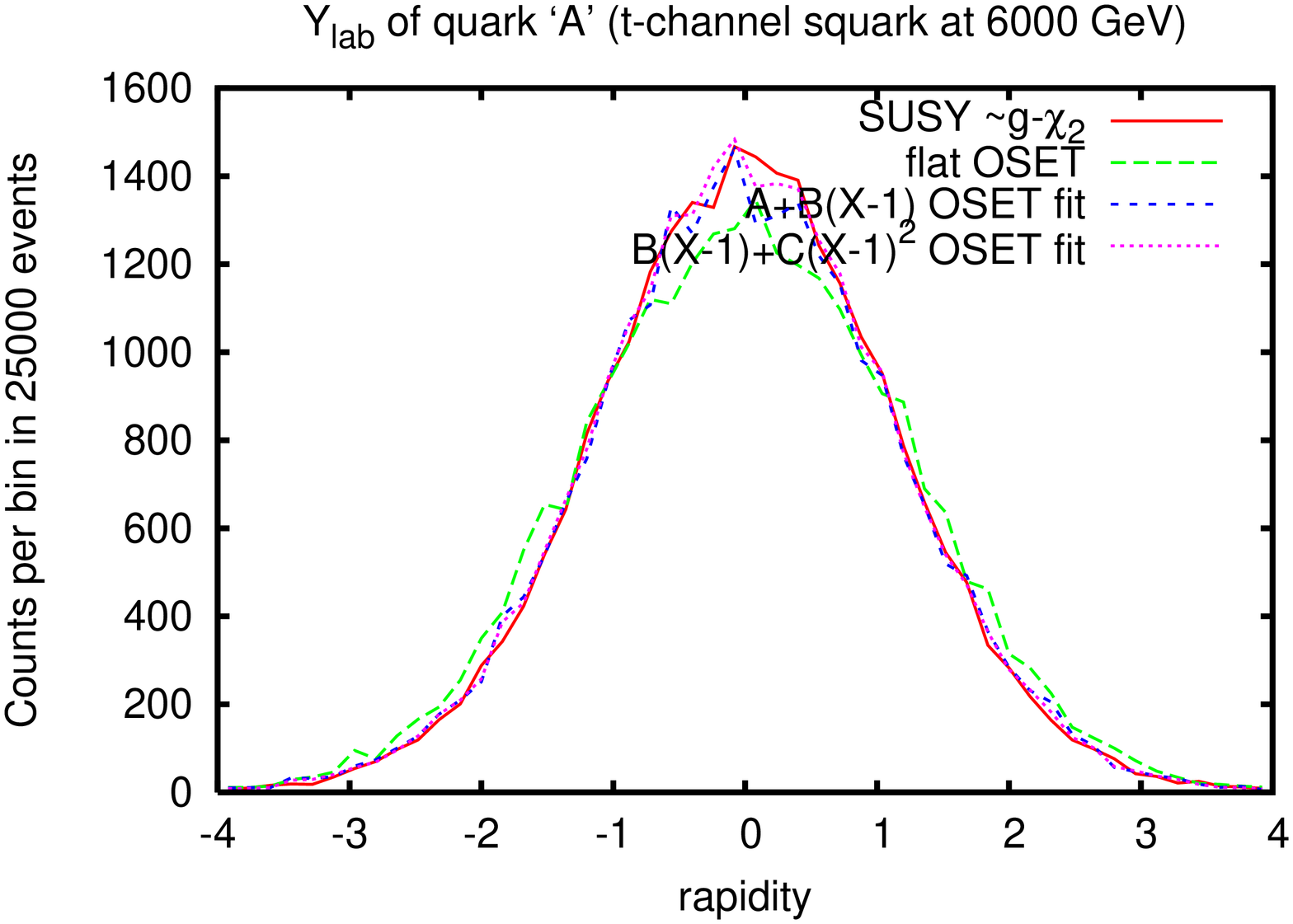}
\caption{Inclusive $p_T$ and rapidity distributions for SUSY $\tilde
\chi_2 \tilde g$ associated production with a heavy (6000 GeV) $t$-channel
propagator, compared with an OSET with a flat matrix element
$|\mathcal{M}|^2 \propto 1$ and an OSET fit using our leading order
shape parametrization (LOSP). In this case, the LOSP consists of
$\Msq=B(X-1)+C(X-1)^2$. The fit is obtained by fitting to the shape of the
gluino $p_T$ distribution. In all cases, the correct $s_0$ and
$\Delta$ are chosen. Top left: $p_T$ of partonic gluino. Top right:
$p_T$ of quark ``A'' (note unlike Figure \ref{fig:ex1and2}, gluino
decay is directly 3-body). Middle left: Missing $E_T$. Middle right:
$p_T$ of $Z$ boson.  Bottom: Rapidity distributions for gluino (left)
and quark from gluino decay (right).}
\label{fig:FitChi2GlPt6000}
\end{center}
\end{figure}

The first example (Figures \ref{fig:FitChi2GlPt} and
\ref{fig:FitChi2GlRap}) comes from the SUSY $\tilde{\chi}_2 \tilde{g}$
associated production channel from Figure \ref{fig:ex1and2}. Despite
the fact that this process exhibits $p$-wave suppression, there are
pieces of the amplitude proportional to powers of $\xi$, so the
transverse shape invariance of $\frac{d \sigma}{d x_T}$ means that
there is an effective $\Msq \propto 1$ contribution to the amplitude.
As shown in Figures \ref{fig:FitChi2GlPt} and \ref{fig:FitChi2GlRap},
a constant amplitude gives a reasonable characterization of most
kinematic quantities.  To better model, say, the $H_T$ distribution,
we can account for the p-wave suppression of near-threshold events by
adding an additional $\Msq \propto (1- 1/X)$ piece.  With a leading
order shape parametrization (LOSP) of $\Msq=A+B(1-1/X)$, all single
object observables are well-described.

This process has qualitatively different behavior if the $t$-channel
(squark) mass $m_{t-\mbox{chan}}$ is much larger than the threshold
value of $\sqrt{\hat s}$---in this case, for $s \lesssim
m_{t-\mbox{chan}}^2$, $\Msq$ grows with $t^2$ like a contact
interaction; because of $\xi$ shape-invariance, this is well modeled
by a LOSP that grows as a quadratic polynomial in $s$: $\Msq=A+B(X-1)$ or
$\Msq=A+B(X-1)+C(X-1)^2$.  We illustrate this behavior and the success
of these fits in Figures \ref{fig:FitChi2GlPt2700} and
\ref{fig:FitChi2GlPt6000}.  In both cases, the constant term $A$ in
$\Msq$ is negligible; in the former, linear growth describes $\Msq$
well, while in the latter including a quadratic term improves the
fit. 

\begin{figure}[tbp]
\begin{center}
\includegraphics[width=2in,angle=-90]{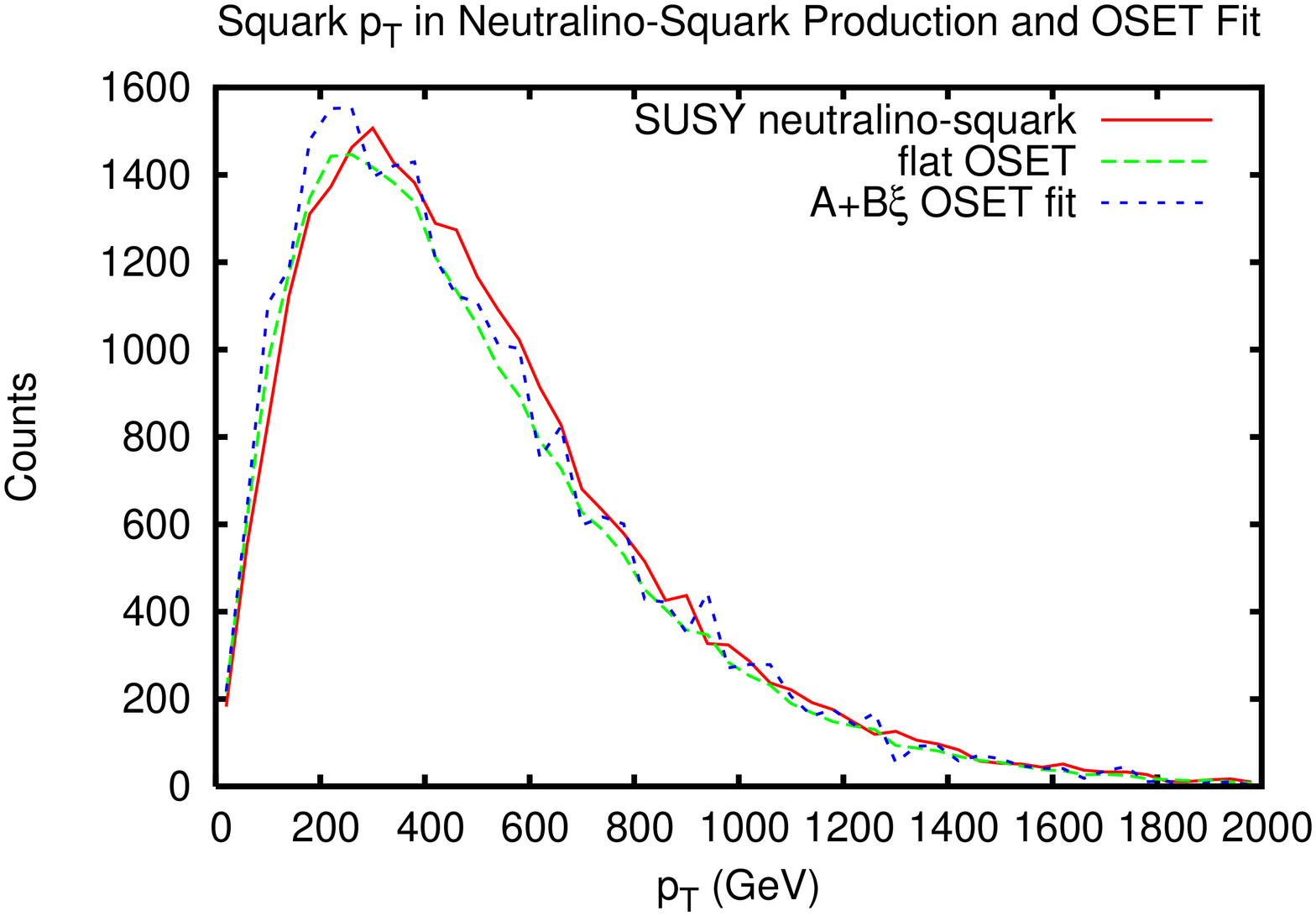}
\includegraphics[width=2in,angle=-90]{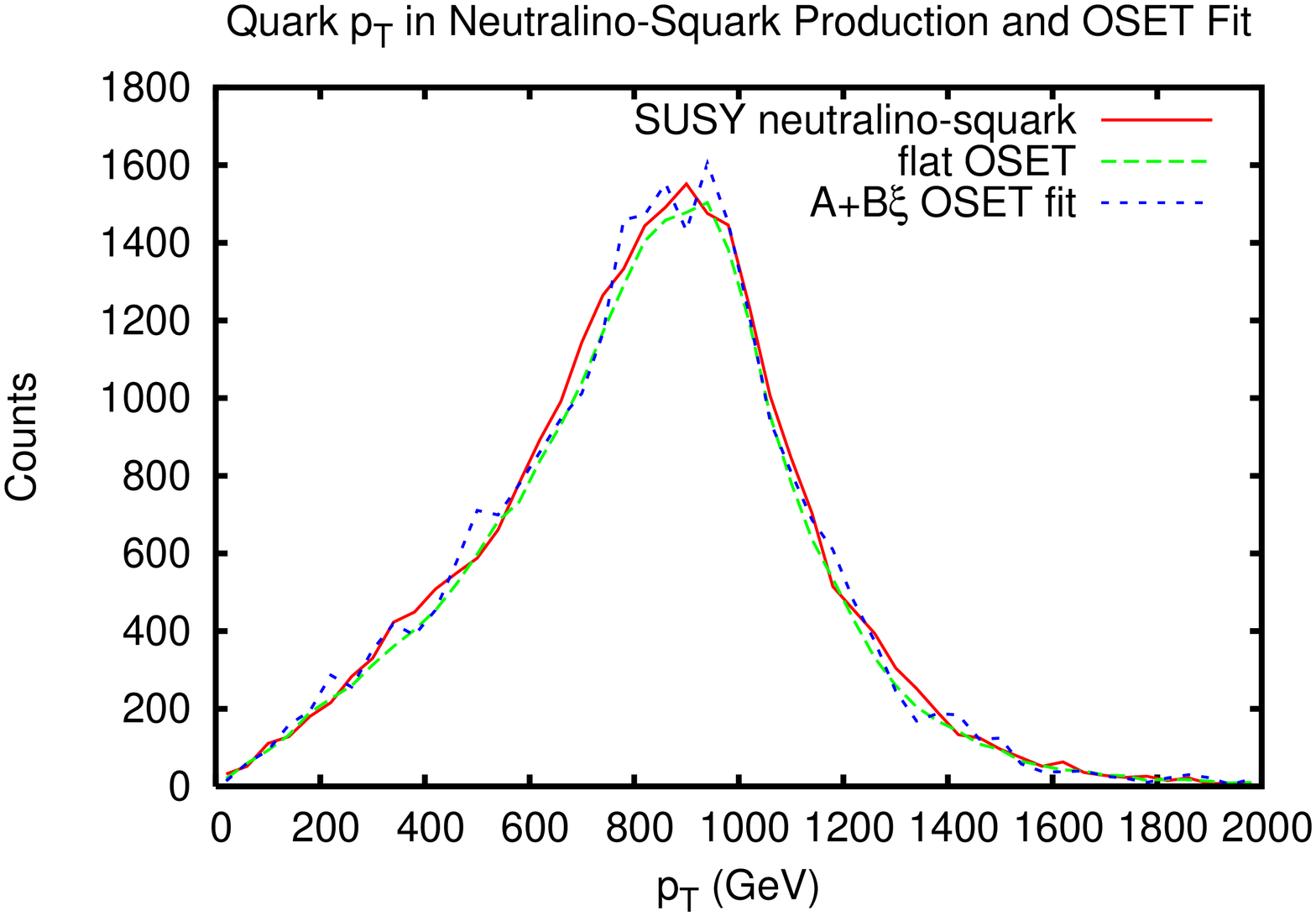}
\includegraphics[width=2in,angle=-90]{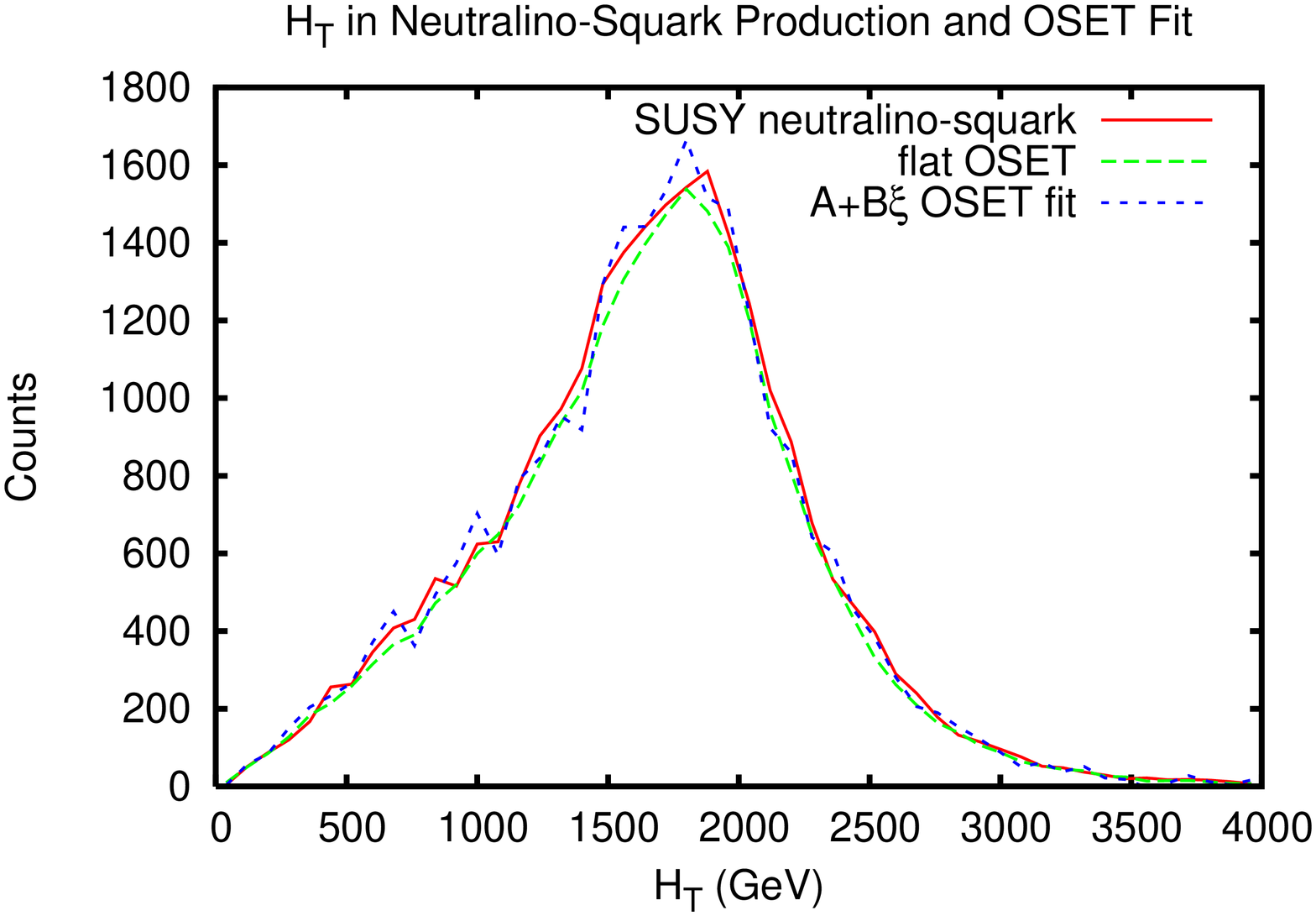}
\includegraphics[width=2in,angle=-90]{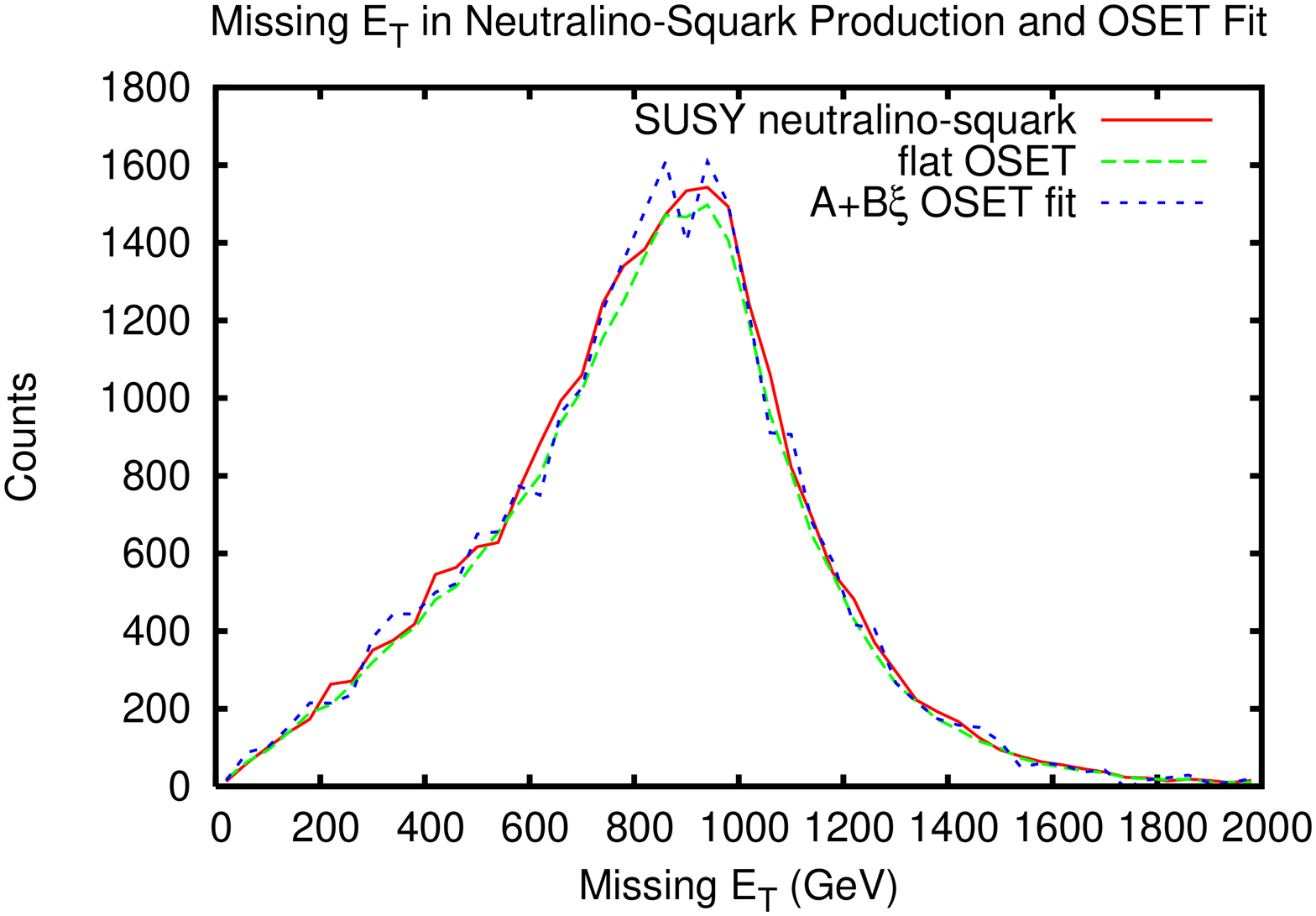}
\caption{Inclusive $p_T$ and $H_T$ distributions for the SUSY
$\tilde \chi_1 \tilde q_L$ associated production process in Figure
\ref{fig:ex3and4} compared with an OSET with a flat matrix element
$\Msq \propto 1$, and an OSET fit using our leading order shape
parametrization (LOSP). In this case, the LOSP consists of
$\Msq=A+B\xi$. The fit is obtained by fitting to the shape of the
primary jet $p_T$ distribution. In all cases, the correct $s_0$ and
$\Delta$ are chosen. Top left: $p_T$ of squark. Top right: $p_T$ of
primary jet. Bottom left: $H_T$ distribution. Bottom right: Missing
$E_T$ distribution.} \label{fig:FitChi1SqPt}
\end{center}
\end{figure}
\begin{figure}[tbp]
\begin{center}
\includegraphics[width=2in,angle=-90]{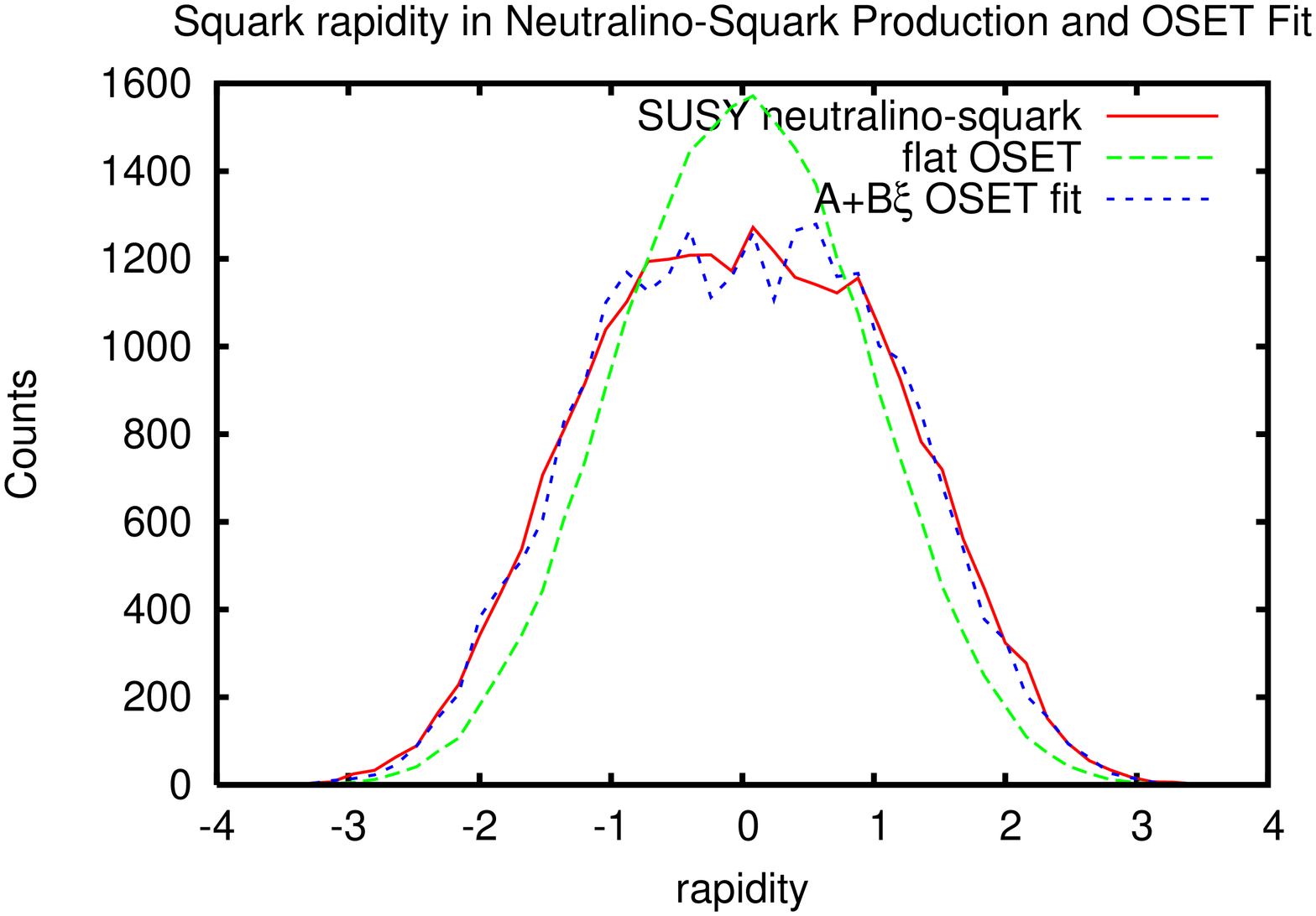}
\includegraphics[width=2in,angle=-90]{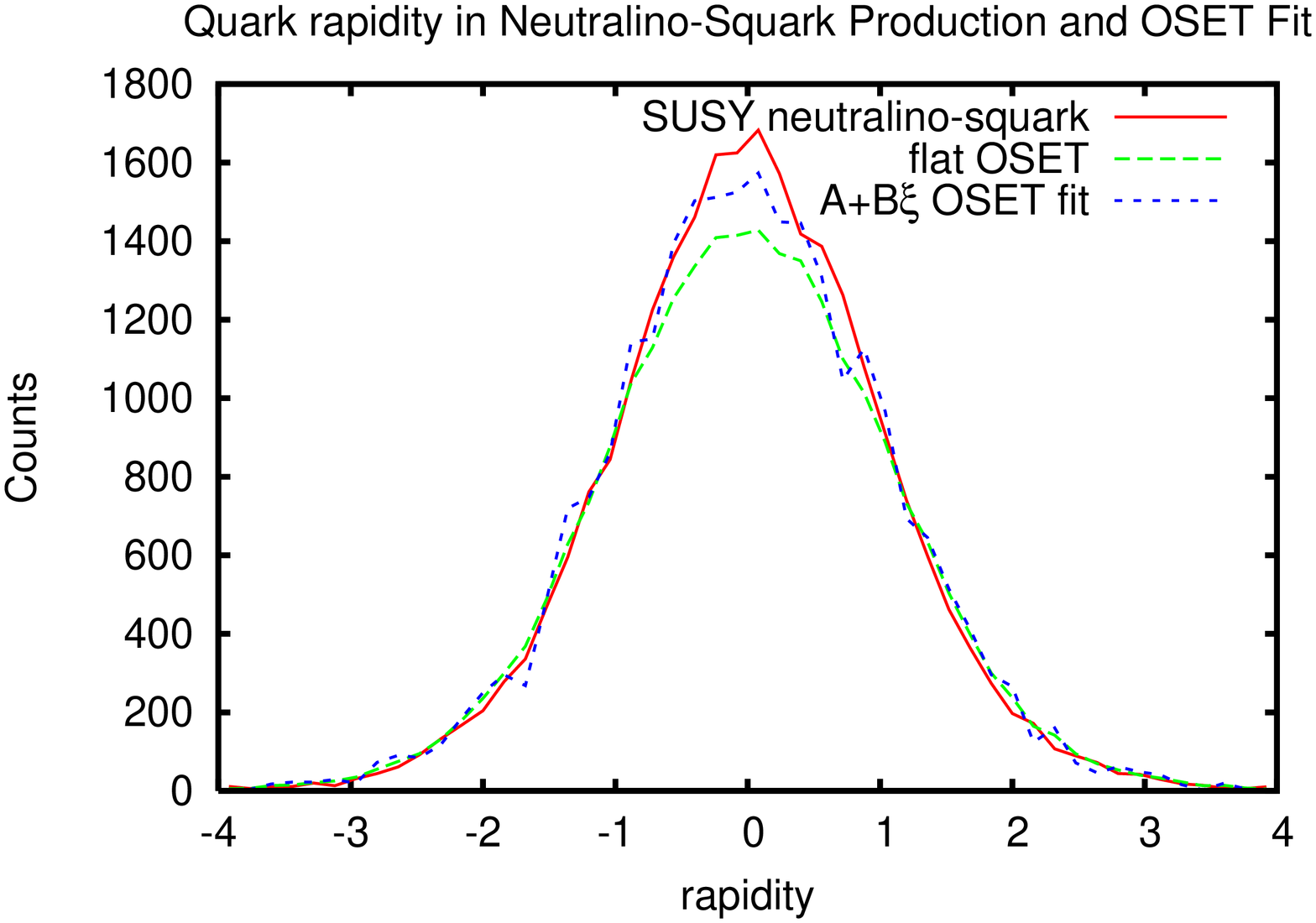}
\caption{Inclusive rapidity distributions for the SUSY $\tilde
  \chi_1 \tilde q_L$ associated production process in Figure
  \ref{fig:ex3and4} compared with an OSET with a flat matrix element
  $\Msq \propto 1$, and an OSET fit using our leading order shape
  parametrization (LOSP). The fit is obtained by fitting to the shape
  of the primary jet $p_T$ distribution. In this case, the LOSP
  consists of $\Msq=A+B\xi$. In all cases, the correct $s_0$ and
  $\Delta$ are chosen. Top left: Rapidity of squark. Top right:
  rapidity of primary jet.}\label{fig:FitChi1SqRap}
\end{center}
\end{figure}

The second example is SUSY $\tilde{\chi}_1 \tilde q_L$ associated
production from Figure \ref{fig:ex3and4}.  This diagram has a strong
$t$-channel pole, and despite the fact that this pole is regulated
by the squark mass, we expect the amplitude to contain high powers
of $\xi$.  Because of the transverse shape invariance, though, we
can capture the $p_T$ distributions in Figure \ref{fig:FitChi1SqPt}
using just a flat matrix element!  This indicates the success of
transverse/rapidity factorization in matrix element modeling.  At
leading order, the $\xi$ dependence affects the squark rapidity
distribution in Figure \ref{fig:FitChi1SqRap}, but after the squark
decays, we see that the rapidity of the resulting jet is
more-or-less consistent with the flat ansatz, though a LOSP of
$\Msq=A+B \xi$ does give a better fit.

\subsection{Washout of Spin Correlations}\label{app:spinwash}

In Section \ref{sec2:Correlations}, we saw that in cascade decays,
di-object invariant mass distributions can retain information about
the spin structure of the underlying model.  As is well known, at
Lepton Colliders one can use production information to also glean
information about spin, but at Hadron Colliders spin information in
production is often lost.  Here, we show two examples comparing
SUSY, UED, and flat matrix elements to see how PDFs and subsequent
two-body decays wash out production spin information.

\begin{figure}[tbp]
\begin{center}
\includegraphics[width=2.2in,angle=-90]{sec2plots/glgl_glpt.eps}
\includegraphics[width=2.2in,angle=-90]{sec2plots/glgl_Apt.eps}

\includegraphics[width=2.2in,angle=-90]{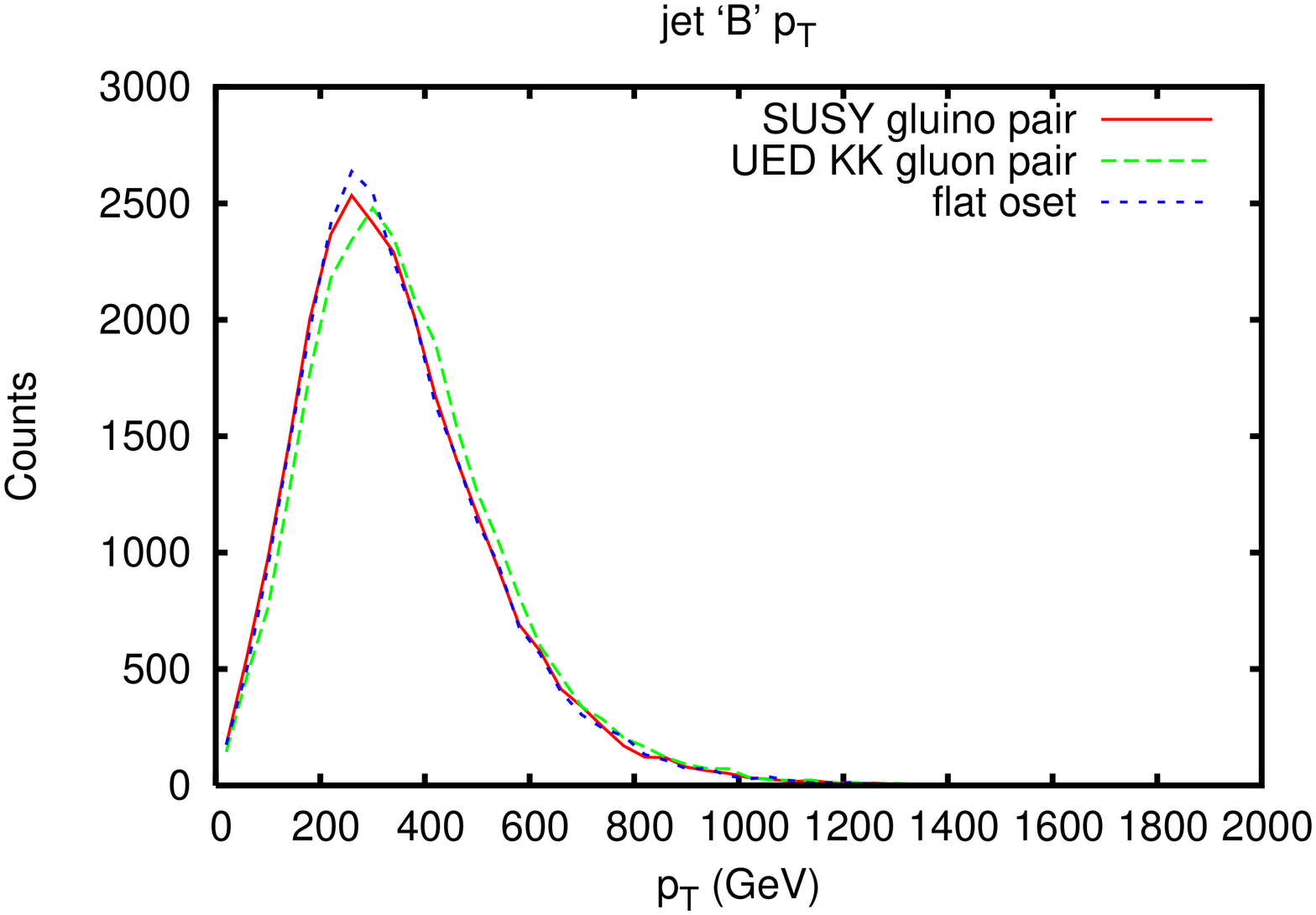}
\includegraphics[width=2.2in,angle=-90]{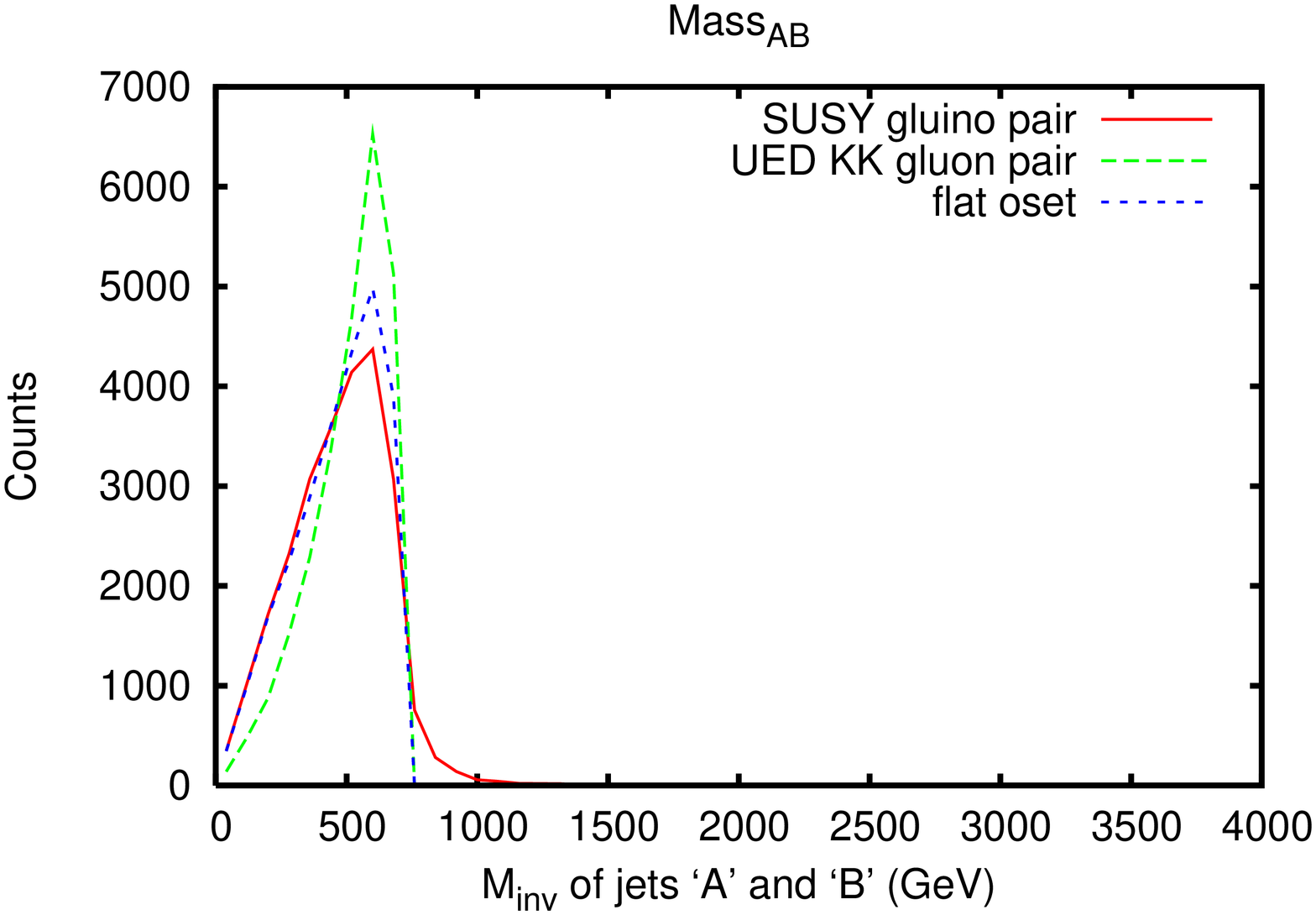}

\includegraphics[width=2.2in,angle=-90]{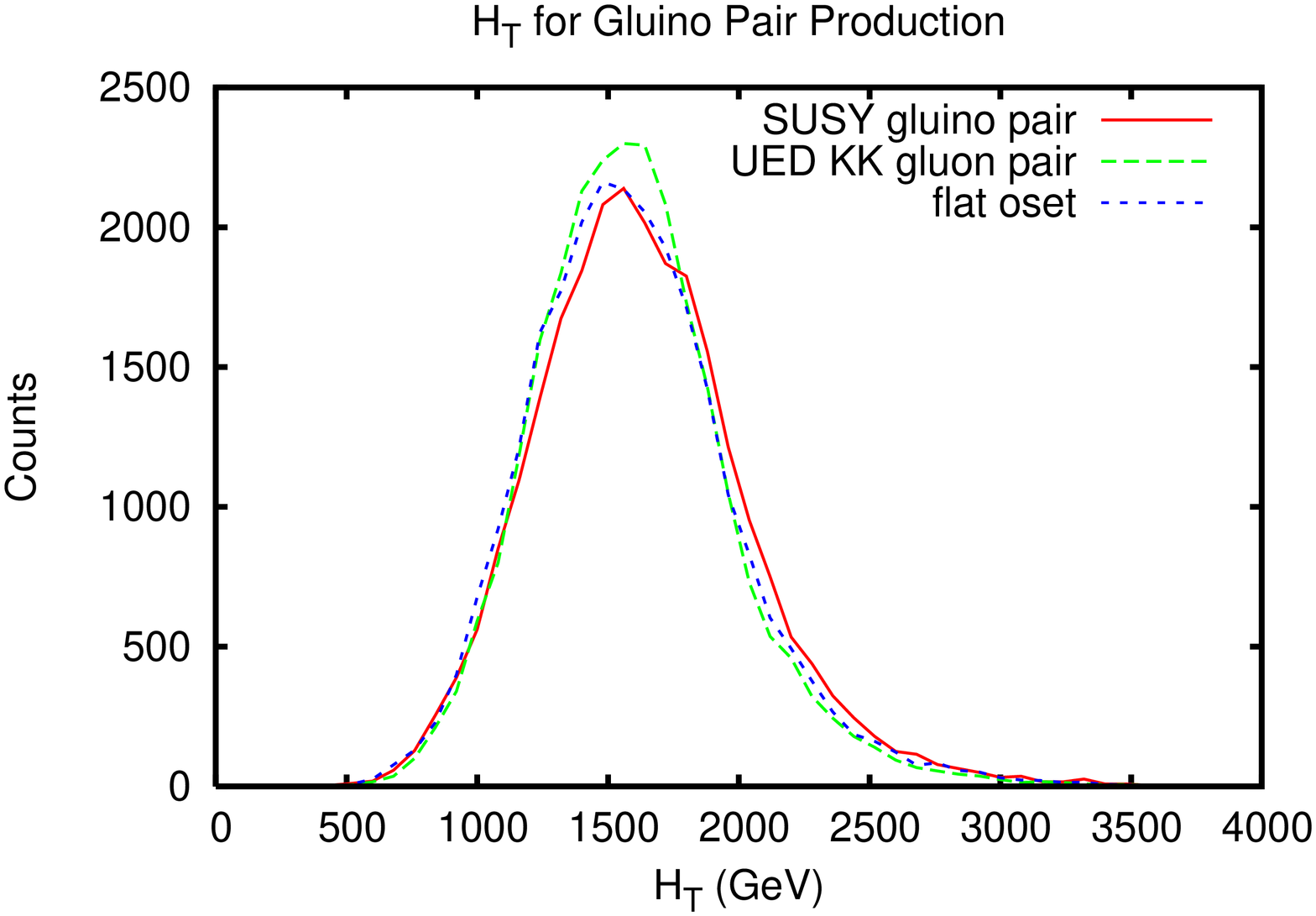}
\includegraphics[width=2.2in,angle=-90]{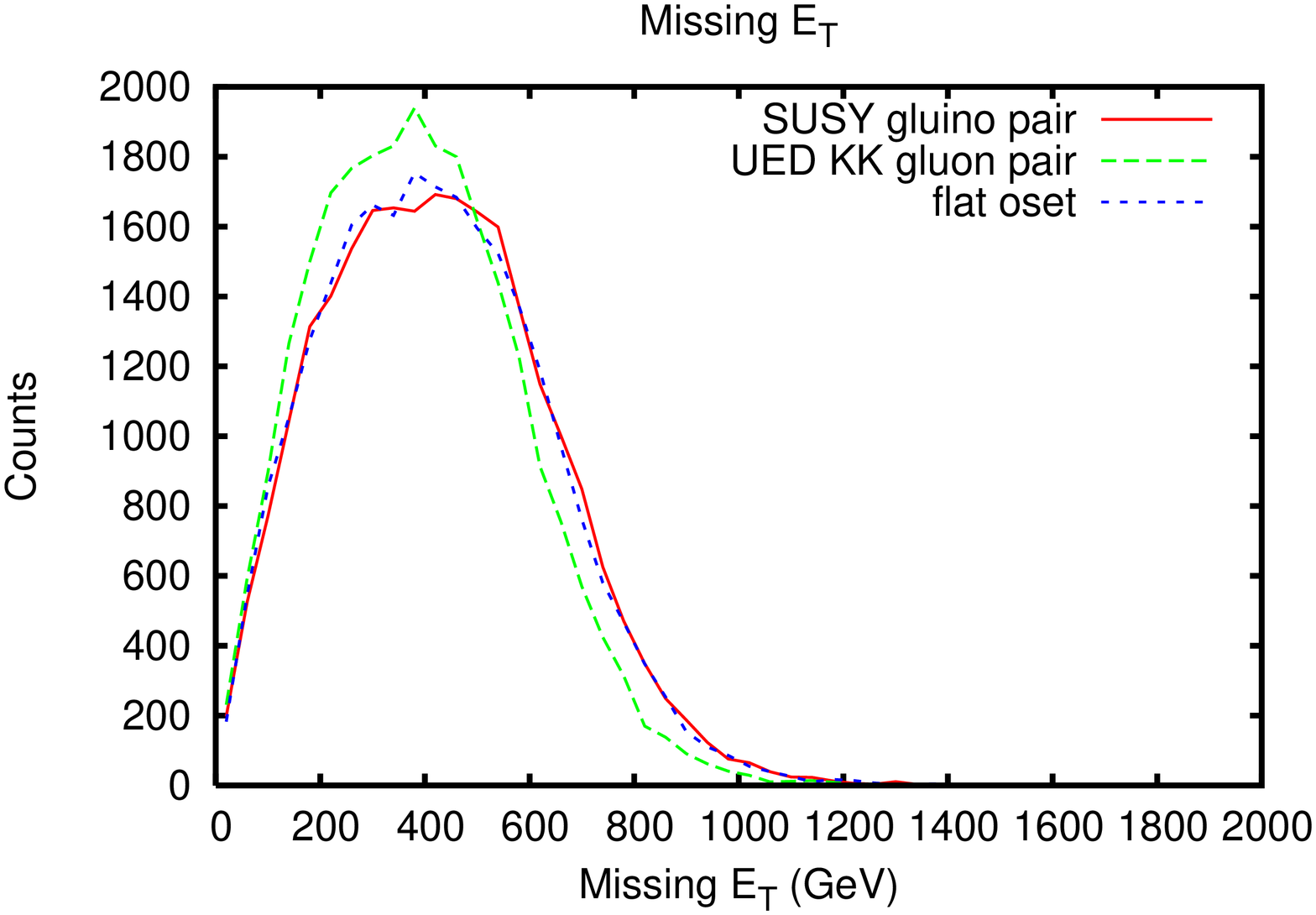}
\caption{Inclusive $p_T$ distributions for SUSY gluino pair-production
  simulated with Pythia, UED KK gluon pair-production simulated with
  HERWIG, and OSET adjoint production.  The SUSY and KK processes have
  the same spectra, illustrated in Figure \ref{fig:ex1and2}.
  The OSET also has the same masses, and uses
  $\Msq \propto 1$.  The scale on the $y$-axis is arbitrary, and
  all processes are normalized to the same total number of events.  Top
  left: partonic gluino $p_T$. Top right: $p_T$ of quark ``A'' (first-stage decay product).
  Middle left: $p_T$ of quark ``B''  (second-stage
  decay product).  Middle right: $E_T^{\rm miss}$.  Bottom left: $H_T$.
  Bottom right: Edge in invariant mass of the two quarks ``A'' and ``B'' from one side
  of process.  The shape of the invariant mass distribution is more
  dependent on vertex structure, and differs from the OSET
  parameterization, but the edge location is uniform as it depends only
  on kinematics.}
\label{fig:FitGlPairPT}
\end{center}
\end{figure}
\begin{figure}[tbp]
\begin{center}
\includegraphics[width=2.2in,angle=-90]{sec2plots/glgl_rapgl.eps}
\includegraphics[width=2.2in,angle=-90]{sec2plots/glgl_rapA.eps}
\includegraphics[width=2.2in,angle=-90]{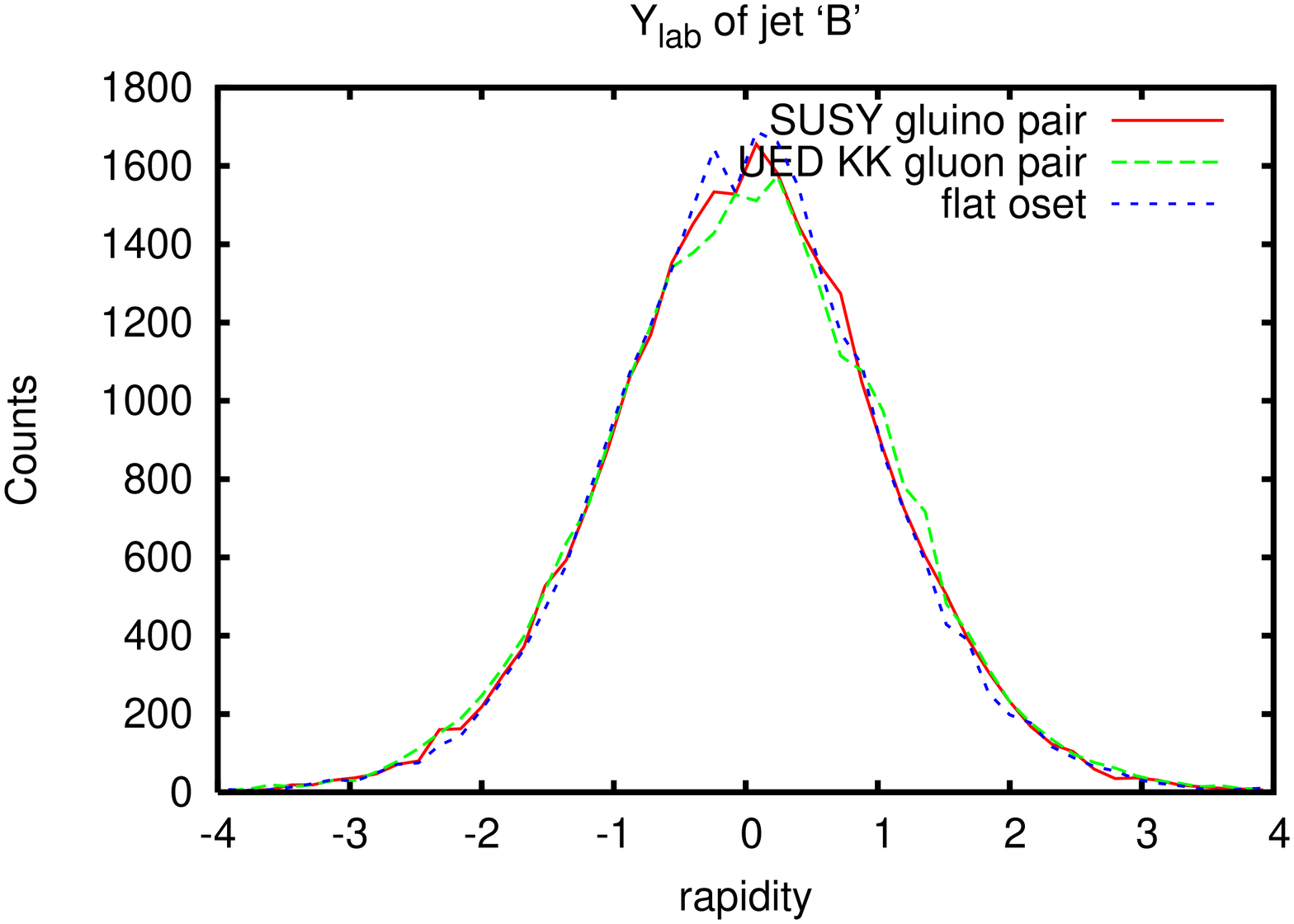}
\includegraphics[width=2.2in,angle=-90]{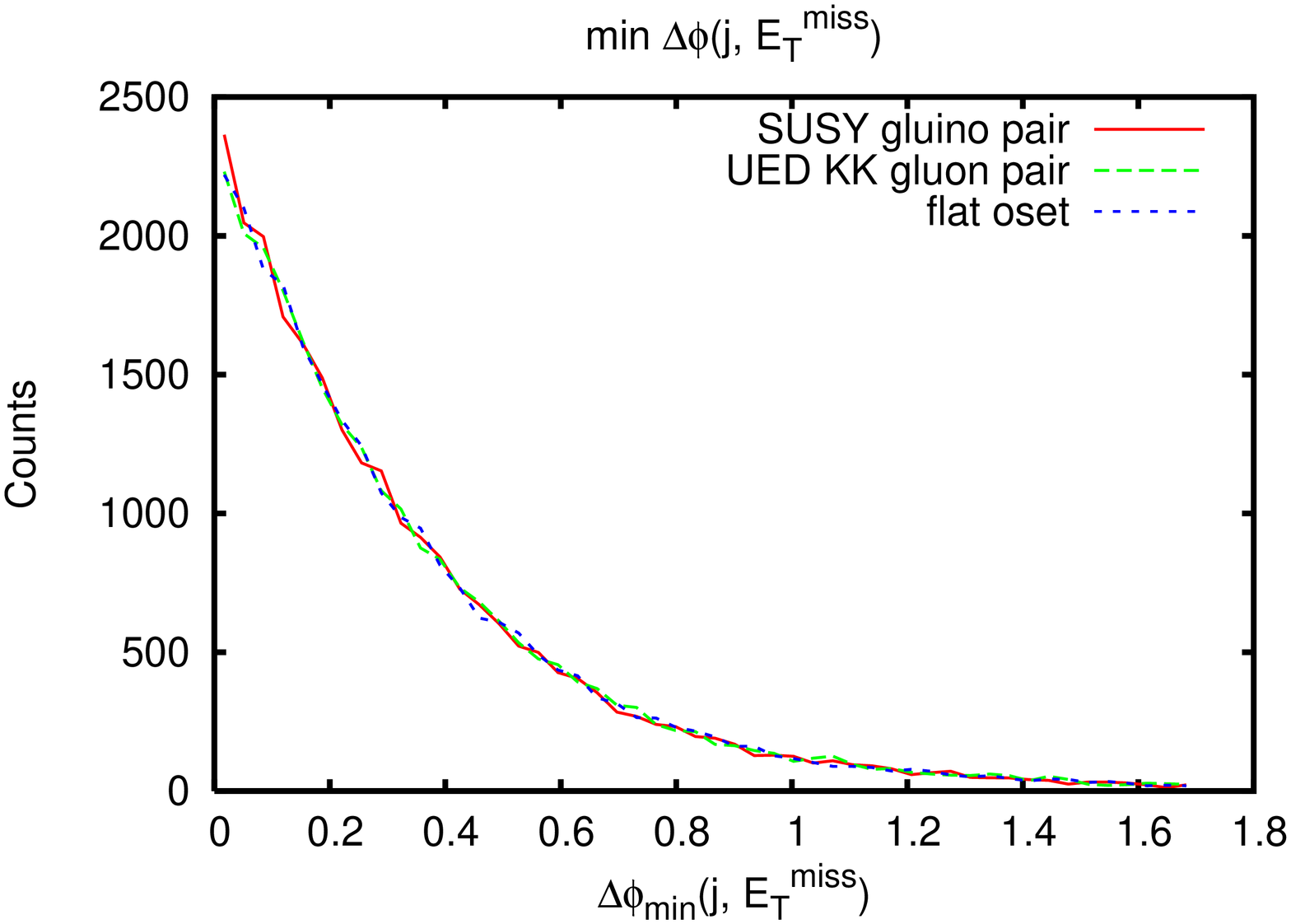}
\caption{Rapidity distributions for the SUSY gluino production, UED KK-gluon
  production, and OSET production processes of Figure
  \ref{fig:FitGlPairPT}.  Top left: lab-frame rapidity of a gluino.
  Top right: lab-frame rapidity of the quark ``A''.  Bottom left: lab-frame rapidity of the quark
  labeled ``B''.  Bottom right: minimum $\delta \phi$ between a quark
  and the partonic missing momentum (sum of transverse momenta of
  the two invisible decay products).}\label{fig:FitGLPairRap}
\end{center}
\end{figure}

The first example involves color adjoint pair production, i.e.\
gluino pair production or KK-gluon pair production as in Figure
\ref{fig:ex1and2}.  Here, we make no attempt to model the matrix
elements beyond a flat ansatz, because both gluino and KK-gluon
production have strong $\Msq \propto 1$ pieces which are well
modeled in Figures \ref{fig:FitGlPairPT} and \ref{fig:FitGLPairRap}.
We observe that while the near threshold behavior of the gluino is
more suppressed than the KK-gluon due to a $\beta^2$ contribution to
the gluino amplitude, the effect is small.  As expected from Section
\ref{sec2:Correlations}, the biggest difference is in the $m_{AB}$
distribution, which measures spin correlations from gluino decays
through on-shell squarks, as discussed further in Appendix \ref{app:decay}.  (The extreme sharpness of the KK-gluon
edge is due to our not properly taking into account the KK-gluon
width.  The difference in the shape of the approach is real.)  Also
noticeable is the small decrease in missing energy in the UED case,
due to non-negligible final state correlations.

\begin{figure}[tbp]
\begin{center}
\includegraphics[width=2.2in,angle=-90]{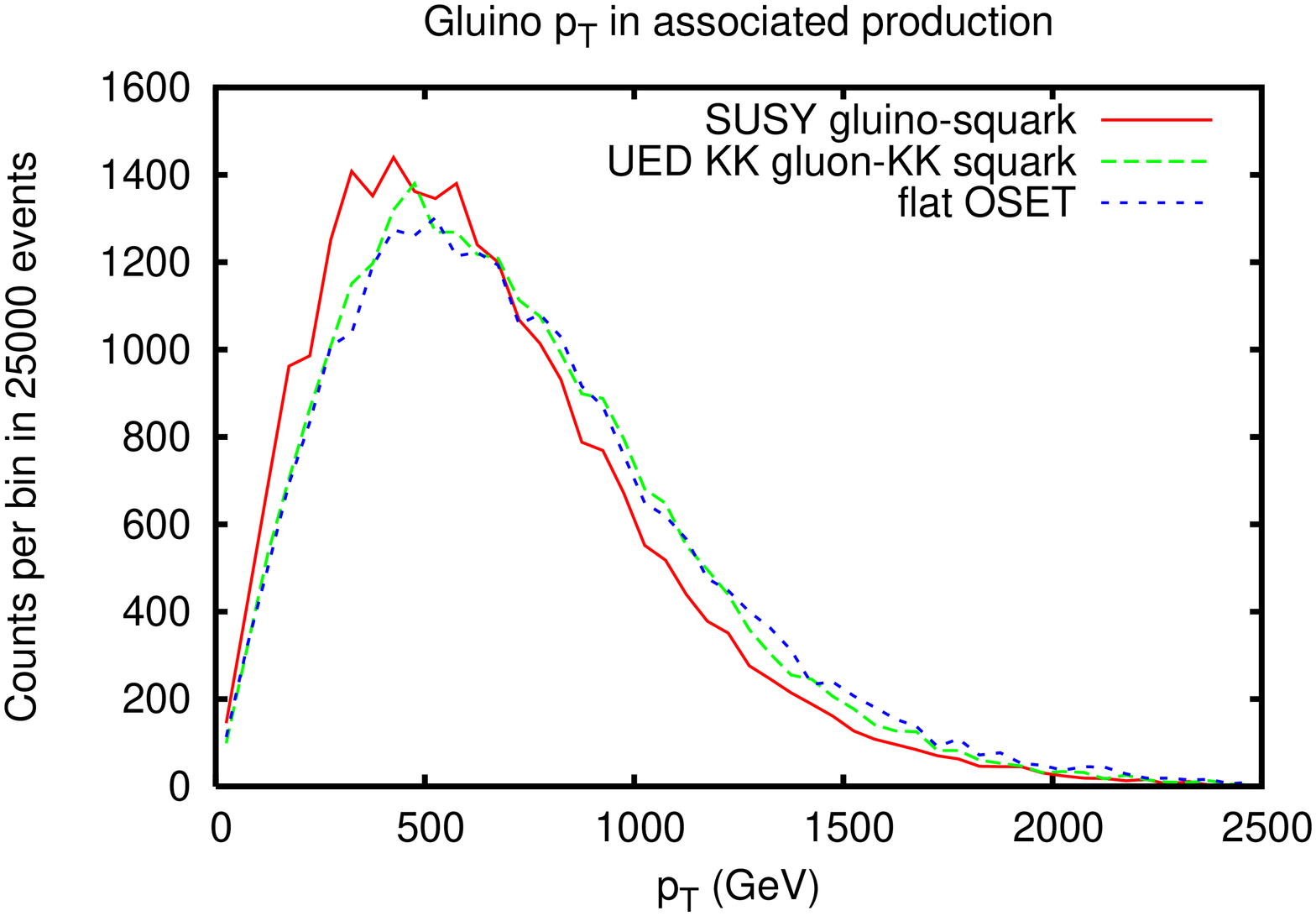}
\includegraphics[width=2.2in,angle=-90]{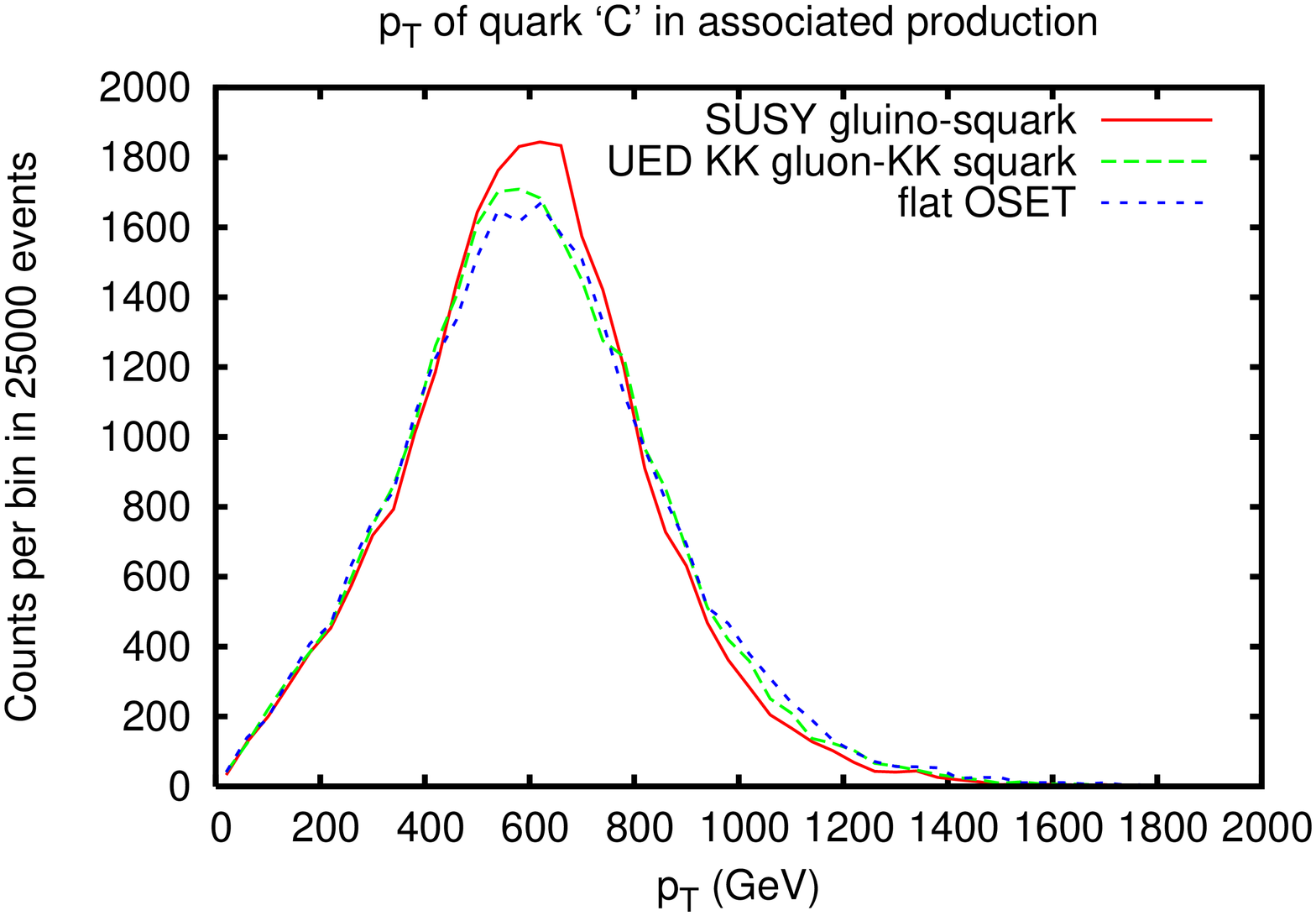}
\includegraphics[width=2.2in,angle=-90]{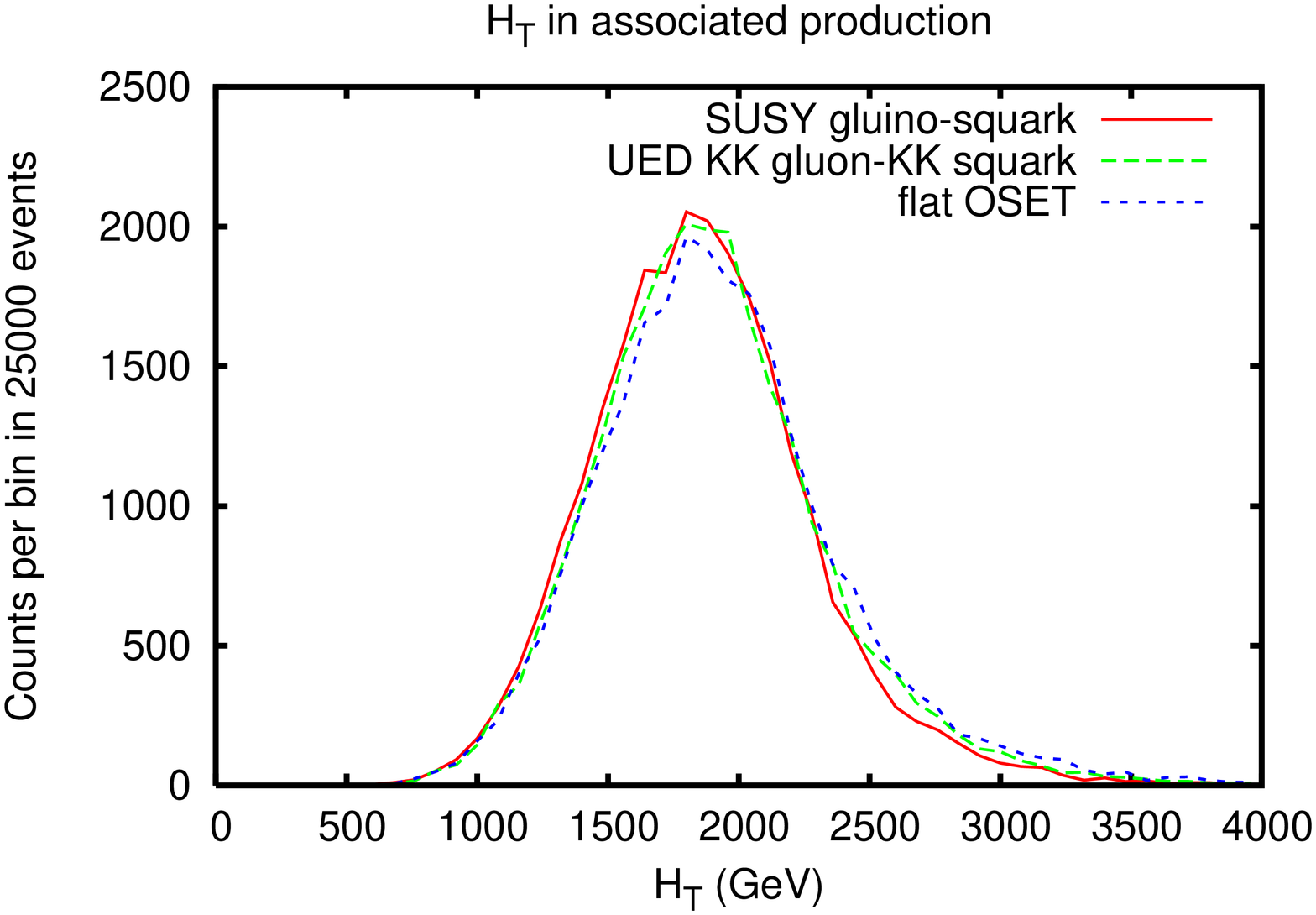}
\includegraphics[width=2.2in,angle=-90]{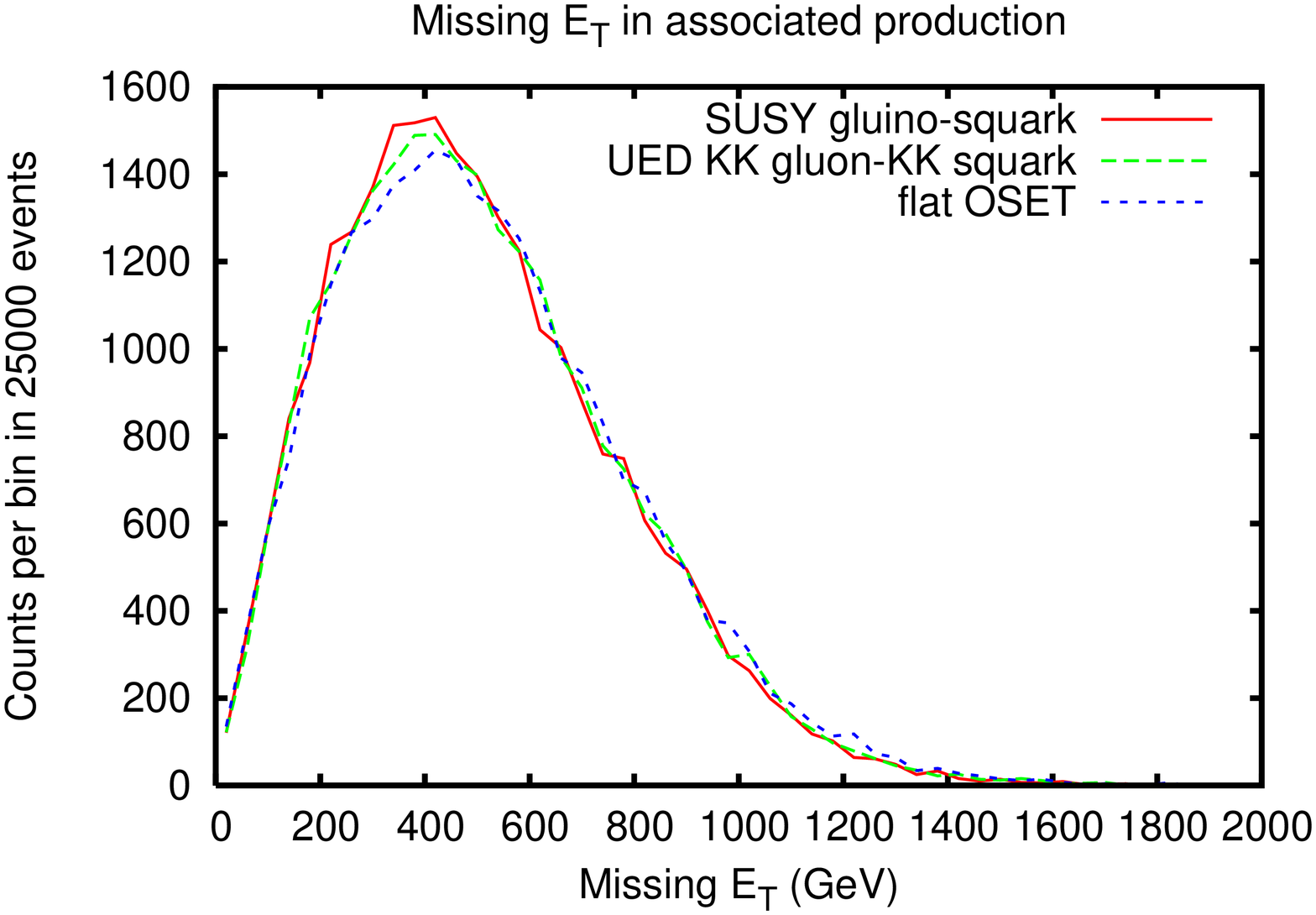}
\caption{Inclusive $p_T$ distributions for SUSY gluino-squark
production simulated with Pythia, UED KK gluon-KK quark production
simulated with HERWIG, and an analogous OSET process.  The SUSY and KK
processes have the same spectra, illustrated in Figure \ref{fig:ex3and4}; the
OSET has the same masses, and uses $\Msq \propto 1$.  The scale
in the $y$-axis is arbitrary, and all processes are normalized to the
same total number of events.  Top left: partonic gluino $p_T$. Top
right: $p_T$ of quark ``C''
(first-stage decay product).  Bottom left: $H_T$.  Bottom right:
$E_T^{\rm miss}$.}
\label{fig:FitSqGlprimaryPT}
\end{center}
\end{figure}
\begin{figure}[tbp]
\begin{center}
\includegraphics[width=2.2in,angle=-90]{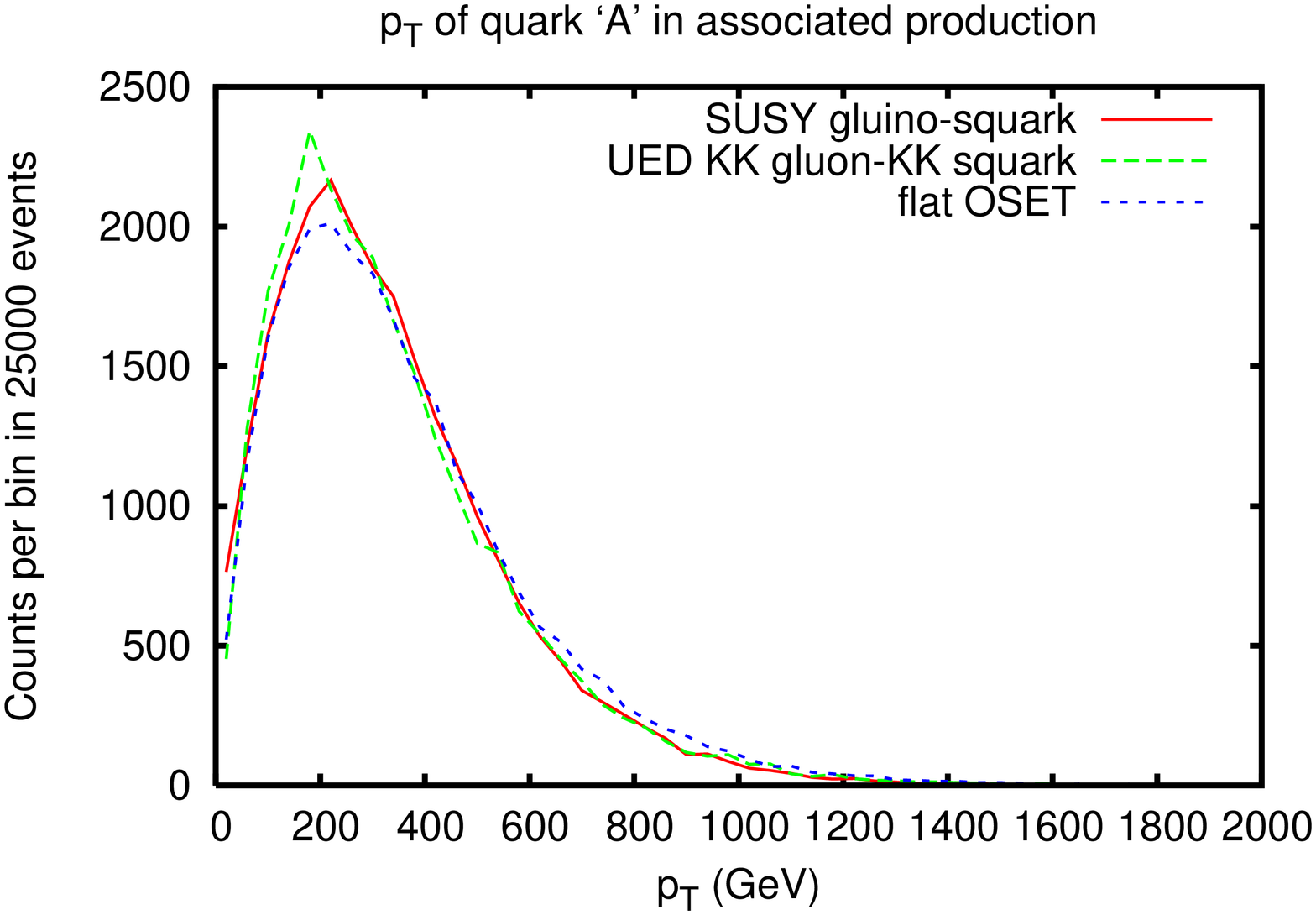}
\includegraphics[width=2.2in,angle=-90]{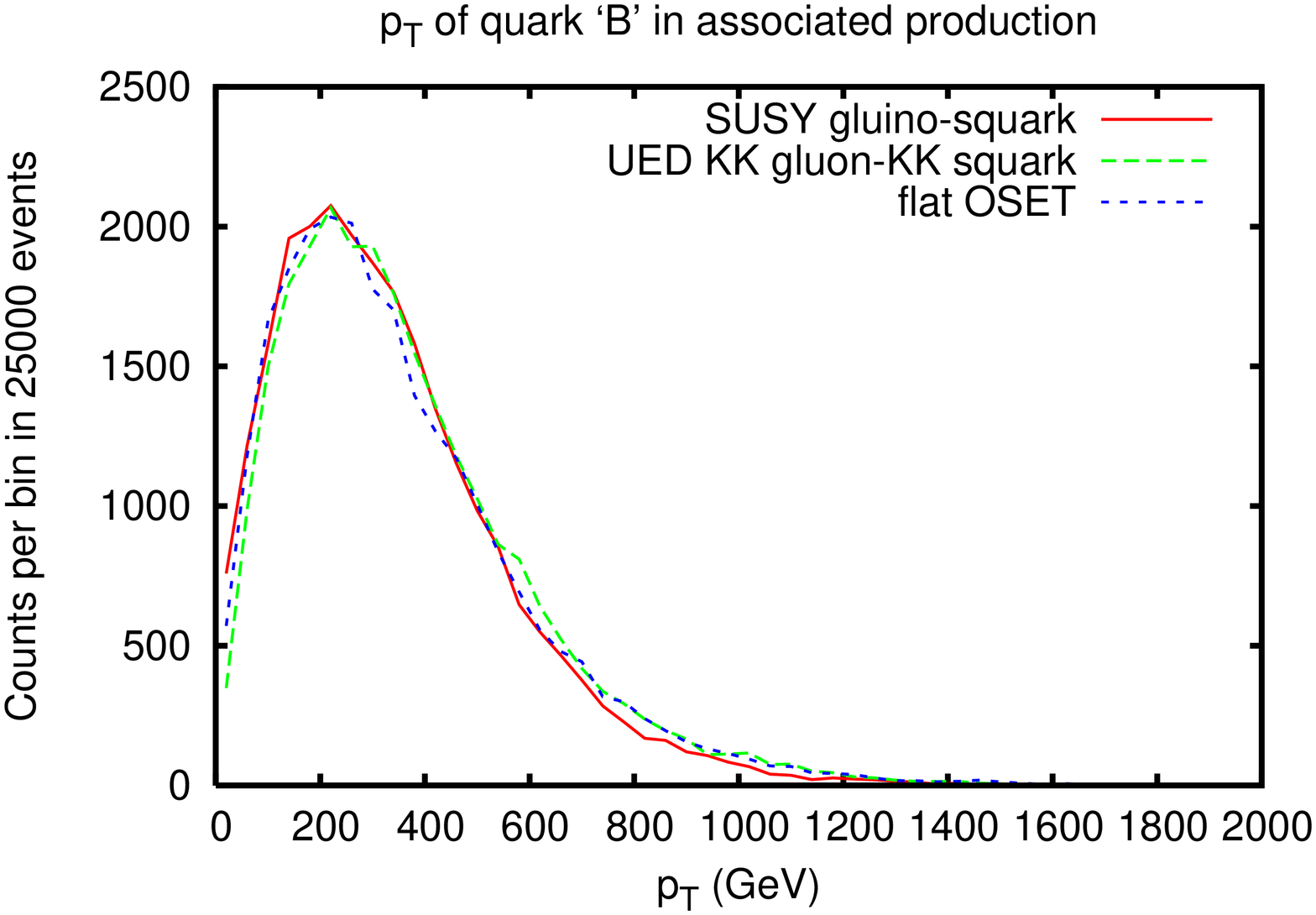}
\includegraphics[width=2.2in,angle=-90]{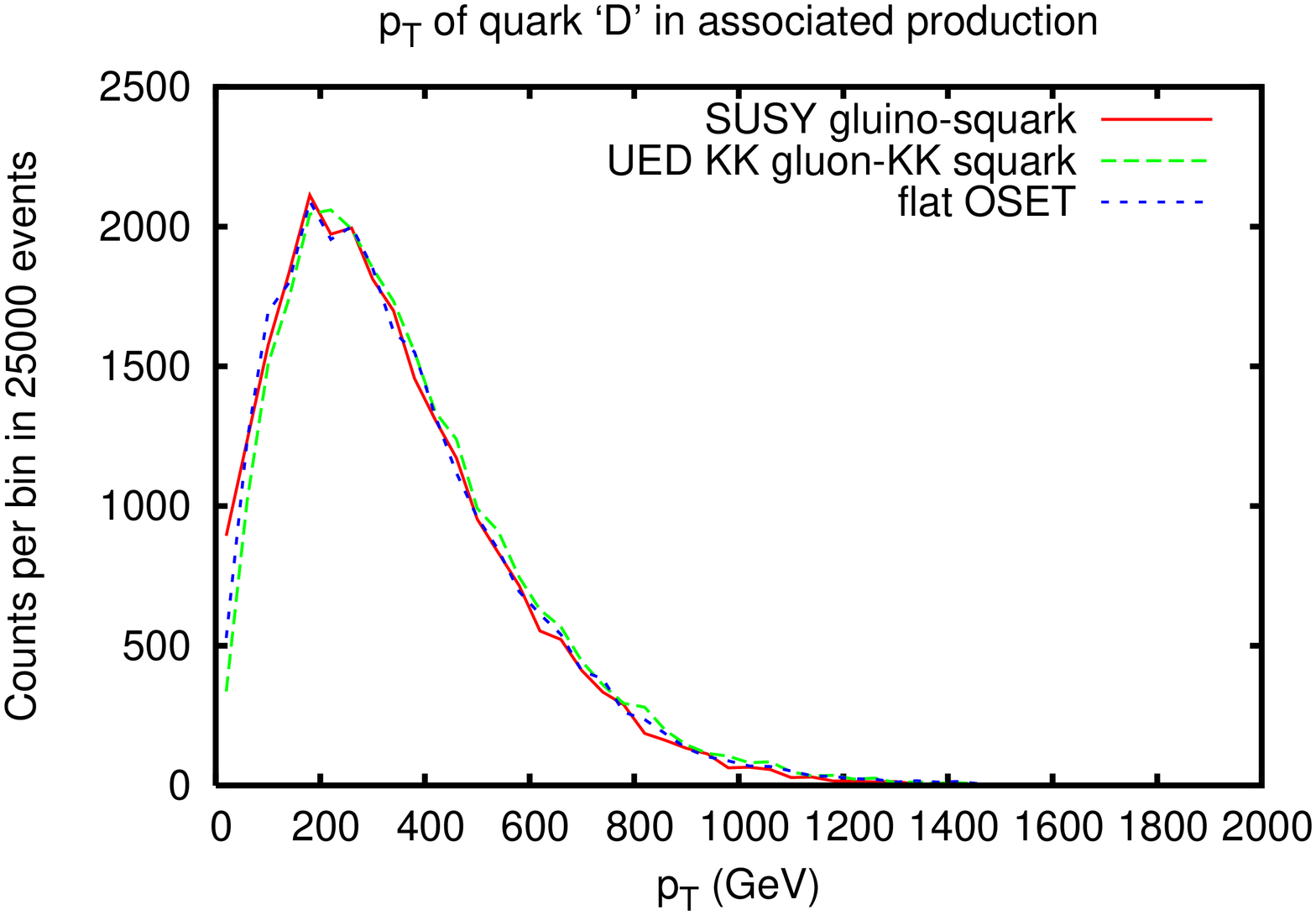}
\includegraphics[width=2.2in,angle=-90]{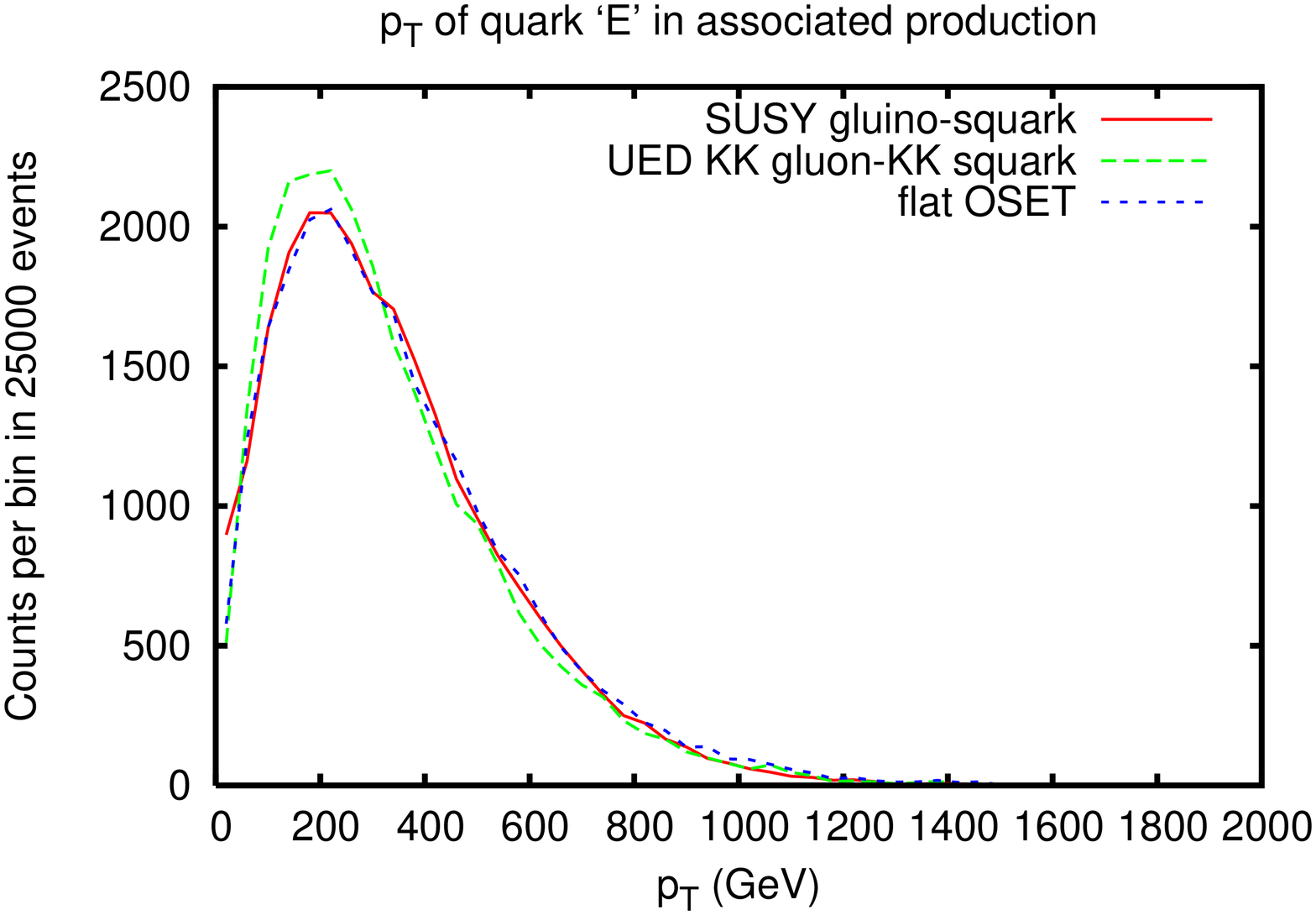}
\caption{Distributions of final state $p_T$s comparing  SUSY gluino-squark
production, UED KK gluon-KK quark production, and an analogous OSET process.
Top left:  $p_T$ of ``A'' quark.  Top right:  $p_T$ of ``B'' quark.
Bottom left:  $p_T$ of ``D'' quark.  Bottom Right:  $p_T$ of ``E'' quark.
Despite the different underlying dynamics among the three models, after
long cascade decays, phase space dominates the transverse structure.
} \label{fig:FitSqGlsecondaryPT}
\end{center}
\end{figure}
\begin{figure}[tbp]
\begin{center}
\includegraphics[width=2.2in,angle=-90]{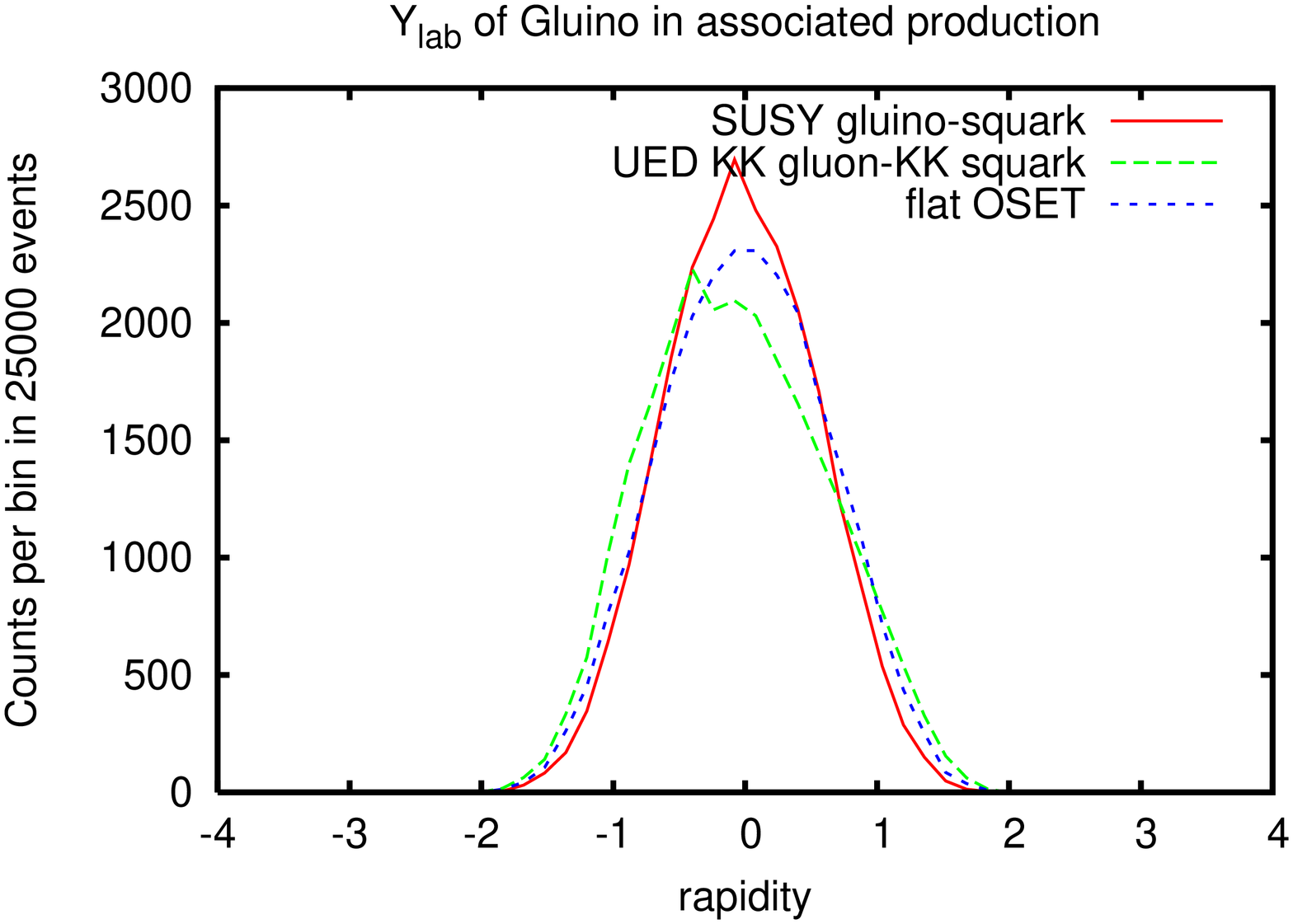}
\includegraphics[width=2.2in,angle=-90]{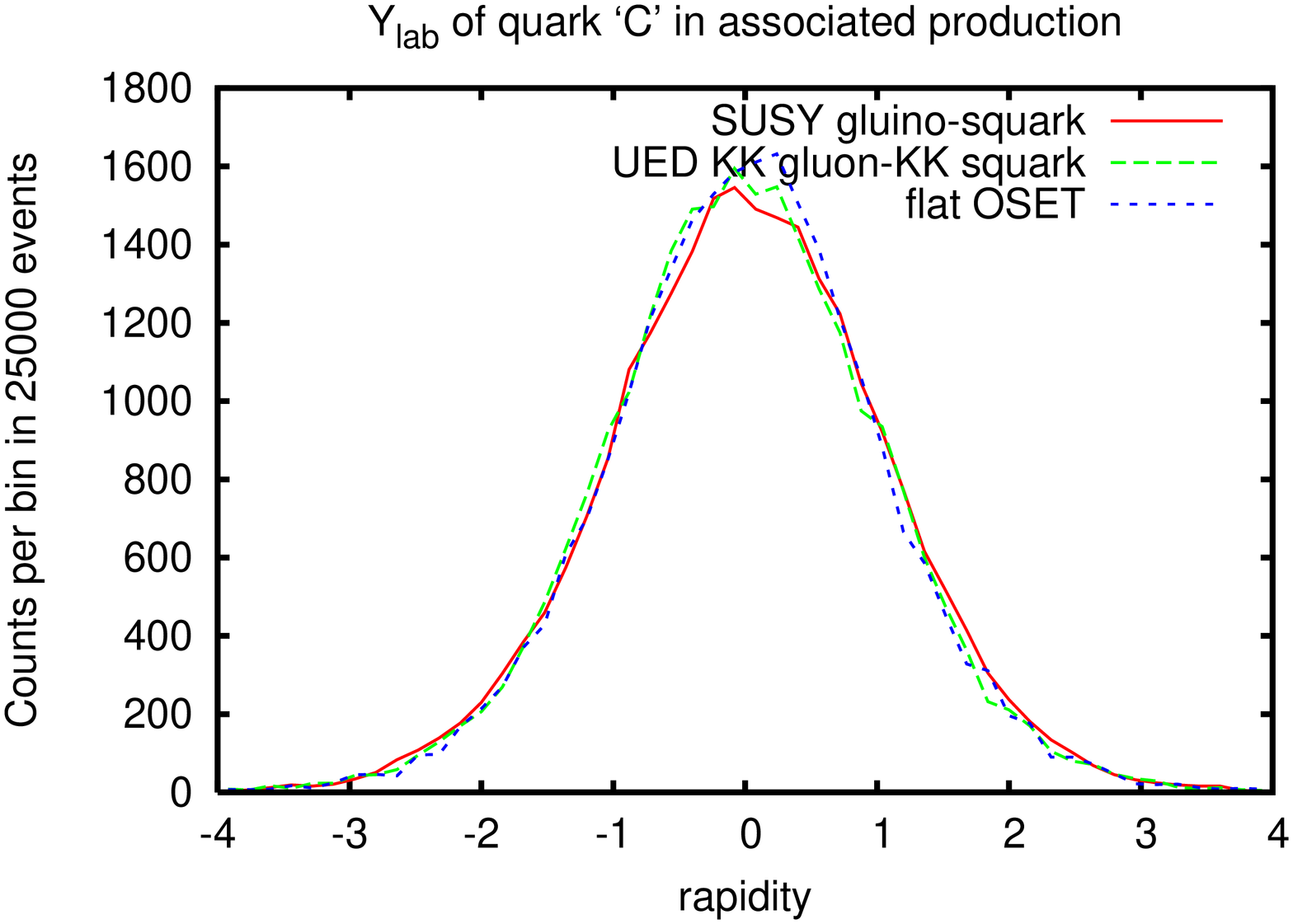}
\includegraphics[width=2.2in,angle=-90]{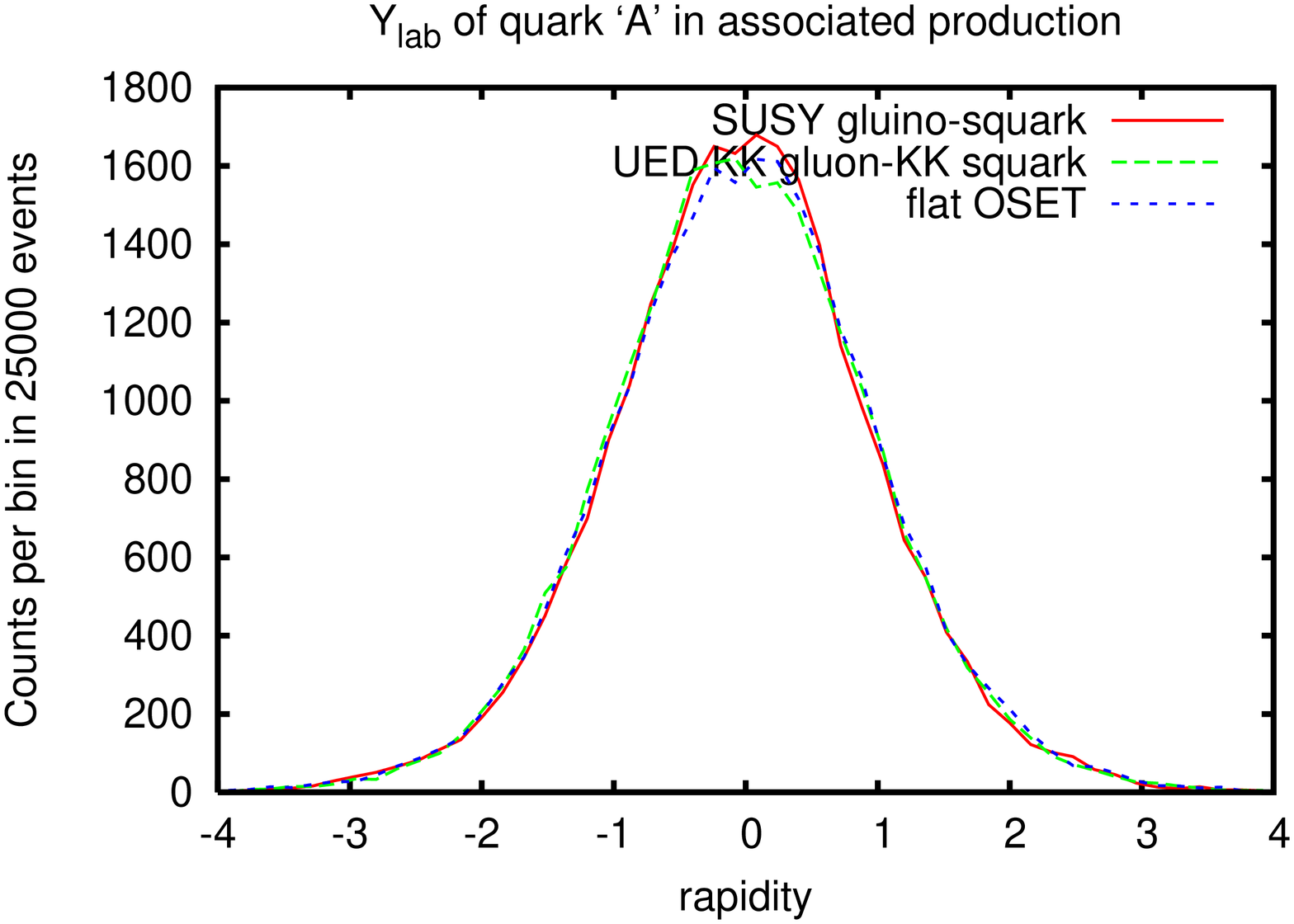}
\includegraphics[width=2.2in,angle=-90]{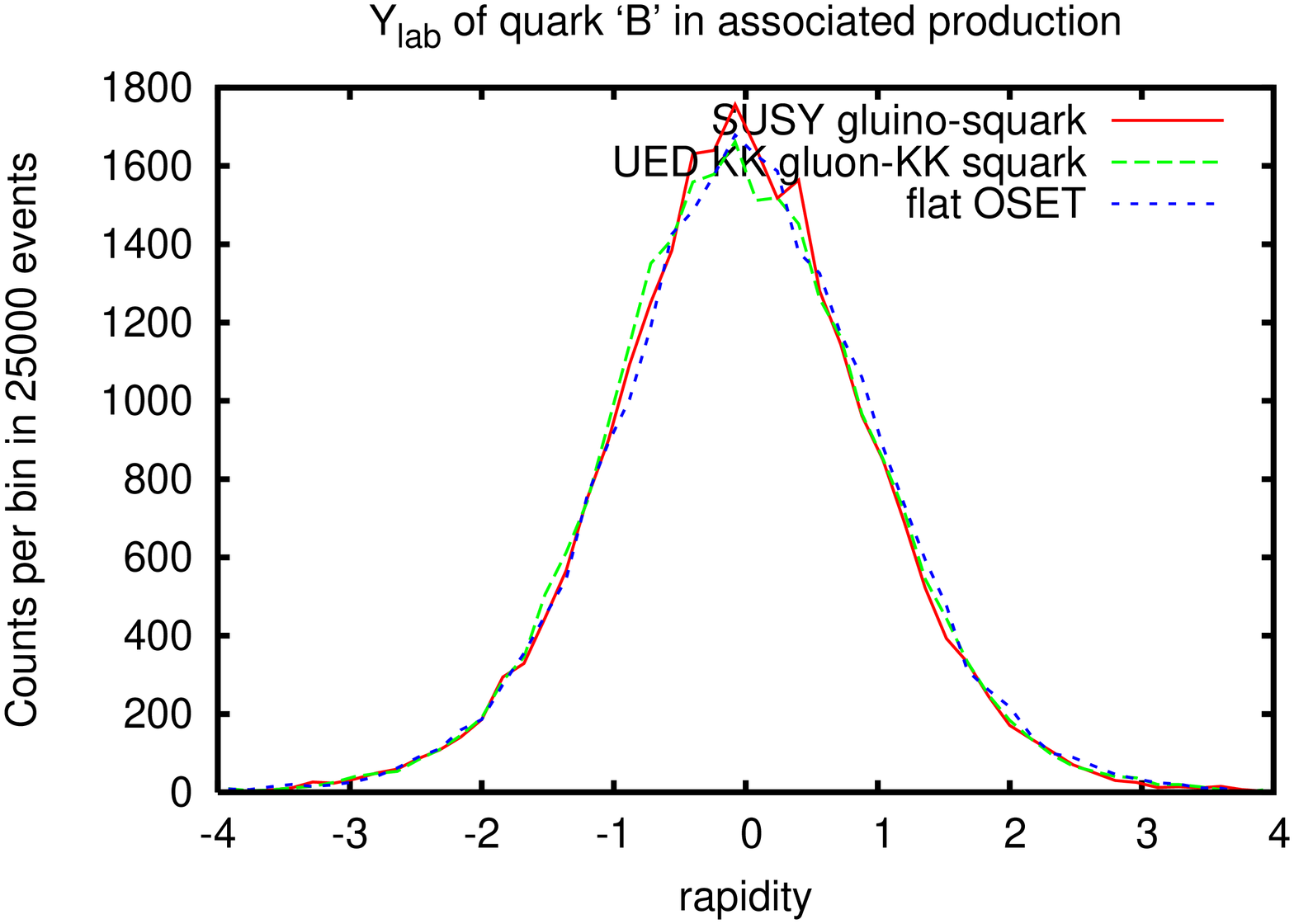}
\includegraphics[width=2.2in,angle=-90]{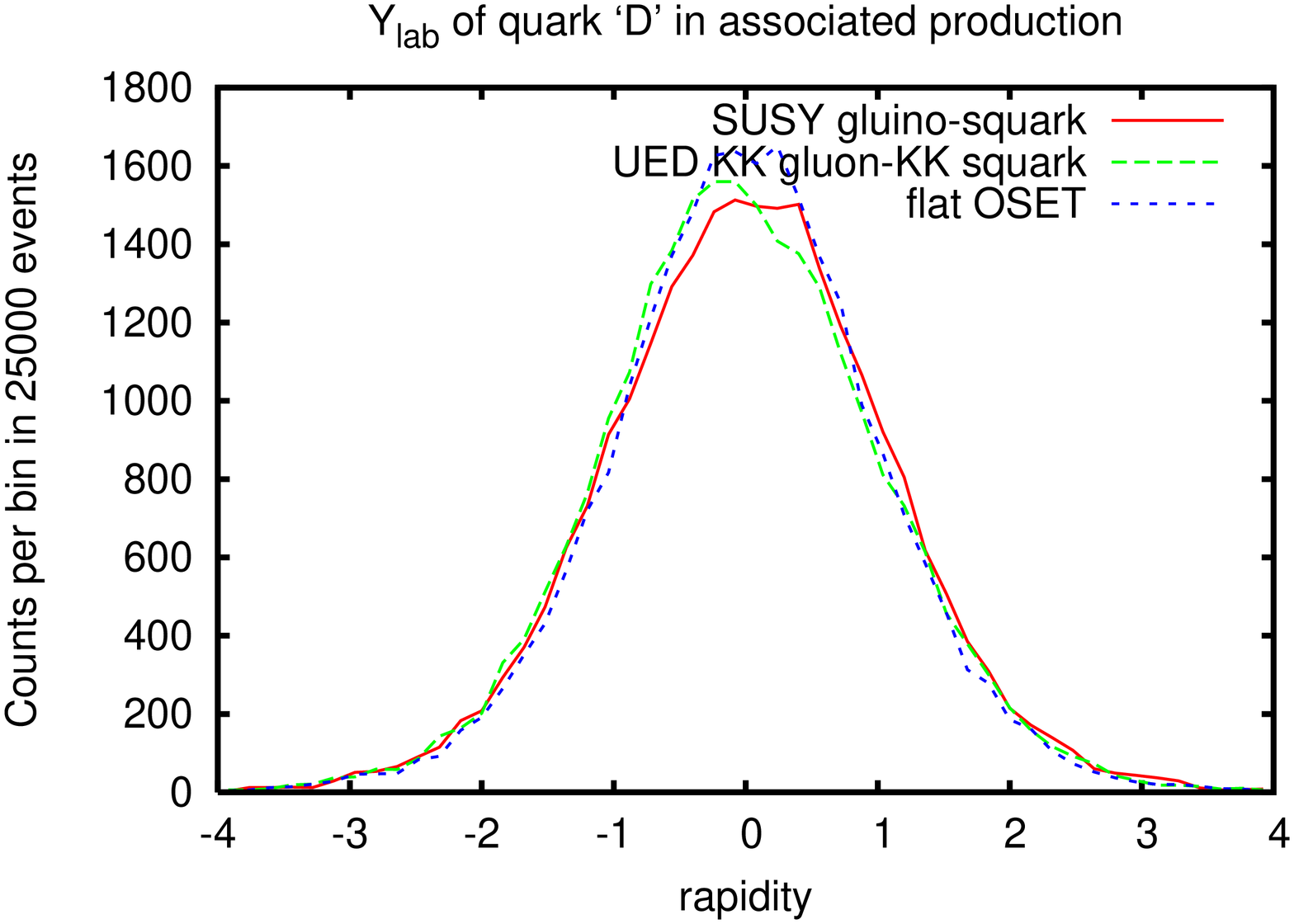}
\includegraphics[width=2.2in,angle=-90]{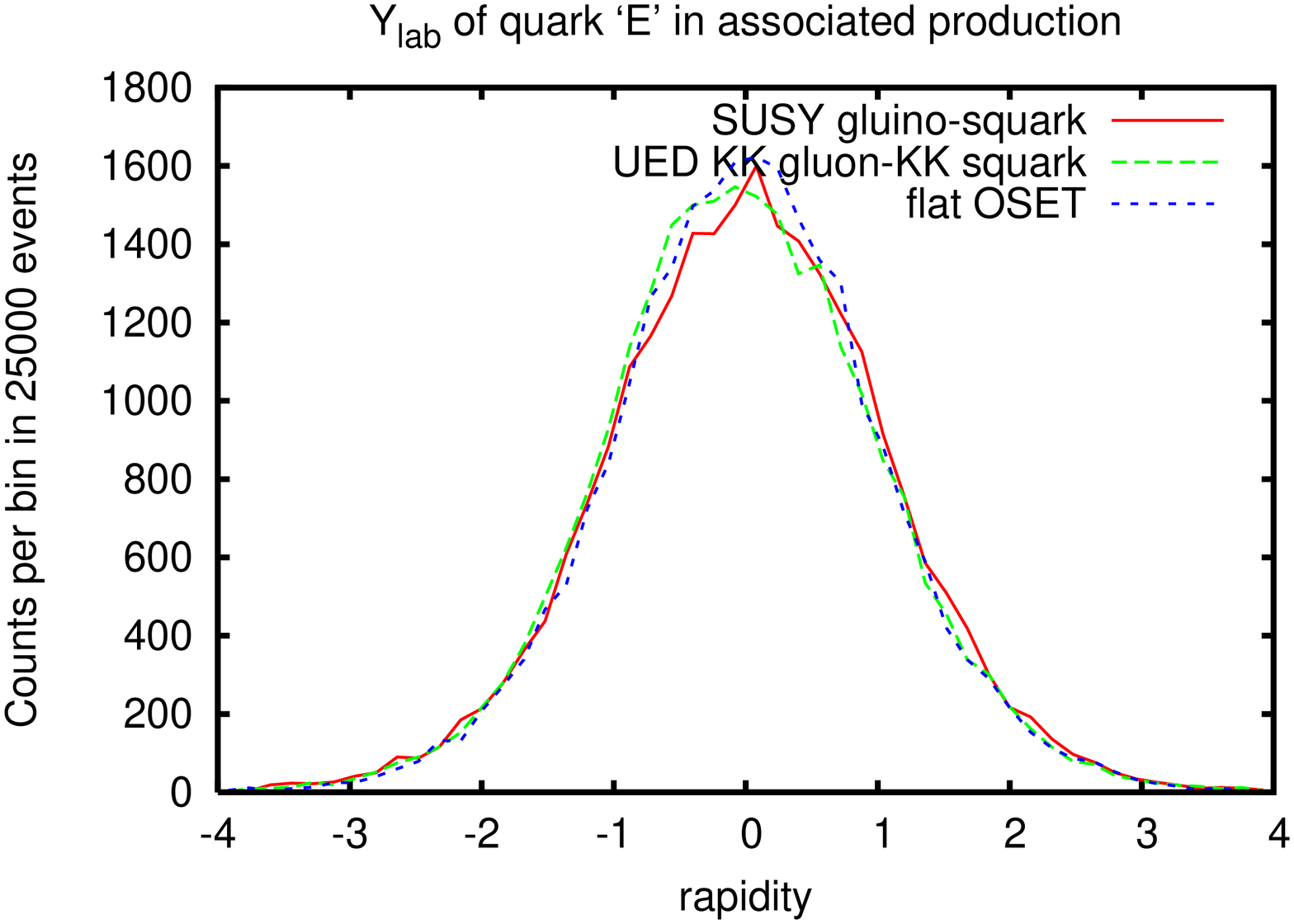}
\caption{Distributions of rapidities comparing  SUSY gluino-squark
production, UED KK gluon-KK quark production, and an analogous OSET process.
Top left:  gluino rapidity.  Top right:   ``C'' quark rapidity.
Middle left: ``A'' quark rapidity. Middle right: ``B'' quark rapidity.
Bottom left:  ``D'' quark rapidity.  Bottom Right:  ``E'' quark rapidity.
Despite the different underlying dynamics among the three models, after
long cascade decays, phase space dominates the rapidity structure.}\label{fig:FitSqGlRap}
\end{center}
\end{figure}

Finally, we look at gluino/squark vs.\ KK-gluon/KK-quark associated
production from Figure \ref{fig:ex3and4}.  In Figure
\ref{fig:FitSqGlprimaryPT}, we see a difference in the squark $p_T$
threshold behavior, but it is washed out in the $p_T$ distribution
in the resulting quark from the the squark decay.  There is little
if anything to distinquish SUSY, UED, and a flat ansatz in single
particle kinematic distributions in Figures
\ref{fig:FitSqGlsecondaryPT} and \ref{fig:FitSqGlRap}.  After the
long cascade decays, all distributions are dominated by phase space
considerations alone.


%% file: AppendixDecay.tex
\section{Spin Correlations and Decay Kinematics: Examples}\label{app:decay}

In this appendix, we develop the discussion in Section
\ref{sec2:Correlations} on the inclusion of spin correlations and
the resulting impact on final state kinematics. This discussion is
only meant to highlight the basic physics. A more detailed study of
this topic is presented in \cite{LW-spin2}.

We have argued that for variables such as $p_T$ and $\eta$, our
approximation captures the leading features. In the rest frame of
the decaying particle, there are correlations between the various
decay products and the direction of the polarization of the decaying
particle.  In many cases however, we do not have enough information
to reconstruct the rest frame of the decaying particle. Moreover,
the directional information is always projected onto the transverse
plane of the decay. Therefore, such correlations will tend to be
washed out after boosting to the lab frame, which depends on the
transverse velocity.

A good example of such an effect is top quark
decay. Consider  a top quark with some definite helicity. In the
rest frame of the top quark, there is a correlation between the
direction of the decay products, such as the charge lepton, and the
direction of top polarization. On the other hand, if we can only
measure the transverse momentum of the lepton, the only effect is
that the $p_T$ of the lepton will be somewhat harder (softer) if the
top quark is right-handed (left-handed) polarized. On the other
hand, this is purely an effect of transverse boost from the top rest
frame to the lab frame.  If the top quark if produced far away from
threshold, such as decaying from a heavy resonance
\cite{Lillie:2007yh}, this effect can be significant and carries
useful information.

\begin{figure}[tbp]
\begin{tabular}{cc}
\includegraphics[angle=270,scale=0.3]{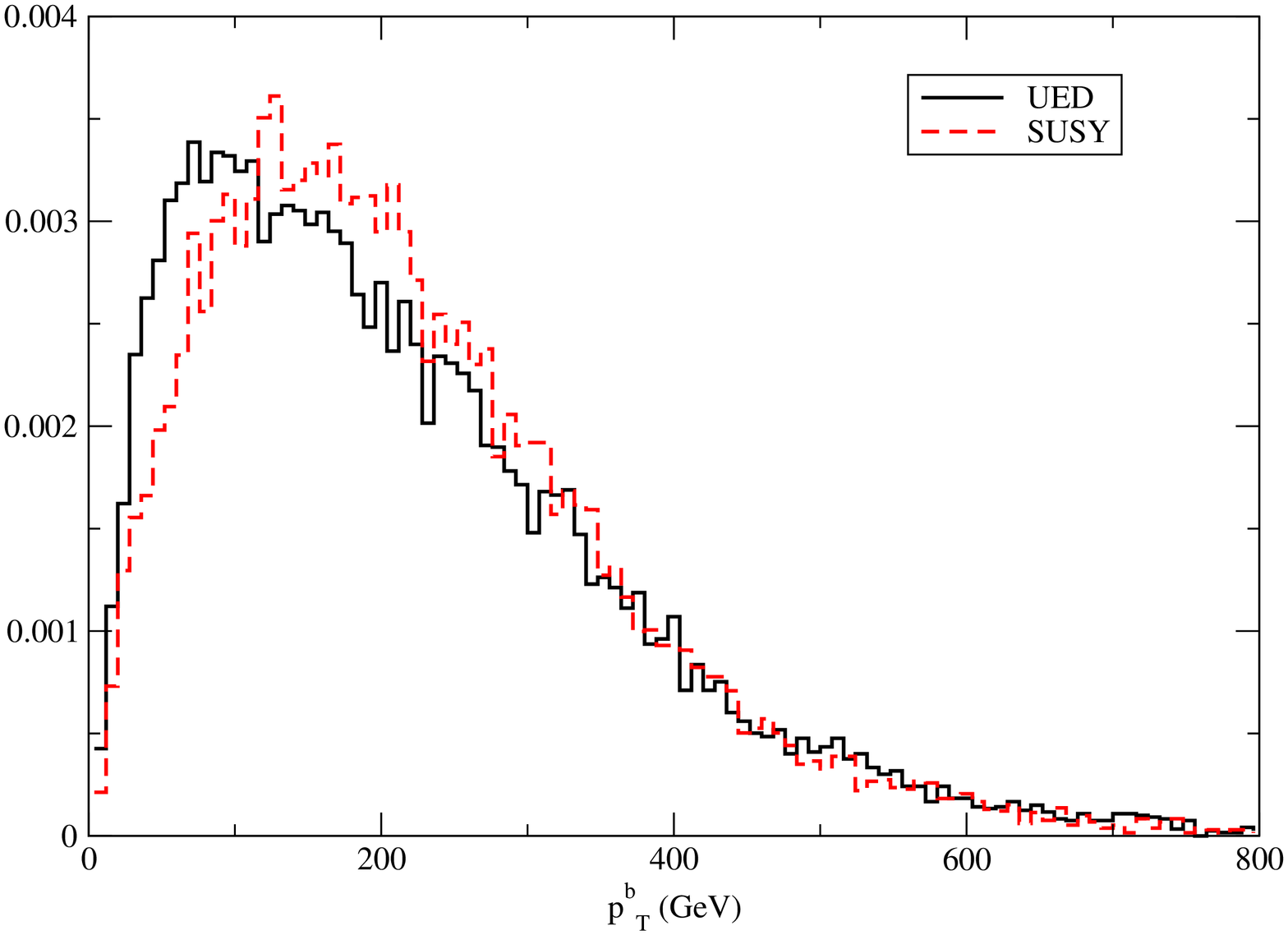}&
\includegraphics[angle=270,scale=0.3]{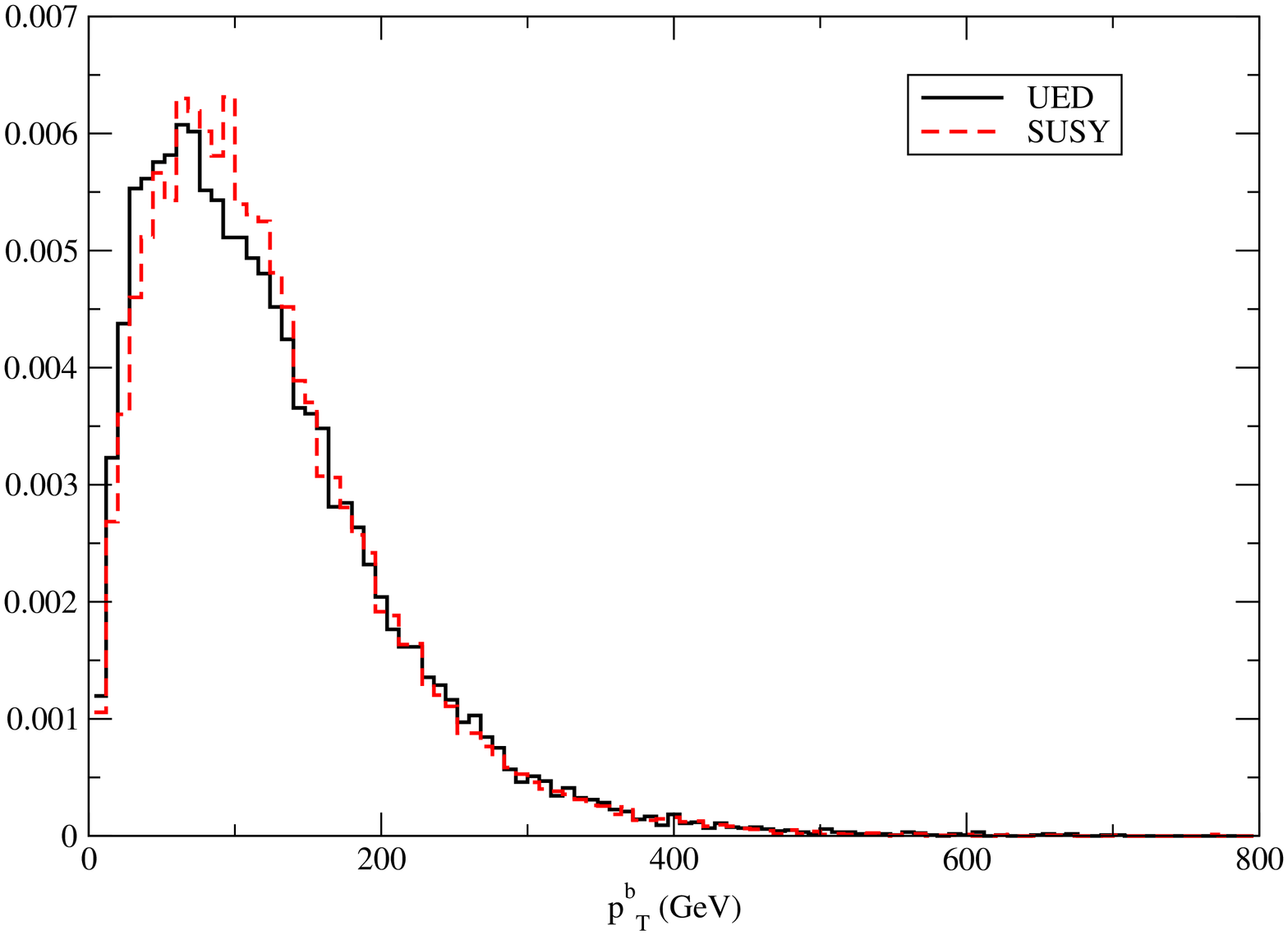}
\end{tabular}
\caption{Comparison of jet-$P_T$ distribution in SUSY and UED-like
decay chains where there is colored adjoint pair production followed by
$Adj \rightarrow b \bar{b} Ne$.  Left:  gluino/KK-gluon at 700 GeV,
sbottom/KK-bottom at 900 GeV,  and bino/B' at 100 GeV.   Right:  Same
process with gluino/KK-gluon at 400 GeV and sbottom/KK-bottom at 500 GeV. \label{pT_comp}}
\end{figure}

\begin{figure}[tbp]
\begin{tabular}{cc}
\includegraphics[angle=270,scale=0.3]{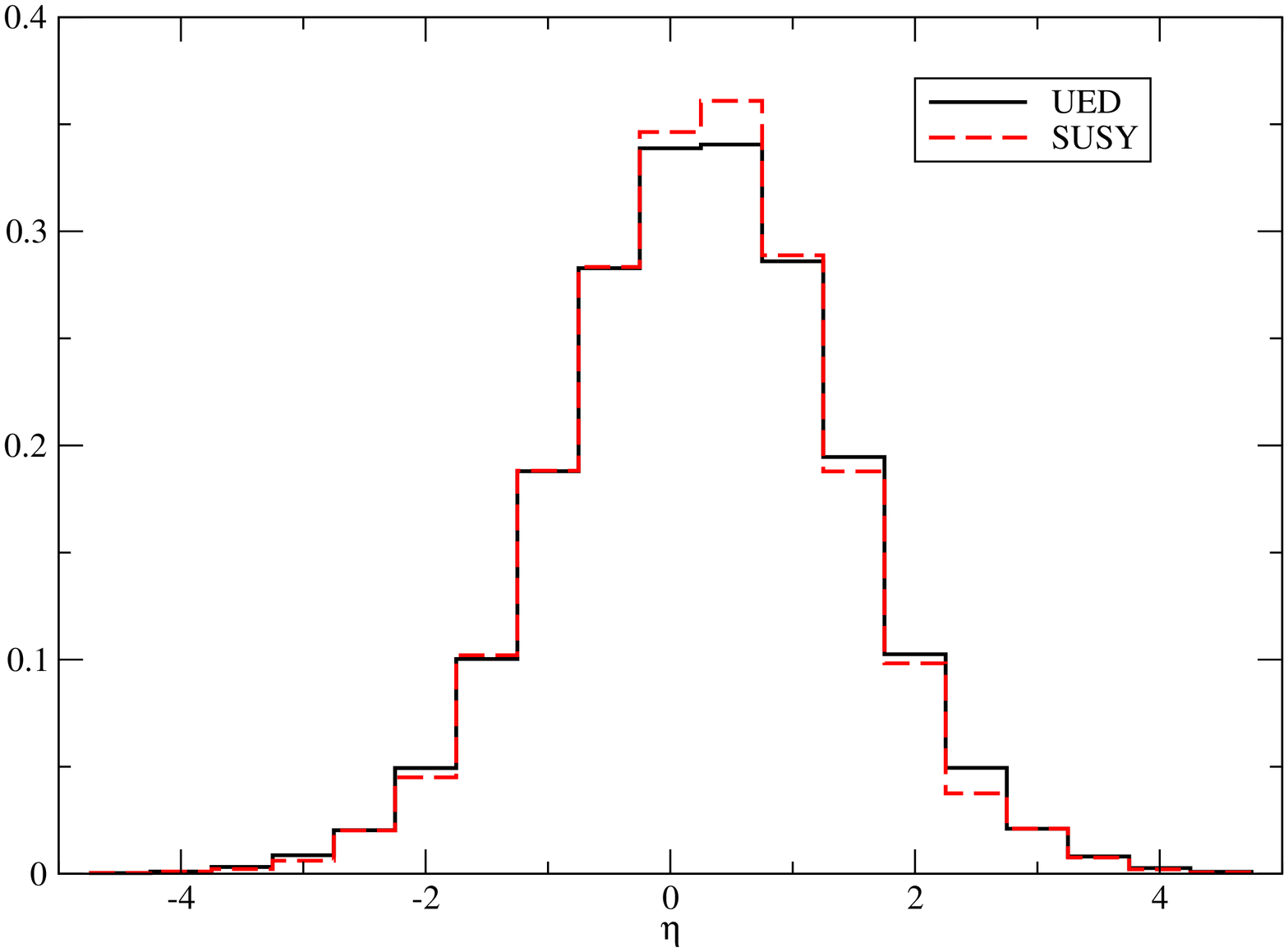}&
\includegraphics[angle=270,scale=0.3]{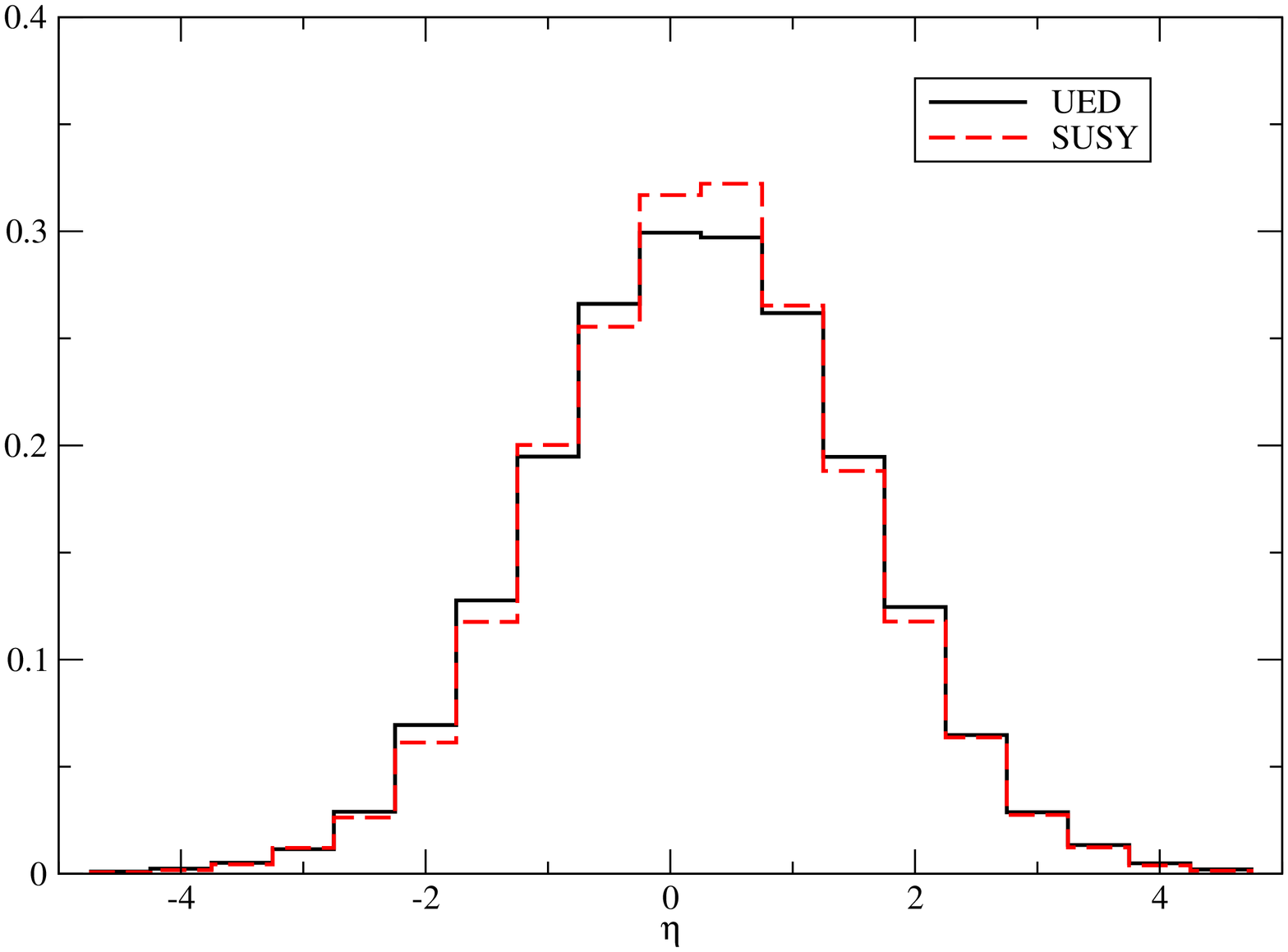}
\end{tabular}
\caption{Comparison of rapidity observables in SUSY and UED-like
decay chains using the same processes as Figure \ref{pT_comp}. \label{eta_comp}}
\end{figure}

On a case by case basis, the origin of such correlation effects is
subtle as it depends on the spin of the particles as well as the
chirality of their couplings. For most new physics scenarios that we
have in mind at the LHC, new particles are mostly produced close to
threshold. Therefore, without an extreme boost, we do not expect
the $p_T$ spectrum to be severely distorted in our approximation.
Similarly, we expect our approximation will be accurate for $\eta$
variables as well. As an example of such a comparison, we compared
the jet-$P_T$ distribution in the decay chain $Adj \rightarrow b
\bar{b} Ne$. We have used two choices of mass parameters.  For
supersymmetry, we have chosen the gluino mass to be 700 (400) GeV,
the sbottom mass to be 900 (500) GeV, and the LSP (Bino) mass to be 100
GeV.  All the other superpartners are decoupled. For comparison, we
adopted an UED-like scenario where KK-gluon, KK-bottom, and KK-$B'$
have the same masses as gluino, sbottom and Bino, respectively. The
result is shown in Figure~\ref{pT_comp}. The difference between the
two scenarios are not significant.  A comparison the rapidity
distribution is shown in Figure~\ref{eta_comp}.  Although we
have presented studies here on the off-shell 3-body decay, we have
checked that on-shell decay does not make a qualitative difference to
our conclusion (see, for example, Figures \ref{fig:FitGlPairPT} and \ref{fig:FitGLPairRap}).

\begin{figure}[tbp]
\begin{tabular}{cc}
\includegraphics[angle=270,scale=0.3]{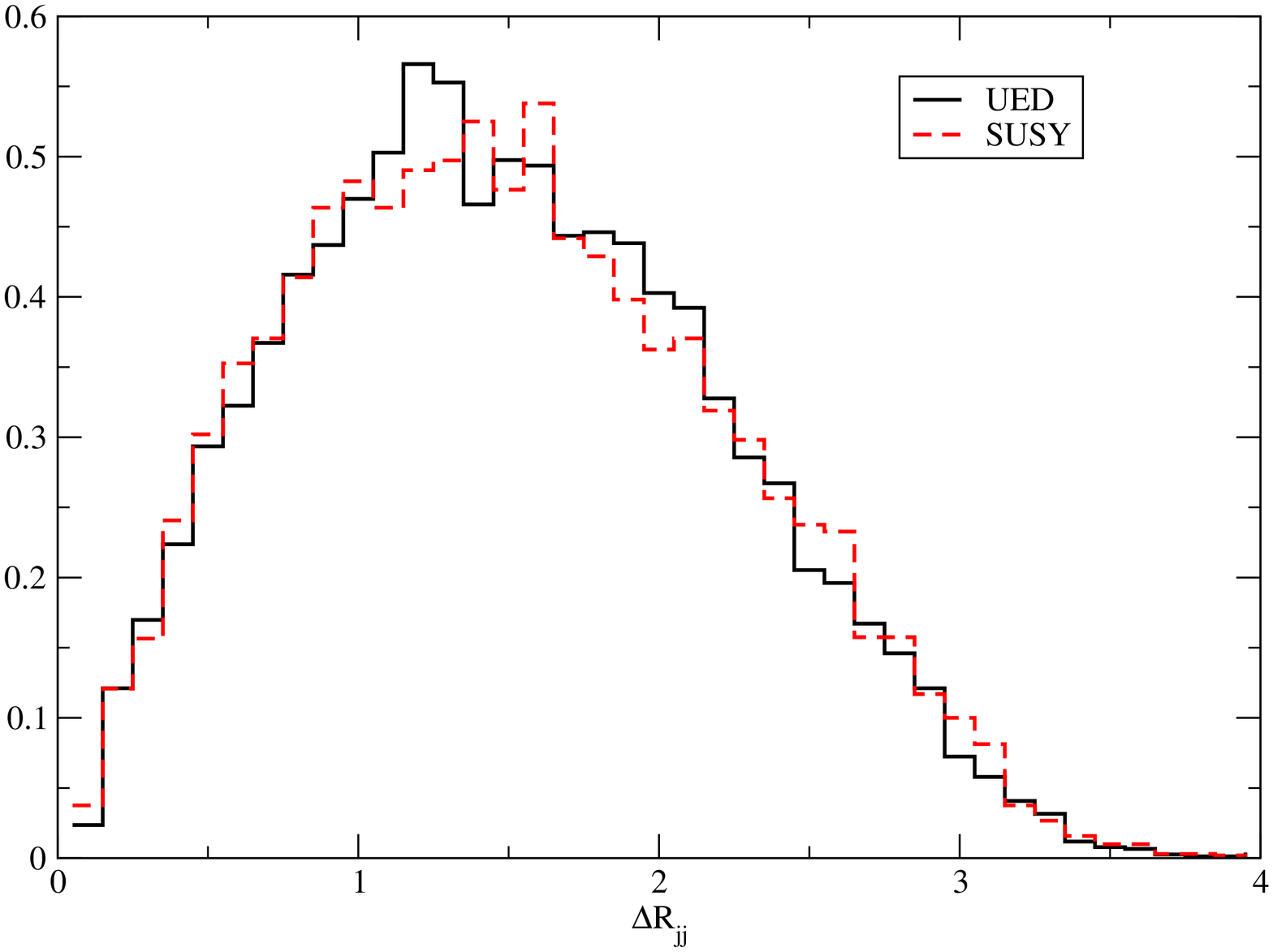}&
\includegraphics[angle=270,scale=0.3]{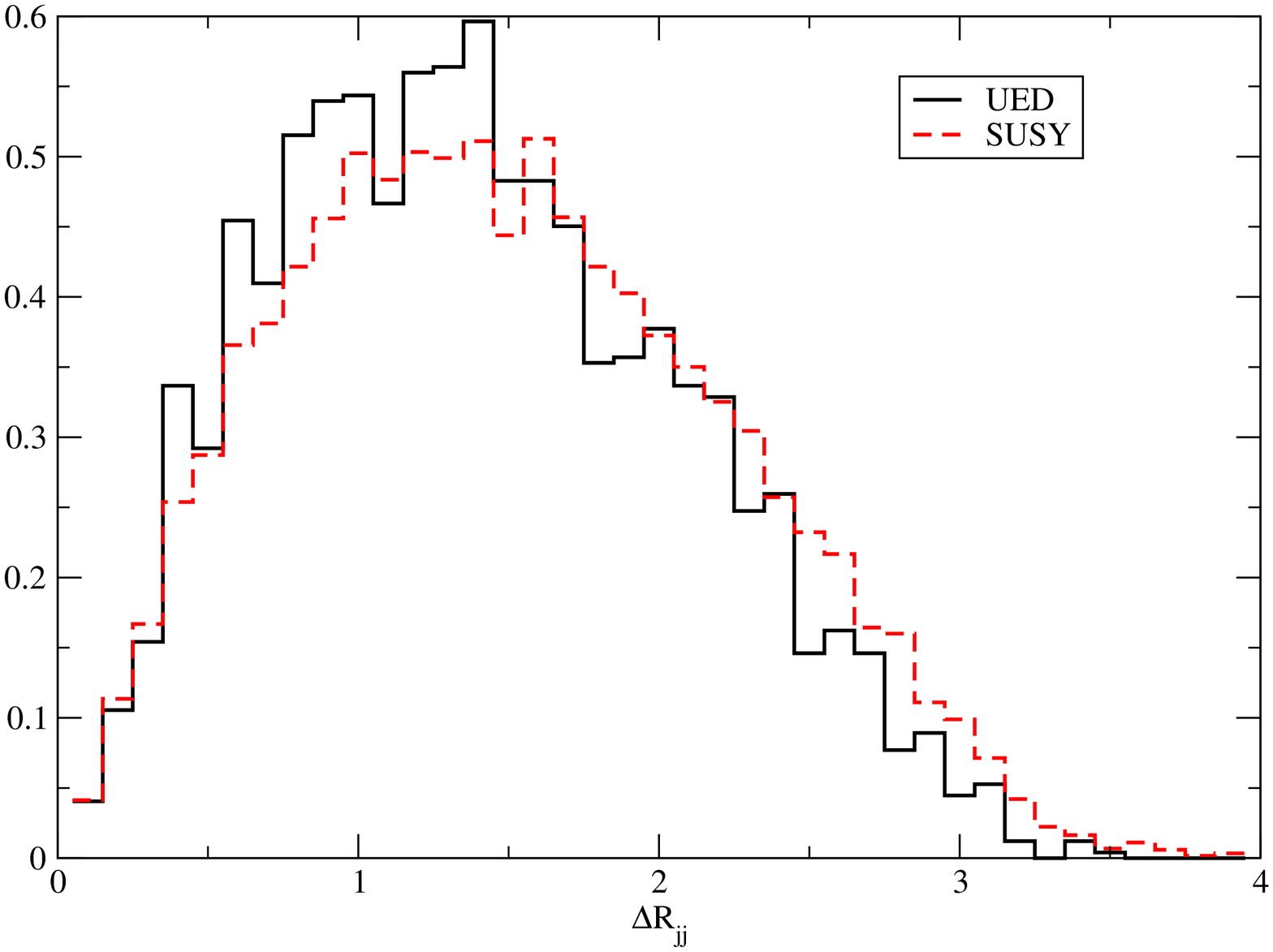}
\end{tabular}
\caption{Comparison of $\Delta R$ between two $b$-jets  in SUSY and UED-like
decay chains using the same processes as Figure \ref{pT_comp}. \label{dR_comp}}
\end{figure}

Another important two object correlation variable is $\Delta R$,
which is important for isolation cuts. We have compared this
variable in the example described above with the same parameter
choice, shown in Figure~\ref{dR_comp}. We see that there are
virtually no important differences.

\begin{figure}tbp]
\begin{tabular}{cc}
\includegraphics[angle=270,scale=0.3]{sec2plots/glgl_minDPhi.eps}&
\includegraphics[angle=270,scale=0.3]{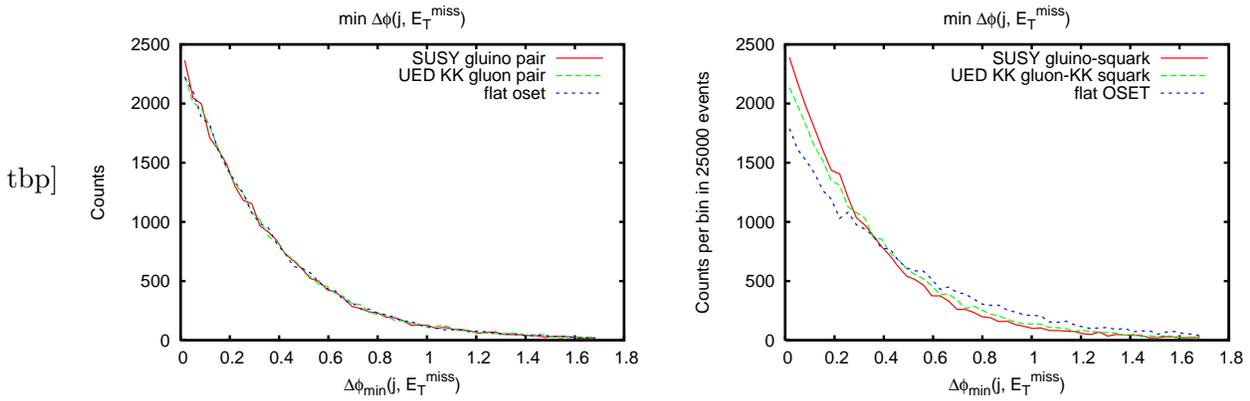}
\end{tabular}
\caption{Comparison of $\delta \phi(j, E_T^{\rm miss})$ for
adjoint pair-production and adjoint-triplet associated production.}\label{fig:minDPhi}
\end{figure}

We end our discussion on transverse variables with a study of
observables that correlate the direction of missing energy and the
$P_T$ of any jet. Such variables, $\delta \phi(j, E_T^{\rm miss})$, are
used to remove QCD background from searches for new physics in
hadronic channels.  At the partonic level, these distributions are
virtually identical for SUSY, UED, and the OSET processes with the
same topologies (see Figure \ref{fig:minDPhi}, again for the $\tilde
g \tilde g$ and $\tilde q \tilde g$ production process).

We emphasize here that the difference caused by spin correlations
are generally sub-dominant to phase space considerations. The
detailed magnitude and structure of the difference will depend on
parameter choices and couplings. What we have shown are by no means
the best or worst possible cases. More studies are necessary to
fully assess the range of possible variations.

Next, we consider invariant mass distributions. To this end, we
again consider the decay chain in Figure~\ref{decay}. As we have
pointed out in the text, the shape depends on spin identities of new
physics particles in the decay chain, the chirality of the couplings
and the mass hierarchy.

As an example, we consider the case when $A$ is a fermion and both
the $X-1-A$ and $A-2-Y$ couples are left-handed (or right-handed).
In this case, the invariant mass distribution has the following form
\begin{equation}
\frac{\mbox{d}}{\mbox{d} t_{12}} \Gamma = a_0 + a_1 t_{12},
\end{equation}
where $a_1$ depends on the spins of the external particles. When both $X$
and $Y$ are spin-1 vectors (transversely polarized), the slope is
positive. On the other hand, if one of them is a scalar, the slope
is negative. This fact can be simply understood. In the rest frame
of the intermediate particle $A$, the preference of colinearity or
anti-colinearity of $\mathbf{p}_1$ and $\mathbf{p}_2$ is fixed by
angular momentum conservation.

\begin{figure}[tbp]
\begin{tabular}{cc}
\includegraphics[angle=270,scale=0.3]{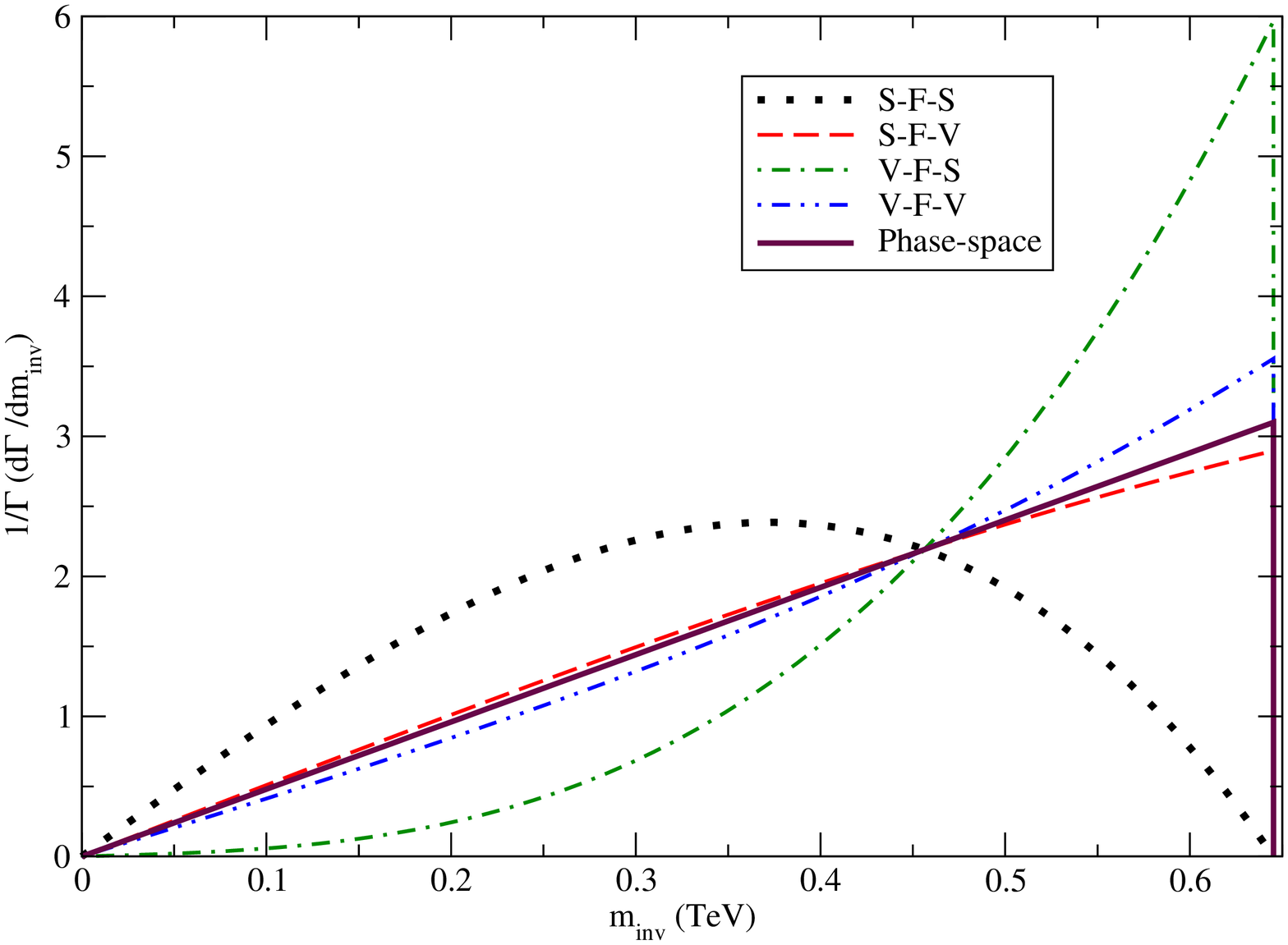}&
\includegraphics[angle=270,scale=0.3]{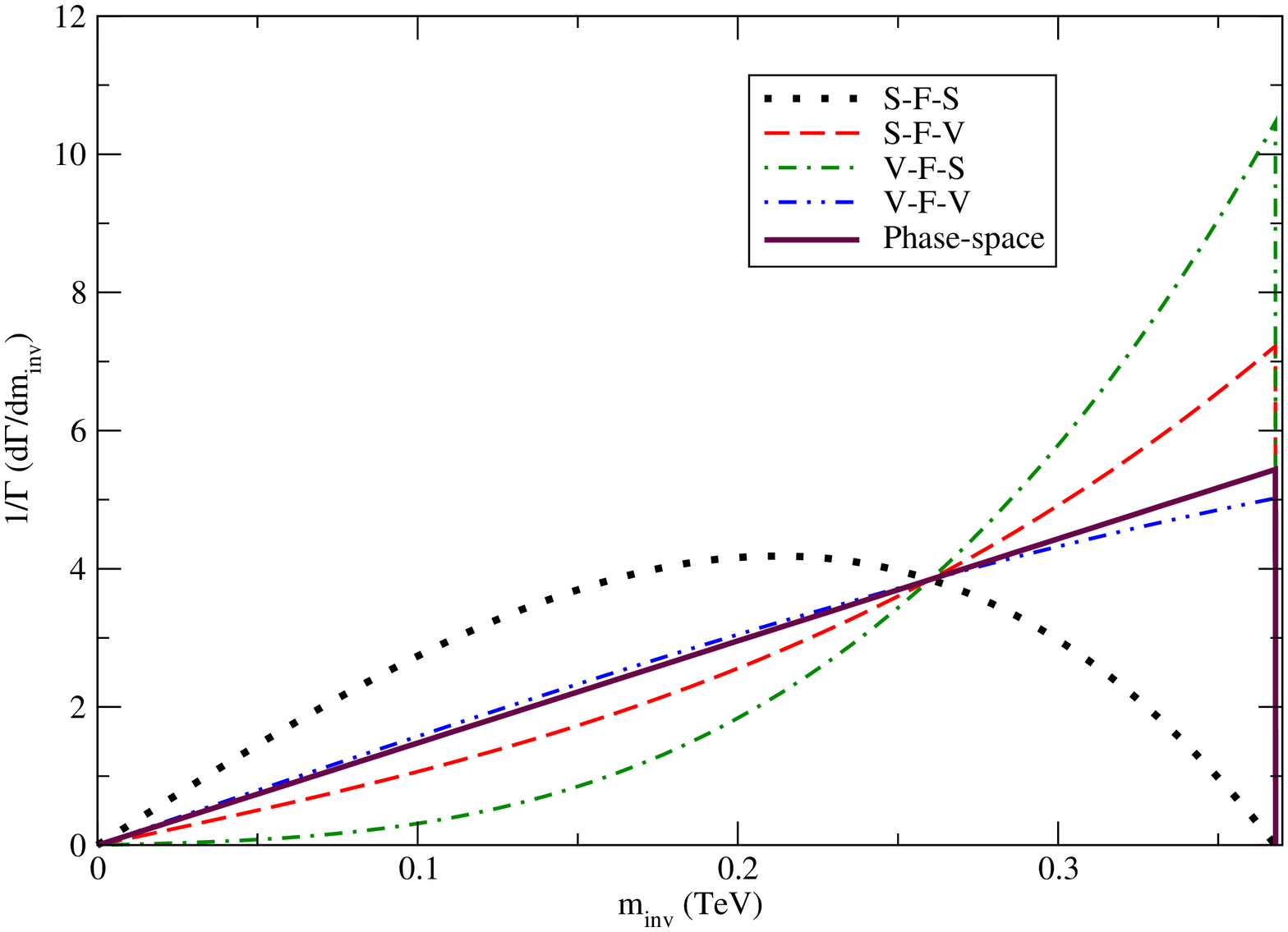}
\end{tabular}
\caption{Comparison of shapes invariant mass distributions. We focus
  on the decay chain of Figure~\ref{decay} in which $A$ is a fermion. We vary the spin of
  outside particles as well as mass hierarchy. \label{minv_comp}}
\end{figure}

Moreover, $a_1$ also depends on the mass hierarchy of the particles
involved in the decay chain. For example, in the case that $X$ and
$Y$ are spin-1 vectors, the slope will become negative once $M_A^2 >
2 M_Y^2 $. In this case, decaying into $Y$ will be longitudinally
dominated and more similar to decaying into a scalar. Such
ambiguities in interpretation of the shapes demonstrate perfectly
the fact the we must have some information about the mass hierarchy
before we can definitely determine the spin of new physics
particles.

Therefore, the shape of the two-body invariant mass distributions
could vary significantly as a result of the spin correlations. Of
course, realistically, we do not expect all two-body invariant mass
distributions to carry such clean information, as we are clearly
limited by statistics, resolution, and trigger/cuts effects to
measure the shape of distributions. Moreover, it is difficult to
identify unambiguously the correct objects to pair, in particular if
such a combination involve jet object(s). Such combinatorics tends
to reduce the statistical significance of any feature in the
invariant mass distribution.

Next, we consider the case when the intermediate particle is
off-shell and therefore the decay is 3-body. We expect that the
effects of virtual particle exchange will be enhanced if it is close
to on-shell.  Interference between several processes will also be
important here. Consider the process in Figure~\ref{decay-2}. If
process $A$ dominates, we have a propagator $(q^2 - M_A^2)^{-2}$,
where $q=p_2 + p_{\rm Y}$. The presence of such an propagator
generically pushes the distribution towards smaller $t_{12}$. On the
other hand, if process $B$ dominates, the presence of propagator
$(t_{12} - M_B^2)^{-2}$ will push the distribution to higher
$t_{12}$. Such effects are well-known in electroweak-ino decays when
both $W/Z$ and the slepton are off-shell
\cite{Baer:1995va,Hinchliffe:1996iu}. In principle, such
deformations could be parameterized by a polynomial of $t_{12}$.
Given the uncertainties of the shape measurements, we do not expect
that we need many terms in this expansion to produce a good fit.

\begin{figure}[tbp]
\begin{tabular}{cc}
\includegraphics[angle=270,scale=0.3]{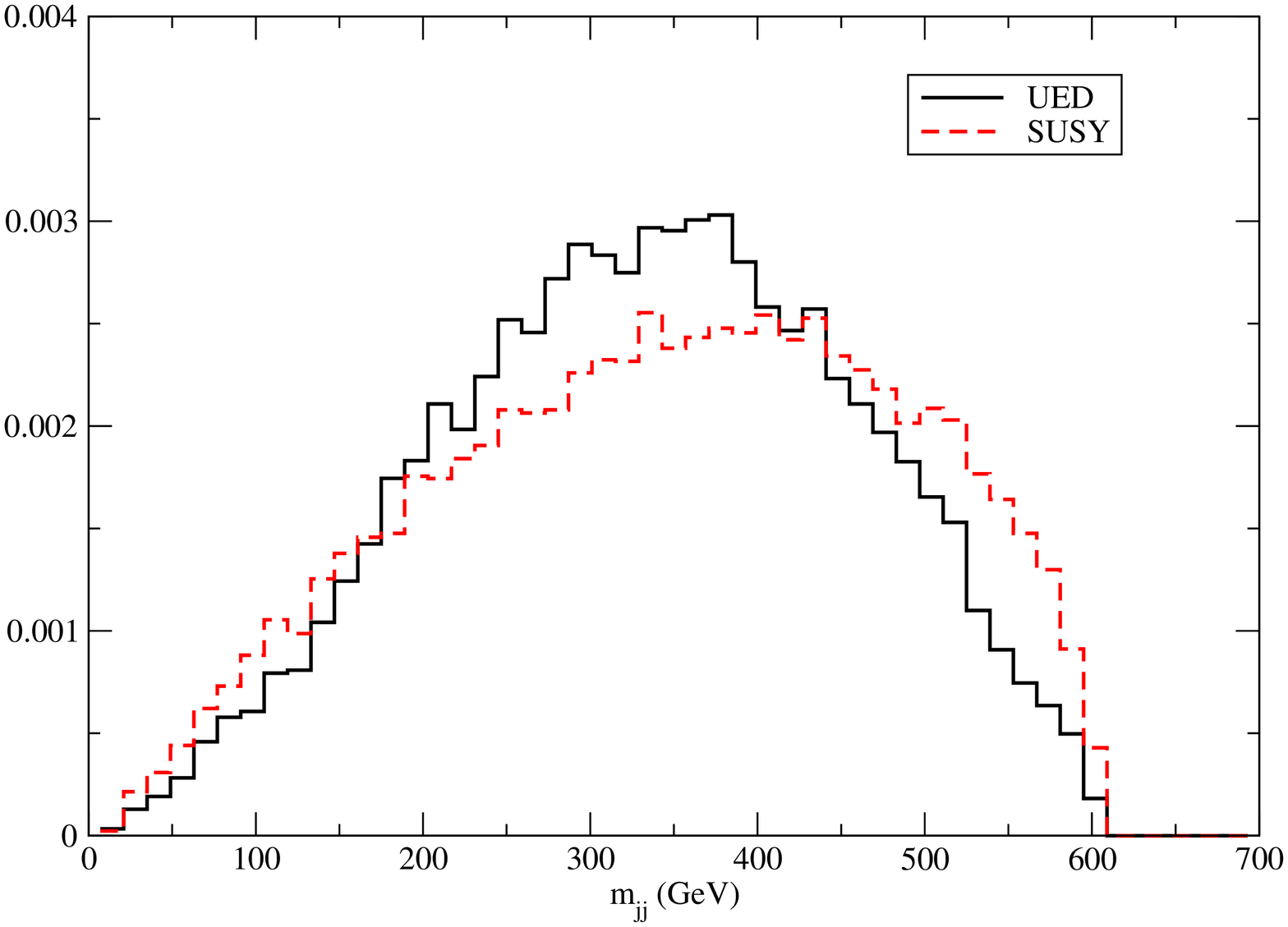}&
\includegraphics[angle=270,scale=0.3]{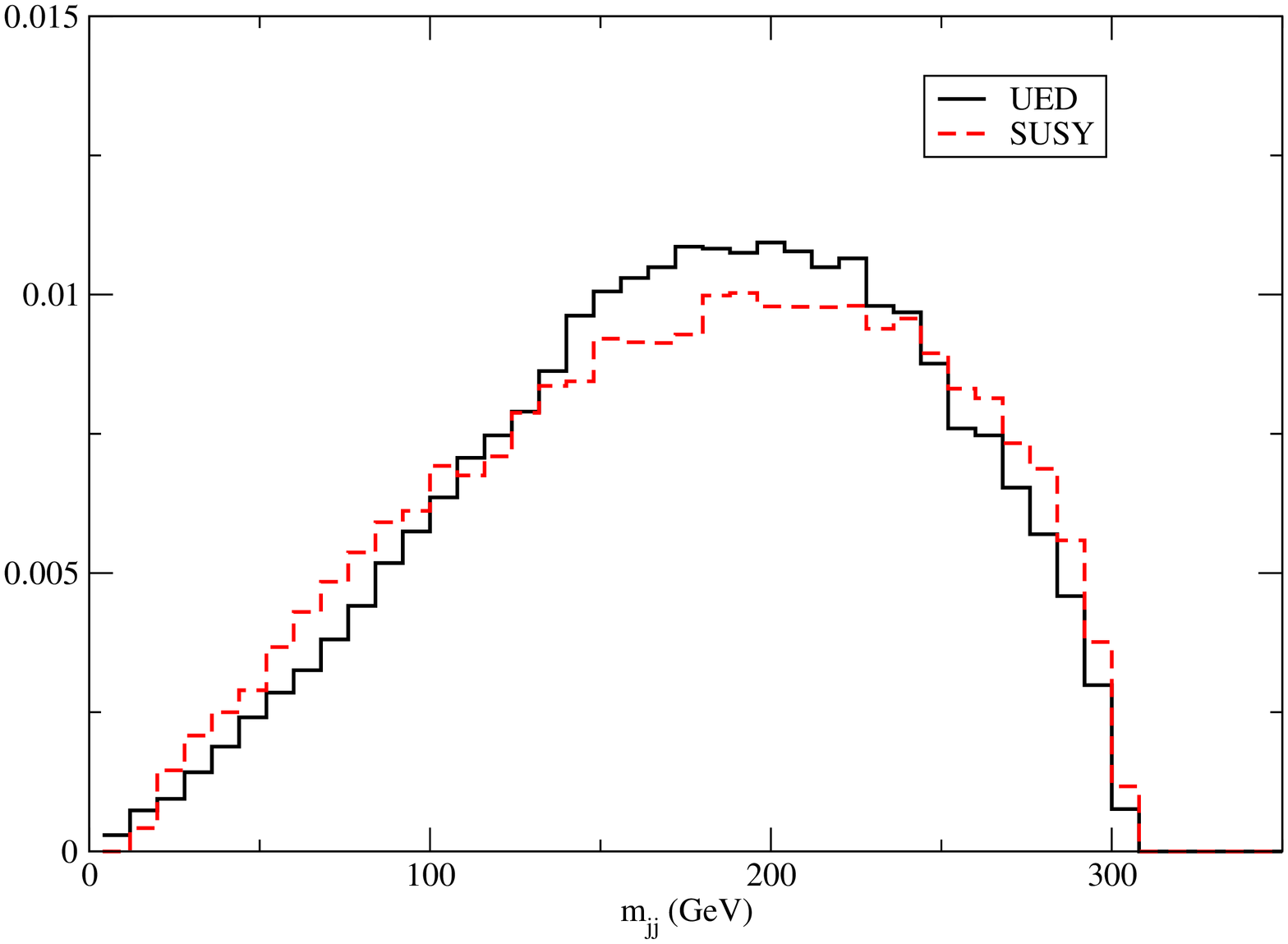}
\end{tabular}
\caption{Comparison of di-object invariant mass distributions for
3-body decays. The mass splitting is $600$ GeV and $300$ Gev on the
left and right respectively.\label{fig:InvMassCorr}}
\end{figure}

Spin correlation will have an important effect here as well. For
example, consider the case where process $A$ dominates. If $X$ and
$Y$ are fermions and $A$ is a scalar, we have no spin correlation
between 1 and 2. The propagator forces the distribution to peak at
the lower end. On the other hand, consider the case where $A$ is a
fermion and $X$ and $Y$ are transeversely polarized spin-1 vectors,
which has identical couplings as the on-shell example discussed
above.  A completely parallel argument of angular momentum
conservation tell us the distribution is forced to peak more towards
the higher $t_{12}$. Several examples of such effects are shown in
Figure~\ref{fig:InvMassCorr}.

